\documentclass[11pt,titlepage]{thesis}

\title{Exact and Perturbed\vspace{1cm}\\Friedmann-Lema\^{i}tre Cosmologies}

\author{Paul Ullrich}
\degree{Master of Mathematics}
\discipline{Applied Mathematics}
\date{2007}

\usepackage{amsmath}
\usepackage{amsfonts}
\usepackage{amssymb}
\usepackage{graphicx}
\usepackage{mathrsfs}

\newtheorem{proposition}{Proposition}[section]

\newtheorem{theorem}{Theorem}[section]

\begin{document}

\maketitle
\prepages
\edeclarationpage

\begin{abstract}
In this thesis we first apply the $1+3$ covariant description of general relativity to analyze $n$-fluid Friedmann-Lema\^{i}tre (FL) cosmologies; that is, homogeneous and isotropic cosmologies whose matter-energy content consists of $n$ non-interacting fluids.  We are motivated to study FL models of this type as observations suggest the physical universe is closely described by a FL model with a matter content consisting of radiation, dust and a cosmological constant.  Secondly, we use the $1+3$ covariant description to analyse scalar, vector and tensor perturbations of FL cosmologies containing a perfect fluid and a cosmological constant.  In particular, we provide a thorough discussion of the behaviour of perturbations in the physically interesting cases of a dust or radiation background.
\end{abstract}

\begin{acknowledgements}
First and foremost, I would like to thank Dr. John Wainwright for his patience, guidance and support throughout the preparation of this thesis.  I also would like to thank Dr. Achim Kempf and Dr. C. G. Hewitt for their time and comments.
\end{acknowledgements}

\tableofcontents
\listoffigures
\listoftables

\mainbody
\chapter{Introduction} \label{chap:geometricalbackground}

\section{Cosmological Models} \label{sec:cosmologicalmodels}

The goal of cosmology is to describe the large-scale structure, origins and evolution of the universe as a whole.  Within the framework of general relativity, a \textit{cosmological model}, denoted \mbox{($\mathcal{M}$, $\mathbf{g}$, $\mathbf{u}$)} consists of a differential manifold $\mathcal{M}$, a Lorentzian metric $\mathbf{g}$ (which describes the geometry of the space-time) and a \textit{unit timelike vector field} $\mathbf{u}$,
\begin{equation} \label{eq:normalized4velocity}
g_{ij} u^i u^j = -1,
\end{equation} that defines the world-lines of a family of fundamental observers.  In cosmology one represents the matter content as a smooth distribution (a continuum) whose particles represent galaxies in the present epoch.  The time-like vector field $\mathbf{u}$ is the 4-velocity of the particles in this continuum.  One member of the congruence of time-like curves defined by $\mathbf{u}$, namely the world-line of our galaxy, plays a particularly important role, since it is the position from which we make observations.

The evolution of the universe is assumed to be governed by the \textit{Einstein field equations (EFEs)},
\begin{equation} \label{eq:einsteinfieldequations}
R_{ab} - \tfrac{1}{2} R g_{ab} = T_{ab},
\end{equation} where $R_{ab}$ is the \textit{Ricci tensor}, $R$ is the \textit{Ricci scalar} ($R = g^{ab} R_{ab}$) and $T_{ab}$ is the \textit{energy-momentum tensor} of the distribution of matter and energy in the universe.  Here and throughout we will use geometerized units so that $c = 1$ and $8 \pi G = 1$, where $c$ is the speed of light in vacuum and $G$ is the gravitational constant.

There are two fundamental assumptions that are used to construct models of the universe:

\begin{itemize}
\item[I.] \textit{The universe is isotropic about our galaxy.}  More precisely, at any event on the world-line of our galaxy, all directions orthogonal to the world-line are physically and geometrically equivalent, when the universe is viewed on a sufficiently large scale.

A variety of cosmological observations (\textit{e.g.} number counts of distant galaxies and peculiar velocity surveys) provide support for assumption I in the present epoch.  The strongest support is provided by the temperature of the cosmic microwave background radiation (CMBR) which is isotropic to within $10^{-5}$ (\textit{i.e.} \mbox{$\frac{\Delta T}{T} \sim 10^{-5}$}).

\item[II.] \textit{The Copernican Principle.}  This principle states that our galaxy is not in a preferred position in the universe.  Unlike assumption I, this assumption is philosophical, rather than observational.
\end{itemize}

Assumptions I and II together imply that the universe satisfies the so-called \textit{Cosmological Principle}, namely
\begin{itemize}
\item[i)]  the universe is isotropic about all fundamental world-lines,
and
\item[ii)]  the universe is spatially homogeneous, \textit{i.e.} all physical and geometric quantities do not vary in space.
\end{itemize}

It is a well-known consequence of assumptions I and II (we give a concise proof in Chapter \ref{chap:friedmannlemaitre}; see Theorem \ref{thm:derivationoftheRWmetric}) that the metric $g_{ab}$ can be written in the form of the \textit{Robertson-Walker (RW) metric},\footnote{Here we use Greek indices to denote spatial indices.  They can take on the values 1, 2 or 3.}
\begin{equation} \label{eq:RWmetric}
ds^2 = -dt^2 + \ell^2(t) \tilde{g}_{\alpha \beta}(x^{\lambda}) dx^{\alpha} dx^{\beta},
\end{equation} where $\tilde{g}_{\alpha \beta}$ is a 3-metric of constant curvature.  The preferred timelike vector field $\mathbf{u}$ is given by
\begin{equation} \label{eq:RWfourvelocity}
\mathbf{u} = \frac{\partial}{\partial t}.
\end{equation}

The Robertson-Walker metric provides the basis for the Friedmann-Lema\^{i}tre (FL) cosmological models that play a central role in modern cosmology.  A \textit{Friedmann-Lema\^{i}tre cosmology}\footnote{The first explicit FL cosmological models were first discovered by Friedmann in 1929 (for some historical details, see Ellis (1989)).} is a cosmological model ($\mathcal{M}$, $\mathbf{g}$, $\mathbf{u}$) that satisfies the Einstein field equations (\ref{eq:einsteinfieldequations}) with a suitable energy-momentum tensor $T_{ab}$, with metric $\mathbf{g}$ and fundamental congruence $\mathbf{u}$ having the Robertson-Walker form (\ref{eq:RWmetric}) and (\ref{eq:RWfourvelocity}).

In this thesis we study FL cosmologies and their perturbations within a general mathematical framework.  In section \ref{sec:geometricalbackground} we provide the necessary geometrical background for this study.

\section{Geometrical Background} \label{sec:geometricalbackground}

In this thesis we will make use of the so-called \textit{$1+3$ covariant description of general relativity}\footnote{See, for example, Ellis (1973), Wainwright and Ellis (1997), p1-30, or Ellis and van Elst (1998), Chapter 2.} which has been developed for use in spacetimes in which there is a preferred timelike congruence $\mathbf{u}$, as in the case of a cosmological model ($\mathcal{M}$, $\mathbf{g}$, $\mathbf{u}$).  The ``1+3'' refers to the fact that one performs a ``time + space'' decomposition relative to $\mathbf{u}$ by projecting tensors and tensorial equations parallel to $\mathbf{u}$ and orthogonal to $\mathbf{u}$.  The second aspect of the $1+3$ description is to write a tensor as a sum of algebraically simpler parts, \textit{i.e.} to give an algebraic decomposition.  For example, an arbitrary rank 2 contravariant tensor can be written as the sum of a symmetric tracefree tensor, an anti-symmetric tensor and the trace multiplied by the metric tensor.

The first step is to introduce the basic variables in the $1+3$ description, which are defined by applying projection relative to $\mathbf{u}$ and algebraically decomposing the covariant derivative $\nabla_{b} u_{a}$, the energy-momentum tensor $T_{ab}$ and the Riemann-Christoffel tensor as follows:

\begin{enumerate}
\item[i)]  The \textit{covariant derivative of the fundamental 4-velocity}, $\nabla_b u_a$, describes the kinematics of the continuum described by the vector field $\mathbf{u}$.  The so-called \textit{kinematic quantities} are defined by writing $\nabla_{b} u_{a}$ as a sum of simpler parts and projecting relative to $\mathbf{u}$.  We define these quantities in section \ref{sec:kinematicquantities}.

\item[ii)]  The \textit{energy-momentum tensor} $T_{ab}$ describes the matter and energy content of the universe and so gives rise to a set of quantities known as \textit{matter variables} when decomposed relative to $\mathbf{u}$.  We discuss this decomposition in section \ref{sec:energymomentumtensor}.

\item[iii)]  The \textit{Weyl curvature tensor} $C_{abcd}$, which is defined as the tracefree part of the Riemann curvature tensor, is decomposed relative to $\mathbf{u}$ into two tracefree symmetric tensors, the so-called\footnote{There is a formal analogy with the decomposition of the electromagnetic field tensor $F_{ab}$ into an electric field $E_{a}$ and a magnetic field $H_{a}$.} electric part $E_{ab}$ and magnetic part $H_{ab}$.  We discuss this decomposition in section \ref{sec:weylcurvaturetensor}.
\end{enumerate}

The second step is to derive evolution and constraint equations for the basic variables by using the Ricci identities and Bianchi identities in conjunction with the Einstein field equations.  We will discuss this process in section \ref{sec:evolutionandconstraintequations}.

\subsection{Projection and Differentiation}

Projection parallel to $\mathbf{u}$ is achieved by simply contracting a vector or tensor with $u^{a}$.  Projection into the 3-space orthogonal to $\mathbf{u}$ is defined by the \textit{projection tensor}
\begin{equation} \label{eq:projectiontensor}
h_{ab} = g_{ab} + u_a u_b,
\end{equation} which satisfies the identities
\begin{displaymath}
h^{\ c}_{a} h^{\ b}_{c} = h^{\ b}_{a}, \qquad h^{\ b}_{a} u_b = 0, \qquad h^{\ a}_{a} = 3.
\end{displaymath}  Corresponding to these two types of projection, two derivatives are defined.

The overdot $\dot{\ }$ denotes differentiation along the fundamental congruence, defined for scalars as
\begin{equation}
\dot{f} = f_{,a} u^a,
\end{equation} and for tensors as
\begin{equation} \label{eq:fundamentalcongderiv}
\dot{T}_{a b \cdots c} = u^{d} \nabla_{d} T_{a b \cdots c},
\end{equation} where $\nabla_{a}$ denotes covariant differentiation.

Following Maartens (1997), we use angle brackets to denote the orthogonal projections of vectors, the orthogonally projected symmetric trace-free part of tensors, and their time derivatives, \textit{i.e.}
\begin{align}
X^{\langle a \rangle} &= h^{a}_{\ b} X^{b}, & T^{\langle ab \rangle} &= \left[ h^{(a}_{\ \ c} h^{b)}_{\ \ d} - \tfrac{1}{3} h^{ab} h_{cd} \right] T^{cd}, \\
\dot{X}^{\langle a \rangle} &= h^{a}_{\ b} \dot{X}^{b}, & \dot{T}^{\langle ab \rangle} &= \left[ h^{(a}_{\ \ c} h^{b)}_{\ \ d} - \tfrac{1}{3} h^{ab} h_{cd} \right] \dot{T}^{cd}.
\end{align}  Further, we will use the projection symbol $\perp$ to identify orthogonal projection over all tensor indices, \textit{i.e.}
\begin{equation}
(T^{ab}_{\phantom{ab}cd})_{\perp} = h^{a}_{\ p} h^{b}_{\ q} h^{r}_{\ c} h^{s}_{\ d} T^{pq}_{\phantom{pq}rs}.
\end{equation}  The \textit{spatial covariant derivative} $\hat{\nabla}_a$ is defined by projecting all free indices of the covariant derivative (compare Ellis (1973), Ellis and Bruni (1989)).  For example,
\begin{equation} \label{eq:orthogonalcovariantderiv}
\hat{\nabla}_{a} f = h^{\ b}_{a} \nabla_{b} f, \qquad \hat{\nabla}_a T^{bc}\ _{\!de} = h^{\ p}_{a} h^{\ b}_{q} h^{\ c}_{r} h^{\ s}_{d} h^{\ t}_{e} \nabla_{p} T^{qr}_{\phantom{qr}st}.
\end{equation}

The covariant derivative of a scalar can then be decomposed into a derivative along the fundamental congruence $\mathbf{u}$ and a spatial covariant derivative via
\begin{equation}
\nabla_a f = \hat{\nabla}_a f - u_a \dot{f}.
\end{equation}

The totally skew pseudotensor is defined by
\begin{equation}
\eta^{abcd} = \eta^{[abcd]}, \qquad \eta^{0123} = (-g)^{-1/2},
\end{equation} where
\begin{equation}
g = \det(g_{ab}).
\end{equation}  The \textit{projected permutation tensor} is then defined by (compare Maartens et al. (1995)),
\begin{equation}
\epsilon_{abc} = \eta_{abcd} u^d.
\end{equation} and is used to define the \textit{spatial curl of a vector} by
\begin{equation} \label{eq:covariantvectorcurl}
\mathsf{curl}\ X_a = \epsilon_{abc} \hat{\nabla}^b X^c.
\end{equation} and the \textit{spatial curl of a rank 2 tensor} by
\begin{equation} \label{eq:covarianttensorcurl}
\mathsf{curl}\ T_{ab} = \epsilon_{cd(a} \hat{\nabla}^c T_{b)}^{\ \ d}.
\end{equation}

\subsection{Kinematic Quantities} \label{sec:kinematicquantities}

We now focus on the decomposition of the covariant derivative of the fundamental 4-velocity.  Following Ellis (1973, p8-11), we decompose the covariant derivative of a unit timelike vector field $\mathbf{u}$ as
\begin{equation} \label{eq:covudecomposition}
\nabla_{b} u_{a} = \sigma_{ab} + \omega_{ab} + H h_{ab} - \dot{u}_a u_b,
\end{equation} with $\sigma_{ab}$ symmetric and trace-free, $\omega_{ab}$ antisymmetric, and
\begin{equation} \label{eq:orthogonalityproperties}
\sigma_{ab} u^a = 0, \qquad \omega_{ab} u^a = 0, \qquad \dot{u}_a u^a = 0.
\end{equation}
It follows from (\ref{eq:covudecomposition}) and (\ref{eq:orthogonalityproperties}) that
\begin{eqnarray}
\label{eq:hubblescalar} H & = & \tfrac{1}{3} \nabla_{a} u^{a}, \\
\label{eq:kinacceleration} \dot{u}_a & = & u^{b} \nabla_{b} u_{a}, \\
\label{eq:kinshear} \sigma_{ab} & = & \nabla_{(a} u_{b)} - H h_{ab} + \dot{u}_{(a} u_{b)}, \\
\label{eq:kinvorticity} \omega_{ab} & = & \nabla_{[b} u_{a]} + \dot{u}_{[a} u_{b]}.
\end{eqnarray}  Here, $H$ is \textit{the Hubble scalar}\footnote{The \textit{rate of expansion scalar} $\Theta$, defined by $\Theta = 3 H$, is often used in the expansion (\ref{eq:covudecomposition}).}, $\dot{u}_a$ is \textit{the acceleration vector}, $\sigma_{ab}$ is \textit{the rate of shear tensor} and $\omega_{ab}$ is \textit{the vorticity tensor}.  In practice we will usually replace $\omega_{ab}$ by \textit{the vorticity vector}, defined by
\begin{equation} \label{eq:kinvorticityvector}
\omega^a = \tfrac{1}{2} \eta^{abcd} u_b \omega_{cd}
\end{equation} which satisfies $u^a \omega_a = 0$.  We can write the vorticity tensor in terms of the vorticity vector using the relation
\begin{equation} \label{eq:kinvorticityvector2}
\omega^{ab} = \eta^{abcd} \omega_c u_d.
\end{equation}  The magnitude of the rate of shear tensor and the vorticity tensor are defined by
\begin{displaymath}
\sigma^2 = \tfrac{1}{2} \sigma_{ab} \sigma^{ab}, \qquad \omega^2 = \omega_{a} \omega^{a}.
\end{displaymath}

We now state two useful propositions concerning an \textit{irrotational} vector field $u^{a}$, \textit{i.e.} $\omega_{a} = 0$.

\bigskip

\begin{proposition} \label{prop:irrotationalpotential} Given a vector field $\mathbf{u}$, there exist scalar fields $f(x^{a})$ and $t(x^{a})$ such that\footnote{The minus sign ensures that $\dot{t} > 0$, \textit{i.e.} $t$ increases into the future along the fundamental congruence.} \mbox{$u_{a} = - f t_{,a}$} if and only if $\omega_{a} = 0$.
\end{proposition}

\paragraph{Proof:}  Given $u_{a} = - f t_{,a}$, it follows by differentiating that $u_{[a,b} u_{c]} = 0$ and that $\omega_a = 0$, by (\ref{eq:kinvorticityvector}).  The converse is a special case of Frobenius' Theorem, which is usually stated in terms of differential forms (see for example Flanders (1989), see the theorem on page 94).\ $\square$

\bigskip

\paragraph{Comment:}  The significance of this result is that if $\mathbf{\omega} = 0$, then the 3-spaces orthogonal to $u^a$ at each point form spacelike hypersurfaces given by \mbox{$t = \mathrm{constant}$}, \textit{i.e.} there is a one-parameter family of spacelike hypersurfaces orthogonal to the timelike congruence defined by $\mathbf{u}$.

If we can further impose the condition of zero acceleration ($\dot{u}_{a} = 0$), we obtain the following result.

\bigskip

\begin{proposition} \label{prop:irrotationalgeodesicpotential} $\omega_{a} = 0, \dot{u}_{a} = 0\ \Longleftrightarrow\ u_{a} = - t_{,a}$, for some scalar field $t(x^{a})$.
\end{proposition}

\paragraph{Proof:}  Given $\omega_a = \dot{u}_a = 0$, equation (\ref{eq:kinvorticity}) implies \mbox{$u_{[a;b]} = u_{[a,b]} = 0$}.  It then follows from Proposition \ref{prop:irrotationalpotential} that $f_{,[a} t_{,b]} = 0$.  This in turn implies that $f = g(t)$ and hence that one can redefine $t$ to set $f = 1$.\ $\square$

\bigskip

\noindent In this case, the scalar $t$ that labels the hypersurfaces is clock time along the normal congruence.

\subsection{Decomposition of the Energy-Momentum Tensor} \label{sec:energymomentumtensor}

The energy-momentum tensor $T_{ab}$ is a symmetric tensor that specifies the total matter-energy content of a cosmological model.  We use the standard decomposition of the energy-momentum tensor with respect to a timelike vector field $\mathbf{u}$ (see Ellis (1973), p7):
\begin{equation} \label{eq:energymomentumtensor}
T_{ab} = \mu u_a u_b + q_a u_b + q_b u_a + p h_{ab} + \pi_{ab},
\end{equation} where
\begin{equation}
q_a u^a = 0, \qquad \pi_{ab} u^b = 0 \qquad \pi_{\ a}^a = 0, \qquad \pi_{ab} = \pi_{ba}.
\end{equation}  In this equation, $\mu$ is the \textit{total energy density} measured by an observer moving with fluid 4-velocity $u^a$, $q_a$ is the \textit{energy flux relative to $u^a$} (representing heat conduction and diffusion), $p$ is the \textit{isotropic pressure} and $\pi_{ab}$ is the \textit{trace-free anisotropic pressure} (due to processes such as viscosity).

A \textit{perfect fluid} with 4-velocity $\mathbf{u}$ is defined by
\begin{equation} \label{eq:perfectfluidquantities}
q_a = 0, \qquad \pi_{ab} = 0,
\end{equation} in which case the energy-momentum tensor (\ref{eq:energymomentumtensor}) simplifies to
\begin{equation} \label{eq:perfectfluid}
T_{ab} = \mu u_a u_b + p (g_{ab} + u_a u_b).
\end{equation}

\subsection{The Weyl Curvature Tensor} \label{sec:weylcurvaturetensor}

The \textit{Weyl conformal curvature tensor} (or simply the \textit{Weyl tensor}) is defined in terms of the Riemann tensor and the Ricci tensor according to
\begin{equation} \label{eq:weyltensor}
C_{abcd} = R_{abcd} - \tfrac{1}{2} (g_{ac} R_{bd} + g_{bd} R_{ac} - g_{bc} R_{ad} - g_{ad} R_{bc}) + \tfrac{1}{6} R (g_{ac} g_{bd} - g_{ad} g_{bc}).
\end{equation}  (see, for example, Ellis 1973, p7).  We note that the Weyl curvature tensor has all the symmetry properties of the Riemann tensor and is in addition trace-free ($C^{a}_{\ bad} = 0$).

It is helpful to view (\ref{eq:weyltensor}) as a decomposition of the Riemann tensor $R_{abcd}$, with its 20 independent components, into a tracefree part $C_{abcd}$ and the trace $R_{ab}$:
\begin{equation}
R_{abcd}\ \longleftrightarrow\ \{C_{abcd}, R_{ab}\},
\end{equation} each of which have 10 independent components.  The Ricci tensor is determined at each point by the energy-momentum tensor through the field equations (\ref{eq:einsteinfieldequations}).  On the other hand, as we shall see later, the Weyl tensor governs the propagation of changes in the gravitational field, \textit{i.e.} gravitational waves.

Within the $1+3$ formalism the Weyl tensor can be decomposed further into an ``electric part'' $E_{ab}$ and a ``magnetic part'' $H_{ab}$ by projecting $C_{abcd}$ and its dual orthogonal to $\mathbf{u}$:
\begin{equation} \label{eq:weyldecomposition}
E_{ac} = C_{abcd} u^b u^d, \qquad H_{ac} = \tfrac{1}{2} \eta^{\ \ ef}_{ab} C_{efcd} u^{b} u^{d}.
\end{equation}  We note that $E_{ab}$ and $H_{ab}$ are symmetric, trace-free and orthogonal to $\mathbf{u}$:
\begin{equation}
E_{ab} u^b = 0, \qquad H_{ab} u^b = 0.
\end{equation}  In addition, this decomposition describes the Weyl tensor completely since\footnote{The Weyl tensor can, in fact, be written in terms of $E_{ab}$ and $H_{ab}$ (see, for example, Ellis and van Elst (1998) eq. (22)).}
\begin{equation}
E_{ab} = 0 = H_{ab} \Longleftrightarrow C_{abcd} = 0.
\end{equation}  Note that through (\ref{eq:weyltensor}) $E_{ab}$, $H_{ab}$ and $R_{ab}$ provide a full description of the space-time curvature as described by the Riemann tensor.

\subsection{Spatial Curvature} \label{sec:gaussequation}

If $\mathbf{u}$ is irrotational, we have seen that there exists a family of hypersurfaces (\textit{i.e.} 3-dimensional manifolds) orthogonal to $\mathbf{u}$.  In this situation the projection tensor $h_{ab}$ can be viewed as the 3-metric induced on these hypersurfaces and the spatial covariant derivative $\hat{\nabla}_{a}$ can be viewed as the covariant derivative determined by this 3-metric.

Analogous to the spacetime Ricci identities, the 3-space Ricci identities are given by
\begin{equation} \label{eq:ThreeRicciIdentities}
\hat{\nabla}_{c} \hat{\nabla}_{d} X_a - \hat{\nabla}_{d} \hat{\nabla}_{c} X_a =\ ^{(3)}\!R_{abcd} X^b,
\end{equation} where $^{(3)}\!R_{abcd}$ is the \textit{spatial Riemann tensor} and $X_a$ is any vector field in the hypersurfaces, \textit{i.e.} $X_a u^a = 0$.  It can be shown by applying (\ref{eq:ThreeRicciIdentities}) to an arbitrary vector field $\mathbf{X}$ that is orthogonal to $\mathbf{u}$, expanding the left hand side using (\ref{eq:orthogonalcovariantderiv}) and (\ref{eq:projectiontensor}) and introducing the spacetime Riemann tensor via
\begin{equation}
\nabla_{c} \nabla_{d} X_a - \nabla_{d} \nabla_{c} X_a = R_{abcd} X^b,
\end{equation} that $^{(3)}\!R_{abcd}$ is related to $R_{abcd}$ via
\begin{equation} \label{eq:gaussequation}
^{(3)}\!R_{abcd} = (R_{abcd})_{\perp} - \Theta_{ac} \Theta_{bd} + \Theta_{ad} \Theta_{bc},
\end{equation} where $\Theta_{ab}$ is the expansion tensor of the congruence $\mathbf{u}$, defined by
\begin{equation}
\Theta_{ab} = \sigma_{ab} + H h_{ab}.
\end{equation}  Equation (\ref{eq:gaussequation}) is the well-known \textit{Gauss equation} (see, for example, Ellis (1973), p33-4).

The \textit{spatial Ricci tensor} and \textit{spatial Ricci scalar} are then defined in the usual manner:
\begin{equation} \label{eq:spatialricci3tensor}
^{(3)}\!R_{ab} = h^{cd}\ ^{(3)}\!R_{acbd}, \qquad ^{(3)}\!R = h^{ab}\ ^{(3)}\!R_{ab}.
\end{equation}  Further, the \textit{trace-free spatial Ricci tensor} is defined as
\begin{equation} \label{a2eq:spatialricci3tensor}
^{(3)}\!S_{ab} =\ ^{(3)}\!R_{ab} - \tfrac{1}{3}\ ^{(3)}\!R h_{ab}.
\end{equation}  We can obtain an expression for the spatial Ricci scalar and trace-free spatial Ricci tensor in terms of the spacetime quantities from (\ref{eq:GeneralizedGaussEquation}), incorporating the Weyl tensor through (\ref{eq:weyltensor}) and (\ref{eq:weyldecomposition}):
\begin{eqnarray}
\label{a2eq:Generalricci3scalar} ^{(3)}\!R & = & R + 2 R_{bd} u^b u^d - 6 H^2 + 2 \sigma^2, \\
\label{a2eq:Generalspatialricci3tensorfull} ^{(3)}\!S_{ab} & = & E_{ab} + \tfrac{1}{2} R_{\langle ab \rangle} - H \sigma_{ab} + \sigma^{c}_{\ \langle a} \sigma_{b \rangle c}.
\end{eqnarray}  Then using the field equations (\ref{eq:einsteinfieldequations}) in the form (\ref{a1eq:contractedefes1})-(\ref{a1eq:contractedefes4}) to eliminate $R$ and $R_{ab}$, we obtain
\begin{eqnarray}
\label{a2eq:ricci3scalarB} ^{(3)}\!R & = & -6 H^2 + 2 \mu + 2 \sigma^2, \\
\label{a2eq:spatialricci3tensorB} ^{(3)}\!S_{ab} & = & E_{ab} + \tfrac{1}{2} \pi_{ab} - H \sigma_{ab} + \sigma^{c}_{\ \langle a} \sigma_{b \rangle c}.
\end{eqnarray}

We conclude this section with a brief digression on spaces of constant curvature.  A differentiable manifold $\mathcal{M}$ of dimension $n$ with metric $\mathbf{g}$ is said to be a \textit{space of constant curvature} if and only if
\begin{equation} \label{a2eq:constantcurvature3riemanntensor}
R_{abcd} = C (g_{ac} g_{bd} - g_{ad} g_{bc}).
\end{equation}  It follows by contraction that
\begin{equation} \label{a2eq:constantcurvature3ricci}
R = n (n-1) C, \quad \mbox{and} \quad R_{ab} - \tfrac{1}{n} R g_{ab} = 0,
\end{equation} \textit{i.e.} the tracefree Ricci tensor $S_{ab} \equiv R_{ab} - \tfrac{1}{n} R g_{ab}$ is zero.  In addition it follows from the twice contracted Bianchi identities that if $n > 2$ then $R$ is constant (see proposition \ref{prop:constantcurvaturetracefreericci}).  If $n = 3$, the Riemann tensor is completely determined by the Ricci tensor according to\footnote{See, for example, Ellis (1973) eq. (80).}
\begin{equation} \label{a2eq:riemann3tensorbyricci3tensor}
R_{abcd} = 4 h_{[d[b} S_{a]c]} + \tfrac{1}{3} R g_{a[c} g_{b]d}.
\end{equation}  It follows that \textit{if $n = 3$ then $(\mathcal{M}, \mathbf{g})$ is a space of constant curvature if and only if $S_{ab} = 0$}.  Within the framework of the $1+3$ covariant formalism this result can be stated as follows.

\begin{proposition} \label{prop:constantcurvaturetracefreericci} Let $\mathbf{u}$ be an irrotational timelike vector field.  The hypersurfaces orthogonal to $\mathbf{u}$ are spaces of constant curvature if and only if
\begin{equation}
^{(3)}\!S_{ab} = 0.
\end{equation}  Further, if this condition holds, then
\begin{equation}
\hat{\nabla}_{a}\ \!^{(3)}\!R = 0.
\end{equation}
\end{proposition}

\subsection{Evolution and Constraint Equations} \label{sec:evolutionandconstraintequations}

In the $1+3$ covariant formalism, the information contained in the EFEs is displayed in an indirect manner by using the \textit{Ricci identities} applied to the fundamental 4-velocity $\mathbf{u}$,
\begin{equation} \label{eq:ricciidentities}
\nabla_{c} \nabla_{d} u_{a} - \nabla_{d} \nabla_{c} u_{a} = R_{abcd} u^b,
\end{equation} and the \textit{Bianchi identities},
\begin{equation} \label{eq:bianchiidentities}
\nabla_{[a} R_{bc]de} = 0.
\end{equation}  The latter can be written in divergence form by contracting once and then twice and using (\ref{eq:weyltensor}) to introduce the Weyl tensor:\footnote{Also see Wainwright and Ellis (1997), p27.}
\begin{equation} \label{eq:bianchiidentities_a}
\nabla_{d} C_{abc}^{\phantom{abc}d} = - \nabla_{[a} R^{c}_{\phantom{c}b]} - \tfrac{1}{6} \delta^{c}_{\ [a} \nabla_{b]} R,
\end{equation} and
\begin{equation} \label{eq:bianchiidentities_b}
\nabla_{b} (R_{a}^{\phantom{a}b} - \tfrac{1}{2} R \delta_{a}^{\ b}) = 0.
\end{equation}  The Riemann tensor in (\ref{eq:ricciidentities}) is likewise expressed in terms of the Weyl tensor and Ricci tensor using (\ref{eq:weyltensor}).  In all these identities, the Ricci tensor is expressed in terms of the energy-momentum tensor (\ref{eq:energymomentumtensor}) using the EFEs in the form
\begin{equation}
R_{ab} = T_{ab} - \tfrac{1}{2} g_{ab} T^{c}_{\ c},
\end{equation} which introduces the fluid 4-velocity $\mathbf{u}$ into (\ref{eq:bianchiidentities_a}) and (\ref{eq:bianchiidentities_b}).  Equation (\ref{eq:covudecomposition}) is used to replace all occurrences of $\nabla_{b} u_{a}$ by the kinematic quantities and (\ref{eq:weyldecomposition}) is used to replace $C_{abcd}$ by $E_{ab}$ and $H_{ab}$.  Finally, one projects all three identities parallel to and orthogonal to $\mathbf{u}$.  The result is a set of \textit{evolution equations} giving the derivatives of $H$, $\sigma_{ab}$, $\omega_a$, $E_{ab} + \tfrac{1}{2} \pi_{ab}$, $H_{ab}$, $\mu$ and $q_{a}$ along the congruence $\mathbf{u}$, and a set of \textit{constraint equations} that contain only the spatial covariant derivative $\hat{\nabla}_{a}$.  We refer the reader to Appendix \ref{app:evolutionandconstrainteqs} for the complete list of equations that arise from this process.

\section{Overview of the Thesis}

The main goal of this thesis is twofold.  Firstly, we analyze the dynamics of so-called $n$-fluid FL cosmologies, that is, homogeneous and isotropic cosmologies whose matter-energy content consists of $n$ non-interacting fluids.  Secondly, we give a unified coordinate-independent discussion of perturbations of FL cosmologies containing a perfect fluid and a cosmological constant.  Some parts of this thesis are new, while other parts provide a unified presentation of known results for these cosmologies.  Our treatment throughout is completely coordinate-independent since it is based on the so-called $1+3$ covariant description of general relativity and in this respect it differs from much of the current literature.

In Chapter 1, we introduce the $1+3$ covariant description of general relativity and in Appendix \ref{app:evolutionandconstrainteqs} we give the full system of evolution equations and constraint equations.  In this sense the thesis is self-contained.  However, a reader who has no prior knowledge of the $1+3$ covariant description will encounter an initially steep learning curve.  We assume the reader has a basic knowledge of general relativity and some familiarity with dynamical systems theory (see, for example, Wainwright and Ellis (1997), Chapter 4).

Chapter 2 plays a supporting role.  In this chapter we describe the geometrical, \textit{i.e.} coordinate independent properties of the Robertson-Walker metric, and give two characterizations of the Friedmann-Lema\^{i}tre models that highlight their very special nature.

In Chapter 3, we give a complete qualitative description of the dynamics of $n$-fluid FL cosmologies.  A central role is played by a new formulation of the evolution equations for these models as a dynamical system on a compact $n$-dimensional state space.  The most important example is the $3$-fluid model in which the three fluids are radiation, pressure-free matter and a cosmological constant, since it provides a model of the physical universe.  Our discussion of this model is one of the highlights of this chapter.

In Chapter 4, we give a unified derivation of the governing equations for scalar, vector and tensor perturbations of FL cosmologies containing a perfect fluid and a cosmological constant, using the so-called geometrical approach pioneered by Ellis et al (1989).  We then analyze the behaviour of the perturbations in various regimes using both exact and approximate solutions.  In particular we determine how the perturbations affect the so-called observational quantities that determine the extent to which a perturbed model deviates from an exact FL model.

The Appendices contain a wealth of additional material relating to perturbations of FL universes, not all of which is needed for the purposes of this thesis.

\chapter{The Friedmann-Lema\^{i}tre Cosmologies} \label{chap:friedmannlemaitre}

In this chapter we first derive the geometrical (\textit{i.e.} coordinate independent) properties of the Robertson-Walker metric within the framework of the $1+3$ description of general relativity.  We then give two characterizations of the Friedmann-Lema\^{i}tre cosmologies, firstly using the kinematic quantities and secondly using the Weyl tensor.

\section{Properties of the RW Metric}

In this section we show that the high symmetry of the RW line-element leads to strong restrictions on the kinematic quantities of the fundamental congruence, on the Weyl tensor and on the Ricci tensor.  We then derive the RW line-element assuming that the Cosmological Principle holds.

\bigskip
\begin{proposition} \label{prop:kinematiccharacterization1} The kinematic quantities of the fundamental congruence (\ref{eq:RWfourvelocity}) associated with the Robertson-Walker metric (\ref{eq:RWmetric}) satisfy
\begin{equation}
\sigma_{ab} = 0, \qquad \omega_{a} = 0, \qquad \dot{u}_a = 0, \qquad \hat{\nabla}_{a} H = 0.
\end{equation}
\end{proposition}

\paragraph{Proof.}  For the RW line-element (\ref{eq:RWmetric}), the covariant derivative
\begin{equation}
\nabla_{b} u_{a} = u_{a,b} - \Gamma^{c}_{\ ab} u_c
\end{equation} simplifies to
\begin{equation}
\nabla_{b} u_{a} = \tfrac{1}{2} g_{ab,0}.
\end{equation}  Furthermore, it follows from the RW metric (\ref{eq:RWmetric}) that
\begin{equation}
g_{\alpha \beta, 0} = 2 \frac{\dot{\ell}}{\ell} g_{\alpha \beta}, \quad \mbox{and} \quad g_{a0,0} = 0.
\end{equation}  Since the projection tensor $h_{ab}$ has components $h_{\alpha \beta} = g_{\alpha \beta}$, $h_{a 0} = 0$, it follows that
\begin{equation}
\nabla_{b} u_{a} = \frac{\dot{\ell}}{\ell} h_{ab}.
\end{equation}  The decomposition (\ref{eq:covudecomposition}) immediately implies $\sigma_{ab} = 0$, $\omega_{ab} = 0$, $\dot{u}_{a} = 0$ and
\begin{equation}
H = \frac{\dot{\ell}}{\ell}.
\end{equation}  Thus, $H$ is purely a function of $t$, which implies $\hat{\nabla}_a H = 0$.\ $\square$

\bigskip

We now consider the restrictions on the spatial curvature associated with the Robertson-Walker metric.

\begin{proposition} \label{prop:spatialcharacterization}  The spatial curvature of the RW metric satisfies
\begin{equation}
^{(3)}\!S_{ab} = 0, \quad \hat{\nabla}_{a}\!\ ^{(3)}\!R = 0.
\end{equation}
\end{proposition}

\paragraph{Proof:}  Since the hypersurfaces orthogonal to $\mathbf{u}$ are spaces of constant curvature, the result follows immediately from proposition \ref{prop:constantcurvaturetracefreericci}.\ $\square$

\bigskip

We now show that the Weyl tensor of the Robertson-Walker metric is zero.

\bigskip
\begin{proposition} \label{prop:weylcharacterization3} The Robertson-Walker metric (\ref{eq:RWmetric}) has zero Weyl tensor:
\begin{equation}
E_{ab} = 0, \qquad H_{ab} = 0
\end{equation}
\end{proposition}

\paragraph{Proof:}  We make use of the contracted Ricci identities in the form\footnote{These equations are an intermediate step in deriving (\ref{a1eq:ricciidentity2}) and (\ref{a1eq:ricciidentity6}), and follow from the Ricci identities, without use of the field equations.  In fact, on applying the field equations (\ref{eq:einsteinfieldequations}) in the form (\ref{a1eq:contractedefes1})-(\ref{a1eq:contractedefes4}) to (\ref{eq:ricciidentit2_R}) and (\ref{eq:ricciidentit6_R}), we obtain (\ref{a1eq:ricciidentity2}) and (\ref{a1eq:ricciidentity6}).}
\begin{eqnarray}
\label{eq:ricciidentit2_R} \dot{\sigma}_{\langle ab \rangle} & = & -2 H \sigma_{ab} + \nabla_{\langle a} \dot{u}_{b \rangle} + \dot{u}_{\langle a} \dot{u}_{b \rangle} - \sigma_{\langle a}^{\ \ c} \sigma_{b \rangle c} - \omega_{\langle a} \omega_{b \rangle} - (E_{ab} - \tfrac{1}{2} R_{\langle a b \rangle}), \\
\label{eq:ricciidentit6_R} 0 & = & H_{ab} - 2 \dot{u}_{\langle a} \omega_{b \rangle} - \hat{\nabla}_{\langle a} \omega_{b \rangle} - \mathsf{curl}(\sigma_{ab}).
\end{eqnarray}  We note that $H_{ab} = 0$ is an immediate consequence of (\ref{eq:ricciidentit6_R}), on applying proposition \ref{prop:kinematiccharacterization1}.  If we apply proposition \ref{prop:kinematiccharacterization1} to (\ref{eq:ricciidentit2_R}) and \ref{prop:spatialcharacterization} to (\ref{a2eq:Generalspatialricci3tensorfull}), we obtain
\begin{equation} \label{eq:weylcharacterizationequation}
E_{ab} - \tfrac{1}{2} R_{\langle ab \rangle} = 0, \qquad E_{ab} + \tfrac{1}{2} R_{\langle ab \rangle} = 0,
\end{equation} which implies $E_{ab} = 0$.\ $\square$

\bigskip
Finally, we derive the restrictions satisfied by the Ricci tensor of the Robertson-Walker metric (\ref{eq:RWmetric}).

\begin{proposition} \label{prop:energymomentumisotropy}
The Ricci tensor of the Robertson-Walker metric (\ref{eq:RWmetric}) satisfies
\begin{equation}
R_{ab} u^{a} h^{b}_{\ c} = 0, \qquad R_{\langle ab \rangle} = 0.
\end{equation}
\end{proposition}

\paragraph{Proof:}  We note that $R_{\langle ab \rangle} = 0$ follows immediately from (\ref{eq:weylcharacterizationequation}).  In order to show \mbox{$R_{ab} u^{a} h^{b}_{\ c} = 0$}, we make use of the contracted Ricci identities in the form\footnote{This equation is an intermediate step in deriving (\ref{a1eq:ricciidentity4}), and follow from the Ricci identities, without use of the field equations.  In fact, on applying the field equations (\ref{eq:einsteinfieldequations}) in the form (\ref{a1eq:contractedefes1})-(\ref{a1eq:contractedefes4}) to (\ref{eq:ricciidentity4_R}), we obtain (\ref{a1eq:ricciidentity4}).}
\begin{equation}
\label{eq:ricciidentity4_R} 0 = \hat{\nabla}^b \sigma_{ab} - 2 \hat{\nabla}_a H - \mathsf{curl}(\omega_a) - 2 \epsilon_{abc} \dot{u}^b \omega^c + R_{bc} u^{b} h^{c}_{\ a},
\end{equation}  Then on applying proposition \ref{prop:kinematiccharacterization1} to (\ref{eq:ricciidentity4_R}), we obtain \mbox{$R_{ab} u^{a} h^{b}_{\ c} = 0$}.\ $\square$

\bigskip

We conclude this section by showing that the Cosmological Principle, as stated in section \ref{sec:cosmologicalmodels}, leads to the RW metric.

\begin{proposition} \label{prop:RWCharacterization}  If the fundamental congruence of a cosmological model ($\mathcal{M}$, $\mathbf{g}$, $\mathbf{u}$) satisfies
\begin{equation} \label{eq:RWCharacterizationQuantities}
\sigma_{ab} = 0, \quad \omega_{a} = 0, \quad \dot{u}_{a} = 0, \quad \hat{\nabla}_{a} H = 0,
\end{equation} and the hypersurfaces orthogonal to $\mathbf{u}$ satisfy
\begin{equation}
^{(3)}S_{ab} = 0,
\end{equation} then $\mathbf{g}$ and $\mathbf{u}$ can be written in the RW form (\ref{eq:RWmetric}) and (\ref{eq:RWfourvelocity}).
\end{proposition}

\paragraph{Proof:}  By proposition \ref{prop:irrotationalgeodesicpotential}, the fundamental 4-velocity $\mathbf{u}$ can be written as
\begin{equation} \label{eq:RWfundamental4velocityfield}
u_{a} = - t_{,a},
\end{equation} where $t$ is a scalar field.  We then introduce \textit{co-moving coordinates} $x^{\alpha}$, \textit{i.e.} coordinates $x^{\alpha}$ that are constant along the world lines of the fundamental congruence.  Then the metric and 4-velocity assume the form
\begin{equation} \label{eq:characterizationmetric}
ds^2 = -dt^2 + g_{\alpha \beta}(t, x^{\lambda}) dx^{\alpha} dx^{\beta}, \qquad \mathbf{u} = \frac{\partial}{\partial t}.
\end{equation}  Next, using the metric (\ref{eq:characterizationmetric}), it can be quickly shown that
\begin{equation}
\nabla_{b} u_{a} = \tfrac{1}{2} g_{ab,0}.
\end{equation}  This equation, in conjunction with (\ref{eq:projectiontensor}), (\ref{eq:kinshear}), (\ref{eq:RWCharacterizationQuantities}) and (\ref{eq:RWfundamental4velocityfield}), leads to
\begin{equation} \label{eq:kinematiccharacterizationshear}
g_{\alpha \beta,0} = 2 H(t) g_{\alpha \beta}.
\end{equation}  We introduce the length scale function $\ell(t)$ via
\begin{equation}
H(t) = \frac{\dot{\ell}}{\ell},
\end{equation} and thus can integrate (\ref{eq:kinematiccharacterizationshear}) to obtain
\begin{equation}
g_{\alpha \beta}(t, x^{\lambda}) = \ell(t)^2 \tilde{g}_{\alpha \beta} (x^{\lambda}).
\end{equation}  Since we have that $^{(3)}\!S_{ab} = 0$, it follows from proposition \ref{prop:constantcurvaturetracefreericci} that $\tilde{g}_{\alpha \beta}$ describes a 3-space of constant curvature.  Thus the metric (\ref{eq:characterizationmetric}) assumes the Robertson-Walker form (\ref{eq:RWmetric}).\ $\square$

\bigskip

\begin{theorem}[Derivation of the RW Metric] \label{thm:derivationoftheRWmetric}  If a cosmological model ($\mathcal{M}$, $\mathbf{g}$, $\mathbf{u}$) satisfies the Cosmological Principle relative to $\mathbf{u}$, then $\mathbf{g}$ and $\mathbf{u}$ can be written in the RW form (\ref{eq:RWmetric}) and (\ref{eq:RWfourvelocity}).
\end{theorem}

\paragraph{Proof:}  The Cosmological Principle implies that any geometrically defined vector orthogonal to $\mathbf{u}$ must be zero, since otherwise it would define a preferred direction orthogonal to $\mathbf{u}$.  In particular, we have $\omega_a = \dot{u}_{a} = \hat{\nabla}_{a} H = 0$.  Likewise, any geometrically defined rank two symmetric tracefree tensor $X_{ab}$ that is orthogonal to $\mathbf{u}$ ($X_{ab} u^{b} = 0$) must be zero, since otherwise its eigenvectors would define a preferred direction orthogonal to $\mathbf{u}$.  In particular we have $\sigma_{ab} =\ ^{(3)}\!S_{ab} = 0$.  The desired result thus follows from proposition \ref{prop:RWCharacterization}. \ $\square$

\section{Characterizations of the FL Cosmologies}

In this section we present two characterizations of FL cosmologies in terms of the kinematic quantities and Weyl curvature.

\subsection{The Kinematic Characterization} \label{sec:FLkinematiccharacterization}

We now present the \textit{kinematic characterization of the Friedmann-Lema\^{i}tre models}.\footnote{One can replace the assumption $\dot{u}_{a} = 0$ in Theorem \ref{thm:kinematiccharacterization} by the assumption that the perfect fluid has a linear equation of state \mbox{$p = (\gamma - 1) \mu$}.  However, the proof requires a more detailed analysis and is beyond the scope of this thesis (see instead Collins and Wainwright (1983)).}

\bigskip
\begin{theorem} \label{thm:kinematiccharacterization} A cosmological model ($\mathcal{M}$,$\mathbf{g}$,$\mathbf{u}$) which satisfies the Einstein field equations with perfect fluid source is a Friedmann-Lema\^{i}tre cosmology if and only if the kinematic quantities of the fluid 4-velocity $\mathbf{u}$ satisfy
\begin{equation} \label{eq:kinematicrestrictions}
\sigma_{ab} = 0, \qquad \omega_{a} = 0, \qquad \dot{u}_a = 0.
\end{equation}
\end{theorem}

\paragraph{Proof.}\footnote{Krasinski (1997) gives the outline of a proof, referring to other sources for some parts (see page 11).}  If we assume the model is a FL cosmology, the kinematic restrictions (\ref{eq:kinematicrestrictions}) follow immediately from proposition \ref{prop:kinematiccharacterization1}.

Conversely, assume that the restrictions (\ref{eq:kinematicrestrictions}) hold.  Since the source is a perfect fluid with 4-velocity $\mathbf{u}$, equation (\ref{eq:perfectfluidquantities}) implies $\pi_{ab} = 0$ and $q_{a} = 0$.  Then the conditions (\ref{eq:kinematicrestrictions}) in conjunction with (\ref{a1eq:ricciidentity2}) imply $E_{ab} = 0$, which in turn implies $^{(3)}\!S_{ab} = 0$ via (\ref{a2eq:spatialricci3tensorB}).  The conditions (\ref{eq:kinematicrestrictions}) and $q_{a} = 0$ in conjunction with (\ref{a1eq:ricciidentity4}) imply $\hat{\nabla}_{a} H = 0$.  The desired result then follows from proposition \ref{prop:RWCharacterization}. \ $\square$

\subsection{The Weyl Characterization} \label{sec:FLweylcharacterization}

We now present a characterization of the FL models using the Weyl tensor (see section \ref{sec:weylcurvaturetensor}).

\bigskip
\begin{proposition} \label{prop:weylcharacterization1} If a cosmological model ($\mathcal{M}$, $\mathbf{g}$, $\mathbf{u}$) satisfies the Einstein field equations with perfect fluid source having 4-velocity $\mathbf{u}$ and barotropic equation of state $p = p(\mu)$ with $\mu + p \neq 0$, and has zero Weyl tensor,
\begin{equation} \label{eq:zeroweylcomponents}
E_{ab} = 0, \qquad H_{ab} = 0,
\end{equation} then the kinematic quantities satisfy
\begin{equation}
\sigma_{ab} = 0, \qquad \omega_{ab} = 0, \qquad \dot{u}_a = 0.
\end{equation}
\end{proposition}

\paragraph{Proof.}  On account of (\ref{eq:zeroweylcomponents}), equations (\ref{a1eq:fullbianchi1}), (\ref{a1eq:fullbianchi3}) and (\ref{a1eq:fullbianchi4}) simplify to
\begin{equation} \label{eq:zerokinematicsfromzeroweyl}
(\mu + p) \sigma_{ab} = 0, \qquad \hat{\nabla}_{a} \mu = 0, \qquad (\mu + p) \omega_{a} = 0.
\end{equation}  Since $(\mu + p) \neq 0$, (\ref{eq:zerokinematicsfromzeroweyl}) implies $\sigma_{ab} = 0$ and $\omega_{a} = 0$.  We now make use of the chain rule to write
\begin{equation} \label{eq:pressureenergyrelation}
\hat{\nabla}_{a} p = \frac{dp}{d\mu} \hat{\nabla}_{a} \mu = 0.
\end{equation}  Then using the perfect fluid condition, \textit{i.e.} $\pi_{ab} = 0$ and $q_{a} = 0$, it follows from (\ref{a1eq:contractedbianchi2}) and (\ref{eq:pressureenergyrelation}) that $(\mu + p) \dot{u}_{a} = 0$, and hence $\dot{u}_{a} = 0$, which completes the proof.\ $\square$

\bigskip

We can now present the main result of this section, which we shall refer to as the \textit{Weyl characterization of the Friedmann-Lema\^{i}tre models}.

\bigskip
\begin{theorem} \label{thm:weylcharacterization}
A cosmological model ($\mathcal{M}$,$\mathbf{g}$,$\mathbf{u}$), which satisfies the Einstein field equations with perfect fluid source having 4-velocity $\mathbf{u}$ and barotropic equation of state $p = p(\mu)$ with $\mu + p \neq 0$, is a Friedmann-Lema\^{i}tre model if and only if the Weyl tensor is zero:
\begin{displaymath}
E_{ab} = 0, \qquad H_{ab} = 0.
\end{displaymath}
\end{theorem}

\paragraph{Proof.}  This result follows immediately from proposition \ref{prop:weylcharacterization3} and theorem \ref{thm:kinematiccharacterization} in conjunction with proposition \ref{prop:weylcharacterization1}. $\square$

\bigskip

We now present one final result which restricts the gradients of the physical and geometrical scalars in an FL cosmology.

\begin{proposition}  A Friedmann-Lema\^{i}tre cosmology satisfies
\begin{equation}
\hat{\nabla}_{a} \mu = \hat{\nabla}_{a} p = \hat{\nabla}_{a} H = \hat{\nabla}_{a}\!\ ^{(3)}\!R = 0.
\end{equation}
\end{proposition}

\paragraph{Proof:}  We note that \mbox{$\hat{\nabla}_{a} H = \hat{\nabla}_{a}\!\ ^{(3)}\!R = 0$} follows immediately from propositions \ref{prop:kinematiccharacterization1} and \ref{prop:spatialcharacterization}.  Further, it follows from (\ref{a1eq:contractedefes3}), (\ref{a1eq:contractedefes4}) and proposition \ref{prop:energymomentumisotropy} that \mbox{$q_{a} = 0$} and \mbox{$\pi_{ab} = 0$}.  Hence, (\ref{a1eq:contractedbianchi2}) and (\ref{a1eq:fullbianchi3}) along with theorems \ref{thm:kinematiccharacterization} and \ref{thm:weylcharacterization} implies \mbox{$\hat{\nabla}_{a} \mu = \hat{\nabla}_{a} p = 0$}.\ $\square$

\chapter{Dynamics of $n$-Fluid FL Cosmologies}

The main goal of this chapter is a description of the dynamics of $n$-fluid FL cosmologies.  In section \ref{sec:behavclassnfluidFL}, using the results from chapter \ref{chap:geometricalbackground}, we derive the basic evolution equations for $n$-fluid FL cosmologies and describe their behaviour using dynamical systems techniques.  In section \ref{sec:FLParameterization}, we describe the degrees of freedom of the $n$-fluid FL cosmologies in two ways, firstly, using intrinsic parameters and secondly, using observational parameters.  In sections \ref{sec:FL2fluidmodels} and \ref{sec:FL3fluidmodels}, we apply the general analysis to the case of 2-fluid and 3-fluid FL cosmologies.  Finally, in section \ref{sec:FLsolutions} we give a unified presentation of the known solutions of the EFEs for single fluid and 2-fluid FL cosmologies, and relate them to the state space analysis in section \ref{sec:FL2fluidmodels}.

\section{Behaviour and Classification of $n$-Fluid FL Cosmologies} \label{sec:behavclassnfluidFL}

In this section we derive the evolution equations for $n$-fluid FL cosmologies without introducing local coordinates and a line-element, by specializing the general evolution and constraint equations (appendix \ref{app:evolutionandconstrainteqs}) using the restrictions derived in chapter \ref{chap:friedmannlemaitre}.  We formulate the evolution equations as ODEs for a set of dimensionless variables in a compact state, which enables us to use dynamical systems methods to describe the evolution of $n$-fluid FL cosmologies qualitatively.

\subsection{Basic Evolution Equations} \label{sec:FLevolutioneqs}

We know from theorems \ref{thm:kinematiccharacterization} and \ref{thm:weylcharacterization} and proposition \ref{prop:energymomentumisotropy} that FL models have the following properties:

\begin{enumerate}
\item[i)] the kinematic quantities of the fundamental congruence $\mathbf{u}$ satisfy
\begin{equation}
\sigma_{ab} = 0, \quad \omega_a = 0, \quad \dot{u}_a = 0,
\end{equation}
\item[ii)] the Weyl curvature is zero:
\begin{equation}
E_{ab} = 0, \quad H_{ab} = 0,
\end{equation}
and
\item[iii)] the stress-energy tensor satisfies
\begin{equation}
\pi_{ab} = 0, \quad q_a = 0,
\end{equation} \textit{i.e.} it has the form of a perfect fluid.
\end{enumerate}

The remaining non-zero physical and geometric variables that appear in the general evolution and constraint equations in Appendix \ref{app:evolutionandconstrainteqs} (\ref{a1eq:ricciidentity1})-(\ref{a1eq:fullbianchi4}) are the Hubble scalar $H$, the energy density $\mu$ and the pressure $p$ of the matter content.  Propositions \ref{prop:kinematiccharacterization1} and \ref{prop:energymomentumisotropy} show that these scalars have zero spatial gradient
\begin{equation}
h_{a}^{\ b} H_{,b} = 0, \quad h_{a}^{\ b} \mu_{,b} = 0, \quad h_{a}^{\ b} p_{,b} = 0,
\end{equation} and hence are purely functions of clock time $t$ along the fundamental congruence.  With the above restrictions, the general evolution and constraint equations reduce to two evolution equations for $\mu$ and $H$, namely (\ref{a1eq:ricciidentity1}) and (\ref{a1eq:contractedbianchi1}), which read
\begin{eqnarray}
\label{eveq:consenergy1} \dot{\mu} & = & -3 H (\mu + p), \\
\label{eveq:raychaudhurieq1} \dot{H} & = & - H^2 - \tfrac{1}{6} (\mu + 3 p),
\end{eqnarray} where the overdot denotes differentiation with respect to $t$.  The spatial geometry of the model is determined by the Gauss equation in the form (\ref{a2eq:ricci3scalarB}) and (\ref{a2eq:spatialricci3tensorB}).  It follows that the tracefree Ricci tensor is zero,
\begin{equation}
^{(3)}\!S_{ab} = 0,
\end{equation} and that the spatial Ricci scalar is determined by $H$ and $\mu$ according to
\begin{equation} \label{eveq:friedmanneq1}
^{(3)}\!R = -6 H^2 + 2 \mu.
\end{equation}

In order to obtain the standard evolution equations, we introduce the length scale function $\ell(t)$, which is defined by $H$, up to a constant multiple, according to
\begin{equation} \label{eq:hubblelengthscale}
H = \frac{\dot{\ell}}{\ell}.
\end{equation}  This definition is consistent with the length scale defined by the RW metric (\ref{eq:RWmetric}).

To proceed we need the following result:

\bigskip
\begin{proposition}
In a Friedmann-Lema\^{i}tre model the quantity
\begin{equation} \label{eq:curvaturelengthscale}
^{(3)}\!R(t) \ell(t)^2
\end{equation}
is constant in time.
\end{proposition}

\paragraph{Proof.}  A straightforward calculation using (\ref{eveq:consenergy1})-(\ref{eq:hubblelengthscale}) yields $(^{(3)}\!R \ell^2)^{\bullet} = 0$. $\square$
\bigskip

Since $\ell$ is only defined by (\ref{eq:hubblelengthscale}) up to a multiplicative constant, we may choose it to satisfy\footnote{The factor of $\tfrac{1}{6}$ is chosen for compatibility with the RW metric.}
\begin{equation} \label{eq:curvatureparameter}
\tfrac{1}{6}\ ^{(3)}\!R \ell^2 = K = \pm 1,\ \mbox{or}\ \ 0,
\end{equation} where $K$ is the \textit{spatial curvature parameter}.  The value for $K$ indicates whether $^{(3)}R$ is positive, negative or zero.  Models where $K = -1, 0$ and $+1$ are then referred to, respectively, as \textit{open FL models}, \textit{flat FL models} and \textit{closed FL models}.

It is important to note that if $K = \pm 1$, then $\ell(t)$ is completely determined via eq (\ref{eq:curvatureparameter}), by the spatial geometry, represented by $^{(3)}\!R$.  On the other hand, if $K = 0$, equation (\ref{eq:curvatureparameter}) does not restrict $\ell(t)$, which means that \textit{if $K = 0$ then $\ell(t)$ can be rescaled with a multiplicative constant}.

We now present the standard form of the evolution equations for FL models (see, for example, Wainwright and Ellis 1997, p52).  Using (\ref{eq:hubblelengthscale}) and (\ref{eq:curvatureparameter}), equations (\ref{eveq:consenergy1}), (\ref{eveq:raychaudhurieq1}) and (\ref{eveq:friedmanneq1}) are rewritten as the \textit{conservation equation},
\begin{equation} \label{eq:FLconservation}
\dot{\mu} = -3 \frac{\dot{\ell}}{\ell} (\mu + p),
\end{equation} the \textit{Raychaudhuri equation},
\begin{equation} \label{eq:FLraychaudhuri}
\frac{\ddot{\ell}}{\ell} = -\tfrac{1}{6} (\mu + 3 p),
\end{equation} and the \textit{Friedmann equation},
\begin{equation} \label{eq:FLfriedmann}
3 \dot{\ell}^2 = \mu \ell^2 - 3 K.
\end{equation}  Note that (\ref{eq:FLraychaudhuri}) is a consequence of (\ref{eq:FLconservation}) and (\ref{eq:FLfriedmann}) provided that $\dot{\ell} \neq 0$, as follows by differentiating (\ref{eq:FLfriedmann}).

\subsection{$n$-Fluid FL Cosmologies} \label{sec:nFluidFLCosmologies}

In this chapter, we consider FL cosmological models whose matter content consists of $n$ fluids with the same four-velocity $\mathbf{u}$.  The total energy density and pressure are equal to a sum of the energy density and pressure of the individual components:
\begin{equation} \label{eq:FLtotalmattercontent}
\mu = \sum_{i=1}^{n} \mu_i, \quad p = \sum_{i=1}^{n} p_i.
\end{equation}  The fluids are assumed to be non-interacting, which means that each fluid component individually satisfies the conservation equation (\ref{eq:FLconservation}), \textit{i.e.}
\begin{equation} \label{eq:FLconservationNFluidA}
\dot{\mu}_i = -3 \frac{\dot{\ell}}{\ell} (\mu_i + p_i).
\end{equation}  We further assume that each fluid obeys a linear equation of state with equation of state parameters $\gamma_i$ ($i = 1, \ldots, n$), \textit{i.e.}
\begin{equation} \label{eq:FLequationofstateNFluid}
p_i = (\gamma_i - 1) \mu_i,
\end{equation} with
\begin{equation}
0 \leq \gamma_i \leq 2.
\end{equation}  The most important physical values are $\gamma_i = 1$ for pressure-free matter and $\gamma_i = \frac{4}{3}$ for radiation.  The value $\gamma_i = 0$ can be used to represent a cosmological constant and $\gamma_i = 2$ is sometimes considered to correspond to a so-called stiff fluid.  For definiteness we order the equation of state parameters according to
\begin{equation} \label{eq:FLgammaordering}
\gamma_1 > \gamma_2 > \cdots > \gamma_n,
\end{equation} and assume that $\gamma_1 \geq 1$, \textit{i.e.} at least one matter component is matter with a non-negative pressure.  The conservation equations (\ref{eq:FLconservationNFluidA}) then assume the form
\begin{equation} \label{eq:FLconservationNFluid}
\dot{\mu}_i = -3 \frac{\dot{\ell}}{\ell} \gamma_i \mu_i.
\end{equation}  As well, the Friedmann equation (\ref{eq:FLfriedmann}) and Raychaudhuri equation (\ref{eq:FLraychaudhuri}) assume the form
\begin{equation} \label{eq:FLfriedmannNFluid}
\dot{\ell}^2 = \sum_{i=1}^{n} \tfrac{1}{3} \mu_i \ell^2 - K,
\end{equation} and
\begin{equation} \label{eq:FLraychaudhuriNFluid}
\frac{\ddot{\ell}}{\ell} = - \tfrac{1}{6} \sum_{i=1}^{n} (3 \gamma_i  - 2) \mu_i.
\end{equation}


In summary, we will use the term \textit{$n$-fluid FL cosmology} to refer to an FL cosmological model whose matter content consists of $n$ non-interacting fluids with the same 4-velocity $\mathbf{u}$ and equations of state given by (\ref{eq:FLequationofstateNFluid})-(\ref{eq:FLgammaordering}).  Following Jantzen and Uggla (1992), we will use abbreviations such as $RDC$-universe and $RC\Lambda$-universe to identify $n$-fluid FL cosmologies, where $R$ denotes radiation, $D$ denotes dust, $C$ denotes curvature and $\Lambda$ denotes a cosmological constant.

We have shown that the state of an $n$-fluid FL cosmology as a function of time $t$ is determined by the functions ($\ell$, $\mu_1$, \ldots, $\mu_n$) which satisfy the system of ODEs (\ref{eq:FLconservationNFluid}) and (\ref{eq:FLfriedmannNFluid}).  The state of an $n$-fluid cosmology can equivalently be described by the variables ($H$, $\mu_1$, \ldots, $\mu_n$), which satisfy the evolution equations
\begin{eqnarray}
\label{eq:FLconservationH} \dot{\mu}_i & = & -3 H \gamma_i \mu_i, \\
\label{eq:FLraychaudhuriH} \dot{H} & = & - H^2 - \tfrac{1}{6} \sum_{i=1}^{n} (3 \gamma_i - 2) \mu_i.
\end{eqnarray}  The first of these is a restatement of (\ref{eq:FLconservationNFluid}) using the definition (\ref{eq:hubblelengthscale}) of $H$.  The second follows from (\ref{eveq:raychaudhurieq1}) on using (\ref{eq:FLtotalmattercontent}) and (\ref{eq:FLequationofstateNFluid}).  Further, upon applying (\ref{eq:hubblelengthscale}), the Friedmann equation (\ref{eq:FLfriedmannNFluid}) gives an additional constraint that must be satisfied, namely
\begin{equation} \label{eq:FLfriedmannH}
H^2 = \tfrac{1}{3} \sum_{i=1}^{n} \mu_i - \frac{K}{\ell^2}.
\end{equation}  We note that this equation explicitly determines $\ell$ in terms of $H$ and $\mu$ when $K \neq 0$.

At this stage, we present four simple and well-known solutions of equations (\ref{eq:FLconservationNFluid})-(\ref{eq:FLraychaudhuriNFluid}) that play an important role when we formulate the evolution equations (\ref{eq:FLconservationH})-(\ref{eq:FLraychaudhuriH})for the  $n$-fluid FL cosmologies as a dynamical system.

\begin{enumerate}
\item \textit{Flat FL universe $(K = 0, 0 < \gamma \leq 2)$.}  For a single fluid, the length scale, energy density and pressure are
\begin{equation} \label{eq:FLsolFlat}
\ell = \ell_0 \left(\frac{t}{t_0}\right)^{\frac{2}{3 \gamma}}, \qquad \mu = \frac{4}{3 \gamma^2 t^2}, \qquad p = (\gamma - 1) \mu.
\end{equation}  In the case of dust ($\gamma = 1$), this solution is the \textit{Einstein-de Sitter universe} (Einstein and de Sitter (1932)).

\item \textit{de Sitter universe $(K = 0)$.}  The de Sitter universe can be thought of as the special case of the flat FL universe when the single fluid is a cosmological constant (\textit{i.e.} when $\gamma = 0$).  The length scale, energy density and pressure are
\begin{equation} \label{eq:FLsolDeSitter}
\ell = \ell_0 \exp \left( \sqrt{\frac{\Lambda}{3}} t \right), \qquad \mu = \Lambda, \qquad p = - \Lambda.
\end{equation}

\item \textit{Milne universe $(K = -1)$.}  The length scale, energy density and pressure are
\begin{equation} \label{eq:FLsolMilne}
\ell = t, \qquad \mu = 0, \qquad p = 0.
\end{equation}  The Milne universe can be thought of as the open FL universe with zero matter content.

\item \textit{Einstein static universe $(K = +1)$.}  This solution arises when we require that the length scale is constant.  The length scale, total energy density and total pressure are
\begin{equation} \label{eq:FLsolEinsteinStatic}
\ell = \ell_0, \qquad \mu = 3 \ell_0^{-2}, \qquad p = - \ell_0^{-2},
\end{equation} where $\ell_0 > 0$ is constant.\footnote{The values of $\mu$ and $p$ follow (\ref{eq:FLraychaudhuri}) and (\ref{eq:FLfriedmann}).}  The matter-energy content can be interpreted as two, or more generally, $n$ non-interacting fluids.  For example, for a 2-fluid model, the energy densities are given by \footnote{These expressions are obtained by solving the linear system
\begin{displaymath}
\mu_1 + \mu_2 = 3 \ell^2, \qquad (3 \gamma_1 - 2) \mu_1 + (3 \gamma_2 - 2) \mu_2 = 0,
\end{displaymath} which follows from (\ref{eq:FLfriedmannNFluid}) and (\ref{eq:FLraychaudhuriNFluid}) with $\ell = \ell_0$ and $n = 2$.}
\begin{equation}
\mu_1 = \left( \frac{2 - 3 \gamma_2}{\gamma_1 - \gamma_2} \right) \ell_0^{-2}, \qquad \mu_2 = \left( \frac{3 \gamma_1 - 2}{\gamma_1 - \gamma_2} \right) \ell_0^{-2},
\end{equation} where $0 \leq \gamma_2 < \tfrac{2}{3} < \gamma_1 \leq 2$.   Einstein's choice was $\gamma_1 = 1$ (pressure-free matter) and $\gamma_2 = 0$ (a cosmological constant) so that $\mu_1 = 2 \ell_0^{-2}$, $p_1 = 0$ and $\mu_2 = \ell_0^{-2} = \Lambda$, $p_2 = - \Lambda$, where $\Lambda$ is the cosmological constant.
\end{enumerate}

\subsection{Hubble-Normalized Scalars} \label{sec:FLnondimensionalsystem}

In this section we introduce so-called \textit{Hubble-normalized scalars}, which form a set of dimensionless variables that, together with the Hubble scalar $H$, describe the state of $n$-fluid FL cosmologies.  These dimensionless scalars also play an important role in relating the models to observations (see section \ref{ssec:ObservationalParameters}).

Firstly, the \textit{density parameters} $\Omega_i$, which describe the dynamical effect of the matter densities, are defined by
\begin{equation} \label{eq:FLdensityparameter}
\Omega_i = \frac{\mu_i}{3 H^2}.
\end{equation}  The \textit{total density parameter} $\Omega$, which describes the dynamical effect of the total matter content, is defined by
\begin{equation} \label{eq:FLtotaldensityparameter}
\Omega = \frac{\mu}{3 H^2}.
\end{equation}  It follows from (\ref{eq:FLtotalmattercontent}) and (\ref{eq:FLdensityparameter}) that
\begin{equation} \label{eq:FLtotaldensityparameterS}
\Omega = \sum_{i=1}^{n} \Omega_i.
\end{equation}  Lastly, the dynamical effect of the spatial curvature is described by $\Omega_k$, defined by
\begin{equation} \label{eq:FLdynamicalcurvatureR}
\Omega_k = - \frac{^{(3)}\!R}{6 H^2}.
\end{equation}  It follows from (\ref{eq:curvatureparameter}) that
\begin{equation} \label{eq:FLdynamicalcurvature}
\Omega_k = - \frac{K}{H^2 \ell^2}.
\end{equation}  In terms of the parameters $\Omega$ and $\Omega_k$, the Friedmann equation (\ref{eq:FLfriedmannNFluid}) assumes the simple form
\begin{equation} \label{eq:FLomegarelation}
\Omega + \Omega_k = 1,
\end{equation} which may be further expanded in terms of $\Omega_i$ as
\begin{equation} \label{eq:FLomegaAomegaKrelation}
\sum_{i=1}^{n} \Omega_i + \Omega_k = 1.
\end{equation}  We note that (\ref{eq:FLomegarelation}) implies that $\Omega$ determines the spatial curvature as follows:
\begin{eqnarray}
\label{eq:OmegaCurvatureRelation1} \Omega < 1 & \Longleftrightarrow & \mbox{open FL,} \\
\label{eq:OmegaCurvatureRelation2} \Omega = 1 & \Longleftrightarrow & \mbox{flat FL,} \\
\label{eq:OmegaCurvatureRelation3} \Omega > 1 & \Longleftrightarrow & \mbox{closed FL.}
\end{eqnarray}

There is an additional dimensionless variable that plays a fundamental role, namely the \textit{deceleration parameter} $q$, defined by
\begin{equation} \label{eq:FLq}
q = - \frac{\ddot{\ell} \ell}{(\dot{\ell})^2}.
\end{equation}  Clearly, if $q < 0$ then the expansion is accelerating ($\ddot{\ell} > 0$) and if $q > 0$ then the expansion is decelerating ($\ddot{\ell} < 0$).  In an $n$-fluid FL model, $q$ can be expressed in terms of the density parameters by using the Raychaudhuri equation (\ref{eq:FLraychaudhuriNFluid}) with (\ref{eq:hubblelengthscale}) and (\ref{eq:FLdensityparameter}).  We write
\begin{equation} \label{eq:qomegarelation}
q = \tfrac{1}{2} \sum_{i=1}^{n} (3 \gamma_i - 2) \Omega_i.
\end{equation}

The quantities $\Omega_i$, $\Omega$, $\Omega_k$ and $q$ are dimensionless, whereas $H$ has dimensions of $(\mathit{length})^{-1}$.  Together, these variables are fundamental quantities in the FL models and more general models (see, for example, Wainwright and Ellis (1997), section 5.2 and chapters 12 and 13).

The evolution equations for $\Omega_i$ and $\Omega_k$ can be obtained by differentiating (\ref{eq:FLdensityparameter}) and (\ref{eq:FLdynamicalcurvature}) and applying (\ref{eq:FLfriedmann}), (\ref{eq:FLconservationH}), (\ref{eq:FLraychaudhuriH}) and (\ref{eq:qomegarelation}).  On performing this procedure, we obtain
\begin{equation}
\frac{d\Omega_i}{dt} = (2 q - (3 \gamma_i - 2)) \Omega_i H, \quad \mbox{and} \quad \frac{d\Omega_k}{dt} = 2 q \Omega_k H.
\end{equation}

\subsection{A Compact State Space for $n$-fluid FL Cosmologies} \label{ssec:compactifiedstatespace}

The one drawback of using H-normalized variables is that they cannot give a unified description of models that have both an expanding ($H > 0$) and a contracting ($H < 0$) epoch.  In particular, when $H = 0$ the density parameters (\ref{eq:FLdensityparameter}) and (\ref{eq:FLdynamicalcurvatureR}) are undefined (\textit{i.e.} \mbox{$\Omega_i \rightarrow + \infty$} as \mbox{$H \rightarrow 0$}).  In order to circumvent this difficulty, we introduce a set of variables which are bounded throughout the evolution of the model.

For an $n$-fluid FL cosmology, with total energy density
\begin{equation} \label{eq:FL2totalmattercontent}
\mu = \sum_{i=1}^{n} \mu_i > 0,
\end{equation} we define the dimensionless \textit{matter variables}, which describe the relative significance of the different matter components:
\begin{equation} \label{eq:FLchi}
\chi_i = \frac{\mu_i}{\mu}, \quad \mu_i \geq 0.
\end{equation}  Then by (\ref{eq:FL2totalmattercontent}) we have
\begin{equation} \label{eq:FLchirestriction}
\sum_{i=1}^{n} \chi_i = 1, \qquad 0 \leq \chi_i \leq 1,
\end{equation} \textit{i.e.} the $\chi_i$ are bounded.

Following Wainwright (1996, p124-5), we now introduce a \textit{curvature parameter} $\tilde{\Omega}$ so as to compactify the state space:
\begin{equation} \label{eq:FLtildeOmega}
\tilde{\Omega} = \mathrm{arctan} \left( \frac{\sqrt{3} H}{\sqrt{\mu}} \right).
\end{equation}  It follows that $\tilde{\Omega}$ satisfies
\begin{equation} \label{eq:FLtildeOmegaConstraint}
- \frac{\pi}{2} < \tilde{\Omega} < \frac{\pi}{2},
\end{equation} and that
\begin{equation} \label{eq:FLtildeOmegaHrelation}
\mathrm{sign}(\tilde{\Omega}) = \mathrm{sign}(H).
\end{equation}  This variable thus describes the state of expansion of a model.  In addition, it describes the spatial curvature, \textit{i.e.} the deviation from flatness, as a consequence of (\ref{eq:FLtotaldensityparameter}) and (\ref{eq:OmegaCurvatureRelation1})-(\ref{eq:OmegaCurvatureRelation3}).  In particular, it follows from (\ref{eq:OmegaCurvatureRelation2}) and (\ref{eq:FLtotaldensityparameter}) that \mbox{$\tilde{\Omega} = \tfrac{\pi}{4}$} gives the flat FL models.

We now derive evolution equations for $\tilde{\Omega}$ and the $\chi_i$.  It is convenient to introduce a quantity $\tilde{q}$ that plays the role of the deceleration parameter $q$, but unlike $q$, is bounded when $H = 0$.  Motivated by (\ref{eq:qomegarelation}), we define
\begin{equation} \label{eq:FLqtilde}
\tilde{q} = \tfrac{1}{2} \sum_{i=1}^{n} (3 \gamma_i - 2) \chi_i.
\end{equation}  Together, (\ref{eq:FLgammaordering}) and (\ref{eq:FLqtilde}) imply that $\tilde{q}$ is restricted by
\begin{equation} \label{eq:FLqtildeConstraint}
\tfrac{1}{2} (3 \gamma_n - 2) \leq \tilde{q} \leq \tfrac{1}{2} (3 \gamma_1 - 2).
\end{equation}  It follows from (\ref{eq:FLtotalmattercontent}), (\ref{eq:FLconservationH}), (\ref{eq:FLchi}), (\ref{eq:FLchirestriction}) and (\ref{eq:FLqtilde}) that the total energy density satisfies
\begin{equation} \label{eq:FLTotalenergyEvolutionA}
\frac{d\mu}{dt} = - 2 (\tilde{q} + 1) H \mu.
\end{equation}  The evolution equation for $\chi_i$ then follows from (\ref{eq:FLconservationH}), (\ref{eq:FLchi}), (\ref{eq:FLqtilde}) and (\ref{eq:FLTotalenergyEvolutionA}) and is given by
\begin{equation} \label{eq:FLChiEvolutionA}
\frac{d\chi_i}{dt} = [2 \tilde{q} - (3 \gamma_i - 2)] H \chi_i.
\end{equation}  Upon noting that (\ref{eq:FLtildeOmega}) implies
\begin{equation} \label{eq:FLHubbleitoMuOmega}
H = \sqrt{\frac{\mu}{3}} \tan \tilde{\Omega},
\end{equation} we can rewrite (\ref{eq:FLChiEvolutionA}) as
\begin{equation} \label{eq:FLChiEvolutionB}
\frac{d\chi_i}{dt} = \sqrt{\frac{\mu}{3}} \frac{1}{\cos \tilde{\Omega}} [2 \tilde{q} - (3 \gamma_i - 2)] (\sin \tilde{\Omega}) \chi_i.
\end{equation}  The form of (\ref{eq:FLChiEvolutionB}) suggests defining a new time variable $\tilde{\tau}$ by
\begin{equation} \label{eq:FLtildetime}
\frac{dt}{d\tilde{\tau}} = \sqrt{\frac{3}{\mu}} \cos \tilde{\Omega}.
\end{equation}  It is essential to note that this change of variable is well-defined, since $\cos \tilde{\Omega} > 0$ on account of (\ref{eq:FLtildeOmegaConstraint}).  In terms of $\tilde{\tau}$, (\ref{eq:FLChiEvolutionB}) reads
\begin{equation} \label{eq:FLChiEvolution}
\frac{d\chi_i}{d\tilde{\tau}} = [2 \tilde{q} - (3 \gamma_i - 2)] (\sin \tilde{\Omega}) \chi_i.
\end{equation}  In order to obtain the evolution equation for $\tilde{\Omega}$ we need the evolution equations for $\mu$ and $H$.  First, it follows from (\ref{eq:FLTotalenergyEvolutionA}) and (\ref{eq:FLtildetime}) that
\begin{equation} \label{eq:FLmuevolutiontilde}
\frac{d\mu}{d\tilde{\tau}} = - 2 \sqrt{3 \mu} H (\tilde{q} + 1) \cos \tilde{\Omega}.
\end{equation}  Secondly, equations (\ref{eq:FLraychaudhuriH}), (\ref{eq:FLchi}), (\ref{eq:FLqtilde}), (\ref{eq:FLHubbleitoMuOmega}) and (\ref{eq:FLtildetime}) lead to
\begin{equation} \label{eq:FLHubbleEvolution}
\frac{dH}{d\tilde{\tau}} = - \sqrt{\frac{\mu}{3}} ( \tan^2 \tilde{\Omega} + \tilde{q} ) \cos \tilde{\Omega}.
\end{equation}  We differentiate (\ref{eq:FLHubbleitoMuOmega}) with respect to $\tilde{\tau}$ and use (\ref{eq:FLmuevolutiontilde}).  After simplifying using (\ref{eq:FLHubbleitoMuOmega}), we obtain
\begin{equation} \label{eq:FLtildeOmegaEvolution}
\frac{d \tilde{\Omega}}{d \tilde{\tau}} = - \tilde{q} \cos 2 \tilde{\Omega} \cos \tilde{\Omega}.
\end{equation}

Equations (\ref{eq:FLChiEvolution}) and (\ref{eq:FLtildeOmegaEvolution}) are the desired evolution equations.  The state space is the bounded $n$-dimensional subset of $\mathbb{R}^{n+1}$ described by \mbox{($\tilde{\Omega}$, $\chi_1$, \ldots, $\chi_n$)} and subject to the restrictions (\ref{eq:FLchirestriction}) and (\ref{eq:FLtildeOmegaConstraint}).  The state space is not compact due to the strict inequality in (\ref{eq:FLtildeOmegaConstraint}) and so we seek to compactify this space, since then we can apply theorems from the theory of dynamical systems.

In order to compactify the space, we simply observe that the right hand sides of the ODEs (\ref{eq:FLChiEvolution}) and (\ref{eq:FLtildeOmegaEvolution}) are well-defined, in fact analytic, at the points $\tilde{\Omega} = \pm \tfrac{\pi}{2}$, which form part of the boundary of the state space.  We can thus extend the state space to include the points with $\tilde{\Omega} = \pm \tfrac{\pi}{2}$, thereby compactifying it.  Since
\begin{equation}
\lim_{\vert \tilde{\Omega} \vert \to \frac{\pi}{2}^{-}} \mu = 0,
\end{equation} as follows from (\ref{eq:FLHubbleitoMuOmega}), we interpret the sets $\tilde{\Omega} = \pm \frac{\pi}{2}$ as representing FL cosmologies with $\mu = 0$. It follows from (\ref{eq:FLfriedmann}) that the set $\tilde{\Omega} = \pm \frac{\pi}{2}$ represents the Milne universe, given by (\ref{eq:FLsolMilne}).

\vspace{1cm}

\noindent \begin{tabular}{|p{\textwidth}|}
\hline \ \\
\textbf{State Space Representation of $n$-Fluid FL Cosmologies} \\ \ \\
\textbf{\ \ \ State Space:} \hspace{3.7cm} $(\tilde{\Omega}, \chi_1, \ldots, \chi_n) \in \mathbb{R}^{n+1}$ \\
\begin{equation}
\label{eq:T_FLcompactstatespace} \sum_{i=1}^{n} \chi_i = 1, \quad 0 \leq \chi_i \leq 1, \quad - \tfrac{\pi}{2} \leq \tilde{\Omega} \leq \tfrac{\pi}{2}.
\end{equation} \\
\textbf{\ \ \ Evolution Equations:}
\begin{eqnarray}
\label{eq:T_FLtildeOmegaEvolution} \frac{d \tilde{\Omega}}{d \tilde{\tau}} & = & - \tilde{q} \cos 2 \tilde{\Omega} \cos \tilde{\Omega}, \\
\label{eq:T_FLChiEvolution} \frac{d \chi_i}{d \tilde{\tau}} & = & \left[ 2 \tilde{q} - (3 \gamma_i - 2) \right] (\sin \tilde{\Omega}) \chi_i,
\end{eqnarray}
\qquad with
\begin{equation} \label{eq:T_tildeQ}
\tilde{q} = \tfrac{1}{2} \sum_{i=1}^{n} (3 \gamma_i - 2) \chi_i.
\end{equation} \\
\hline
\end{tabular}

\paragraph{Comments:}
\begin{enumerate}
\item[i)]  The ODEs (\ref{eq:T_FLtildeOmegaEvolution}) and (\ref{eq:T_FLChiEvolution}) describe the essential dynamics of $n$-fluid FL cosmologies in terms of the dimensionless variables \mbox{$(\tilde{\Omega}, \chi_1, \ldots, \chi_n)$} and the dimensionless time $\tilde{\tau}$.  Since the state space, as defined by (\ref{eq:T_FLcompactstatespace}) is compact, \textit{the solutions of the ODEs are defined for all $\tilde{\tau} \in \mathbb{R}$}, and as a result the ODEs define a dynamical system on the state space (see, for example, Wainwright and Ellis (1997), p87, Corollary 4.1).  The evolution of an $n$-fluid FL cosmology is thus described by an orbit of the dynamical system (a solution curve of the ODEs).  As a consequence, we are guaranteed the existence of a past attractor $\mathcal{A}^{-}$ of the ODEs (\textit{i.e.} an invariant set to which all orbits, except possibly a set of measure zero, are past asymptotic to as $\tilde{\tau} \to -\infty$) and a future attractor $\mathcal{A}^{+}$ (\textit{i.e.} as a past attractor, except as $\tilde{\tau} \to +\infty$).  The past attractor describes the asymptotic regime at early times and the future attractor describes the asymptotic regime at late times.

\item[ii)]  This description of the evolution of $n$-fluid FL cosmologies as a dynamical system on a compact state space is new.  It generalizes the formulation given by Wainwright (1996) for the case $n = 2$.

\item[iii)]  The physical state of an $n$-fluid FL cosmology is determined by the variables \mbox{$(H, \mu_1, \ldots, \mu_n)$,} which are related to the dimensionless variables via the total density $\mu$, according to
\begin{equation} \label{eq:compactstatespacerelations}
H = \sqrt{\frac{\mu}{3}} \tan \tilde{\Omega}, \qquad \mu_i = \mu \chi_i,
\end{equation} (see (\ref{eq:FLHubbleitoMuOmega}) and (\ref{eq:FLchi})).  On using (\ref{eq:compactstatespacerelations}) to eliminate $H$, (\ref{eq:FLmuevolutiontilde}) assumes the form
\begin{equation} \label{eq:FLmuevolutiontilde2}
\frac{d \mu}{d \tilde{\tau}} = - 2 (\tilde{q} + 1) (\sin \tilde{\Omega}) \mu.
\end{equation}  Then (\ref{eq:FLmuevolutiontilde2}) determines $\mu$ as an integral in terms of $\tilde{\Omega}(\tilde{\tau})$ and $\chi_i(\tilde{\tau})$, up to an arbitrary constant.  This freedom implies that \textit{each orbit determines a 1-parameter family of cosmological models}, whose physical variables are related by a multiplicative constant.  Since an $n$-dimensional dynamical system has an $(n-1)$-parameter family of orbits, it follows that \textit{the class of $n$-fluid FL cosmologies is labelled by $n$ parameters}.  We will specify these parameters explicitly in section \ref{sec:FLParameterization}.

\item[iv)]  Clock time $t$ is related to $\tilde{\tau}$ via the ODE (\ref{eq:FLtildetime}) and can thus be expressed as an integral involving $\tilde{\Omega}(\tilde{\tau})$ and $\mu(\tilde{\tau})$, the latter determined from (\ref{eq:FLmuevolutiontilde2}).  Clock time will take on values in an interval
\begin{equation}
t_i < t < t_f,
\end{equation} where $t_i$ is finite or $- \infty$ and $t_f$ is finite or $+ \infty$, depending on the model.

\item[v)]  The dimensionless variables $\tilde{\Omega}$, $\chi_i$ and $\tilde{q}$ are closely related to the Hubble-normalized scalars $\Omega_i$ and $q$ when $H \neq 0$.  From (\ref{eq:FLtotaldensityparameter}) and (\ref{eq:FLtildeOmega}) we have
\begin{eqnarray} \label{eq:FLOmegaTildeOmegaRelation}
\tan^2 \tilde{\Omega} = \frac{1}{\Omega}, \quad - \tfrac{\pi}{2} < \tilde{\Omega} < \tfrac{\pi}{2}, \quad \tilde{\Omega} \neq 0.
\end{eqnarray}  Similarly, from (\ref{eq:FLdensityparameter}), (\ref{eq:FLtotaldensityparameter}) and (\ref{eq:FLchi}) we have
\begin{equation} \label{eq:FLChiOmegaRelation}
\chi_i = \frac{\Omega_i}{\Omega}.
\end{equation}  Finally, from (\ref{eq:qomegarelation}), (\ref{eq:FLqtilde}) and (\ref{eq:FLChiOmegaRelation}) we have
\begin{equation} \label{eq:FLQTildeQRelation}
\tilde{q} = \frac{q}{\Omega}.
\end{equation}

\end{enumerate}


\subsection{Invariant Sets and Equilibrium Points in the Compact State Space} \label{ssec:FLCompactifiedStructure}

Our goal in the rest of this section is to describe the qualitative behaviour of the orbits in the compact state space, thereby understanding the dynamics of $n$-fluid FL cosmologies.  Since the state space is compact, it follows that every orbit has a past and future attractor (see, for example, Wainwright and Ellis (1997), p99-100).

In general the attractors of a dynamical system are complicated invariant sets.  But in the present situation we will show that they are determined by the equilibrium points of the dynamical system (section \ref{ssec:FLAsymptoticBehaviour}), which we now present.

The form of the ODEs (\ref{eq:T_FLtildeOmegaEvolution}) and (\ref{eq:T_FLChiEvolution}) determines three physically important invariant sets, given as follows:

\begin{enumerate}
\item[1)] \textit{Flat FL set.}  This set is the $n-1$ dimensional subset given by $\tilde{\Omega} = \pm \tfrac{\pi}{4}$ (which corresponds to $\Omega = 1$, $\Omega_k = 0$ on account of (\ref{eq:FLOmegaTildeOmegaRelation}) and (\ref{eq:FLomegarelation})).  The orbits in this set describe flat $n$-fluid FL cosmologies, expanding if $\tilde{\Omega} = \tfrac{\pi}{4}$ and contracting if $\tilde{\Omega} = - \tfrac{\pi}{4}$.

\item[2)] \textit{Milne set.}  This set is the $n-1$ dimensional subset given by $\tilde{\Omega} = \pm \tfrac{\pi}{2}$ (which corresponds to $\Omega = 0$ and $\Omega_k = 1$ on account of (\ref{eq:FLOmegaTildeOmegaRelation}) and (\ref{eq:FLomegarelation})).  The orbits in this set describe the Milne vacuum solution.

\item[3)] \textit{$m$-fluid sets ($1 \leq m < n$).}  These sets are determined by setting $n-m$ of the $\chi_i$ variables to be zero, with the $m$ remaining $\chi_i$ variables being non-zero.  The orbits in these sets describe $m$-fluid FL cosmologies.
\end{enumerate}

\noindent The equilibrium points are determined by setting the right hand side of equations (\ref{eq:T_FLtildeOmegaEvolution}) and (\ref{eq:T_FLChiEvolution}) to zero.  If $\tilde{\Omega} \neq 0$, then (\ref{eq:T_FLChiEvolution}) and (\ref{eq:T_tildeQ}) implies $\chi_i = 1$, for one value of $i$, $\chi_j = 0$, for $j \neq i$, and \mbox{$\tilde{q} = \tfrac{1}{2} (3 \gamma_i - 2) \neq 0$}.  Then (\ref{eq:T_FLtildeOmegaEvolution}) implies \mbox{$\cos \tilde{\Omega} = 0$} (\textit{i.e.} \mbox{$\tilde{\Omega} = \pm \tfrac{\pi}{2}$}) or \mbox{$\cos 2 \tilde{\Omega} = 0$} (\textit{i.e.} \mbox{$\tilde{\Omega} = \pm \tfrac{\pi}{4}$}).  If $\tilde{\Omega} = 0$ then (\ref{eq:T_FLtildeOmegaEvolution}) implies $\tilde{q} = 0$, with $\tilde{q}$ given by (\ref{eq:T_tildeQ}).  In this way, we obtain the following sets of equilibrium points:

\begin{enumerate}
\item[1)] \textit{Flat FL points ($F_i^{\pm}$)} are given by
\begin{equation} \label{eq:FlatFLEquilibriumPoints}
\tilde{\Omega} = \pm \pi / 4, \qquad \chi_i = 1, \qquad \chi_j = 0\ \mathrm{if}\ j \neq i.
\end{equation}  These equilibrium points lie in the intersection of the flat FL set and the various 1-fluid sets.

\item[2)] \textit{Milne points ($M_i^{\pm}$)} are given by
\begin{equation} \label{eq:MilneEquilibriumPoints}
\tilde{\Omega} = \pm \tfrac{\pi}{2}, \qquad \chi_i = 1, \chi_j = 0\ \mathrm{if}\ j \neq i.
\end{equation}  These equilibrium points lie in the intersection of the Milne set and the various 1-fluid sets.

\item[3)] \textit{Einstein static points ($E$)} are given by
\begin{equation} \label{eq:FLEinsteinStaticConstriant}
\tilde{\Omega} = 0, \qquad \sum_{i=1}^{n} (3 \gamma_i - 2) \chi_i = 0,
\end{equation} and so constitute an ($n-2$)-dimensional set of equilibrium points.  These equilibrium points correspond to the $n$-fluid interpretation of the Einstein static model (\ref{eq:FLsolEinsteinStatic}) and only exist when $\gamma_n < 2/3$.
\end{enumerate}

\subsection{Classification of $n$-Fluid FL Cosmologies} \label{ssec:FLClassificationTheorem}

In this section, we will prove that the $n$-fluid FL cosmologies that have an epoch of expansion (\textit{i.e.} $H > 0$ in some time interval) form three qualitatively different generic\footnote{Invariant sets of dimension $n$.} subclasses, defined as follows:

\begin{enumerate}
\item[1)] \textit{Ever-expanding models} satisfy $H > 0$ for all $t$.

\item[2)] \textit{Recollapsing models} satisfy $H(t_{\ast}) = 0$, $H > 0$ for $t < t_{\ast}$ and $H < 0$ for $t > t_{\ast}$ for some $t_{\ast}$.

\item[3)] \textit{Bouncing models} satisfy $H(t_{\ast}) = 0$, $H < 0$ for $t < t_{\ast}$ and $H > 0$ for $t > t_{\ast}$ for some $t_{\ast}$.
\end{enumerate}

The variables $\tilde{\Omega}$ and $\tilde{q}$ are particularly important in this analysis.  We begin with the following proposition on the behaviour of $\tilde{q}$.

\bigskip

\begin{proposition} \label{prop:FLQProp} In any $n$-fluid FL cosmology with $n > 1$, $\tilde{q}$ is strictly decreasing (increasing) in an epoch of expansion (contraction).  Further, if $n = 1$ then $\tilde{q}$ is constant.
\end{proposition}

\paragraph{Proof:}  By differentiating (\ref{eq:FLqtilde}) with respect to $\tilde{\tau}$ and using (\ref{eq:FLChiEvolution}), we obtain\footnote{This equation is equivalent to (14.15) in Wainwright and Ellis (1997), since $\tilde{q} = \tfrac{1}{2} (3 w + 1)$.  The proposition is equivalent to $c_s^2 - w > 0$, if $n > 1$.}
\begin{equation} \label{eq:QFLfrac1}
\frac{d \tilde{q}}{d \tilde{\tau}} = \tfrac{1}{2} \left[ 4 \tilde{q}^2 - \sum_{i=1}^{n} (3 \gamma_i - 2)^2 \chi_i \right] \sin \tilde{\Omega}.
\end{equation}  By the Cauchy-Schwarz inequality and (\ref{eq:FLchirestriction}) we have
\begin{equation} \label{eq:QFLfrac2}
4 \tilde{q}^2 = \left[ \sum_{i=1}^{n} (3 \gamma_i - 2) \chi_i \right]^2 \leq \left[ \sum_{i=1}^{n} (3 \gamma_i - 2)^2 \chi_i \right] \left[ \sum_{i=1}^{n} \chi_i \right] = \sum_{i=1}^{n} (3 \gamma_i - 2)^2 \chi_i,
\end{equation} with equality holding if and only if $n = 1$. \footnote{Equality holds in (\ref{eq:QFLfrac2}) if and only if there exists $c$ such that
\begin{displaymath}
(3 \gamma_i - 2) \sqrt{\chi_i} = c \sqrt{\chi_i}, \qquad i = 1, 2, \ldots, n,
\end{displaymath} which is possible if and only if $n = 1$.}

On recalling (\ref{eq:FLtildeOmegaHrelation}), it follows that
\begin{equation}
\mathrm{sign} \left( \frac{d\tilde{q}}{d\tilde{\tau}} \right) = - \mathrm{sign} ( H ),
\end{equation} which establishes the result.  $\square$

\bigskip

This proposition leads to a second result, presented here:

\begin{proposition} \label{prop:FLNoCross} For any non-static $n$-fluid FL cosmology, $\tilde{\Omega}(\tilde{\tau}_\ast) = 0$ is satisfied for at most one $\tilde{\tau}_\ast$, and in this situation $\frac{d\tilde{\Omega}}{d\tilde{\tau}} (\tilde{\tau}_{\ast}) \neq 0$.
\end{proposition}

\paragraph{Proof:}  We first show, by contradiction, that $\tilde{\Omega}(\tilde{\tau}_\ast) = 0$ is satisfied for at most one $\tilde{\tau}_\ast$.  Suppose \mbox{$\tilde{\Omega}(\tilde{\tau}_1) = 0 = \tilde{\Omega}(\tilde{\tau}_2)$} and $\tilde{\Omega} \neq 0$ for $\tilde{\tau} \in (\tilde{\tau}_1, \tilde{\tau}_2)$.  Without loss of generality, we assume that $\tilde{\Omega} > 0$ for $\tilde{\tau} \in (\tilde{\tau}_1, \tilde{\tau}_2)$.  Then by proposition \ref{prop:FLQProp} it follows that $\tilde{q}(\tilde{\tau}_2) \leq \tilde{q}(\tilde{\tau}_1)$ since $\tilde{q}$ is strictly decreasing whenever $\tilde{\Omega} > 0$.  Further, by (\ref{eq:FLtildeOmegaEvolution}), we have that $\tilde{\Omega}$ satisfies
\begin{equation}
\tilde{\Omega}^{\prime}(\tilde{\tau}_1) = - \tilde{q}(\tilde{\tau}_1), \quad \mbox{and} \quad \tilde{\Omega}^{\prime}(\tilde{\tau}_2) = - \tilde{q}(\tilde{\tau}_2),
\end{equation} where the prime ($\prime$) denotes differentiation with respect to $\tilde{\tau}$.  But $\tilde{\Omega}(\tilde{\tau}) > 0$ for $\tilde{\tau} \in (\tilde{\tau}_1, \tilde{\tau}_2)$ implies that $\tilde{\Omega}^{\prime}(\tilde{\tau}_1) > 0$ (\textit{i.e.} $\tilde{q}(\tilde{\tau}_1) < 0$) and $\tilde{\Omega}^{\prime}(\tilde{\tau}_2) < 0$ (\textit{i.e.} $\tilde{q}(\tilde{\tau}_2) > 0$) and so $\tilde{q}(\tilde{\tau}_2) > \tilde{q}(\tilde{\tau}_1)$, which is a contradiction.

Finally, it follows from (\ref{eq:T_FLtildeOmegaEvolution}) that if $\tilde{\Omega} = 0$, $\frac{d\tilde{\Omega}}{d\tilde{\tau}} = 0$ is satisfied only when $\tilde{q} = 0$, which is disallowed for non-static models.\ $\square$


\bigskip

The main result of this section now follows immediately.

\begin{theorem} \label{thm:NFluidClassification1} There are exactly three generic classes of $n$-fluid FL cosmologies ($n > 1$) that have an expanding epoch, namely ever-expanding models, recollapsing models and bouncing models.  Bouncing models occur if and only if $\gamma_n < \tfrac{2}{3}$.
\end{theorem}

\paragraph{Proof:}  Since $H = 0$ if and only if $\tilde{\Omega} = 0$ on account of (\ref{eq:FLtildeOmega}), proposition \ref{prop:FLNoCross} states that the Hubble scalar can be zero at most once, and that $H$ changes sign if it becomes zero.  Thus models with more than one expanding epoch are excluded, which implies that the only possible classes are those given.  If $\gamma_n > \tfrac{2}{3}$, it follows from (\ref{eq:FLgammaordering}) and (\ref{eq:FLqtilde}) that \mbox{$\tilde{q} > \tfrac{1}{2} (3 \gamma_n - 2) > 0$}, and hence from (\ref{eq:T_FLtildeOmegaEvolution}) that bouncing models are disallowed.\ $\square$

To the best of the author's knowledge, the classification theorem is new for $n > 2$, although it is a well known result for $n = 2$, when the matter content is a perfect fluid ($\gamma_1 > \tfrac{2}{3}$) and a cosmological constant ($\gamma_2 = 0$) (see Rindler (1977), p234-8).

\subsection{Asymptotic Behaviour of Solutions} \label{ssec:FLAsymptoticBehaviour}

In this section we determine the asymptotic behaviour of the three generic classes of $n$-fluid FL cosmologies identified in the previous section.  We will first describe the behaviour of the length scale $\ell(t)$ as a function of clock time $t$ in each of the three generic classes of FL cosmologies, and then use these results to determine the past attractor and future attractor.

Important in this analysis is the set of orbits asymptotic to an Einstein static solution.  In the following proposition, we demonstrate that orbits of this type are non-generic.

\begin{proposition} \label{prop:FLEinsteinStaticDim}  The set of orbits in the compact state space that are asymptotic to an Einstein static equilibrium point has dimension $n-1$.
\end{proposition}

\paragraph{Proof:}  We can write the evolution equations (\ref{eq:FLChiEvolution}) and (\ref{eq:FLtildeOmegaEvolution}) as a dynamical system of the form
\begin{equation}
\frac{dX_i}{d\tilde{\tau}} = F_i(X_0, X_1, \ldots, X_n), \qquad i = 0, \ldots, n,
\end{equation} on $\mathbb{R}^{n+1}$, where $X_0 = \tilde{\Omega}$ and $X_j = \chi_j$ ($j = 1, \ldots, n$).  The physical state space is the hyperplane given by (\ref{eq:T_FLcompactstatespace}), and hence is $n$ dimensional.  Consider the Jacobian matrix $J_{ij} = \frac{\partial F_j}{\partial X_i}$.  Upon evaluating $J_{ij}$ on the Einstein static set, given by (\ref{eq:FLEinsteinStaticConstriant}), we obtain
\begin{equation}
J_{00} = 0, \quad J_{0i} = -\tfrac{1}{2} (3 \gamma_i - 2), \quad J_{i0} = -(3 \gamma_i - 2) \chi_{i,E}, \quad J_{ij} = 0,
\end{equation} where $i,j = 1, \ldots, n$ and $\chi_{i,E}$ denotes the value of $\chi_i$ at the Einstein equilibrium point.  The Jacobian matrix, when restricted to the physical state space, has two non-zero eigenvalues
\begin{equation}
\lambda_{\pm} = \pm \sqrt{\tfrac{1}{2} \sum_{i=1}^{n} (3 \gamma_i - 2)^2 \chi_i} > 0,
\end{equation} and $n-2$ zero eigenvalues.\footnote{The zero eigenvalues reflect the fact that the set of Einstein static equilibrium points has dimension $n-2$.}  It follows\footnote{Using the Stable/Unstable Manifold Theorem.  See, for example, Perko (1996, p107).} that each Einstein static equilibrium point has a 1-dimensional stable manifold and a 1-dimensional unstable manifold.  Thus the set of orbits future asymptotic to the Einstein static set has dimension exactly one larger than the dimension of the Einstein static manifold, \textit{i.e.} dimension $n-1$. $\square$

\bigskip

In the next proposition, we establish the behaviour of the length scale $\ell$ as a function of time $\tilde{\tau}$, for ever-expanding models.

\bigskip

\begin{proposition} \label{prop:DimensionlessTimeTildeTime}  If $\tilde{\Omega} > 0$ for all $\tilde{\tau} \in \mathbb{R}$ and $\tilde{\Omega} \not\to 0$ as $\tilde{\tau} \to \pm \infty$, then
\begin{equation}
\lim_{\tilde{\tau} \to - \infty} \ell(\tilde{\tau}) = -\infty, \quad \mbox{and} \quad \lim_{\tilde{\tau} \to +\infty} \ell(\tilde{\tau}) = +\infty.
\end{equation}
\end{proposition}

\paragraph{Proof:}  The length scale $\ell$ defines a time variable $\tau$ according to
\begin{equation} \label{eq:dimensionlesstime}
\ell = \ell_0 e^{\tau}.
\end{equation}  It follows from (\ref{eq:hubblelengthscale}) and (\ref{eq:dimensionlesstime}) that
\begin{equation} \label{eq:dimensionlesstimeH}
\frac{dt}{d\tau} = \frac{1}{H}.
\end{equation}  Then, using (\ref{eq:FLHubbleitoMuOmega}), (\ref{eq:FLtildetime}) and (\ref{eq:dimensionlesstimeH}), and the chain rule, we obtain
\begin{equation} \label{eq:dimensionlesstimetildetau}
\frac{d\tau}{d\tilde{\tau}} = \sin \tilde{\Omega}.
\end{equation}  Since $\tilde{\Omega} > 0$ for all $\tilde{\tau} \in \mathbb{R}$ and $\tilde{\Omega} \not\to 0$ as $\tilde{\tau} \to \pm \infty$, (\ref{eq:dimensionlesstimetildetau}) then implies
\begin{equation}
\lim_{\tilde{\tau} \to - \infty} \tau(\tilde{\tau}) = -\infty, \quad \mbox{and} \quad \lim_{\tilde{\tau} \to +\infty} \tau(\tilde{\tau}) = +\infty.
\end{equation}  This completes the proof, on applying (\ref{eq:dimensionlesstime}).\ $\square$

\bigskip

\begin{proposition} \label{prop:RecollapsingBouncingQ} For a recollapsing model, $\tilde{q}(\tilde{\tau}) > 0$ for all $\tilde{\tau}$, and for a bouncing model, $\tilde{q}(\tilde{\tau}) < 0$ for all $\tilde{\tau}$.
\end{proposition}

\paragraph{Proof:}  For a recollapsing model, it follows from the definition that there exists $\tilde{\tau}_{\ast}$ so that $\tilde{\Omega}(\tilde{\tau}_{\ast}) = 0$ and that $\tilde{\Omega}(\tilde{\tau})$ changes sign from positive to negative at $\tilde{\tau}_{\ast}$, which implies $\frac{d\tilde{\Omega}}{d\tilde{\tau}}(\tilde{\tau}_{\ast}) < 0$ (strict inequality follows from Proposition \ref{prop:FLNoCross}).  The evolution equation (\ref{eq:T_FLtildeOmegaEvolution}) now implies $q(\tilde{\tau}_{\ast}) > 0$.  By Proposition \ref{prop:FLQProp}, $\tilde{q}$ has a global minimum at $\tilde{\tau}_{\ast}$, \textit{i.e.} \mbox{$\tilde{q}(\tilde{\tau}) \geq \tilde{q}(\tilde{\tau}_{\ast})$} for all $\tilde{\tau}$, which gives the desired result.  The proof for bouncing models is similar.\ $\square$

\bigskip

We now determine the asymptotic behaviour of clock time $t$ (\textit{i.e.} as \mbox{$\tilde{\tau} \to \pm \infty$}).  Let
\begin{equation}
t_i = \lim_{\tilde{\tau} \to -\infty} t(\tilde{\tau}), \qquad t_f = \lim_{\tilde{\tau} \to +\infty} t(\tilde{\tau}),
\end{equation} where the dependence of $t$ on $\tilde{\tau}$ is determined via equation (\ref{eq:FLtildetime}).  Then $t_i$ and $t_f$ are finite or infinite depending on the type of universe and during the evolution $t_i < t < t_f$.  Clock time is related to the length scale via the Friedmann equation.  In particular, for any $n$-fluid FL cosmology, we can integrate (\ref{eq:FLconservationNFluid}) to obtain\footnote{See p46.}
\begin{equation} \label{eq:FLConservationIntegral}
\mu_i \ell^2 = 3 \left( \lambda_i \ell \right)^{-3 \gamma_i + 2}.
\end{equation} where $\lambda_i$ is an integration constant whose inverse can be interpreted as the characteristic length scale of fluid $i$.  On substituting (\ref{eq:FLConservationIntegral}) into the Friedmann equation (\ref{eq:FLfriedmannNFluid}), we obtain
\begin{equation} \label{eq:FLFriedmannTheoremForm}
\left( \frac{d\ell}{dt} \right)^2 = \sum_{i=1}^{n} \left( \lambda_i \ell \right)^{-3 \gamma_i + 2} - K.
\end{equation}



\begin{theorem} \label{thm:FLSingularityThm}  The following statements are true:
\begin{enumerate}
\item[i)] For an ever-expanding $n$-fluid FL model that is not asymptotic to the Einstein static universe, $t_i$ is finite, $t_f = \infty$ and the length scale satisfies \mbox{$\ell(t_i) = 0$} for some $t_i$ and \mbox{$\ell \to + \infty$} as \mbox{$t \to +\infty$}.
\item[ii)] For a recollapsing $n$-fluid FL model, $t_i$ and $t_f$ are finite, \mbox{$\ell(t_i) = 0$} and \mbox{$\ell(t_f) = 0$}.
\item[iii)] For a bouncing $n$-fluid FL model, $t_i$ and $t_f$ are infinite and $\ell(t)$ satisfies \mbox{$\ell(t) \to +\infty$} as \mbox{$t \to \pm \infty$}.
\end{enumerate}
\end{theorem}

\paragraph{Proof:  Case i)}  In this case, we will make use of the following elementary results from calculus:
\begin{enumerate}
\item[(I)] If $f(a) = O(a^r)$ as $a \to 0$, and $r > -1$, then $\displaystyle \lim_{a \to 0} \int_a^1 f(\tilde{a}) d\tilde{a} < \infty$, \textit{i.e.} the integral converges.
\item[(II)] If $f(a) \in O(a^r)$, as $a \to \infty$, and $r \geq -1$ then $\displaystyle \lim_{a \to \infty} \int_1^a f(\tilde{a}) d\tilde{a} = \infty$, \textit{i.e.} the integral diverges.
\end{enumerate}

Recall that ever-expanding models satisfy $H(\tilde{\tau}) > 0\ \forall\ \tilde{\tau}$.  Thus by Proposition \ref{prop:DimensionlessTimeTildeTime}, solutions which are not asymptotic to the Einstein static solution satisfy $\ell \to 0$ as $\tilde{\tau} \to - \infty$ and $\ell \to \infty$ as $\tilde{\tau} \to \infty$.  Let
\begin{equation} \label{eq:DenomF}
f(\ell) = \frac{1}{\sqrt{\sum_{i=1}^{n} \left( \lambda_i \ell \right)^{-3 \gamma_i + 2} - K}}.
\end{equation}  For an ever-expanding model, this function is defined and continuous (for $0 < a < \infty$) since $H > 0$ implies that the denominator is strictly positive (see (\ref{eq:FLFriedmannTheoremForm})).  Hence, we can integrate (\ref{eq:FLFriedmannTheoremForm}) to obtain
\begin{equation} \label{eq:FLFriedmannIntegral}
t(\ell) = t(1) - \int_\ell^1 f(\tilde{\ell}) d\tilde{\ell}, \quad \mbox{for $0 < \ell < 1$}.
\end{equation}  Further, the integrand satisfies
\begin{equation}
f(\ell) = O\left( \ell^{\frac{3 \gamma_1 - 2}{2}} \right), \quad \mbox{as $\ell \to 0$.}
\end{equation}  Since $\gamma_1 > \tfrac{2}{3}$ it follows from result (I) that there exists some finite time $t_i$, defined by $t_i = \lim_{\ell \to 0} t(\ell)$, such that $\ell(t_i) = 0$.

Similarly, we can integrate (\ref{eq:FLFriedmannTheoremForm}) to obtain
\begin{equation} \label{eq:FriedmannIntegral2}
t(\ell) = t(1) + \int_1^\ell f(\tilde{\ell}) d\tilde{\ell}, \quad \mbox{for $\ell > 1$}.
\end{equation}  Further, the integrand satisfies
\begin{equation}
f(a) = O\left( \ell^{\frac{3 \gamma_n - 2}{2}} \right), \quad \mbox{as $\ell \to \infty$}.
\end{equation}  Since $\gamma_n \geq 0$ it follows from result (II) $t \to \infty$ as $\ell \to \infty$.

For the following two cases we make use of the fact that $\tilde{q}$ and $\ddot{\ell}$ are related by
\begin{equation} \label{eq:FLsignQRelation}
\frac{\ddot{\ell}}{\ell} = - \tfrac{1}{3} \mu \tilde{q},
\end{equation} as follows from (\ref{eq:FLraychaudhuriNFluid}), (\ref{eq:FLchi}) and (\ref{eq:FLqtilde}).

\paragraph{Case ii)}  For a recollapsing model, Proposition \ref{prop:RecollapsingBouncingQ} implies $\tilde{q}(\tilde{\tau}) > 0$ for all $\tilde{\tau} \in \mathbb{R}$.  Then (\ref{eq:FLsignQRelation}) implies that $\ell(t)$ is concave down on its domain and so we can conclude there exists $t_i$ and $t_f$ such that \mbox{$\ell(t_i) = 0$} and \mbox{$\ell(t_f) = 0$}.

\paragraph{Case iii)} For a bouncing model, Proposition \ref{prop:RecollapsingBouncingQ} implies $\tilde{q}(\tilde{\tau}) < 0$ for all $\tilde{\tau} \in \mathbb{R}$.  Then (\ref{eq:FLsignQRelation}) implies that $\ell(t)$ is concave up on its domain and hence $\ell \to \infty$ as $\tilde{\tau} \to \pm \infty$.  Since $\mathrm{sign}(\dot{\ell}) = \mathrm{sign}(H)$, it follows that $\ell$ is minimal at $\tilde{\tau}_\ast$.  If we choose $\ell_{min} = \ell(\tilde{\tau}_{\ast})$, we have that $\ell(\tilde{\tau}) \geq \ell_{min} > 0$ for all $\tilde{\tau} \in \mathbb{R}$.

We must now show that $\ell$ does not go to infinity in finite time.  Choose $\ell_0$ so that $H > 0$ for $\ell \geq \ell_0$.  Then $f(\ell)$ in (\ref{eq:DenomF}) is defined and continuous for $\ell \geq 1$.  We can integrate (\ref{eq:FLFriedmannTheoremForm}) from $\ell = 1$ to obtain (\ref{eq:FriedmannIntegral2}).  Since the integrand satisfies
\begin{equation}
f(a) = O\left( \ell^{\frac{3 \gamma_n - 2}{2}} \right), \quad \mbox{as $a \to \infty$},
\end{equation} and $\gamma_n \geq 0$, it follows from result (II) that (\ref{eq:FriedmannIntegral2}) diverges, and so these models satisfy $\ell \to \infty$ as $t \to \infty$.  By symmetry, we have that $\ell \to \infty$ as $t \to - \infty$.\ $\square$

\bigskip

The past and future attractors of the dynamical system on the compact state space can now be determined using theorem \ref{thm:FLSingularityThm}.

\begin{theorem} \label{thm:FLasymptoticbehaviour} The past and future attractors in the state space of $n$-fluid FL cosmologies with an expanding epoch are
\begin{equation}
\mathcal{A}^{-} = F_1^{+} \cup F_n^{-},\ \mbox{and}\ \mathcal{A}^{+} = F_1^{-} \cup F_n^{+}, \quad \mbox{for $\gamma_n < \tfrac{2}{3}$,}
\end{equation} and
\begin{equation}
\mathcal{A}^{-} = F_1^{+},\ \mbox{and}\ \mathcal{A}^{+} = F_1^{-} \cup M_{n}^{+} \quad \mbox{for $\gamma_n > \tfrac{2}{3}$,}.
\end{equation}
\end{theorem}

\paragraph{Proof:}  We begin by obtaining algebraic expressions for $\chi_i$ and $\tilde{\Omega}$.  From (\ref{eq:FLConservationIntegral}) we have
\begin{equation} \label{eq:FLmuCArelation}
\mu_i \ell^2 = 3 \left( \lambda_i \ell \right)^{-3 \gamma_i + 2}.
\end{equation}  Then (\ref{eq:FLchi}), (\ref{eq:FLConservationIntegral}) and (\ref{eq:FLmuCArelation}) lead to
\begin{equation} \label{eq:FLalgebraicchiN}
\chi_i = \frac{(\lambda_i \ell)^{-3 \gamma_i + 2}}{\sum_{j=1}^{n} (\lambda_j \ell)^{-3 \gamma_j + 2}}
\end{equation}  Using (\ref{eq:FLtildeOmega}), we write the Friedmann equation (\ref{eq:FLfriedmannH}) in the form
\begin{equation} \label{eq:FLtildeOmegaitoKmuell}
1 - \tan^2 \tilde{\Omega} = \frac{3 K}{\mu \ell^2}.
\end{equation}  Then using (\ref{eq:FLchi}) and (\ref{eq:FLmuCArelation}) we have
\begin{equation} \label{eq:FLalgebraictildeOmega}
1 - \tan^2 \tilde{\Omega} = \frac{K}{\sum_{j=1}^{n} (\lambda_j \ell)^{-3 \gamma_j + 2}}.
\end{equation}

To determine the limits of (\ref{eq:FLalgebraicchiN}) and (\ref{eq:FLalgebraictildeOmega}) as $\ell \to 0$, we write
\begin{equation} \label{eq:FLalgebraicchi1}
\chi_1 = \frac{\lambda_1^{-3 \gamma_i + 2}}{\sum_{j=1}^{n} \lambda_j^{-3 \gamma_j + 2} \ell^{3 (\gamma_1 - \gamma_j)}.}
\end{equation} and, using (\ref{eq:FL2totalmattercontent}), (\ref{eq:FLmuCArelation}) and (\ref{eq:FLalgebraicchiN}),
\begin{equation}
1 - \tan^2 \tilde{\Omega} = K \chi_1 (\lambda_1 \ell)^{3 \gamma_1 - 2}.
\end{equation}  It follows that
\begin{equation}
\lim_{\ell \to 0^{+}} \chi_1 = 1, \quad \lim_{\ell \to 0^{+}} \tilde{\Omega} = \pm \tfrac{\pi}{4},
\end{equation} since $\gamma_1 - \gamma_j > 0$ and $\gamma_1 > \tfrac{2}{3}$.

Similarly, by writing
\begin{equation}
\chi_n = \frac{\lambda_n^{-3 \gamma_n + 2}}{\sum_{j=1}^{n} \lambda_j^{-3 \gamma_j + 2} \ell^{3 (\gamma_n - \gamma_j)}},
\end{equation} and
\begin{equation}
1 - \tan^2 \tilde{\Omega} = K \chi_n (\lambda_n \ell)^{3 \gamma_n - 2},
\end{equation} we obtain
\begin{equation}
\lim_{\ell \to + \infty} \chi_n = 1, \quad \lim_{\ell \to + \infty} \tilde{\Omega} = \left\{ \begin{array}{cl} 0, & \mbox{if $\gamma_n < \tfrac{2}{3}$} \\[0.3ex] \pm \tfrac{\pi}{2}, & \mbox{if $\gamma_n > \tfrac{2}{3}$ and $K = -1$}. \end{array} \right.
\end{equation}

It now follows from Theorem \ref{thm:FLSingularityThm} and equations (\ref{eq:FlatFLEquilibriumPoints}) and (\ref{eq:MilneEquilibriumPoints}) that orbits describing models in the three generic classes approach the equilibrium points given in table \ref{table:FLPastFutureAttractors}, as \mbox{$\tilde{\tau} \to - \infty$} and as \mbox{$\tilde{\tau} \to + \infty$}.\ $\square$

\begin{table}[p]
\begin{center}
\begin{tabular}{lcc}
\underline{Model Type} & \underline{$\tilde{\tau} \to - \infty$} & \underline{$\tilde{\tau} \to + \infty$} \\[0.6ex]
Ever-expanding models & & \\[0.3ex]
\quad $\gamma_n < \tfrac{2}{3}$ & $F_1^{+}$ & $F_n^{+}$ \\[0.3ex]
\quad $\gamma_n > \tfrac{2}{3}$, $K = -1$ & $F_1^{+}$ & $M_n^{+}$ \\[0.3ex]
Recollapsing models & $F_1^{+}$ & $F_1^{-}$ \\[0.3ex]
Bouncing models ($\gamma_n < \tfrac{2}{3}$ only) & $F_n^{-}$ & $F_n^{+}$
\end{tabular}
\end{center}
\caption{The past and future attractors of generic $n$-fluid FL cosmologies.} \label{table:FLPastFutureAttractors}
\end{table}

\paragraph{Comment:}  Since the orbits describing models in the three generic classes approach the equilibrium points in table \ref{table:FLPastFutureAttractors} as $\tilde{\tau} \to \pm \infty$, it follows that the density parameters $\Omega_i$ and $\Omega_k$ approach their values at the equilibrium points, which we give in table \ref{table:FLHubbleAsymptoticBehaviour}.

\begin{table}[p]
\begin{center}
\begin{tabular}{ccll}
\underline{Equilibrium Point} & \quad & \multicolumn{2}{l}{\underline{Behaviour \hspace{2.5in}}} \\[0.6ex]
$F_1^{\pm}$ & & $\Omega_i \to 0$ for $i \neq 1$, \quad $\Omega_1 \to 1$, & $\Omega_k \to 0$, \\[0.3ex]
$F_n^{\pm}$ & & $\Omega_i \to 0$ for $i \neq n$, \quad $\Omega_n \to 1$, & $\Omega_k \to 0$, \\[0.3ex]
$M_i^{\pm}$ & & $\Omega_i \to 0$ for $i = 1, \ldots, n$, & $\Omega_k \to 1$.
\end{tabular}
\end{center}
\caption{Asymptotic behaviour of Hubble-normalized quantities in generic $n$-fluid FL cosmologies.} \label{table:FLHubbleAsymptoticBehaviour}
\end{table}

\clearpage

\section{Parameterization of FL Models} \label{sec:FLParameterization}

In this section we introduce the $n$ essential parameters that label the family of $n$-fluid FL cosmologies, namely a conformal parameter $\lambda$ and $n-1$ dimensionless mass parameters $m_1, \ldots, m_{n-1}$, that we shall refer to as \textit{intrinsic parameters}.  In section \ref{ssec:ObservationalParameters}, we introduce an equivalent set of parameters, namely the observational parameters $\Omega_{i,0}$ and $H_0$ and show how they determine the intrinsic parameters.  Finally, we show that the evolution equations (\ref{eq:FLChiEvolution}) and (\ref{eq:FLtildeOmegaEvolution}) admit $n-1$ independent conserved quantities, which characterize the orbits in the $n$-dimensional state space.  In particular, we will show that the values of the conserved quantities are determined in terms of the intrinsic parameters.

\subsection{The Intrinsic Parameters} \label{ssec:FLIntrinsicParameters}

We now integrate the conservation equation and demonstrate that the constants of integration of this equation can be used to parameterize $n$-fluid FL cosmologies.  As we have seen previously in (\ref{eq:FLConservationIntegral}), on integrating (\ref{eq:FLconservationNFluid}), one obtains
\begin{equation} \label{eq:FLenergydensityNFluid}
\mu_i \ell^2 = 3 \left( \lambda_i \ell \right)^{-3 \gamma_i + 2}.
\end{equation}  The constants of integration $\lambda_i$ have dimension $(\mbox{length})^{-1}$ and so each can be thought of as defining a characteristic length scale for each fluid via $\lambda_i^{-1}$.  When (\ref{eq:FLenergydensityNFluid}) is substituted into (\ref{eq:FLfriedmannNFluid}), there results a first order ODE for $\ell(t)$:
\begin{equation} \label{eq:FLellDE}
\left( \frac{d\ell}{dt} \right)^2 = \sum_{i=1}^{n} \left( \lambda_i \ell \right)^{-3 \gamma_i + 2} - K.
\end{equation}

In order to define a dimensionless length scale and time variable for $n$-fluid FL cosmologies, one needs to choose a preferred $\lambda_i$ that will play the role of an overall conformal factor.  Typically we will choose $\lambda$ to be $\lambda_n$, which is a natural choice if $\gamma_n = 0$.  In this case, it follows from (\ref{eq:FLenergydensityNFluid}) that
\begin{equation} \label{eq:FLpreferredlambda}
\lambda^2 = \tfrac{1}{3} \Lambda.
\end{equation}
The dimensionless length $L$ and time $T$ are then defined as
\begin{equation} \label{eq:FLDimensionlessLengthTime}
L = \lambda \ell, \quad \mbox{and} \quad T = \lambda t.
\end{equation}  Equation (\ref{eq:FLenergydensityNFluid}) suggests defining a set of dimensionless parameters $m_1, \ldots, m_{n-1}$ via\footnote{For $i = n$, this equation states that $m_n = 1$.}
\begin{equation} \label{eq:FLM}
m_i = \left( \frac{\lambda_i}{\lambda} \right)^{-3 \gamma_i + 2},
\end{equation} so that
\begin{equation} \label{eq:FLmuitoL}
\mu_i = 3 \lambda^2 m_i L^{-3 \gamma_i}.
\end{equation}  In dimensionless form, the Friedmann equation (\ref{eq:FLellDE}) then reads
\begin{equation} \label{eq:FLDimlessLDE}
\left( \frac{dL}{dT} \right)^2 = \sum_{i=1}^{n} m_i L^{-3 \gamma_i + 2} - K.
\end{equation}  For any choice of parameters $m_i > 0$, $i = 1, \ldots, n-1$, along with the conditions
\begin{equation}
L(0) = 0, \quad \mbox{and} \quad \frac{dL}{dT} > 0,
\end{equation} the DE (\ref{eq:FLDimlessLDE}) has unique solution (as shown in section \ref{sec:FLexistenceuniqueness}), given by
\begin{equation} \label{eq:FLDimlessLFunction}
L = L(T, m_1, \ldots, m_{n-1}).
\end{equation}  It then follows from (\ref{eq:FLDimensionlessLengthTime}) that $\ell(t)$ can be written as
\begin{equation} \label{eq:FLGeneralLengthScale}
\ell(t) = \lambda^{-1} L(\lambda t, m_1, \ldots, m_{n-1}).
\end{equation}  The line element then takes the form
\begin{equation} \label{eq:FLLineElementClock}
ds^2 = \lambda^{-2} \big( -dT^2 + L^2 d\Sigma^2 \big),
\end{equation} where $L$ is given by (\ref{eq:FLDimlessLFunction}) and $d\Sigma^2$ is a metric of constant curvature. \footnote{Equations (\ref{eq:FLmuitoL}) and (\ref{eq:FLLineElementClock}) illustrate the well-known scale-invariance of Einstein's field equations:  if $g_{ij}$ is a solution, then so is $\tilde{g}_{ij} = \lambda^{-2} g_{ij}$ for any constant $\lambda$, with the matter terms scaling appropriately, \textit{e.g.} $\tilde{\mu} = \lambda^2 \mu$.}  From (\ref{eq:hubblelengthscale}), (\ref{eq:FLDimensionlessLengthTime}) and (\ref{eq:FLDimlessLDE}) we obtain
\begin{equation} \label{eq:FLHitoL}
H^2 = \lambda^2 \left( \sum_{i=1}^{n} m_i L^{-3 \gamma_i} - K L^{-2} \right).
\end{equation}  Further, using (\ref{eq:FLdensityparameter}), (\ref{eq:FLmuitoL}) and (\ref{eq:FLHitoL}) the density parameters $\Omega_i$ are written as
\begin{equation} \label{eq:FLOmegaitoM}
\Omega_i = \frac{m_i L^{-3 \gamma_i + 2}}{\sum_{j=1}^{n} m_j L^{-3 \gamma_j + 2} - K}.
\end{equation}

In summary, equations (\ref{eq:FLmuitoL}) and (\ref{eq:FLGeneralLengthScale}) show that the $n$-fluid FL cosmologies form an $n$-parameter family, labelled by a conformal factor $\lambda$ and $n-1$ mass parameters $m_1, \ldots, m_{n-1}$.  We will refer to these parameters as the \textit{intrinsic parameters}.

\subsection{The Observational Parameters} \label{ssec:ObservationalParameters}

In this section we introduce the observational parameters associated with $n$-fluid FL cosmologies, based on the current epoch $t_0$, which represents the age of the universe.  We denote the values of $H$, $q$, $\Omega_i$ and $\Omega_k$ at $t_0$ by
\begin{displaymath}
H_0 = H(t_0), \quad q_0 = q(t_0), \quad \Omega_{i, 0} = \Omega_i(t_0), \quad \Omega_{k, 0} = \Omega_k(t_0),
\end{displaymath} where $H_0$ is the \textit{Hubble constant}.  Then, the set of constants
\begin{equation} \label{eq:FLObservationalParams}
\left\{ t_0, H_0, q_0, \Omega_{A,0}, \Omega_{k,0} \right\}, \qquad A = 1, \ldots, n,
\end{equation} are collectively referred to as the \textit{observational parameters} (see, for example, Wainwright and Ellis (1997), p55-8).

In addition, we require the parameter $\ell_0$, which denotes the length scale at the present time, \textit{i.e} $\ell_0 = \ell(t_0)$.  This quantity is not an observational parameter, but if $K \neq 0$, $\ell_0$ can be determined in terms of observational parameters using (\ref{eq:FLdynamicalcurvature}) and (\ref{eq:FLomegarelation}), \textit{i.e.} via
\begin{equation} \label{eq:FLell0itoObs1}
\ell_0^2 = \frac{-K}{H_0^2 \Omega_{k,0}} = \frac{-K}{H_0^2 (1 - \sum_{i=1}^{n} \Omega_{i,0})}.
\end{equation}  In the case of zero curvature ($K = 0$), $\ell_0$ can be chosen arbitrarily in order to fix the multiplicative factor that determines $\ell$ (also see section \ref{sec:FLevolutioneqs}).

We now present the observational formulation of the $n$-fluid FL cosmologies.  In this formulation, the dimensionless length scale and dimensionless time variable are given by
\begin{equation} \label{eq:FLObservationalComponents}
a = \frac{\ell}{\ell_0}, \quad \mbox{and} \quad T = H_0 t.
\end{equation}  From (\ref{eq:FLConservationIntegral}) we have
\begin{equation} \label{eq:FLObservationalEnergyDensity}
\mu_i = 3 H_0^2 \Omega_{i,0} a^{-3 \gamma_i}.
\end{equation}  Using (\ref{eq:FLObservationalComponents}) and (\ref{eq:FLObservationalEnergyDensity}), the Friedmann equation (\ref{eq:FLfriedmann}) assumes the form
\begin{equation} \label{eq:FLObservationalDE}
\left( \frac{da}{dT} \right)^2 = \sum_{i=1}^{n} \Omega_{i,0} a^{-3 \gamma_i + 2} + \Omega_{k,0}.
\end{equation}  We use uniqueness of the solution to (\ref{eq:FLObservationalDE}) under some appropriate initial conditions (see section \ref{sec:FLexistenceuniqueness}) to conclude that in an expanding epoch the length scale is given by
\begin{equation}
\ell = \ell_0\ a(H_0 t, \Omega_{1,0}, \ldots, \Omega_{n,0}).
\end{equation}  It follows from (\ref{eq:hubblelengthscale}), (\ref{eq:FLObservationalComponents}) and (\ref{eq:FLObservationalDE}) that the Hubble scalar assumes the form\footnote{An expression of this form is given by Peacock (1999), eq. (3.8).}
\begin{equation} \label{eq:FLObservationalH}
H^2 = H_0^2 \left[ \sum_{i=1}^{n} \Omega_{i,0} a^{-3 \gamma_i} + \Omega_{k,0} a^{-2} \right],
\end{equation} and further that the density parameters (\ref{eq:FLdensityparameter}) and (\ref{eq:FLchi}) assume the form
\begin{equation} \label{eq:FLObservationalHubbleDensity}
\Omega_i = \frac{\Omega_{i,0} a^{-3 \gamma_i + 2}}{\sum_{j=1}^{n} \Omega_{j,0} a^{-3 \gamma_j + 2} + \Omega_{k,0}},
\end{equation} and
\begin{equation} \label{eq:FLObservationalChi}
\chi_i = \frac{\Omega_{i,0} a^{-3 \gamma_i + 2}}{\sum_{j=1}^{n} \Omega_{j,0} a^{-3 \gamma_j + 2}}.
\end{equation}

The parameters $\Omega_{k,0}$, $q_0$, $t_0$ in (\ref{eq:FLObservationalParams}) can be related to the parameters $\{\Omega_{i,0}, H_0\}$, as follows:  First, $\Omega_{k,0}$ is determined in terms of $\Omega_{i,0}$ via the Friedmann equation (\ref{eq:FLomegaAomegaKrelation}):
\begin{equation} \label{eq:OmegaKObservational}
\Omega_{k,0} = 1 - \sum_{i=1}^{n} \Omega_{i,0}.
\end{equation}  Second, $q_0$ is determined in terms of $\Omega_{i,0}$ via (\ref{eq:qomegarelation}):
\begin{equation} \label{eq:Q0Observational}
q_0 = \tfrac{1}{2} \sum_{i=1}^{n} (3 \gamma_i - 2) \Omega_{i,0}.
\end{equation}  Third, knowing $H_0$, we can obtain $t_0$ by integrating (\ref{eq:FLObservationalDE}) as
\begin{equation} \label{eq:T0H0Observational}
t_0 H_0 = \int_0^1 \frac{1}{\sqrt{F(a, \Omega_{i,0})}} da,
\end{equation} where
\begin{equation}
F(y, \Omega_{i,0}) = 1 + \sum_{i=1}^{n} \Omega_{i,0} (a^{2 - 3 \gamma_i} - 1),
\end{equation} and $\Omega_{k,0}$ has been eliminated using (\ref{eq:OmegaKObservational}).  As a consequence, \textit{the only independent observational parameters are the $n+1$ quantities defined by}
\begin{equation} \label{eq:FLIndObservationalParameters}
\{\Omega_{i,0}, H_0\}.
\end{equation}

It is useful to note that the normalized scale factor can be expressed in terms of the redshift $z$, defined by
\begin{equation}
1 + z = \frac{\ell_0}{\ell} = \frac{1}{a},
\end{equation} (see Wainwright and Ellis (1997), eq. (2.41)).  One can thus use (\ref{eq:FLObservationalH}) and (\ref{eq:FLObservationalHubbleDensity}) to express $H$ and $\Omega_i$ in terms of $z$.

We now present the link between the observational parameters and the intrinsic parameters.  On substituting \mbox{$\mu_i = 3 H^2 \Omega_i$} (see (\ref{eq:FLdensityparameter})) into (\ref{eq:FLenergydensityNFluid}) at $t = t_0$, we obtain
\begin{equation} \label{eq:FLDensityObservational}
\lambda_i^{-1} = \ell_0^{-1} (\Omega_{i,0} H_0^2 \ell_0^2)^{\frac{1}{3 \gamma_i - 2}}.
\end{equation}

\subsection{Conserved Quantities} \label{ssec:FLConservedQuantities}

We now show that the evolution equations (\ref{eq:FLChiEvolution}) and (\ref{eq:FLtildeOmegaEvolution}) admit $n-1$ conserved quantities.

From (\ref{eq:FLchi}) and (\ref{eq:FLmuitoL}) we have that
\begin{equation} \label{eq:FLFirstInt1}
\chi_i = \frac{3 \lambda^2}{\mu} m_i L^{-3 \gamma_i}.
\end{equation}  We now eliminate $\lambda^2/\mu$ and $L$ by taking products of powers of three distinct matter variables (we require $n \geq 3$).  Indeed, it follows from (\ref{eq:FLFirstInt1}) that
\begin{equation} \label{eq:FLFirstIntMatter1}
\chi_i^{\gamma_j - \gamma_k} \chi_j^{\gamma_k - \gamma_i} \chi_k^{\gamma_i - \gamma_j} = m_i^{\gamma_j - \gamma_k} m_j^{\gamma_k - \gamma_i} m_k^{\gamma_i - \gamma_j}.
\end{equation}  In the case of $K \neq 0$, we also obtain a conserved quantity describing the curvature of a model.  We use (\ref{eq:hubblelengthscale}) and (\ref{eq:FLtildeOmega}) to rewrite the Friedmann equation (\ref{eq:FLfriedmann}) as
\begin{equation} \label{eq:FLFirstInt2}
\vert 1 - \tan^2 \tilde{\Omega} \vert = \frac{3}{\mu \ell^2}.
\end{equation}  Then from (\ref{eq:FLFirstInt1}) and (\ref{eq:FLFirstInt2}), we have
\begin{equation} \label{eq:FLFirstIntCurvature1}
\chi_i^{3 \gamma_j - 2} \chi_j^{- 3 \gamma_i + 2} \vert 1 - \tan^2 \tilde{\Omega} \vert^{3 (\gamma_i - \gamma_j)} = m_i^{3 \gamma_j - 2} m_j^{-3 \gamma_i + 2}.
\end{equation}

Since there can be only $n-1$ independent conserved quantities, we must fix some of the values of $i$, $j$ and $k$ in (\ref{eq:FLFirstIntMatter1}) and (\ref{eq:FLFirstIntCurvature1}) to obtain a minimal independent set.  We choose $i = 1$, $k = n$ in (\ref{eq:FLFirstIntMatter1}) (with $2 \leq j \leq n-1$) and $i = 1$ and $j = n$ in (\ref{eq:FLFirstIntCurvature1}) and so define \footnote{Observe that if we choose $j = 1$ or $j = n$ in (\ref{eq:FLFirstIntMatter}) we simply obtain $\mathcal{M}_1 = 1$ and $\mathcal{M}_n = 1$.}
\begin{equation} \label{eq:FLFirstIntMatter}
\mathcal{M}_j(\chi_1, \ldots, \chi_n) \equiv \chi_1^{\gamma_j - \gamma_n} \chi_j^{\gamma_n - \gamma_1} \chi_n^{\gamma_1 - \gamma_j},
\end{equation} and
\begin{equation} \label{eq:FLFirstIntCurvature}
\mathcal{K}(\chi_1, \chi_n, \tilde{\Omega}) \equiv \chi_1^{\frac{3 \gamma_n - 2}{3 (\gamma_1 - \gamma_n)}} \chi_n^{- \frac{(3 \gamma_1 - 2)}{3 (\gamma_1 - \gamma_n)}} (1 - \tan^2 \tilde{\Omega}).
\end{equation}  Upon recalling that $m_n = 1$, the conserved quantities are then determined by the intrinsic parameters $m_1, \ldots, m_{n-1}$ according to
\begin{equation} \label{eq:FLFirstIntMatterConstants}
\mathcal{M}_j(\chi_1, \ldots, \chi_n) = m_1^{\gamma_j - \gamma_n} m_j^{\gamma_n - \gamma_1}, \qquad j = 2, \ldots, n-1,
\end{equation} and
\begin{equation} \label{eq:FLFirstIntCurvatureConstants}
\mathcal{K}(\chi_1, \chi_n, \tilde{\Omega}) = K m_1^{\frac{3 \gamma_n - 2}{3 (\gamma_1 - \gamma_n)}}.
\end{equation}

\paragraph{Comment:}  We have derived these conserved quantities algebraically, and it follows immediately from (\ref{eq:FLFirstIntMatterConstants}) and (\ref{eq:FLFirstIntCurvatureConstants}) that
\begin{equation} \label{eq:FLConservedQuantitiesZeroDeriv}
\frac{d \mathcal{M}_{j}}{d \tilde{\tau}} = 0, \qquad \frac{d \mathcal{K}}{d \tilde{\tau}} = 0.
\end{equation}  It involves a lengthy calculation to verify (\ref{eq:FLConservedQuantitiesZeroDeriv}) using the ODEs (\ref{eq:T_FLtildeOmegaEvolution}) and (\ref{eq:T_FLChiEvolution}).


If $K = 0$ (\textit{i.e.} $\tilde{\Omega} = \tfrac{\pi}{4}$) then $\mathcal{K} = 0$.  As described in section \ref{ssec:ObservationalParameters}, we can use the freedom to rescale $\ell_0$ to set
\begin{equation}
\lambda_1 = \lambda_n,
\end{equation} which implies $m_1 = 1$.  Then (\ref{eq:FLFirstIntMatterConstants}) gives
\begin{equation}
\mathcal{M}_j = m_j^{\gamma_n - \gamma_1}, \qquad j = 2, \ldots, n-1.
\end{equation}

The conserved quantities (\ref{eq:FLFirstIntMatter}) and (\ref{eq:FLFirstIntCurvature}) can also be expressed in terms of the density parameters, as follows:
\begin{eqnarray}
\label{eq:FirstIntMatterItoOmega} \mathcal{M}_j(\Omega_1, \ldots, \Omega_n) & = & \Omega_1^{\gamma_j - \gamma_n} \Omega_j^{\gamma_n - \gamma_1} \Omega_n^{\gamma_1 - \gamma_j}, \qquad j = 2, \ldots, n-1, \\
\label{eq:FirstIntCurvatureItoOmega} \mathcal{K}(\Omega_1, \Omega_n, \Omega_k) & = & - \Omega_1^{\frac{3 \gamma_n - 2}{3 (\gamma_1 - \gamma_n)}} \Omega_n^{- \frac{3 \gamma_1 - 2}{3 (\gamma_1 - \gamma_n)}} \Omega_k,
\end{eqnarray}  The conserved quantities are thus related to the observational parameters by
\begin{equation} \label{eq:FirstIntMatterItoObservations}
\mathcal{M}_j(\Omega_1, \ldots, \Omega_n) = \Omega_{1,0}^{(\gamma_j - \gamma_n)} \Omega_{j,0}^{(\gamma_n - \gamma_1)} \Omega_{n,0}^{(\gamma_1 - \gamma_j)},
\end{equation} and
\begin{equation} \label{eq:FirstIntCurvatureItoObservations}
\mathcal{K}(\Omega_1, \Omega_n, \Omega_k) = - \Omega_{1,0}^{\frac{3 \gamma_n - 2}{3 (\gamma_1 - \gamma_n)}} \Omega_{n,0}^{- \frac{(3 \gamma_1 - 2)}{3 (\gamma_1 - \gamma_n)}} \Omega_{k,0},
\end{equation} upon evaluating (\ref{eq:FirstIntMatterItoOmega}) and (\ref{eq:FirstIntCurvatureItoOmega}) at $t = t_0$.  We note that conserved quantities of this type have been previously discovered, for example, by Ehlers and Rindler (1989) for $RDC\Lambda$ FL universes, and more generally by Lake (2006) for $n$-fluid FL universes, using a different method.

\section{2-Fluid FL Cosmologies - Qualitative Analysis} \label{sec:FL2fluidmodels}

We now use the results of sections \ref{sec:behavclassnfluidFL} and \ref{sec:FLParameterization} to give a qualitative analysis of the dynamics of 2-fluid FL cosmologies.  The results of this section, in particular figures \ref{fig:FL2fluidmodel-DR}-\ref{fig:FLsolutions}, are contained in Wainwright (1996), although the use of the conserved quantity $\mathcal{K}$ is new.

\subsection{General Features} \label{ssec:FL2FluidGeneralFeatures}

The state space is two-dimensional, being described by the variables \mbox{($\tilde{\Omega}$, $\chi_1$, $\chi_2$)} subject to the restrictions (\ref{eq:FLchirestriction}).  We use (\ref{eq:FLchirestriction}) to eliminate $\chi_1$ via
\begin{equation} \label{eq:FL2fluidchirestriction}
\chi_1 = 1 - \chi_2.
\end{equation}  The state space is then the rectangle in the ($\tilde{\Omega}$, $\chi_2$)-plane given by
\begin{equation} \label{eq:FL2fluidStateSpace}
- \pi / 2 \leq \tilde{\Omega} \leq \pi / 2, \quad \mbox{and} \quad 0 \leq \chi_2 \leq 1.
\end{equation}

In accordance with theorem \ref{thm:NFluidClassification1}, there are two qualitatively different classes of $2$-fluid FL cosmologies, depending on whether $\gamma_2 > \tfrac{2}{3}$ or $\gamma_2 < \tfrac{2}{3}$ (recall that $\gamma_1$ and $\gamma_2$ are assumed to satisfy \mbox{$0 \leq \gamma_2 < \gamma_1 \leq 2$}).  We will discuss these two classes in sections \ref{ssec:FL2FluidCosmClass1} and \ref{ssec:FL2FluidCosmClass2} using the two most important examples from a physical point of view:\footnote{Other choices are qualitatively similar}

\begin{enumerate}
\item[i)] Universes with radiation and dust ($RDC$-universes, $\gamma_1 = \tfrac{4}{3}$, $\gamma_2 = 1$),
\item[ii)] Universes with dust and cosmological constant ($DC\Lambda$-universes, $\gamma_1 = 1$, $\gamma_2 = 0$).
\end{enumerate}

An important difference between these two cases is that if $\gamma_2 < \tfrac{2}{3}$ the Einstein static universe, represented by an equilibrium point $E$, is included in the family of models.  The point $E$ is given by\footnote{The restriction $\gamma_2 < \tfrac{2}{3}$ ensures that $\chi_{2,E} < 1$, as required.}
\begin{equation} \label{eq:EinsteinStaticChi2}
\chi_{2,E} = \frac{3 \gamma_1 - 2}{3 (\gamma_1 - \gamma_2)}, \quad \tilde{\Omega} = 0, \quad K = +1,
\end{equation} (see equation (\ref{eq:FLEinsteinStaticConstriant})).  A second related difference is the behaviour of the compact deceleration parameter $\tilde{q}$ as given by (\ref{eq:T_tildeQ}).  If $\gamma_2 > \tfrac{2}{3}$ then $\tilde{q}$ is positive on the state space (\ref{eq:T_FLcompactstatespace}) and all models are decelerating.  If $\gamma_2 < \tfrac{2}{3}$ then $\tilde{q}$ is negative on a subset of the state space.  Figure \ref{fig:FL2fluidmodel-HQ} shows that the orbits of ever-expanding models eventually enter this subset, and as a result the models are accelerating at late times.

The portraits of the orbits are drawn by first sketching the various invariant sets and equilibrium points, as given in section \ref{ssec:FLCompactifiedStructure}, in particular the boundary of the state space (\ref{eq:FL2fluidStateSpace}) and the flat FL invariant set $\tilde{\Omega} = \pm \tfrac{\pi}{4}$.  In the case $\gamma_2 < \tfrac{2}{3}$, the orbits that are past and future asymptotic to the Einstein static point $E$ should be drawn.  At this stage, the orbits that have been drawn represent exceptional models and form a ``skeleton'' for the state space.  The orbits of typical models can now be drawn by noting the past and future attractors as given in theorem \ref{thm:FLasymptoticbehaviour}.  In drawing the portraits it is helpful to note that the evolution equations (\ref{eq:T_FLtildeOmegaEvolution}) and (\ref{eq:T_FLChiEvolution}) are invariant under the interchange 
\begin{equation} \label{eq:EvolutionEquationInvariantInterchange}
(\tilde{\Omega}, \tilde{\tau}) \to (-\tilde{\Omega}, -\tilde{\tau}),
\end{equation}
\textit{i.e.} changing the sign of $\tilde{\Omega}$ reverses the direction of time.  In particular, the invariant set \linebreak \mbox{$-\tfrac{\pi}{2} \leq \tilde{\Omega} \leq -\tfrac{\pi}{4}$} represents cosmologies that evolve in the same way as those with \mbox{$\tfrac{\pi}{4} \leq \tilde{\Omega} \leq \tfrac{\pi}{2}$}, with the difference that the direction of time is reversed.  These models are contracting throughout their evolution ($\tilde{\Omega} < 0$ implies $H < 0$) and hence are not potential models of the real universe.

In practice the state space can be sketched numerically\footnote{For example, using the \texttt{plots[implicitplot]} command in Maple.} using the conserved quantity $\mathcal{K}$ that is defined by (\ref{eq:FLFirstIntCurvature}) for the family of 2-fluid cosmologies.  On substituting (\ref{eq:FL2fluidchirestriction}) into (\ref{eq:EinsteinStaticChi2}) and choosing $n=2$, we obtain
\begin{equation} \label{eq:FL2fluidEvolConstraint}
\mathcal{K}(\tilde{\Omega}, \chi_2) = (1 - \chi_2)^{\frac{3 \gamma_2 - 2}{3 (\gamma_1 - \gamma_2)}} \chi_2^{- \frac{(3 \gamma_1 - 2)}{3 (\gamma_1 - \gamma_2)}} (1 - \tan^2 \tilde{\Omega}).
\end{equation}  The orbits are given by
\begin{equation}
\mathcal{K}(\tilde{\Omega}, \chi_2) = \mbox{constant.}
\end{equation}  The value of $\mathcal{K}$ at $E$ is denoted $\mathcal{K}_{E}$ and is given by
\begin{equation} \label{eq:FL2FluidCriticalK}
\mathcal{K}_E = (1 - \chi_{2,E})^{\frac{3 \gamma_2 - 2}{3 (\gamma_1 - \gamma_2)}} (\chi_{2,E})^{- \frac{(3 \gamma_1 - 2)}{3 (\gamma_1 - \gamma_2)}}.
\end{equation}  It follows that the equation
\begin{equation}
\mathcal{K}(\tilde{\Omega}, \chi_2) = \mathcal{K}_E
\end{equation} describes the orbits that are past and future asymptotic to $E$, \textit{i.e.} it describes the stable and unstable manifolds of $E$.

The value of $\mathcal{K}$ determines whether a model is ever-expanding or not, as shown in the following table.

\bigskip

\begin{center}
\begin{tabular}{lccl}
\hspace{0.3in} & \underline{Restriction on $\mathcal{K}$} & & \underline{Qualitative Behaviour} \\[0.5ex]
\multicolumn{2}{l}{\underline{Case $\gamma_2 > \tfrac{2}{3}$}} & & \\[1.0ex]
& $- \infty < \mathcal{K} \leq 0 \phantom{\infty}$ & & Ever-expanding \\[0.3ex]
& $\phantom{\infty} 0 < \mathcal{K} < +\infty$ & & Recollapsing \\
& & & \\
\multicolumn{2}{l}{\underline{Case $\gamma_2 < \tfrac{2}{3}$}} & & \\[1.0ex]
& $- \infty < \mathcal{K} < \mathcal{K}_{E}\ $ & & Ever-expanding \\[0.3ex]
& $\ \mathcal{K}_{E} < \mathcal{K} < +\infty$ & & Recollapsing or bouncing
\end{tabular}
\end{center}

\bigskip

\noindent Further, the sign of $\mathcal{K}$ determines whether the model is open, flat or closed.  We note that all generic classes of FL models are covered by the given values of $\mathcal{K}$.

In the case $\gamma_2 > \tfrac{2}{3}$ the critical value of $\mathcal{K}$ is $\mathcal{K}_{crit} = 0$, indicating that the flat FL models play an exceptional role as regards to future evolution.  In the case $\gamma_2 < \tfrac{2}{3}$, $\mathcal{K}_{crit} = \mathcal{K}_{E}$ and the exceptional models are those that are past or future asymptotic to the Einstein static universe.  This distinction is clearly indicated by comparing figure \ref{fig:FL2fluidmodel-DR} ($RDC$-universes) with figure \ref{fig:FL2fluidmodel-HQ} ($DC\Lambda$-universes).

We note that the conserved quantity $\mathcal{K}$ can also be expressed in terms of the density parameters.  It follows from (\ref{eq:FirstIntCurvatureItoOmega}) with $n = 2$ that
\begin{equation} \label{eq:FL2fluidEvolConstraintDensityParams}
\mathcal{K}(\Omega_1, \Omega_2, \Omega_k) =  - \Omega_{1}^{\frac{3 \gamma_2 - 2}{3 (\gamma_1 - \gamma_2)}} \Omega_{2}^{- \frac{(3 \gamma_1 - 2)}{3 (\gamma_1 - \gamma_2)}} \Omega_{k},
\end{equation} with
\begin{equation}
\Omega_k = 1 - \Omega_1 - \Omega_2.
\end{equation}

\subsection{FL Cosmologies with Radiation and Dust ($RDC$-universes)} \label{ssec:FL2FluidCosmClass1}

\begin{figure}[bt]
\begin{center}
\includegraphics[height=240pt]{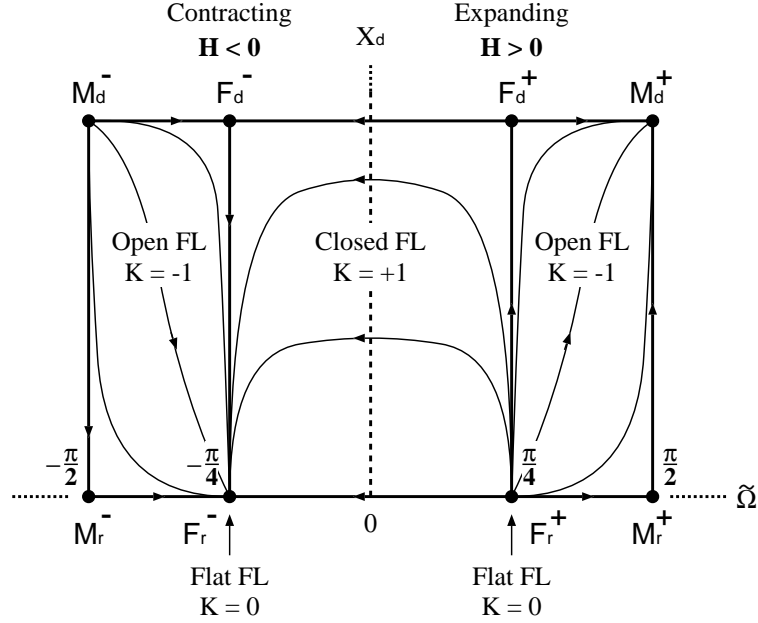}
\end{center}
\caption[The 2-fluid FL state space for $RDC$-universes]{The ($\tilde{\Omega}$, $\chi_2$) state space for 2-fluid FL cosmologies containing radiation ($\gamma_1 = \tfrac{4}{3}$) and dust ($\gamma_2 = 1$), showing regions of expansion ($H > 0$, \textit{i.e.} $\tilde{\Omega} > 0$) and regions of contraction ($H < 0$, \textit{i.e.} $\tilde{\Omega} < 0$).} \label{fig:FL2fluidmodel-DR}
\end{figure}

We now consider $2$-fluid FL cosmologies containing radiation ($\gamma_1 = \tfrac{4}{3}$) and dust ($\gamma_2 = 1$).  The portrait of the orbits, shown in figure \ref{fig:FL2fluidmodel-DR} is drawn as described in the previous subsection.

For the $RDC$-universes, the expression (\ref{eq:FL2fluidEvolConstraint}) for the conserved quantity $\mathcal{K}$ becomes
\begin{equation} \label{eq:RadiationDustK}
\mathcal{K} = \frac{\chi_r}{\chi_d^2} (1 - \tan^2 \tilde{\Omega}).
\end{equation}  Equivalently, the expression (\ref{eq:FL2fluidEvolConstraintDensityParams}) gives $\mathcal{K}$ in terms of the density parameters as
\begin{equation}
\mathcal{K} = - \frac{\Omega_r \Omega_k}{\Omega_d^2}.
\end{equation}

The value of $\mathcal{K}$ is infinite or zero on the skeleton of the state space:

\smallskip

\begin{center}
\begin{tabular}{ll}
$\mathcal{K} = + \infty$ & if $\chi_d = 0$ and $\vert \tilde{\Omega} \vert < \tfrac{\pi}{4}$, \\[0.5ex]
$\mathcal{K} = 0$ & if $\chi_d = 1$ or $\vert \tilde{\Omega} \vert = \tfrac{\pi}{4}$, \\[0.5ex]
$\mathcal{K} = - \infty$ & if $\tilde{\Omega} = \pm \tfrac{\pi}{2}$, or $\chi_d = 0$ and $\vert \tilde{\Omega} \vert > \tfrac{\pi}{4}$.
\end{tabular}
\end{center}

\smallskip

Note that $\mathcal{K}$ is indeterminate at the past attractor $\mathcal{A}^{-} = F_r^{+}$ ($\chi_d = 0$, $\tilde{\Omega} = \tfrac{\pi}{4}$) and at the future attractor $\mathcal{A}^{+} = F_r^{-} \cup M_d^{+}$ ($\chi_d = 0$, $\tilde{\Omega} = \tfrac{\pi}{4}$; $\chi_d = 1$, $\tilde{\Omega} = \tfrac{\pi}{2}$).  This indeterminacy arises since infinitely many orbits meet at $\mathcal{A}^{-}$ and $\mathcal{A}^{+}$.

\paragraph{Discussion:}  Typical models differ qualitatively depending on the value of $\mathcal{K}$, which determines where the orbit lies in the state space.  Firstly, for ever-expanding models if the orbit passes close to $M_r^{+}$, the dust component will never be significant dynamically (\textit{i.e.} $\Omega_d \ll 1$) and this will correspond to $\mathcal{K}$ being large and negative.  On the other hand if the orbit passes close to $F^{+}_{d}$ then $\Omega_d$ will attain a maximum value close to 1, corresponding to $\mathcal{K}$ being negative and close to zero.  Secondly, recollapsing models differ quantitatively by how close the orbit comes to $F_d^{\pm}$, \textit{i.e.} by the maximum value of $\chi_d$, which in turn will depend on $\mathcal{K}$.  One can determine the dependence on $\mathcal{K}$ explicitly, as follows:

\begin{enumerate}
\item[i)] For ever-expanding models, we obtain
\begin{equation}
\left. \Omega_d \right\vert_{max} = \frac{1}{1 + 2 \sqrt{- \mathcal{K}}}, \qquad \mathcal{K} < 0.
\end{equation}

\item[ii)] For recollapsing models an expression for $\chi_r \vert_{min}$ and $\chi_d \vert_{max}$ can be obtained by setting $\tilde{\Omega} = 0$ in (\ref{eq:RadiationDustK}) and solving the resulting second degree polynomial.  We obtain
\begin{equation}
\left. \chi_d \right\vert_{max} = \frac{2}{1 + \sqrt{1 + 4 \mathcal{K}}}, \qquad \left. \chi_r \right\vert_{min} = 1 - \left. \chi_d \right\vert_{max}, \qquad \mathcal{K} > 0.
\end{equation}
\end{enumerate}

\subsection{FL Cosmologies with Dust and Cosmological Constant} \label{ssec:FL2FluidCosmClass2}

\begin{figure}[hbp]
\begin{center}
\includegraphics[height=240pt]{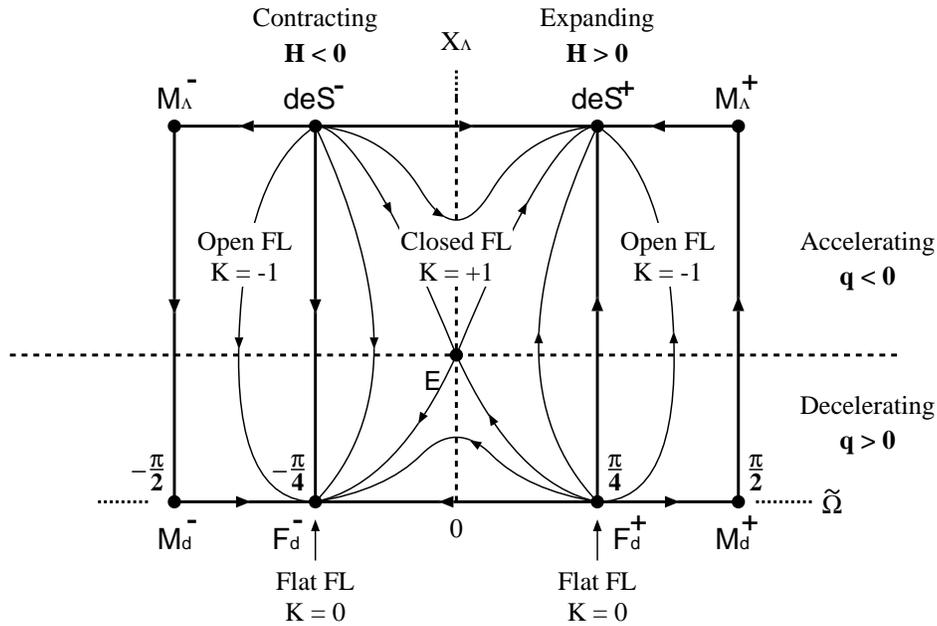}
\end{center}
\caption[The 2-fluid FL state space for $DC\Lambda$-universes]{The ($\tilde{\Omega}$, $\chi_2$) state space for 2-fluid FL cosmologies containing dust ($\gamma_1 = 1$) and cosmological constant ($\gamma_2 = 0$), showing regions of expansion ($H > 0$, \textit{i.e.} $\tilde{\Omega} > 0$), regions of contraction ($H < 0$, \textit{i.e.} $\tilde{\Omega} < 0$), regions of acceleration ($\tilde{q} < 0$) and regions of deceleration ($\tilde{q} > 0$) in relation to the Einstein static solution $E$, which satisfies $H = 0$ and $\tilde{q} = 0$.} \label{fig:FL2fluidmodel-HQ}
\end{figure}

We now consider $2$-fluid FL cosmologies containing radiation ($\gamma_1 = \tfrac{4}{3}$) and dust ($\gamma_2 = 1$).  The portrait of the orbits, shown in figure \ref{fig:FL2fluidmodel-HQ} is drawn as described in section \ref{ssec:FL2FluidGeneralFeatures}.

For the $DC\Lambda$-universes, the expression (\ref{eq:FL2fluidEvolConstraint}) for the conserved quantity $\mathcal{K}$ becomes
\begin{equation} \label{eq:DustCosmologicalConstantK}
\mathcal{K}(\tilde{\Omega}, \chi_\Lambda) = \chi_d^{-2/3} \chi_\Lambda^{-1/3} (1 - \tan^2 \tilde{\Omega}).
\end{equation}  Equivalently, the expression (\ref{eq:FL2fluidEvolConstraintDensityParams}) gives $\mathcal{K}$ in terms of the density parameters as
\begin{equation}
\mathcal{K} = - \Omega_d^{-2/3} \Omega_\Lambda^{-1/3} \Omega_k.
\end{equation}  The critical value of $\chi_\Lambda$, \textit{i.e.} the value of $\chi_\Lambda$ at the Einstein static equilibrium point, is given by (\ref{eq:EinsteinStaticChi2}) as
\begin{equation}
\chi_{\Lambda,E} = \tfrac{1}{3}.
\end{equation}  It then follows from (\ref{eq:FL2FluidCriticalK}) that the critical value of $\mathcal{K}$ is
\begin{equation}
\mathcal{K}_{E} = \left( \tfrac{4}{27} \right)^{-1/3}.
\end{equation}  The value of $\mathcal{K}$ is infinite or zero on the skeleton of the state space:

\smallskip

\begin{center}
\begin{tabular}{ll}
$\mathcal{K} = + \infty$ & if $\vert \tilde{\Omega} \vert < \tfrac{\pi}{4}$ and $\chi_\Lambda = 0\ \mbox{or}\ 1$, \\[0.5ex]
$\mathcal{K} = 0$ & if $\tilde{\Omega} = \pm \tfrac{\pi}{4}$, \\[0.5ex]
$\mathcal{K} = - \infty$ & if $\tilde{\Omega} = \pm \tfrac{\pi}{2}$ or $\vert \tilde{\Omega} \vert > \tfrac{\pi}{4}$ and $\chi_\Lambda = 0\ \mbox{or}\ 1$.
\end{tabular}
\end{center}

\smallskip

Note that $\mathcal{K}$ is indeterminate at the past attractor $\mathcal{A}^{-} = F_d^{+}$ ($\chi_\Lambda = 0$, $\tilde{\Omega} = \tfrac{\pi}{4}$) and at the future attractor $\mathcal{A}^{+} = deS^{+} \cup F_d^{-}$ ($\chi_\Lambda = 1$, $\tilde{\Omega} = \tfrac{\pi}{4}$; $\chi_\Lambda = 0$, $\tilde{\Omega} = - \tfrac{\pi}{4}$).

\begin{figure}[p]
\begin{center}
\includegraphics[height=160pt]{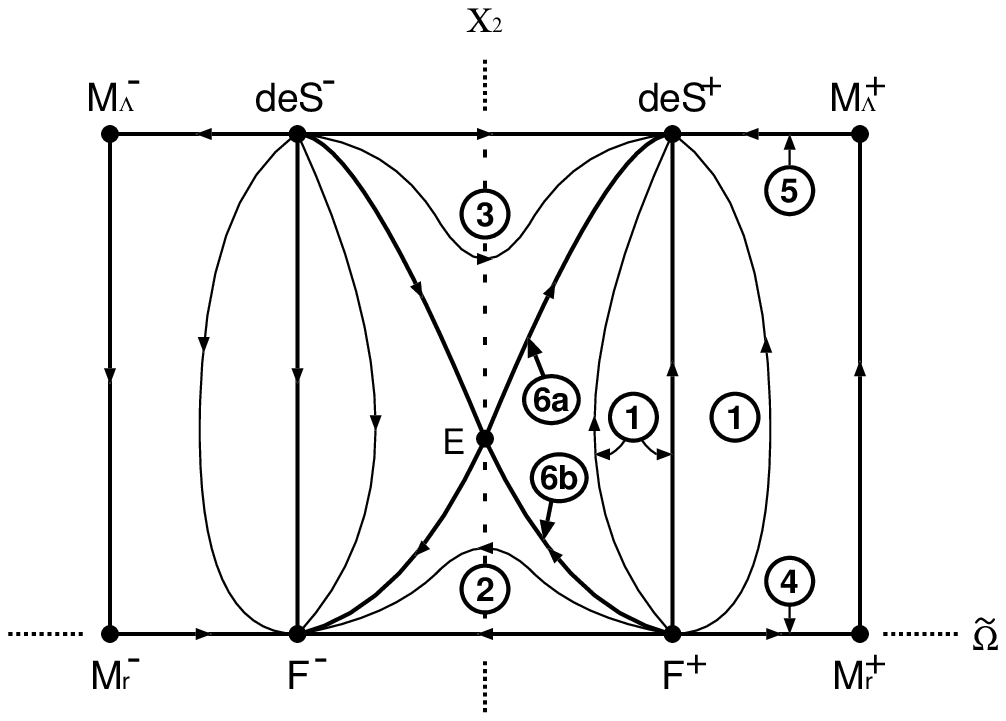}
\end{center}
\begin{center}
\includegraphics[width=60pt]{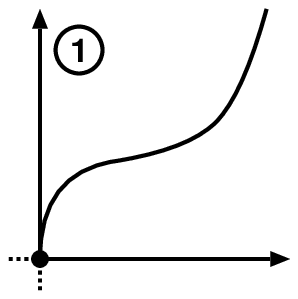}
\includegraphics[width=60pt]{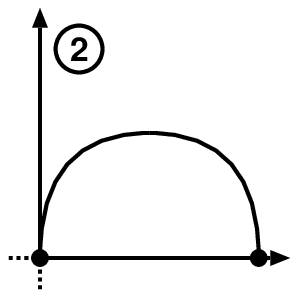}
\includegraphics[width=60pt]{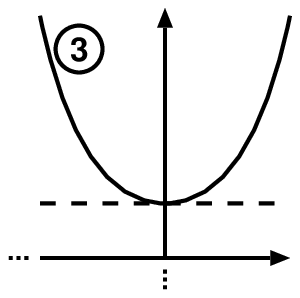}
\end{center}
\begin{center}
\includegraphics[width=60pt]{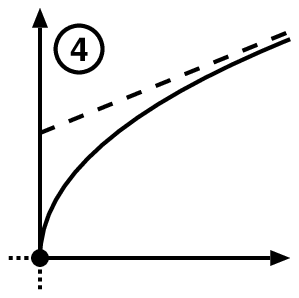}
\includegraphics[width=60pt]{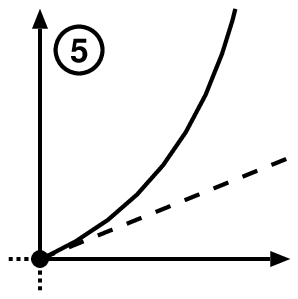}
\includegraphics[width=60pt]{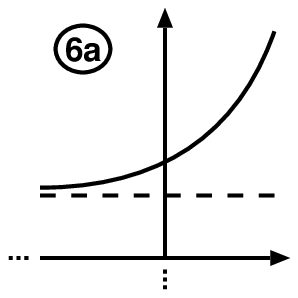}
\includegraphics[width=60pt]{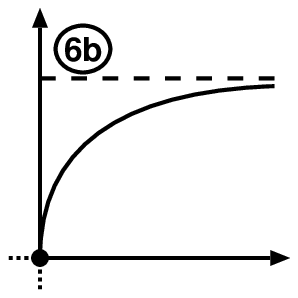}
\end{center}
\begin{center}
\includegraphics[width=60pt]{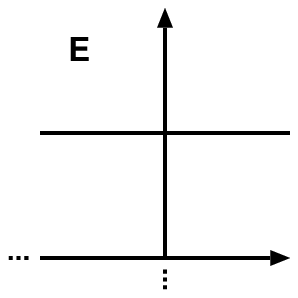}
\includegraphics[width=60pt]{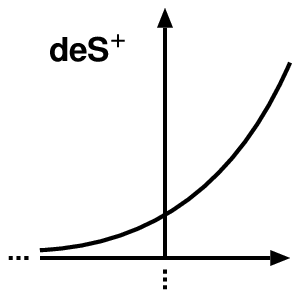}
\includegraphics[width=60pt]{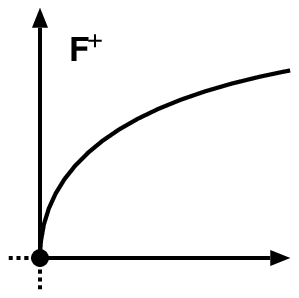}
\includegraphics[width=60pt]{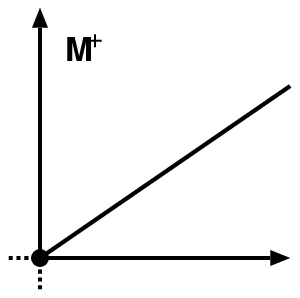}
\end{center}
\caption[Thumbnail plots depicting behaviour of the length scale in FL cosmologies]{The ($\tilde{\Omega}$, $\chi$) representation of the state space for the FL models with matter, a cosmological constant and spatial curvature.  The thumbnail plots depict the length scale $\ell(t)$ of FL models corresponding to orbits in each region of the state space and to each equilibrium point.  Cases (1), (2) and (3) represent the three generic families of FL models, \textit{i.e.} their orbits fill an open subset of the state space (2-parameter families).  They represent, respectively, ever-expanding models, recollapsing models and bouncing models.  On the other hand, cases (4), (5), (6a) and (6b) correspond to a single orbit in the state space (a 1-parameter family), while the cases in the third row represent the unique solutions corresponding to the equilibrium points.} \label{fig:FLsolutions}
\end{figure}

\clearpage

\paragraph{Discussion:}  Firstly, ever-expanding models differ qualitatively depending on the magnitude and the sign of the curvature scalar $\Omega_k$.  If the orbit stays close to flat FL ($\tilde{\Omega} = \tfrac{\pi}{4}$), the curvature will never be significant dynamically (\textit{i.e.} $\Omega_k \ll 1$) and this will correspond to $\mathcal{K}$ being close to zero.  Secondly, recollapsing models (and bouncing models) differ qualitatively by how close the orbit comes to the Einstein static model, \textit{i.e.} by the maximal (or minimal) value of $\chi_\Lambda$.  Since $\chi_\Lambda < \chi_{\Lambda,E}$ in recollapsing models (see (\ref{eq:EinsteinStaticChi2})), the cosmological constant will never be the dominant fluid.  Similarly, since $\chi_\Lambda > \chi_{\Lambda,E}$ in bouncing models, there will be no dust dominant epoch, \textit{i.e.} $\chi_d \approx 1$.  One can determine the dependence on $\mathcal{K}$ explicitly, as follows:

\begin{enumerate}
\item[i)] For ever-expanding models, we obtain
\begin{equation}
\left\vert \Omega_k \right\vert_{max} = \frac{\vert \mathcal{K} \vert}{\mathcal{K}_{E} - \mathcal{K}}, \qquad \mathcal{K} < \mathcal{K}_{E}.
\end{equation}

\item[ii)] For recollapsing and bouncing models, an expression for $\chi_\Lambda \vert_{max}$ and $\chi_\Lambda \vert_{min}$ can be obtained by setting $\tilde{\Omega} = 0$ in (\ref{eq:DustCosmologicalConstantK}) and solving the resulting third degree polynomial.
\end{enumerate}

\section{3-Fluid FL Cosmologies - Qualitative Analysis} \label{sec:FL3fluidmodels}

We now use the results of sections \ref{sec:behavclassnfluidFL}-\ref{sec:FL2fluidmodels} to give a qualitative analysis of the dynamics of 3-fluid FL cosmologies in which the matter content is radiation, dust and a cosmological constant ($\Lambda > 0$) (labelling the $\chi_i$ as $\chi_r$, $\chi_d$ and $\chi_\Lambda$).  The analysis given in this section is new.

\subsection{General Features}

The state space is three-dimensional, being described by the variables ($\tilde{\Omega}$, $\chi_r$, $\chi_d$, $\chi_\Lambda$) subject to the restriction (\ref{eq:FLchirestriction}).  We use (\ref{eq:FLchirestriction}) to eliminate $\chi_r$ via
\begin{equation} \label{eq:ThreeFluidFLChiRelation}
\chi_r = 1 - \chi_d - \chi_\Lambda.
\end{equation}  The physical state space is then a solid triangular prism in $\mathbb{R}^3$ given by
\begin{equation} \label{eq:ThreeFluidStateSpace}
-\frac{\pi}{2} \leq \tilde{\Omega} \leq \frac{\pi}{2}, \qquad 0 \leq \chi_d, \chi_\Lambda \leq 1, \qquad \chi_d + \chi_\Lambda \leq 1.
\end{equation}

\begin{figure}[p]
\begin{center}
\includegraphics[height=240pt]{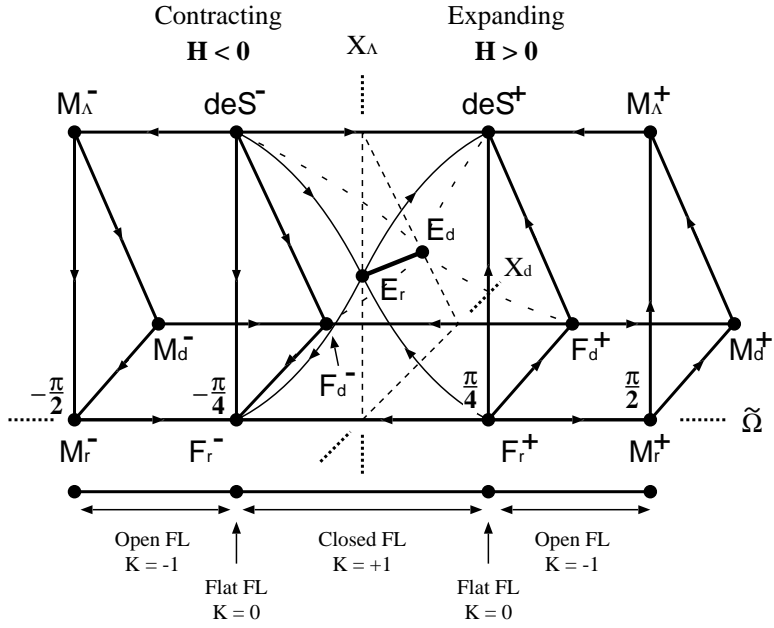}
\end{center}
\caption[The 3-fluid FL state space for radiation, dust and cosmological constant.]{The skeleton of the state space for $RDC\Lambda$ cosmologies, using $\tilde{\Omega}$, $\chi_d$ and $\chi_\Lambda$ as variables, with \mbox{$\chi_r = 1 - \chi_d - \chi_\Lambda$}.} \label{fig:FL3fluidmodel}
\end{figure}

We can construct a ``skeleton'' for the state space using various invariant subsets, as shown in figure \ref{fig:FL3fluidmodel}.  The choice of $\chi_i = 0$ for any one fluid gives a 2-fluid FL cosmology, identified as follows:

\smallskip

\begin{center}
\begin{tabular}{ll}
$\chi_\Lambda = 0$ & $RDC$-universes, \\[0.5ex]
$\chi_d = 0$ & $RC\Lambda$-universes, \\[0.5ex]
$\chi_r = 0$ & $DC\Lambda$-universes.
\end{tabular}
\end{center}

\smallskip

\noindent Further, the orbits in the planes $\tilde{\Omega} = \pm \tfrac{\pi}{4}$ represent the flat 3-fluid universes.

The Einstein static universe is represented by a line segment of equilibrium points denoted $E_r E_d$, which is the intersection of the planes $\tilde{\Omega} = 0$ and $\tilde{q} = 0$.  The stable manifold $\mathcal{S}^{+}$ of the set of Einstein static points is a surface through the line segment $E_r E_d$ terminating at $deS^{-}$ when $\chi_\Lambda = 1$ and at the line segment $F_r^{+} F_d^{+}$ when $\chi_\Lambda = 0$.  This surface can be described by an equation of the form
\begin{equation}
\tilde{\Omega} = f(\chi_d, \chi_\Lambda),
\end{equation} with $\chi_d$ and $\chi_\Lambda$ restricted as in (\ref{eq:ThreeFluidStateSpace}).  Because of the symmetry $(\tilde{\Omega}, \tilde{\tau}) \to (-\tilde{\Omega}, -\tilde{\tau})$ (see (\ref{eq:EvolutionEquationInvariantInterchange})), the unstable manifold $\mathcal{S}^{-}$ is the surface 
\begin{equation}
\tilde{\Omega} = - f(\chi_d, \chi_\Lambda).
\end{equation}

We know from theorem \ref{thm:NFluidClassification1} that there are three generic classes of $RDC\Lambda$ FL universes.  Their orbits are contained in invariant subsets bounded by the surfaces $\mathcal{S}^{\pm}$, as follows:

\smallskip

\begin{center}
\begin{tabular}{ll}
Ever-expanding universes & $\tilde{\Omega} > \vert f(\chi_d, \chi_\Lambda) \vert$ (the subset to the right of $\mathcal{S}^{+}$ and $\mathcal{S}^{-}$), \\[0.5ex]
Recollapsing universes & (the subset between $\mathcal{S}^{+}$ and $\mathcal{S}^{-}$ and below the plane $\tilde{q} = 0$), \\[0.5ex]
Bouncing universes & (the subset between $\mathcal{S}^{+}$ and $\mathcal{S}^{-}$ and above the plane $\tilde{q} = 0$).
\end{tabular}
\end{center}

\smallskip

The planes $\tilde{\Omega} = 0$ and $\tilde{q} = 0$, although not invariant sets, play an important role in describing the dynamics.  The plane $\tilde{\Omega} = 0$ divides the state space into a region of expansion ($\tilde{\Omega} > 0$) and a region of contraction ($\tilde{\Omega} < 0$).  The variable $\tilde{q}$ is given by (\ref{eq:T_tildeQ}) which here specializes to
\begin{equation} \label{eq:RDCLQTilde}
\tilde{q} = \chi_r + \tfrac{1}{2} \chi_d - \chi_\Lambda.
\end{equation}  On using (\ref{eq:ThreeFluidFLChiRelation}) to eliminate $\chi_r$, we find that $\tilde{q} = 0$ on the plane
\begin{equation} \label{eq:ThreeFluidFLChiDChiLRelation}
\chi_d + 4 \chi_\Lambda = 2.
\end{equation}  Equation (\ref{eq:ThreeFluidFLChiDChiLRelation}) then divides the state space into a region of acceleration ($\tilde{q} < 0$) and a region of deceleration ($\tilde{q} > 0$) as follows:

\smallskip

\begin{center}
\begin{tabular}{ll}
Region of acceleration & $\chi_\Lambda > \tfrac{1}{4} (2 - \chi_d)$, \\[0.5ex]
Region of deceleration & $\chi_\Lambda < \tfrac{1}{4} (2 - \chi_d)$.
\end{tabular}
\end{center}

\smallskip

\begin{figure}[p]
\begin{center}
\includegraphics[height=140pt]{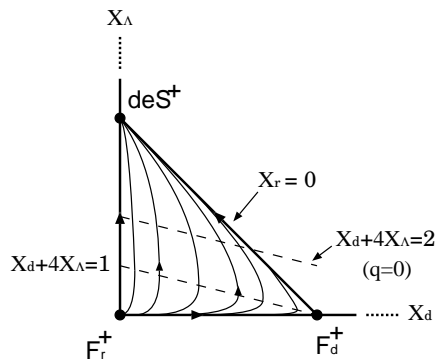}
\end{center}
\caption[3-fluid FL state space cross-section along the expanding flat FL manifold]{The invariant set $\tilde{\Omega} = \tfrac{\pi}{4}$ in the state space for the $RDC\Lambda$ universes, describing the expanding flat FL models.} \label{fig:FL3fluidmodelflat}
\end{figure}

It follows from the general discussion in section \ref{ssec:FLConservedQuantities} that there are two conserved quantities for the $RDC\Lambda$ FL universes, defined by
\begin{eqnarray} \label{eq:ThreeFluidConservedQuantities}
\mathcal{M} = \left( \frac{\chi_r^3 \chi_\Lambda}{\chi_d^4} \right)^{1/3}, \quad \mbox{and} \quad \mathcal{K} = \frac{1 - \tan^2 \tilde{\Omega}}{(\chi_r \chi_\Lambda)^{1/2}}.
\end{eqnarray}  Using (\ref{eq:ThreeFluidFLChiRelation}) we can eliminate $\chi_r$ from these equations and hence obtain $\mathcal{M}$ and $\mathcal{K}$ in terms of $\tilde{\Omega}$, $\chi_d$ and $\chi_\Lambda$.  The values of $\mathcal{M}$ and $\mathcal{K}$ at the Einstein static equilibrium points $E$ are denoted $\mathcal{M}_{E}$ and $\mathcal{K}_{E}$ and are obtained on substituting (\ref{eq:ThreeFluidFLChiDChiLRelation}) into (\ref{eq:ThreeFluidConservedQuantities}).  We have
\begin{align}
\mathcal{M}_{E} & = \left[ \frac{(3 \chi_\Lambda - 1)^3 \chi_\Lambda}{16 (1 - 2 \chi_\Lambda)^4} \right]^{1/3}, & & 0 \leq \mathcal{M}_{E} < \infty, \\[0.5ex]
\mathcal{K}_{E} & = \left[ (3 \chi_\Lambda - 1) \chi_\Lambda \right]^{-1/2}, & & 2 \leq \mathcal{K}_{E} < \infty.
\end{align}  The orbits of typical models are described by the equations
\begin{equation}
\mathcal{K}(\tilde{\Omega}, \chi_d, \chi_\Lambda) = \mbox{constant}, \quad \mbox{and} \quad \mathcal{M}(\chi_d, \chi_\Lambda) = \mbox{constant},
\end{equation} \textit{i.e.} the orbits are the intersections of these two families of surfaces in the state space.  As a special case the flat universes satisfy $\mathcal{K} = 0$ (\textit{i.e.} $\tilde{\Omega} = \tfrac{\pi}{4}$) and then the equation
\begin{equation}
\mathcal{M}(\tilde{\Omega}, \chi_d, \chi_\Lambda) = \mbox{constant}
\end{equation} describes the orbits in the invariant set $\tilde{\Omega} = \tfrac{\pi}{4}$ and can be used to numerically sketch them, as in figure \ref{fig:FL3fluidmodelflat}.

\paragraph{Discussion:}  Knowing the past attractor $\mathcal{A}^{-} = F_r^{+}$ and the future attractor $\mathcal{A}^{+} = F_r^{-} \cup deS^{+}$ one can visualize the orbits of typical universes.  For example, the orbits of ever-expanding models join $F_r^{+}$ to $deS^{+}$.  The models will differ qualitatively depending on whether they come close to $F_d^{+}$ ($\Omega_d \approx 1$ in a neighbourhood of $F_d^{+}$) and whether they come close to the Milne set ($\tilde{\Omega} = \tfrac{\pi}{2}$) or to the surfaces $\mathcal{S}^{\pm}$, in which case there will be an epoch in which spatial curvature is dominant (\textit{i.e.} $\vert \Omega_k \vert \approx 1$).  One can determine the points in state space at which $\vert \Omega_k \vert$ and $\chi_d$ attain a maximum value, as follows.  If $K \neq 0$, it follows from (\ref{eq:FLomegarelation}) and (\ref{eq:FLOmegaTildeOmegaRelation}) that $\tilde{\Omega}$ and $\vert \Omega_k \vert$ are simultaneously extremal.  Since in an ever-expanding model $\tilde{\Omega}$ is extremal when $\tilde{q} = 0$, we conclude that $\vert \Omega_k \vert$ is extremal on the plane given by (\ref{eq:ThreeFluidFLChiDChiLRelation}).  The matter variable $\chi_d$ will attain a maximum when $\chi_d^{\prime} = 0$, which implies $\tilde{q} = \tfrac{1}{2}$, as follows from (\ref{eq:T_FLChiEvolution}) and (\ref{eq:T_tildeQ}).  After substituting (\ref{eq:ThreeFluidFLChiRelation}) to eliminate the $\chi_r$ term in (\ref{eq:RDCLQTilde}), we conclude that $\chi_d$ will attain a maximum on the plane
\begin{equation} \label{eq:FL3FluidMaximalDustPlane}
\chi_d + 4 \chi_\Lambda = 1.
\end{equation}  One can then determine the dependence of $\chi_d \vert_{max}$ on $\mathcal{M}$ explicitly from (\ref{eq:ThreeFluidFLChiRelation}), (\ref{eq:ThreeFluidConservedQuantities}) and (\ref{eq:FL3FluidMaximalDustPlane}), as follows:
\begin{equation} \label{eq:RDCLMaxChiD}
\chi_d \vert_{max} = \frac{3^{3/4}}{4 \mathcal{M}^{3/4} + 3^{3/4}}.
\end{equation}

\subsection{A Model of the Physical Universe}

In this section we discuss the $RDC\Lambda$ FL universes as viable models for the large scale dynamical behaviour of the physical universe.  Current observations of distant galaxies, type $I_a$ supernovae and the cosmic microwave background radiation support the following conclusions:

\begin{enumerate}
\item[i)]  The universe is expanding in the present epoch ($H_0 > 0$).
\item[ii)]  The universe is accelerating in the present epoch ($q_0 < 0$).
\item[iii)]  The early universe was radiation-dominated (hot big-bang).
\item[iv)]  Pressure-free matter (baryonic matter and dark matter) is dynamically significant in the present epoch.
\item[v)]  The spatial geometry is close to flatness in the present epoch.
\end{enumerate}

These conclusions suggest modelling the physical universe as an $RDC\Lambda$ FL cosmology, with observational parameters
\begin{equation} \label{eq:ConcordantActiveParameters}
\Omega_{r,0}, \quad \Omega_{d,0}, \quad \Omega_{k,0}, \quad \Omega_{\Lambda,0}, \quad q_0, \quad H_0, \quad \mbox{and} \quad t_0.
\end{equation}  Equations (\ref{eq:OmegaKObservational})-(\ref{eq:T0H0Observational}) limit the number of independent quantities in this set to four, as expected for an $RDC\Lambda$ model.  Values for these parameters have been determined using observations, such as high redshift galaxy surveys (for example, by 2dF and SDSS; see Spergel et al. (2006) for detailed references) and analyses of the power spectrum of the cosmic microwave background (for example, by WMAP).  We refer the reader to Spergel et al. (2006) and Tegmark et al. (2004) for an analysis of recent observational data, leading to the following values:

\begin{eqnarray}
\label{eq:ConcordantH} H_0 & = & 71.0 \pm 4.0 km\ s^{-1}\ Mpc^{-1}, \\
\label{eq:ConcordantOr} \Omega_{r,0} & = & (4.9 \pm 0.5) \times 10^{-5}, \\
\label{eq:ConcordantOd} \Omega_{d,0} & = & 0.27 \pm 0.03, \\
\label{eq:ConcordantOl} \Omega_{\Lambda,0} & = & 0.73 \pm 0.04, \\
\label{eq:ConcordantOk} \Omega_{k,0} & = & -0.010 \pm 0.014.
\end{eqnarray}

\begin{figure}[p]
\begin{center}
\includegraphics[width=\textwidth]{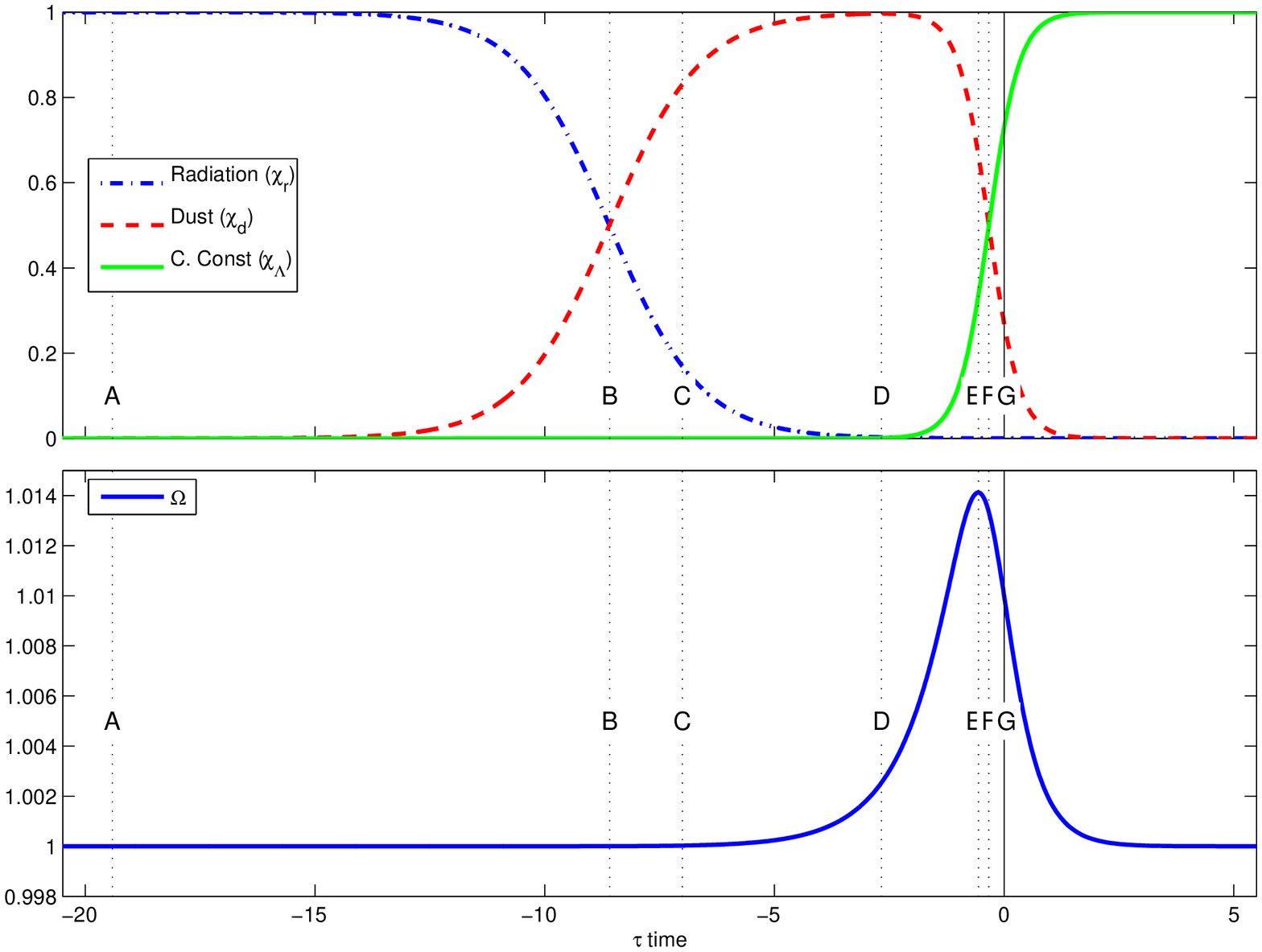}
\end{center}
\begin{center}
\begin{tabular}{clcc}
& \underline{Epoch} & \underline{Redshift ($z$)} & \underline{Temperature ($T$)} \\
\textbf{A} & Big-bang nucleosynthesis & $3.71 \times 10^8$ & $1.01 \times 10^9 K$ \\
\textbf{B} & Matter-radiation equality & $5380$ & $9700 K$ \\
\textbf{C} & Last Scattering & $1100$ & $3010 K$ \\
\textbf{D} & Maximal matter density & $13.5$ & $39.6 K$ \\
\textbf{E} & Maximal spatial curvature & $0.773$ & $4.83 K$ \\
\textbf{F} & Matter-$\Lambda$ equality & $0.374$ & $3.50 K$ \\
\textbf{G} & Present day & $0.0$ & $2.73 K$
\end{tabular}
\end{center}
\caption[Past and future evolution of the $RDC\Lambda$ model of the physical universe]{Past and future evolution of the $RDC\Lambda$ model of the physical universe, depicting the matter variables (top), Hubble-normalized energy density (bottom) and epochs of primary interest.} \label{fig:ConcordantModel}
\end{figure}

Using the observational values (\ref{eq:ConcordantH})-(\ref{eq:ConcordantOk}), one can apply (\ref{eq:T_FLtildeOmegaEvolution})-(\ref{eq:T_tildeQ}) and (\ref{eq:FLOmegaTildeOmegaRelation})-(\ref{eq:FLQTildeQRelation}) to numerically obtain the past and future behaviour of all matter quantities $\chi_i$ and the curvature indicator $\Omega$.  We present the results of one such numerical simulation in figure \ref{fig:ConcordantModel}, along with computed epochs of note.

The conserved quantities in the $RDC\Lambda$ universe, as given by (\ref{eq:ThreeFluidConservedQuantities}), can be expressed in terms of the density parameters as
\begin{equation} \label{eq:ConcordantMatterQuantity}
\mathcal{M} = \left( \frac{\Omega_r^3\ \Omega_\Lambda}{\Omega_d^{4}} \right)^{\tfrac{1}{3}},
\end{equation} and
\begin{equation}
\mathcal{K} = \Omega_r^{-\frac{1}{2}} \Omega_\Lambda^{- \frac{1}{2}} \Omega_k,
\end{equation} (see (\ref{eq:FirstIntMatterItoOmega}) and (\ref{eq:FirstIntCurvatureItoOmega})).  Then using (\ref{eq:ConcordantOr})-(\ref{eq:ConcordantOk}), we obtain
\begin{equation}
\mathcal{M} = (2.63 \pm 0.07) \times 10^{-3},
\end{equation} and
\begin{equation}
\mathcal{K} = -1.72 \pm 1.91.
\end{equation}  The values of these conserved quantities, which were fixed at the end of the inflationary epoch, determine a small subset of $RDC\Lambda$ universes that can potentially describe the real universe.  The value of $\mathcal{M}$ ensures the occurrence of a matter-dominated epoch that is sufficiently long, but not too long (see equation (\ref{eq:RDCLMaxChiD})).  The value of $\mathcal{K}$ ensures that $\Omega_k$ is close to $0$ in the past and also into the future.

\clearpage
\section{Explicit FL Cosmologies} \label{sec:FLsolutions}

In this section we give a unified description of the known solutions of the Friedmann DE that can be expressed in terms of elementary functions, for 1-fluid and 2-fluid FL cosmologies.  In these cases, the Friedmann DE can be transformed to the form
\begin{equation} \label{eq:FLquadraticform}
\left( \frac{d\mathcal{L}}{d\mathcal{T}} \right)^2 = Q(\mathcal{L}, m, K),
\end{equation} where $Q$ is a \textit{quadratic function} in $\mathcal{L}$, $m$ is a dimensionless mass parameter and $K$ is the curvature indicator.  Here $\mathcal{L}$ is a function of the length scale factor $\ell$ and the variable $\mathcal{T}$ is either proportional to conformal time $\eta$ (in the case $\Lambda = 0$) or proportional to clock time $t$ (in the case $\Lambda > 0$).

Recall that the $n$-fluid Friedmann equation can be written as a DE of the form (\ref{eq:FLellDE}), reproduced here for convenience:
\begin{equation} \label{eq:FLfriedmannDE}
\left( \frac{d\ell}{dt} \right)^2 = \sum_{i=1}^{n} ( \lambda_i \ell )^{-3 \gamma_i + 2} - K.
\end{equation}  We are primarily interested in expanding models, which begin at a big-bang singularity, \textit{i.e.} the length scale satisfies $\ell(t_s) = 0$ at some time $t_s$, and $\frac{d\ell}{dt} > 0$ in some interval.  Since the DE (\ref{eq:FLfriedmannDE}) is autonomous, it is invariant under the change $t \rightarrow t+C$.  We will use this freedom to set $t_s = 0$, thereby fixing the constant of integration that arises in solving (\ref{eq:FLfriedmannDE}).  We will thus solve (\ref{eq:FLfriedmannDE}) subject to the requirement $\frac{d\ell}{dt} > 0$ and the initial condition
\begin{equation} \label{eq:FLinitialconds}
\ell(0) = 0.
\end{equation}

Using conformal time (\ref{eq:conformaltime}), equation (\ref{eq:FLfriedmannDE}) becomes a first order ODE for $\ell(\eta)$:
\begin{equation} \label{eq:FLfriedmann_conf}
\frac{1}{\ell^2} \left( \frac{d\ell}{d\eta} \right)^2 = \sum_{i=1}^{n} ( \lambda_i \ell )^{-3 \gamma_i + 2} - K.
\end{equation}  The solution $\ell(\eta)$ can in principle be expressed in terms of clock time $t$ using the equation (\ref{eq:conformaltime}).  As with (\ref{eq:FLfriedmannDE}), when solving (\ref{eq:FLfriedmann_conf}) for $\ell(\eta)$, we will impose the requirement $\frac{d\ell}{d\eta} > 0$ and the initial condition
\begin{equation} \label{eq:FLinitialconds_conf}
\ell(0) = 0.
\end{equation}

In the following sections we consider five classes of models in which either (\ref{eq:FLfriedmannDE}) or (\ref{eq:FLfriedmann_conf}) can be transformed into (\ref{eq:FLquadraticform}).  The solution of (\ref{eq:FLquadraticform}) subject to the requirement $\frac{d\mathcal{L}}{d\mathcal{T}} > 0$ and the initial condition $\mathcal{L}(0) = 0$ is given in Appendix \ref{app:FriedmannSolutions}.  In each case, we give the solution for $\ell(t)$ or $\ell(\eta)$ and note that the corresponding expressions for the energy densities and pressure are given by equation (\ref{eq:FLtotalmattercontent}), (\ref{eq:FLequationofstateNFluid}) and (\ref{eq:FLmuitoL}).

\subsection{Single Fluid FL Cosmologies} \label{ssec:FLFCUniverses}

In this case we consider solutions to the Friedmann equation with $n = 1$, \mbox{$\tfrac{2}{3} < \gamma \leq 2$} and arbitrary curvature.  We choose the conformal factor to be $\lambda = \lambda_1$.  The Friedmann DE (\ref{eq:FLfriedmann_conf}) assumes the form
\begin{equation}
\frac{1}{\ell^2} \left( \frac{d\ell}{d\eta} \right)^2 = ( \lambda \ell )^{-3 \gamma + 2} - K.
\end{equation}  We perform the change of variable
\begin{equation}
\mathcal{L} = ( \lambda \ell )^{1 / \beta}, \qquad \mathcal{T} = \frac{\eta}{\beta},
\end{equation} with
\begin{equation}
\beta = \frac{2}{3 \gamma - 2},
\end{equation} so as to obtain
\begin{equation}
\left( \frac{d\mathcal{L}}{d\mathcal{T}} \right)^2 = 1 - K \mathcal{L}^2.
\end{equation}  Then from (\ref{aFLsol:case1}) and (\ref{aFLsol:case2}), the solution is
\begin{equation} \label{eq:FLsolutionszerolambda1}
\mathcal{L} = S_K( \mathcal{T} ),
\end{equation} where
\begin{equation} \label{eq:FLsolutionS_K}
S_K(\mathcal{T}) = \left\{ \begin{array}{cl} \sinh(\mathcal{T}), & \quad \mbox{open FL}\ \ (K = -1), \\[0.3ex] \mathcal{T}, & \quad \mbox{flat FL}\ \ (K = 0), \\[0.3ex] \sin(\mathcal{T}), & \quad \mbox{closed FL}\ \ (K = +1). \end{array} \right.
\end{equation}  Thus,
\begin{equation} \label{eq:FLsolution-FC-recollapse}
\ell(\eta) = \lambda^{-1} \left[ S_K \left( \tfrac{\eta}{\beta} \right) \right]^{\beta}.
\end{equation}  Using (\ref{eq:FLmuitoL}), the energy density is given by
\begin{equation} \label{eq:FLsolution-FC-EnergyDensity}
\mu = 3 \lambda^2 L^{-3 \gamma}, \qquad L = \lambda \ell.
\end{equation}

This solution, for arbitrary $\gamma$, was first given by Harrison (1967) (see equations (152), (18) and (32) in Harrison (1967), for open, flat and closed models, respectively).  Specific cases were discovered earlier, however.   Friedmann (1922) gave this solution in the case of dust and positive curvature (\textit{i.e.} $\gamma = 1$ and $K = +1$) and Tolman (1931) gave this solution in the case of radiation and positive curvature (\textit{i.e.} $\gamma = \tfrac{4}{3}$ and $K = +1$).

These solutions describe two qualitatively different classes of models: ever-expanding models if $K = -1$ or $0$, and recollapsing models if $K = +1$.  In this case, it follows from (\ref{eq:FLsolutionS_K}) and (\ref{eq:FLsolution-FC-recollapse}) that recollapse (\textit{i.e.} $\ell \to 0$ as $\eta \to \eta_f$) will occur at
\begin{equation}
\eta_{f} = \beta \pi.
\end{equation}


For flat FL ($K = 0$) we can obtain $\ell = \ell(t)$ explicitly by integrating (\ref{eq:FLfriedmannDE}) with $n = 1$ and $K = 0$.  The result is
\begin{equation} \label{eq:FLsolution-F}
\ell(t) = \lambda^{-1} \left( \frac{3 \gamma}{2} \lambda t \right)^{\tfrac{2}{3 \gamma}}.
\end{equation}  The parameter $\lambda$ is not essential, since by rescaling $\ell$ it can be assigned any value.  Equation (\ref{eq:FLsolution-F}) thus represents the unique flat FL universe with a single fluid.\footnote{That (\ref{eq:FLsolution-F}) does define a unique solution is confirmed by the fact that the physical quantities $H$ and $\mu$ do not depend on $\lambda$, being given by
\begin{displaymath} 
H = \frac{2}{3 \gamma} t^{-1}, \quad \mbox{and} \quad \mu = \frac{4}{3 \gamma^2} t^{-2},
\end{displaymath} on using (\ref{eq:hubblelengthscale}), (\ref{eq:FLtotaldensityparameter}) and the fact that $\Omega = 1$.}

\bigskip
In the case of radiation ($\gamma = \tfrac{4}{3}$, $\beta = 1$) one can integrate (\ref{eq:conformaltime}) to obtain,
\begin{equation}
t = \left\{ \begin{array}{ll} \vspace{0.1in} \lambda^{-1} (\cosh \eta - 1), & \quad \mbox{open FL}\ \ (K = -1), \\ \vspace{0.1in} \lambda^{-1} (1 - \cos \eta), & \quad \mbox{closed FL}\ \ (K = +1). \end{array} \right.
\end{equation}  Substituting back into (\ref{eq:FLsolutionszerolambda1}) and rearranging yields
\begin{equation} \label{eq:FLsolution-RC}
\ell(t) = \lambda^{-1} \left[ 2 \lambda t - K (\lambda t)^2 \right]^{1/2}.
\end{equation}  We note that the second term here represents the contribution to the solution from spatial curvature, \textit{i.e.} the drift from flat FL.  If $K = -1$, the limit $\lambda \to \infty$, gives the Milne solution (given in section \ref{sec:nFluidFLCosmologies}, equation (\ref{eq:FLsolMilne})).

\bigskip
In the case of dust ($\gamma = 1$, $\beta = 2$) one can integrate (\ref{eq:conformaltime}) in closed form, but cannot invert the expression in terms of elementary functions.  In this case $t$ and $\eta$ are related by
\begin{equation}
t = \left\{ \begin{array}{ll} \vspace{0.1in} \tfrac{1}{2} \lambda^{-1} ( \sinh \eta - \eta), & \quad \mbox{open FL}\ \ (K = -1), \\ \vspace{0.1in} \tfrac{1}{2} \lambda^{-1} ( \eta - \sin \eta ), & \quad \mbox{closed FL}\ \ (K = +1). \end{array} \right.
\end{equation}

\subsection{FL Vacuum Cosmologies} \label{ssec:FLCLUniverses}

In this case we solve the Friedmann equation where the only fluid component is a cosmological constant, \textit{i.e.} $n = 1$, $\gamma_1 = 0$.  We choose the conformal factor to be $\lambda = \lambda_1$, which leads to
\begin{equation} \label{eq:FLsolution-CL-lambda}
\lambda = \sqrt{\tfrac{1}{3} \Lambda},
\end{equation} (see equation (\ref{eq:FLpreferredlambda})).  The Friedmann equation (\ref{eq:FLfriedmannDE}) in this case is
\begin{equation}
\left( \frac{d\ell}{dt} \right)^2 = ( \lambda \ell)^2 - K.
\end{equation}  We perform the change of variable
\begin{equation}
\mathcal{L} = \lambda \ell, \qquad \mathcal{T} = \lambda t,
\end{equation} so as to obtain
\begin{equation} \label{eq:FLVacuumDE}
\left( \frac{d\mathcal{L}}{d\mathcal{T}} \right)^2 = \mathcal{L}^2 - K.
\end{equation}  This DE can only be solved subject to $\mathcal{L}(0) = 0$ and $\frac{d\mathcal{L}}{d\mathcal{T}} > 0$ in the case of open FL, with the solution given by (\ref{aFLsol:case1}).  For the case of non-negative curvature, we instead require that $\mathcal{L}(0) = 1$ and that $\mathcal{L}$ is increasing for all $\mathcal{T} > 0$.  It follows that (\ref{eq:FLVacuumDE}) has a unique solution in each case, leading to
\begin{equation} \label{eq:FLcosmologicalconstantsoln}
\ell(t) = \left\{ \begin{array}{ll}
\vspace{0.1in} \lambda^{-1} \sinh ( \lambda t ), & \quad \mbox{open FL}\ \ (K = -1), \\
\vspace{0.1in} \lambda^{-1} \exp \left( \lambda t \right), & \quad \mbox{flat FL}\ \ (K = 0), \\
\lambda^{-1} \cosh \left( \lambda t \right), & \quad \mbox{closed FL}\ \ (K = +1),
\end{array} \right.
\end{equation} where $\lambda$ is given by (\ref{eq:FLsolution-CL-lambda}).  The energy density is constant and given by
\begin{equation}
\mu = 3 \lambda^2 = \Lambda.
\end{equation}

These solutions were first given by Robertson (1933), de Sitter (1917) and Lanczos (1922) in the case of negative, zero and positive curvature, respectively.  They correspond to the orbits $M^{+} \to deS^{+}$ for $K = -1$, $deS^{+}$ for $K = 0$ and $deS^{-} \to deS^{+}$ for $K = +1$, along the 1-fluid cosmological constant manifold in figure \ref{fig:FLsolutions}.  In the case of flat FL, (\ref{eq:FLcosmologicalconstantsoln}) gives the \textit{de Sitter} solution, described in section section \ref{sec:nFluidFLCosmologies}, equation (\ref{eq:FLsolDeSitter}).

Often, in the case of the de Sitter solution, it is preferrable to use conformal time instead of clock time.  In particular, when $K = 0$ this choice of time variable is defined on an interval $(-\infty, \eta_f)$ (see Appendix \ref{app:AlternativeTime}).  Upon defining
\begin{equation} \label{eq:FLdeSitterTildeEta}
\tilde{\eta} = \eta_f - \eta,
\end{equation} the length scale (\ref{eq:FLcosmologicalconstantsoln}) for a flat background is simply
\begin{equation} \label{eq:FLdeSittersoln_c}
\ell(\eta) = \frac{1}{\lambda \tilde{\eta}}.
\end{equation}

\subsection{Flat FL with a Single Fluid and $\Lambda$} \label{ssec:FLFLUniverses}

In this case we solve the Friedmann equation with $n = 2$, \mbox{$\gamma_1 = \gamma$} arbitrary, \mbox{$\gamma_2 = 0$} and zero curvature (\textit{i.e.} $K = 0$).  We choose the conformal factor to be $\lambda = \lambda_2$, which leads to (\ref{eq:FLsolution-CL-lambda}).  As described in section \ref{ssec:ObservationalParameters}, we can use the freedom to scale $\ell$ to set $\lambda_1 = \lambda$.  The Friedmann equation (\ref{eq:FLfriedmannDE}) then takes the form
\begin{equation}
\left( \frac{d\ell}{dt} \right)^2 = ( \lambda \ell )^{- 3 \gamma + 2} + ( \lambda \ell )^{2}.
\end{equation}  We apply the change of variable
\begin{equation}
\mathcal{L} = (\lambda \ell)^{\frac{3}{2} \gamma}, \qquad \mathcal{T} = \tfrac{3}{2} \gamma \lambda t
\end{equation} so as to write (\ref{eq:FLfriedmannDE}) as
\begin{equation}
\left( \frac{d\mathcal{L}}{d\mathcal{T}} \right)^{2} = 1 + \mathcal{L}^2.
\end{equation}  Then using (\ref{aFLsol:case1}), we obtain
\begin{equation}
\mathcal{L} = \sinh \mathcal{T},
\end{equation} and so
\begin{equation} \label{eq:FLFlatCosmoSoln}
\ell(t) = \lambda^{-1} \left[ \sinh \left( \frac{3 \gamma}{2} \lambda t \right) \right]^{\frac{2}{3 \gamma}}.
\end{equation}  Using (\ref{eq:FLmuitoL}), the energy density is given by
\begin{equation} \label{eq:FLFlatCosmoEnergyDensity}
\mu = \mu_f + \mu_\Lambda = 3 \lambda^2 L^{-3 \gamma} + \Lambda, \qquad L = \lambda \ell.
\end{equation}  These solutions were first given by Harrison (1967) in equation (23).  They represent ever-expanding models and correspond to the orbit $F^{+} \to deS^{+}$ along the $K = 0$ invariant set, as in figure \ref{fig:FLsolutions}.

\subsection{FL Cosmologies with Radiation and Dust ($RDC$-Universes)} \label{ssec:FLRDCUniverses}

In this case we solve the Friedmann equation with $n = 2$, \mbox{$\gamma_1 = \tfrac{4}{3}$} and \mbox{$\gamma_2 = 1$}.  We choose $\lambda = \lambda_2$ as the conformal parameter.  Then, by (\ref{eq:FLM}), the single mass parameter is given by
\begin{equation}
m = m_r = \left( \frac{\lambda_r}{\lambda} \right)^{-2}.
\end{equation}  The Friedmann equation (\ref{eq:FLfriedmann_conf}) assumes the form
\begin{equation}
\frac{1}{\ell^2} \left( \frac{d\ell}{d\eta} \right)^2 = m ( \lambda \ell )^{-2} + ( \lambda \ell )^{-1} - K.
\end{equation}  We apply the change of variable
\begin{equation}
\mathcal{L} = \frac{1}{\sqrt{m}} (\lambda \ell), \qquad \mathcal{T} = \eta,
\end{equation} which gives
\begin{equation} \label{eq:FL2fluidDRDE}
\left( \frac{d\mathcal{L}}{d\mathcal{T}} \right)^2 = 1 + \frac{1}{\sqrt{m}} \mathcal{L} - K \mathcal{L}^2.
\end{equation}  Using (\ref{aFLsol:case1}) and (\ref{aFLsol:case2}), the solution is
\begin{equation}
\mathcal{L} = S_K(\mathcal{T}) + \frac{1}{\sqrt{m}} S_K(\tfrac{1}{2} \mathcal{T})^2,
\end{equation} and so
\begin{equation} \label{eq:FLdrcSolutions}
\ell(\eta) = \lambda^{-1} \left( \sqrt{m} S_K(\eta) + S_K(\tfrac{1}{2} \eta)^2 \right),
\end{equation} where $S_K$ is given by (\ref{eq:FLsolutionS_K}).  Using (\ref{eq:FLmuitoL}), the energy density is given by
\begin{equation} \label{eq:FLdrcEnergyDensity}
\mu = \mu_r + \mu_d = 3 \lambda^2 m L^{-4} + 3 \lambda^2 L^{-3}, \qquad L = \lambda \ell.
\end{equation}  In this case, one can integrate (\ref{eq:conformaltime}) so as to obtain an expression for the clock time $t = t(\eta)$, but cannot invert this expression to obtain $\ell = \ell(t)$:
\begin{equation}
t = \left\{ \begin{array}{ll}
\vspace{0.1in} \displaystyle \lambda^{-1} \left( \sqrt{m} ( \cosh \eta - 1 ) + \tfrac{1}{2} (\sinh \eta - \eta) \right), & \quad \mbox{open}\ \ (K = -1), \\
\vspace{0.1in} \displaystyle \lambda^{-1} \left( \tfrac{1}{2} \sqrt{m} \eta^2 + \tfrac{1}{12} \eta^3 \right), & \quad \mbox{flat}\ \ (K = 0), \\
\displaystyle \lambda^{-1} \left( \sqrt{m} ( 1 - \cos \eta ) + \tfrac{1}{2} ( \eta - \sin \eta ) \right), & \quad \mbox{closed}\ \ (K = +1).
\end{array} \right.
\end{equation}  The solution (\ref{eq:FLdrcSolutions}) contains the one-fluid dust and radiation solutions given in (\ref{eq:FLsolution-FC-recollapse}) as special cases.  One can see this by writing (\ref{eq:FLdrcSolutions}) in terms of parameters $\lambda_r = \lambda_1$ and $\lambda_d = \lambda_2$,
\begin{equation}
\ell(\eta) = \lambda_r^{-1} S_K(\eta) + \lambda_d^{-1} S_K(\tfrac{1}{2} \eta)^2,
\end{equation} and letting $\lambda_r \to \infty$ or $\lambda_d \to \infty$.

These solutions were first given\footnote{Harrison reports that in the case $K = +1$ this solution was given by Lema\^{i}tre (1927), but we have not verified this reference.} by Chernin (1966).  They describe all orbits in the state space for $RDC$-universes with $H > 0$ in some epoch (see figure \ref{fig:FL2fluidmodel-DR}) and hence describe two qualitatively different classes of models:  ever-expanding models if $K = -1$ (the orbits $F_1^{+} \to M^{+}$) or $K = 0$ (the orbit $F_1^{+} \to F_2^{+}$) and recollapsing models if $K = +1$ (the orbits $F_1^{+} \to F_1^{-}$).  The dimensionless parameter $m$ determines key physical properties of the models.  First, it determines the epoch of matter-radiation equality ($\mu_r = \mu_d$) according to
\begin{equation}
L_{eq} = m,
\end{equation} as follows from (\ref{eq:FLdrcEnergyDensity}).  Second, for recollapsing models, $m$ determines the time of recollapse, according to
\begin{equation}
\eta_f = 2 (\pi - \mathrm{arctan}(2 \sqrt{m})),
\end{equation} as follows from (\ref{eq:FLdrcSolutions}).

\subsection{FL Cosmologies with Radiation and $\Lambda$ ($RC\Lambda$-Universes)} \label{ssec:FLRCLUniverses}

In this case we solve the Friedmann equation with $n = 2$, \mbox{$\gamma_1 = \tfrac{4}{3}$}, \mbox{$\gamma_2 = 0$} and arbitrary curvature.  We choose the conformal factor to be $\lambda = \lambda_2$ so that (\ref{eq:FLsolution-CL-lambda}) holds.  Then, by (\ref{eq:FLM}), the single mass parameter is given by
\begin{equation}
m = m_1 = \left( \frac{\lambda_1}{\lambda} \right)^{-2}.
\end{equation}  The Friedmann equation (\ref{eq:FLfriedmannDE}) then takes the form
\begin{equation}
\left( \frac{d\ell}{dt} \right)^2 = m ( \lambda \ell )^{- 2} + ( \lambda \ell )^{2} - K.
\end{equation}  Upon making a change of variable according to
\begin{equation}
\mathcal{L} = \frac{1}{\sqrt{m}} (\lambda \ell)^2, \qquad \mathcal{T} = 2 \lambda t,
\end{equation} we obtain
\begin{equation}
\left( \frac{d\mathcal{L}}{d\mathcal{T}} \right)^2 = 1 - \frac{1}{\sqrt{m}} K \mathcal{L} + \mathcal{L}^2.
\end{equation}  Using (\ref{aFLsol:case1}), the solution is
\begin{equation}
\mathcal{L} = \sinh \mathcal{T} - \frac{1}{\sqrt{m}} K \sinh^2( \tfrac{1}{2} \mathcal{T} ),
\end{equation} and so
\begin{equation} \label{eq:FLradiationDESolution}
\ell(t) = \lambda^{-1} \left[ \sqrt{m} \sinh(2 \lambda t) - K \sinh^2 (\lambda t) \right]^{1/2}.
\end{equation}  Using (\ref{eq:FLmuitoL}), the energy density is given by
\begin{equation} \label{eq:FLrclEnergyDensity}
\mu = \mu_r + \mu_\Lambda = 3 \lambda^2 m L^{-4} + \Lambda, \qquad L = \lambda \ell.
\end{equation}  The solution (\ref{eq:FLradiationDESolution}) contains the $C\Lambda$ vacuum solutions given in (\ref{eq:FLcosmologicalconstantsoln}) and the one-fluid radiation solution given in (\ref{eq:FLsolution-RC}) as special cases.  One can see this by writing (\ref{eq:FLradiationDESolution}) in terms of parameters $\lambda_r = \lambda_1$ and $\lambda_\Lambda = \lambda_2$,
\begin{equation}
\ell(\eta) = \left[ \lambda_r^{-1} \lambda_\Lambda^{-1} \sinh(2 \lambda_\Lambda t) - K \lambda_\Lambda^{-2} \sinh^2(\lambda_\Lambda t) \right]^{1/2},
\end{equation} and letting $\lambda_r \to \infty$ or $\lambda_\Lambda \to 0$.  Also, in the case of zero curvature ($K = 0$), (\ref{eq:FLradiationDESolution}) reduces to (\ref{eq:FLFlatCosmoSoln}) with $\gamma = \tfrac{4}{3}$.

This solution was first given by Harrison (1967) (see equations (59) and (38), for the case of negative ($K = -1$) and positive ($K = +1$) curvature, respectively).  It describes all orbits in the state space for $RC\Lambda$-universes with $H > 0$ in some epoch (see figure \ref{fig:FL2fluidmodel-HQ}) and an initial singularity.  Hence, this solution describes both ever-expanding models and recollapsing models.  The parameter $m$ determines key physical properties of the models.   First, it determines the epoch of radiation-cosmological constant equality ($\mu_r = \mu_\Lambda$) according to
\begin{equation}
L_{eq} = m^{\frac{1}{4}},
\end{equation} as follows from (\ref{eq:FLsolution-CL-lambda}) and (\ref{eq:FLrclEnergyDensity}).  Second, for recollapsing models, $m$ determines the time of recollapse according to
\begin{equation}
t_f = \lambda^{-1} \mathrm{arctanh}(2 \sqrt{m}).
\end{equation}  This result follows upon rewriting (\ref{eq:FLradiationDESolution}) using standard hyperbolic identities as
\begin{equation}
\ell(t) = \lambda^{-1} \left[ \tfrac{1}{2} \sinh (2 \lambda t) \left( 2 \sqrt{m} - K \tanh ( \lambda t ) \right) \right]^{1/2}.
\end{equation}

We note that at the critical value of $m$, denoted $m^{\ast}$ and given by $m^{\ast} = \tfrac{1}{4}$, the solution (\ref{eq:FLradiationDESolution}) simplifies to
\begin{equation}
\ell(t) = \lambda^{-1} \left[ \tfrac{1}{2} (1 - \exp(-2 \lambda t)) \right]^{\frac{1}{2}}.
\end{equation}  This solution is future asymptotic to the Einstein static solution and corresponds to the orbit $F^{+} \to E$ in figure \ref{fig:FL2fluidmodel-HQ}.


\subsection{Explicit Solutions in Terms of Observational Parameters}

In sections \ref{ssec:FLFCUniverses}-\ref{ssec:FLRCLUniverses}, we have presented the explicit solutions using intrinsic parameters, namely, a conformal factor $\lambda$, the curvature indicator $K$ and, for the 2-fluid solutions with non-zero curvature, a dimensionless mass parameter $m$, which describes the relative significance of the two fluids.

In order to link with observations it is desirable to write the solutions in terms of the observational parameters.  In order to do so we need to use the formula that relates the intrinsic parameters $\lambda_i$ to the observational parameters, derived in section \ref{ssec:ObservationalParameters}.  In the case of $K \neq 0$, we can use equations (\ref{eq:FLell0itoObs1}) and (\ref{eq:FLDensityObservational}) to write
\begin{equation} \label{eq:lambdaitoObs}
\lambda_i^{-2} = \ell_0^2 \left( \frac{\Omega_{i,0}}{ \vert \Omega_{k,0} \vert} \right)^{\frac{2}{3 \gamma_i - 2}}.
\end{equation}  One can think of (\ref{eq:lambdaitoObs}) as relating $\lambda_i^{-1}$, the length scale determined by the $\mathrm{i}^{\mathrm{th}}$ fluid, to $\ell_0$, the length scale determined by the spatial curvature.  In particular, we will need the following special cases of (\ref{eq:lambdaitoObs}):
\begin{align}
\label{eq:lambdaRitoObs} \lambda_r^{-2} &= \ell_0^2 \left( \frac{\Omega_{r,0}}{\vert \Omega_{k,0} \vert} \right), & \mbox{Radiation ($\gamma_r = \tfrac{4}{3}$),} & \\
\label{eq:lambdaDitoObs} \lambda_d^{-2} &= \ell_0^2 \left( \frac{\Omega_{d,0}}{\vert \Omega_{k,0} \vert} \right)^{2}, & \mbox{Dust ($\gamma_d = 1$),} &
\end{align}  Note that the Friedmann equation (\ref{eq:FLomegarelation}), evaluated at $t_0$ gives
\begin{equation} \label{eq:FriedmannOmegaObs}
\Omega_{k,0} = 1 - \Omega_0,
\end{equation} where $\Omega_0$ is the total density parameter evaluated at $t_0$.  In the case of a cosmological constant (\textit{i.e.} $\gamma = 0$), we can obtain an expression for $\lambda_\Lambda$ directly from (\ref{eq:FLDensityObservational}), as follows:
\begin{equation} \label{eq:lambdaLitoObs}
\lambda_\Lambda^{-2} = \frac{1}{\Omega_{\Lambda,0} H_0^2} = \frac{3}{\Lambda}, \qquad \mbox{($\gamma_\Lambda = 0$).}
\end{equation}

In the case of $K = 0$ and $n > 1$, we use the freedom to scale $\ell$ in order to set $\lambda_1$ equal to $\lambda_n$.  Then, as a result of (\ref{eq:FLDensityObservational}), the length scale $\ell_0$ is determined explicitly in terms of $\Omega_{i,0}$ and $H_0$ according to
\begin{equation}
(\ell_0 H_0)^{6 (\gamma_1 - \gamma_n)} = \Omega_{1,0}^{3 \gamma_n - 2} \Omega_{n,0}^{- (3 \gamma_1 - 2)}.
\end{equation}  The intrinsic parameters $\lambda_i$ are then obtained via (\ref{eq:FLDensityObservational}).

There is no work to be done in writing the densities in observational form, since they are given in general by equation (\ref{eq:FLObservationalEnergyDensity}), which we repeat for convenience:
\begin{equation}
\mu_i = 3 H_0^2 \Omega_{i,0} a^{3 \gamma_i},
\end{equation} where $a = \ell / \ell_0$.

In order to illustrate this process we give the length scale function for the $RDC$-universes in terms of the observational parameters.  We first rewrite $\ell(\eta)$ in (\ref{eq:FLdrcSolutions}) in terms of $\lambda_r$ and $\lambda_d$:
\begin{equation}
\ell(\eta) = \lambda_r^{-1} S_K(\eta) + \lambda_d^{-1} S_K(\tfrac{1}{2} \eta)^2.
\end{equation}  Then (\ref{eq:lambdaRitoObs}) and (\ref{eq:lambdaDitoObs}) immediately give
\begin{equation}
\ell(t) = \ell_0 \left[ \left( \frac{\Omega_{r,0}}{\vert 1 - \Omega_0 \vert} \right)^{1/2} S_K(\eta) + \left( \frac{\Omega_{d,0}}{\vert 1 - \Omega_0 \vert} \right) S_K(\tfrac{1}{2} \eta)^2 \right],
\end{equation} where $\Omega_0 = \Omega_{r,0} + \Omega_{d,0}$.  Then $\Omega_{r,0} = 0$ gives the $DC$-universe and $\Omega_{d,0} = 0$ gives the $RC$-universe.  Peacock (1999) refers to the terms
\begin{displaymath}
\frac{\Omega_{d,0}}{2 \vert 1 - \Omega_0 \vert}, \quad \mbox{and} \quad \frac{\Omega_{r,0}}{2 \vert 1 - \Omega_0 \vert}
\end{displaymath} as the ``dimensionless masses'' (p79), when discussing RDC-universes.  Note that these quantities are closely related to those given in (\ref{eq:lambdaRitoObs}) and (\ref{eq:lambdaDitoObs}), upon applying (\ref{eq:FriedmannOmegaObs}).  These quantities also appear in the DC- and RC-universes as given by Coles and Lucchin (1995) (see p39, eq (2.4.2) for dust in terms of $\eta$ and p41, eq (2.5.2) for radiation in terms of $t$).

\section{Discussion}

We summarize the principal features of the dynamics of $n$-fluid FL universes.

\begin{itemize}
\item[i)] $n$-fluid FL universes have at most one epoch of expansion ($H > 0$) and at most one epoch of acceleration ($q < 0$, or equivalently, $w < - \tfrac{1}{3}$).  These results depend crucially on the assumption that the fluids are non-interacting.  We refer to Tolman (1934), p402 and p429-431, and to Clifton and Barrow (2007) for a discussion of models in which this assumption is not made.

\item[ii)] On approach to a singularity (either initial or final, $\ell \to 0$) the $1$-fluid is dominant (\textit{i.e.} $\chi_1 \to 1$) and the matter content dominates the spatial curvature ($\Omega \to 1$, $\Omega_k \to 0$).  In a typical ever-expanding model ($\ell \to \infty$) the $n$-fluid is dominant ($\chi_n \to 1$), and the matter content dominates the spatial curvature if and only if $\gamma_n < \tfrac{2}{3}$.

\item[iii)] The $n$-fluid FL cosmologies admit $n-1$ conserved quantities formed from the density parameters $\Omega_1, \ldots, \Omega_n$.  In particular, they admit one curvature quantity and $n-2$ matter quantities.
\end{itemize}

We now comment on an important difference between $n$-fluid FL cosmologies with $n > 2$ and $2$-fluid cosmologies.  For a $2$-fluid FL model, there are two essential parameters, a conformal parameter $\lambda$ and a mass parameter $m_1$, or equivalently, two observational parameters $\Omega_{1,0}$ and $\Omega_{2,0}$.  There is also one conserved quantity $\mathcal{K}$ that keeps track of the spatial curvature.  For a $3$-fluid FL model there is an additional mass parameter $m_2$, and hence three observational parameters $\Omega_{1,0}$, $\Omega_{2,0}$ and $\Omega_{3,0}$.  There is also a second conserved quantity $\mathcal{M}$.  The new dynamical feature that arises in generalizing from a $2$-fluid model to a $3$-fluid model relates to the intermediate dynamics.  In an ever-expanding 3-fluid model the additional degree of freedom is the maximum value attained by $\chi_2$, which indicates the extent to which fluid-$2$ becomes dominant during the evolution (recall that fluid-$1$ is dominant at early times ($\chi_1 \to 1$) and fluid-$3$ is dominant at late times ($\chi_3 \to 1$)).  In an $RDC\Lambda$-model of the physical universe, the maximum value of $\chi_{dust}$ is close to unity, which ensures that there is an extended matter-dominated epoch.

The dynamical systems analysis of the $n$-fluid FL universes leads to an interesting conclusion regarding the behaviour of the spatial curvature, as follows.  \textit{For any $\epsilon > 0$, there is an open set of $n$-fluid FL universes with $\gamma_n < \tfrac{2}{3}$ whose spatial curvature scalar $\Omega_k$ satisfies $\vert \Omega_k \vert < \epsilon$ throughout the evolution.}  In other words, if $\gamma_n < \tfrac{2}{3}$, in particular, if fluid $n$ is a cosmological constant, the total matter-energy content ``controls'' the spatial curvature throughout the evolution.  This result is a consequence of the fact that generic ever-expanding models satisfy $\Omega_k \to 0$ as $\ell \to 0$ and since $\gamma_n < \tfrac{2}{3}$, $\Omega_k \to 0$ as $\ell \to + \infty$.  It follows that $\vert \Omega_k \vert$ attains a maximum value $\epsilon$ which depends on the orbit in question.  By restricting to orbits that lie within a sufficiently small neighbourhood of the flat FL invariant set $\Omega_k = 0$, the value $\epsilon$ can be made arbitrarily small.  The above result has been found using different methods by Lake (2005), for $DC\Lambda$ FL universes.  He indicates that the result will also hold for $RDC\Lambda$ universes (see also Lake (2006)).  Lake comments on the significance of this result in connection with the so-called flatness problem (Lake (2005)).

One of the main results of this chapter is the analysis of the $RDC\Lambda$ FL universes and the comparisons with observations.  Discussion of these models dates back to the 1930's.  Our analysis is novel in that it gives new insight into the dynamics of these models, by representing the evolution as orbits in a compact state space.  We now give a brief survey of previous work on these models.

Tolman (1934) gives the Friedmann equation for closed $RDC\Lambda$ universes (equation (106.3), p408) and states that de Sitter has studied the integration of this equation (de Sitter (1930), (1931), not readily available).  He gives the special solutions of this DE that represent universes that are past or future asymptotic to the Einstein static universe, in the form $t = f(\ell)$ [equations (161.8) and (161.4)].

Coquereaux and Grossman (1982) give a detailed qualitative discussion of the Friedmann equation for closed $RDC\Lambda$-universes using elliptic and Weierstrass functions.  Debrowski and Stelmach (1986) extend the analysis to open $RDC\Lambda$-universes.  They give an explicit solution in terms of elementary functions when the parameters are suitably restricted.

The work that is closest to ours is that of Ehlers and Rinder (1989), who give a dynamical systems analysis of $RDC\Lambda$-universes, using Hubble-normalized variables.  They give the explicit expressions for the density parameters as functions of the scale factor [equation (19)] and also show the existence of two conserved quantities [equation (29)].  Their state space representation is inevitably incomplete since the Hubble-normalized variables are undefined at the instant of maximum expansion in recollapsing and bouncing models.

\chapter{Perturbations of FL Models}

The high degree of isotropy of the cosmic microwave background (CMB), when combined with the Copernican Principle, provides strong support for the belief that the large scale structure of the observable universe is very well described by the Friedmann-Lema\^{i}tre universes, which are exactly isotropic and spatially homogeneous.  The real universe is, of course, not exactly isotropic and spatially homogeneous, since there is complex structure associated with the observed distribution of galaxies, the overall expansion may not be exactly isotropic, and there may be primordial gravitational waves.  But the current belief of most cosmologists is that the deviations from an exact FL cosmological model are sufficiently small that they can be described by considering linear perturbations of the FL models.

\section{Historical Development}

In this section we briefly discuss the history of the theory of linear perturbations of FL cosmologies.  There are two main approaches, which we shall refer to as the \textit{metric approach} and the \textit{geometrical approach}.

\subsection{Metric Approach}

The metric approach to perturbations of the FL models was introduced by Lifshitz (1946) and subsequently discussed in greater detail by Lifshitz and Khalatnikov (1963).  In the metric approach, one distinguishes a \textit{background spacetime}, which is a FL model, and a \textit{perturbed spacetime}, which represents the physical universe.  The metric for the perturbed spacetime is written in the form
\begin{equation}
g_{ab} = g_{(0) ab} + \delta g_{ab},
\end{equation} where $g_{(0)ab}$ is the metric of the FL model and $\delta g_{ab}$ is a small perturbation, \textit{i.e.} $\vert \delta g_{ab} \vert \ll \vert g_{(0) ab} \vert$.  The fluid 4-velocity and energy density in the perturbed spacetime are likewise written in the form
\begin{eqnarray}
u^{i} & = & u^{i}_{(0)} + \delta u^{i}, \\
\mu & = & \mu_{(0)} + \delta \mu.
\end{eqnarray}  The density contrast is defined by
\begin{equation} \label{eq:DensityContrast}
\delta = \frac{\delta \mu}{\mu_{(0)}}.
\end{equation}

\subsubsection{The problem of gauge}

In defining a perturbation, one is effectively establishing a one-to-one correspondence between points in the physical spacetime and points in the background spacetime.  However, this correspondence is not unique: one may make an infinitesimal change in the correspondence keeping the background coordinates fixed.  Such a transformation is called a \textit{gauge transformation}.  A quantity that is unchanged by an infinitesimal gauge transformation is said to be \textit{gauge-invariant}.  We recommend Ellis and Bruni (1989) for a detailed explanation of gauge-invariance.

The problem of gauge invariance has plagued the study of linear perturbations since the pioneering work of Lifshitz (1946), resulting in authors publishing contradictory predictions on the behaviour of perturbations of FL cosmologies.  The reason is that the metric perturbation $\delta g_{ab}$, the velocity perturbation $\delta u^{i}$ and the density contrast $\delta$ are not gauge-invariant quantities, and hence their time evolution can depend on the choice of gauge.

A result due to Stewart and Walker (1974) gives a useful criteria for identifying gauge-invariant quantities:

\bigskip
\begin{theorem}[Stewart-Walker Lemma] \label{thm:StewartWalkerLemma} Let $T_{(0)}$ be a tensor field on a background spacetime and let $T = T_{(0)} + \Delta T$ be the corresponding tensor on a perturbed spacetime.  If $T_{(0)} = 0$ then $T$ is gauge-invariant.
\end{theorem}
\bigskip

In 1980 Bardeen reformulated the metric approach using gauge-invariant variables.  He did not make use of the Stewart-Walker Lemma in choosing the gauge-invariant variables.  Instead, he wrote out the transformation laws for the perturbations $\delta g_{ab}$, $\delta u^{a}$ and $\delta$ under an infinitesimal gauge transformation and then, by inspection, formed linear combinations of the perturbations and their derivatives that were invariant.

This approach placed the analysis of perturbations of Friedmann-Lema\^{i}tre models on a sounder mathematical foundation and has been used in much subsequent research.\footnote{We refer to Mukhanov (2005) as a recent text that uses a simplified version of the Bardeen approach, restricting considerations to perturbations of flat FL (see chapter 7).}  Nevertheless, the metric approach still possesses several shortcomings:

\begin{enumerate}
\item[i)] First, it is not clear what physical quantities Bardeen's gauge-invariant quantities correspond to.  Indeed, a gauge-invariant quantity can have different physical interpretations in different gauges.  The most important ambiguity concerns the density contrast $\delta$, as defined by (\ref{eq:DensityContrast}).  Bardeen defines two gauge-invariant quantities in terms of $\delta$, denoted by $\epsilon_m$ and $\epsilon_g$.  The first equals the density contrast $\delta$ when one uses the so-called co-moving gauge, while the second equals $\delta$ in the so-called Newtonian gauge (see Bardeen 1980, p22).

\item[ii)] Second, as emphasized by Hawking (1966), the metric tensor is not a physically significant quantity since one cannot measure it directly.

\end{enumerate}



\subsection{Geometrical Approach}

The geometrical approach has its origins in a paper by Hawking (1966).  Instead of using the components of a perturbed metric as basic variables, he proposed using the evolution equations for the kinematic quantities, the Weyl curvature and the matter density to study how perturbations evolve.  Although this paper initiated a significant new approach to studying linear perturbations of FL, the analysis of density perturbations was flawed and the paper had little impact.  Subsequently, motivated by Hawking's paper, Lyth and Mukherjee (1988) used the evolution equations for $\mu$ and $H$ to study density perturbations in an FL cosmology, although they did not do so in a fully gauge-invariant way.  The decisive step was taken by Ellis \& Bruni (1989), when they proposed using the spatial gradient of the matter density ($\hat{\nabla}_{a} \mu$) as the basic variable to describe density perturbations.  Unlike the density contrast $\delta$, the spatial gradient is a gauge-invariant quantity by the Stewart-Walker lemma, since it is zero in any FL model (see proposition \ref{prop:energymomentumisotropy}).  The first comprehensive discussion of the geometrical approach was given by Bruni, Dunsby and Ellis (1992)\footnote{The name ``geometrical approach'' was introduced in this reference.  This approach is also referred to as ``the gauge-invariant and covariant (GIC) approach''.}.  They also related the Bardeen approach to the geometrical approach, and gave clear physical interpretations of his gauge-invariant variables.

In this chapter we give an introduction to the geometrical approach to linear perturbations of FL models.

\section{The Linearized Einstein Field Equations} \label{sec:PertLinearizedEquations}

The geometrical approach is based on the conviction that one should use variables that have a direct physical or geometric meaning, and are both coordinate-independent and gauge-invariant.  One thus uses tensorial quantities that are identically zero in an FL universe, and hence gauge-invariant, on account of the Stewart-Walker lemma.

We shall see that a complete set of tensorial quantities is the following:

\bigskip \noindent
\begin{tabular}{llll}
\ \ \ i) & The density and Hubble gradients: & $\hat{\nabla}_a \mu$, $\hat{\nabla}_a H$ \\
\ \ \ ii) & The fluid kinematic quantities: & $\sigma_{ab}$, $\omega_{a}$ \\
\ \ \ iii) & The Weyl curvature: & $E_{ab}$, $H_{ab}$
\end{tabular}
\bigskip

\noindent These variables have two additional advantages:
\begin{enumerate}
\item[a)] they directly describe the deviation of the physical universe from an idealized FL universe, and
\item[b)] they are, in principle, observable, when normalized to be dimensionless using a power of $H$.  We refer in particular to the fundamental paper of Sachs and Wolfe (1967), who analyzed the observations of distant galaxies from this point of view, and Maartens et al. (1995b), who used the cosmic microwave background observations to bound these quantities.
\end{enumerate}

In the geometrical approach one thus replaces the EFEs for the metric tensor components by the equivalent system of evolution equations and constraints, satisfied by the tensorial quantities i)-iii), as given in Appendix \ref{app:evolutionandconstrainteqs}.

For simplicity we will assume the matter content consists of a perfect fluid with barotropic equation of state $p = p(\mu)$ and a cosmological constant $\Lambda$.  We will use the standard notation,
\begin{equation} \label{eq:soundspeeds}
w = \frac{p}{\mu}, \quad c_s^2 = \frac{dp}{d\mu},
\end{equation} where $c_s$ is the speed of sound.

\subsection{The Linearization Process}

In order to derive the governing equations for linear perturbations of a FL cosmology, we must distinguish between quantities which are non-zero in the background (referred to as \textit{zero-order quantities}) and those which are zero in the background (referred to as \textit{first-order quantities}).  For a FL background,  the zero-order quantities are
\begin{equation} \label{lineq:zeroorderquantities}
\mu, \quad w, \quad c_s^2 \quad \mbox{and} \quad H.
\end{equation}  It follows from propositions \ref{prop:kinematiccharacterization1}-\ref{prop:energymomentumisotropy} that the quantities
\begin{equation} \label{lineq:firstorderquantities}
\hat{\nabla}_a \mu, \quad \hat{\nabla}_a H, \quad \sigma_{ab}, \quad \omega_{a}, \quad E_{ab}, \quad H_{ab},
\end{equation} are first order.  For sake of brevity, we define
\begin{equation} \label{eq:fractionaldensitygradient}
X_a = \hat{\nabla}_a \mu,
\end{equation} and
\begin{equation} \label{eq:spatialgradienthubblescalar}
Z_a = 3 \hat{\nabla}_a H.
\end{equation}

We note that the acceleration $\dot{u}_a$ is not an independent quantity, since it can be expressed in terms of the density gradient $X_{a}$ via (\ref{a1eq:contractedbianchi4}), on using (\ref{eq:soundspeeds}) and (\ref{eq:fractionaldensitygradient}):
\begin{equation} \label{eq:densitygradientaccelerationrelation}
\dot{u}_a = - \frac{c_s^2}{(1 + w) \mu} X_a.
\end{equation}

The evolution equations for the zero-order quantities are simply the evolution equations for a FL cosmology with a single fluid and cosmological constant, given by (\ref{eveq:consenergy1}) and (\ref{eveq:raychaudhurieq1}), with $\mu$ replaced by $\mu + \Lambda$ and $p$ by $p - \Lambda$ to make the cosmological constant explicit.  We also include the evolution equation for $w$, which follows from (\ref{lineq:zeroorder2}) and (\ref{eq:soundspeeds}).

\bigskip
\noindent \begin{tabular}{|p{\textwidth}|}
\hline
\paragraph{Zero-Order Evolution Equations}
\begin{eqnarray}
\label{lineq:zeroorder1} \dot{H} & = & -H^2 - \tfrac{1}{6} (1 + 3 w) \mu + \tfrac{1}{3} \Lambda, \\
\label{lineq:zeroorder2} \dot{\mu} & = & - 3 H (1 + w) \mu, \\
\label{lineq:zeroorder3} \dot{w} & = & - 3 H (1 + w) (c_s^2 - w).
\end{eqnarray} \\
\hline
\end{tabular}

\paragraph{Note:}  The Friedmann equation (\ref{eveq:friedmanneq1}) plays an auxiliary role in that if $K \neq 0$, it serves to express $\ell(t)$ in terms of the zero-order quantities.  We use (\ref{eq:curvatureparameter}) to write (\ref{eveq:friedmanneq1}) as follows:
\begin{equation} \label{lineq:zeroorder0}
H^2 = \tfrac{1}{3} (\mu + \Lambda) - \frac{K}{\ell^2}.
\end{equation}

The evolution equations for the first order quantities are obtained from the general system of evolution equations and constraints in Appendix \ref{app:evolutionandconstrainteqs}, by a process of linearization.  There are three aspects to the \textit{linearization procedure}:

\begin{enumerate}
\item[$L_1$)]  Drop all products of the first order quantities (\ref{lineq:firstorderquantities}) and their spatial derivatives in the general non-linear evolution equations.
\item[$L_2$)]  The resulting equations form a coupled system of linear PDEs for the first order quantities, whose coefficients depend on the zero order quantities.  For example, there are terms such as $H X_a$, $H E_{ab}$ and $\mu \sigma_{ab}$.  The linearization process involves replacing the zeroth order quantity by its value in the background FL model in each of these products.
\item[$L_3$)]  The resulting equations also contain the projected spatial covariant derivative operator $\hat{\nabla}_a$ acting on first order quantities.  The linearization process involves replacing $\hat{\nabla}_a$, which is defined on the physical spacetime, by the corresponding operator on the background FL spacetime.
\end{enumerate}

\subsection{The Linearized Evolution and Constraint Equations}

We now derive the governing equations for linearized perturbations by applying the linearization procedure $L_1$ to the general system of evolution equations and constraint equations in Appendix \ref{app:evolutionandconstrainteqs}.

One aspect of the linearization process, namely dealing with the terms $\hat{\nabla}_{\langle a} \dot{u}_{b \rangle}$, $\mathsf{curl}(\dot{u}_{a})$ and $\hat{\nabla}_{a} (\hat{\nabla}^{b} \dot{u}_{b})$ involving the spatial derivative of $\dot{u}_{a}$, requires particular attention.  First, it follows from (\ref{eq:densitygradientaccelerationrelation}) and $L_1$ that\footnote{Since $p = p(\mu)$, we can use (\ref{eq:soundspeeds}) to write \begin{displaymath} \hat{\nabla}_{a} \frac{c_s^2}{(1 + w) \mu} = F(\mu) \hat{\nabla}_{a} \mu = F(\mu) X_{a}, \end{displaymath} where $F(\mu)$ is a function of $\mu$ whose specific form is unimportant.  Then the term $F(\mu) X_{a} X_{b}$ is dropped.}
\begin{equation} \label{eq:accelerationsymmetrizationderiv}
\hat{\nabla}_{\langle a} \dot{u}_{b \rangle} = \frac{- c_s^2}{(1 + w) \mu} \hat{\nabla}_{\langle a} X_{b \rangle}.
\end{equation}  Second, it follows in a similar manner from (\ref{eq:densitygradientaccelerationrelation}), (\ref{eq:fractionaldensitygradient}) and $L_1$ that
\begin{equation} \label{eq:accelerationcurlterm_a}
\mathsf{curl}(\dot{u}_{a}) = \frac{- c_s^2}{(1 + w) \mu}\ \mathsf{curl}(\hat{\nabla}_{a} \mu).
\end{equation}  Applying (\ref{adiff:curl1}) with $f = \mu$ and using (\ref{a1eq:contractedbianchi3}) gives
\begin{equation} \label{eq:accelerationcurlterm}
\mathsf{curl}(\dot{u}_{a}) = 6 c_s^2 H \omega_{a}.
\end{equation}  Finally, it again follows from (\ref{eq:densitygradientaccelerationrelation}) and repeated application of $L_1$ that
\begin{equation}
\hat{\nabla}_{a} (\hat{\nabla}^{b} \dot{u}_{b}) = \frac{- c_s^2}{(1 + w) \mu} \hat{\nabla}_{a} (\hat{\nabla}^{2} \mu).
\end{equation}  Then the commutation property (\ref{adiff:laplace1}) with $f = \mu$ and the evolution equation (\ref{a1eq:contractedbianchi3}) for $\mu$ leads to
\begin{equation} \label{eq:accelerationdoubledivergenceterm}
\hat{\nabla}_{a} (\hat{\nabla}^{b} \dot{u}_{b}) = \frac{- c_s^2}{(1 + w) \mu} \left( \hat{\nabla}^2 - \frac{2K}{\ell^2} \right) X_{a} - 6 c_s^2 H \mathsf{curl}(\omega_a).
\end{equation}

The evolution equations for the first order quantities are now obtained from (\ref{a1eq:ricciidentity2}) (\ref{a1eq:ricciidentity3}), (\ref{a1eq:fullbianchi1}), (\ref{a1eq:fullbianchi2}), (\ref{eq:fractionaldensitygradient_evol}) and (\ref{eq:spatialgradienthubblescalar_evol}) on applying the linearization procedure $L_1$ and using (\ref{eq:accelerationsymmetrizationderiv}), (\ref{eq:accelerationcurlterm}) and (\ref{eq:accelerationdoubledivergenceterm}) to eliminate terms involving spatial derivatives of $\dot{u}_{a}$.

\bigskip
\noindent \begin{tabular}{|p{\textwidth}|}
\hline
\paragraph{First-Order Evolution Equations}
\begin{eqnarray}
\label{lineq:Xevolution} \dot{X}_{\langle a \rangle} & = & - 4 H X_a - (1 + w) \mu Z_a, \\
\label{lineq:Zevolution} \dot{Z}_{\langle a \rangle} & = & - 3 H Z_a - \tfrac{1}{2} X_a - \frac{c_s^2}{(1 + w) \mu} \left( \hat{\nabla}^2 + \frac{K}{\ell^2} \right) X_a - 6 c_s^2 H\ \mathsf{curl}(\omega_a), \\
\label{lineq:sigmaevolution} \dot{\sigma}_{\langle ab \rangle} & = & -2 H \sigma_{ab} - \frac{c_s^2}{(1 + w) \mu} \hat{\nabla}_{\langle a} X_{b \rangle} - E_{ab}, \\
\label{lineq:omegaevolution} \dot{\omega}_{\langle a \rangle} & = & (3 c_s^2 - 2) H \omega_a, \\
\label{lineq:weylEevolution} \dot{E}_{\langle ab \rangle} & = & - 3 H E_{ab} + \mathsf{curl}(H_{ab}) - \tfrac{1}{2} (1 + w) \mu \sigma_{ab}, \\
\label{lineq:weylHevolution} \dot{H}_{\langle ab \rangle} & = & - 3 H H_{ab} - \mathsf{curl}(E_{ab}).
\end{eqnarray} \\
\hline
\end{tabular}
\bigskip

Similarly, constraints on the first-order quantities are obtained on applying the linearization procedure $L_1$ to (\ref{a1eq:ricciidentity4}), (\ref{a1eq:ricciidentity5}), (\ref{a1eq:ricciidentity6}), (\ref{a1eq:fullbianchi3}) and (\ref{a1eq:fullbianchi4}).

\bigskip
\noindent \begin{tabular}{|p{\textwidth}|}
\hline
\paragraph{First-Order Constraint Equations}
\begin{eqnarray}
\label{lineq:sigmaconstraint} \hat{\nabla}^b \sigma_{ab} & = & \tfrac{2}{3} Z_a +  \mathsf{curl}(\omega_a), \\
\label{lineq:omegaconstraint} \hat{\nabla}^a \omega_a & = & 0, \\
\label{lineq:curlsigmaconstraint} \hat{\nabla}_{\langle a} \omega_{b \rangle} & = & H_{ab} - \mathsf{curl}( \sigma_{ab} ), \\
\label{lineq:Eweylconstraint} \hat{\nabla}^b E_{ab} & = & \tfrac{1}{3} X_a, \\
\label{lineq:Hweylconstraint} \hat{\nabla}^b H_{ab} & = & (1 + w) \mu \omega_a.
\end{eqnarray} \\
\hline
\end{tabular}
\bigskip

There are two additional constraints involving the curls of the spatial gradients $X_a$ and $Z_a$.  Indeed, using (\ref{eq:densitygradientaccelerationrelation}), (\ref{eq:accelerationcurlterm_a}) and (\ref{eq:accelerationcurlterm}) leads to
\begin{equation} \label{eq:densitygradientvorticityrelation}
\mathsf{curl}(X_a) = 6 (1 + w) \mu H \omega_a.
\end{equation}  This equation is in fact a consequence of the main system of evolution equations and constraints and hence we list it as an auxiliary equation.  There is an analogous equation for $\mathsf{curl}(Z_a)$ which we shall not need.

\paragraph{Comments:}
\begin{enumerate}
\item[i)] The evolution equations (\ref{lineq:Xevolution})-(\ref{lineq:weylHevolution}) and constraints (\ref{lineq:sigmaconstraint})-(\ref{lineq:Hweylconstraint}) form a closed system of linear first order  PDEs for the first order quantities (\ref{lineq:firstorderquantities}).  The coefficients $\mu$, $w$, $c_s^2$ and $H$ are zero-order quantities and are evaluated in the background FL model.  Note that if $K \neq 0$ the term $K / \ell^2$ in (\ref{lineq:Zevolution}) is given in terms of zero order quantities by the constraint (\ref{lineq:zeroorder0}).

\item[ii)] It can be shown that the system of evolution equations and constraints is consistent: that is, the time derivatives of each constraint equation (\ref{lineq:sigmaconstraint})-(\ref{lineq:Hweylconstraint}) are identically satisfied as a consequence of the other equations.  As a result, once a set of initial data is chosen to satisfy the constraint equations, these equations will hold at all later times (see Bruni et al. (1992), p42).

\item[iii)]  Equations (\ref{lineq:Xevolution})-(\ref{lineq:weylHevolution}) and (\ref{lineq:sigmaconstraint})-(\ref{lineq:Hweylconstraint}) form the basis for the geometrical approach to the analysis of linear perturbations of FL models with a barotropic perfect fluid and cosmological constant as source.  They can be extracted from Bruni et al. (1992) by specializing their more general equations to the case of a barotropic perfect fluid (set $q_a$, $\pi_{ab}$ and $\mathscr{E}$ to zero in equations (50)-(54), (57)-(62)).  Most authors, however, do not make use of the full system of evolution and constraint equations.
\end{enumerate}

\section{Harmonic Decomposition} \label{ssec:spatialharmonics}

Before proceeding, we must discuss the fundamental role of spatial harmonics in the geometrical theory.  The harmonic decomposition is analogous to a Fourier decomposition of a quantity that is defined along the background 3-spaces.  Hence, harmonics are defined so as to be constant along flow lines (\textit{i.e.}, independent of proper time) and orthogonal to the fluid 4-velocity $\mathbf{u}$.  We refer the reader to Appendix \ref{app:covariantharmonics} for a complete discussion of spatial harmonics in the geometrical approach.

The key idea behind the harmonic decomposition is that a scalar, vector or tracefree symmetric rank 2 tensor orthogonal to $u^a$ can be written as a linear combination of scalar, vector and tensor harmonics with purely time-dependent coefficients.  In the literature it is tacitly assumed that the harmonics are complete in some appropriate space of tensor fields.

The expansions are as follows:
\begin{eqnarray}
\label{eq:harmonicbasis1} T & = & \sum_k T_{(0,k)} Q^{(0)}, \\
\label{eq:harmonicbasis2} T_{a} & = & \sum_k T_{(0,k)} Q^{(0)}_{a} + \sum_k T_{(1,k)} Q^{(1)}_{a}, \\
\label{eq:harmonicbasis3} T_{ab} & = & \sum_k T_{(0,k)} Q^{(0)}_{ab} + \sum_k T_{(1,k)} Q^{(1)}_{ab}  + \sum_k T_{(2,k)} Q^{(2)}_{ab}.
\end{eqnarray}  This decomposition has two consequences:  Firstly, it decouples the temporal and spatial dependence of each field, since all coefficients are purely functions of time and harmonics are constant along flow lines, \textit{i.e.} $T_{(d,k)} = T_{(d,k)}(t)$ and $\dot{Q} = 0$ for all harmonics $Q$ (see Appendix \ref{app:covariantharmonics}).  Second, in the framework of linear perturbation theory, it removes any co-dependence between scalar, vector and tensor harmonics.  As in Bruni et al. (1992, p51), the summation in (\ref{eq:harmonicbasis1})-(\ref{eq:harmonicbasis3}) can be over a discrete set or an integral over a continuously varying index.

We note that the scalar decomposition (\ref{eq:harmonicbasis1}) is analogous to the $3$-dimensional Fourier transform for scalar fields.  The vector decomposition (\ref{eq:harmonicbasis2}) is similarly analogous to the Fourier transform, except applied to the curl-free (scalar) component and the divergence-free (vector) component of the vector field.  The tensor decomposition (\ref{eq:harmonicbasis3}) separates the curl-free divergent (scalar) component, the remaining divergent (vector) component and the divergence-free (tensor) component of the vector field.

Using this decomposition, it is natural to consider the kinematic quantities and Weyl tensor components of the real universe to be the fundamental description of the linear perturbations (since they are each gauge-invariant, each of these components can be expanded using the harmonic basis functions).

The decoupling of the perturbation types is extremely important in the linearized theory since it allows us to consider scalar, vector and tensor perturbations independently.  For each perturbation type, we will proceed as follows:  First, we first define a \textit{basic variable}.  Second, we use the linearized evolution equations (\ref{lineq:zeroorder1})-(\ref{lineq:Hweylconstraint}) to derive a second order differential equation which describes the evolution of the basic variable, which we shall call the \textit{governing DE}.  Finally, we use the linearized evolution and constraint equations to express all kinematic quantities and Weyl tensor components in terms of the basic variable, hence providing a complete solution to the linearized equations for each type of perturbation.

\section{Scalar Perturbations} \label{ssec:scalarperturbations}

We first focus on \textit{scalar perturbations}, which are defined as solutions of the linearized equations (\ref{lineq:zeroorder1})-(\ref{lineq:Hweylconstraint}) that can be expanded in terms of the scalar harmonics $Q^{(0)}$, $Q^{(0)}_{a}$ and $Q^{(0)}_{ab}$, defined by equations (\ref{ahaeq:scaharmdefinition}), (\ref{ahaeq:scavecdef}) and (\ref{ahaeq:scatensdef}).  Following Ellis et al. (1989), we choose the basic quantity for scalar perturbations to be the fractional density gradient $\mathcal{D}_a$, defined as
\begin{equation} \label{eq:fractionaldensitygradientDC}
\mathcal{D}_{a} = \frac{\ell \hat{\nabla}_{a} \mu}{\mu},
\end{equation} or equivalently, using (\ref{eq:fractionaldensitygradient}),
\begin{equation} \label{eq:fractionaldensitygradientD}
\mathcal{D}_{a} = \frac{\ell}{\mu} X_{a}.
\end{equation}  We also rescale $Z_a$ according to \mbox{$\mathcal{Z}_a = \ell Z_{a}$}, \textit{i.e.}
\begin{equation} \label{eq:fractionaldensitygradientZC}
\mathcal{Z}_{a} = 3 \ell \hat{\nabla}_a H.
\end{equation}

\subsection{The Governing DE}

In this section we derive the governing DE for scalar perturbations, following Ellis et al. (1990) (also see Wainwright and Ellis (1997, p290-3)).  Since we are considering scalar harmonics, we can write
\begin{equation} \label{eq:densitygradientscalarharmonics}
\mathcal{D}_{a} = \sum_{k} \mathcal{D}_{(k)} Q^{(0,k)}_{a}.
\end{equation}

Since (\ref{eq:densitygradientscalarharmonics}) and (\ref{ahaeq:scaharmzerocurl}) imply $\mathsf{curl}(\mathcal{D}_{a}) = 0$, and hence $\mathsf{curl}(X_{a}) = 0$, (\ref{eq:densitygradientvorticityrelation}) gives
\begin{equation} \label{eq:scalarzerovorticity}
\omega_a = 0.
\end{equation}  It follows that equations (\ref{lineq:Xevolution}) and (\ref{lineq:Zevolution}) decouple from the remaining linearized evolution equations.  Expressing these equations in terms of the rescaled variables $\mathcal{D}_{a}$ and $\mathcal{Z}_{a}$ leads to
\begin{eqnarray}
\label{lineq:Devolution} \dot{\mathcal{D}}_{\langle a \rangle} & = & 3 w H \mathcal{D}_{a} - (1 + w) \mu \mathcal{Z}_{a}, \\
\label{lineq:newZevolution} \dot{\mathcal{Z}}_{\langle a \rangle} & = & -2 H \mathcal{Z}_{a} - \tfrac{1}{2} \mu \mathcal{D}_{a} - \frac{c_s^2}{(1 + w) \mu} \left( \hat{\nabla}^2 + \frac{K}{\ell^2} \right) \mathcal{D}_{a}.
\end{eqnarray}  These equations can be combined to give a second order evolution equation for $\mathcal{D}_{a}$.  On differentiating (\ref{lineq:Devolution}) and using (\ref{lineq:zeroorder1})-(\ref{lineq:zeroorder3}), (\ref{lineq:zeroorder0}) and (\ref{lineq:newZevolution}), we obtain
\begin{equation} \label{eq:scalarperturbationmasterequation_a}
\ddot{\mathcal{D}}_{\langle a \rangle} + \mathcal{A} H \dot{\mathcal{D}}_{\langle a \rangle} + \mathcal{B} H^2 \mathcal{D}_{a} - c_s^2 \left( \hat{\nabla}^2 - \frac{2K}{\ell^2} \right) \mathcal{D}_{a} = 0,
\end{equation} where
\begin{equation} \label{eq:scalarperturbationmasterequation_a1}
\mathcal{A} = 2 - 3 w - 3 (w - c_s^2),
\end{equation} and
\begin{equation} \label{eq:scalarperturbationmasterequation_a2}
\mathcal{B} = -\tfrac{3}{2} (1 - w)(1 + 3w) \Omega - 6 w \Omega_\Lambda + 3 (w - c_s^2) \bigg( \frac{K}{H^2 \ell^2} - 3 \bigg).
\end{equation}

We now expand each term in terms of scalar harmonics.  The derivatives of $\mathcal{D}_{a}$ along the fundamental congruence are obtained by differentiating (\ref{eq:densitygradientscalarharmonics}) and noting $\dot{Q}_a^{(0,k)} = 0$ (see (\ref{ahaeq:scaharmtimeinvariance})), as follows:
\begin{equation} \label{eq:densitygradientscalarharmonicsderiv}
\dot{\mathcal{D}}_{\langle a \rangle} = \sum_{k} \dot{\mathcal{D}}_{(k)} Q^{(0,k)}_{a}, \qquad \ddot{\mathcal{D}}_{\langle a \rangle} = \sum_{k} \ddot{\mathcal{D}}_{(k)} Q^{(0,k)}_{a}.
\end{equation}  In order to remove the Laplacian term from the evolution equation (\ref{eq:scalarperturbationmasterequation_a}), we use the identity (\ref{ahaeq:scavecprop2}), which can be written as
\begin{equation} \label{eq:scavecprop2x}
\left( \hat{\nabla}^2 - \frac{2 K}{\ell^2} \right) Q^{(0)}_{a} = - \frac{k^2}{\ell^2} Q^{(0)}_a.
\end{equation}  On applying (\ref{eq:densitygradientscalarharmonicsderiv}) and (\ref{eq:scavecprop2x}) to (\ref{eq:scalarperturbationmasterequation_a}), we obtain
\begin{equation} \label{eq:scalarperturbationmasterequation}
\ddot{\mathcal{D}}_{(k)} + \mathcal{A} H \dot{\mathcal{D}}_{(k)} + \left( \mathcal{B} + \frac{c_s^2 k^2}{H^2 \ell^2} \right) H^2 \mathcal{D}_{(k)}  = 0,
\end{equation} where $\mathcal{A}$ and $\mathcal{B}$ are given by (\ref{eq:scalarperturbationmasterequation_a1}) and (\ref{eq:scalarperturbationmasterequation_a2}).  This is the governing equation for scalar perturbations, in terms of clock time $t$.  This equation is equivalent\footnote{Bruni et al. (1992) make use of the scalar quantity \mbox{$\vartriangle = \hat{\nabla}^{a} \mathcal{D}_{a}$} instead of $\mathcal{D}_{(k)}$, defined in this manner so as to restrict considerations to scalar perturbations without it being necessary to resort to a harmonic expansion.  This choice of scalar quantity automatically removes all vector harmonic components from $\mathcal{D}_{a}$ since vector harmonics are divergence-free by definition (\ref{ahaeq:vecharmtimeindependence}).} to (73) in Bruni et al. (1992), upon specializing to the case of a perfect fluid, and (214) in Ellis and van Elst (1998), in the case of $\Lambda = 0$.

\subsection{Kinematic Quantities and Weyl Curvature} \label{ssec:ScalarPertQuantities}

We now write the kinematic quantities and Weyl curvature in terms of the density gradient coefficients $\mathcal{D}_{(k)}$ and the scalar harmonics $Q^{(0)}$.  Since we are considering scalar harmonics, $\mathcal{Z}_{a}$ and $\sigma_{ab}$ can be expanded as
\begin{equation} \label{eq:shearscalarharmonics}
\mathcal{Z}_{a} = \sum_{k} \mathcal{Z}_{(k)} Q^{(0,k)}_{a}, \qquad \sigma_{ab} = \sum_{k} \sigma_{(k)} Q^{(0,k)}_{ab}.
\end{equation}  We can use (\ref{lineq:Devolution}) to express $\mathcal{Z}_{(k)}$ in terms of $\mathcal{D}_{(k)}$ and $\dot{\mathcal{D}}_{(k)}$, and since $\omega_a = 0$ it follows from (\ref{lineq:sigmaconstraint}) and (\ref{ahaeq:scatenprop1}) that
\begin{equation}
\mathcal{Z}_{(k)} = \frac{(k^2 - 3 K)}{k} \sigma_{(k)}.
\end{equation}  Next, it follows from (\ref{eq:shearscalarharmonics}) and (\ref{ahaeq:scaharmzerocurl}) that $\mathsf{curl}(\sigma_{ab}) = 0$.  Hence, using (\ref{lineq:curlsigmaconstraint}) we have that
\begin{equation}
H_{ab} = 0.
\end{equation}  Expanding $E_{ab}$ in terms of scalar harmonics via
\begin{equation} \label{eq:electricweylscalarharmonics}
E_{ab} = \sum_{k} E_{(k)} Q^{(0,k)}_{ab},
\end{equation} and using (\ref{ahaeq:scatenprop1}) leads to
\begin{equation} \label{eq:electricweyldivscalarexpansion}
\ell \hat{\nabla}^b E_{ab} = \sum_{k} \frac{2}{3 k} (k^2 - 3 K) E_{(k)} Q_{a}^{(0,k)}.
\end{equation}  The coefficients $E_{(k)}$ can then be expressed in terms of $\mathcal{D}_{(k)}$ via the constraint (\ref{lineq:Eweylconstraint}).  We thus obtain the following expressions for the harmonic coefficients of the non-zero Hubble-normalized first-order quantities:

\begin{eqnarray}
\label{eq:scalarkinZ} \mathcal{Z}_{(k)} & = & - \frac{1}{(1+w)} \left[ -3 w H \mathcal{D}_{(k)} + \dot{\mathcal{D}}_{(k)} \right], \\
\label{eq:scalarkinS} \sigma_{(k)} & = & - \frac{k}{k^2 - 3K} \mathcal{Z}_{(k)}, \\
\label{eq:scalarkinE} E_{(k)} & = & \frac{k}{2 (k^2 - 3K)} \mu \mathcal{D}_{(k)}.
\end{eqnarray}

At this stage, we have satisfied all of the linearized evolution equations and constraint equations except for (\ref{lineq:sigmaevolution}) and (\ref{lineq:weylEevolution}).  We can now verify that $D_{a}$, $\sigma_{ab}$ and $E_{ab}$, as given by (\ref{eq:densitygradientscalarharmonics}), (\ref{eq:shearscalarharmonics}), (\ref{eq:electricweylscalarharmonics}) and (\ref{eq:scalarkinZ})-(\ref{eq:scalarkinE}) satisfy (\ref{lineq:sigmaevolution}) and (\ref{lineq:weylEevolution}) (via Maple).  We have thus established the following result:

\begin{proposition} \label{prop:ScalarPerturbations}  For any solution $\mathcal{D}_{(k)}$ of the governing DE (\ref{eq:scalarperturbationmasterequation}), the first order quantities given by (\ref{eq:densitygradientscalarharmonics}), (\ref{eq:shearscalarharmonics}), (\ref{eq:electricweylscalarharmonics}) and (\ref{eq:scalarkinZ})-(\ref{eq:scalarkinE}) satisfy the complete set of linearized evolution and constraint equations (\ref{lineq:Xevolution})-(\ref{lineq:Hweylconstraint}).
\end{proposition}

\section{Vector Perturbations}

In this section we focus on \textit{vector perturbations}, defined as solutions of the linearized equations (\ref{lineq:Xevolution})-(\ref{lineq:Hweylconstraint}) that can be expanded in terms of the vector harmonics $Q^{(1)}_{a}$ and $Q^{(1)}_{ab}$.  Vector perturbations are closely associated with vorticity, and so we choose the basic quantity for vector perturbations to be the co-moving vorticity vector, defined by
\begin{equation} \label{eq:dimensionlessvorticityvector}
\mathcal{W}_a = \ell \omega_a.
\end{equation}

\subsection{The Governing DE}

In this section, we derive the governing DE for vector perturbations.  On differentiating (\ref{eq:dimensionlessvorticityvector}) along the fundamental congruence and using (\ref{lineq:omegaevolution}), we obtain
\begin{equation} \label{eq:vorticityevolution}
\dot{\mathcal{W}}_{\langle a \rangle} - (3 c_s^2 - 1) H \mathcal{W}_{a} = 0.
\end{equation}

We expand (\ref{eq:vorticityevolution}) in terms of the associated vector harmonics $P^{(1,k)}_a$ as
\begin{equation} \label{eq:vorticityvectorharmonics}
\mathcal{W}_{a} = \sum_k \mathcal{W}_{(k)} P^{(1,k)}_{a}.
\end{equation}  On substituting (\ref{eq:vorticityvectorharmonics}) into (\ref{eq:vorticityevolution}), we obtain
\begin{equation} \label{eq:vectorperturbationmasterequation}
\dot{\mathcal{W}}_{(k)} - (3 c_s^2 - 1) H \mathcal{W}_{(k)} = 0.
\end{equation}  This equation is the governing DE for vector perturbations, in terms of clock time $t$.

\subsection{Kinematic Quantities and Weyl Curvature} \label{ssec:VectorPertQuantities}

We now write the kinematic quantities and Weyl curvature in terms of the co-moving vorticity coefficients $\mathcal{W}_{(k)}$ and vector harmonics $Q^{(1)}$.  By the definition of vector perturbations, all quantities can be expanded in terms of vector harmonics and associated vector harmonics, as follows:
\begin{align}
\mathcal{D}_{a} & = \sum_{k} \mathcal{D}_{(k)} Q_{a}^{(1,k)}, & \mathcal{Z}_{a} & = \sum_{k} \mathcal{Z}_{(k)} Q_{a}^{(1,k)}, & \sigma_{ab} & = \sum_{k} \sigma_{(k)} Q^{(1,k)}_{a}, \nonumber \\
E_{ab} & = \sum_{k} E_{(k)} Q_{ab}^{(1,k)}, & H_{ab} & = \sum_{k} H_{(k)} P_{ab}^{(1,k)}.  \label{eq:quantitiesvectorharmonics}
\end{align}

From (\ref{eq:densitygradientvorticityrelation}), (\ref{eq:fractionaldensitygradientD}) and (\ref{ahaeq:associatedvecharm}), we can express $\mathcal{D}_{(k)}$ in terms of $\mathcal{W}_{(k)}$, and hence obtain $\mathcal{Z}_{(k)}$ in terms of $\mathcal{W}_{(k)}$ from (\ref{lineq:Xevolution}), (\ref{eq:fractionaldensitygradientD}) and (\ref{eq:fractionaldensitygradientZC}).  Further, using (\ref{ahaeq:tenvecprop1}), the divergences of $\sigma_{ab}$, $E_{ab}$ and $H_{ab}$ can be written as follows:
\begin{eqnarray}
\ell \hat{\nabla}^{b} \sigma_{ab} & = & \sum_{k} \frac{1}{2 k} (k^2 - 2 K) \sigma_{(k)} Q_{a}, \\
\ell \hat{\nabla}^{b} E_{ab} & = & \sum_{k} \frac{1}{2 k} (k^2 - 2 K) E_{(k)} Q_{a}, \\
\ell \hat{\nabla}^{b} H_{ab} & = & \sum_{k} \frac{1}{2 k} (k^2 - 2 K) H_{(k)} Q_{a}.
\end{eqnarray}  We can now use (\ref{lineq:sigmaconstraint}), (\ref{lineq:Eweylconstraint}) and (\ref{lineq:Hweylconstraint}) to determine the coefficients $\sigma_{(k)}$, $E_{(k)}$ and $H_{(k)}$ in terms of $\mathcal{W}_{(k)}$.  We thus obtain the following expressions for the harmonic coefficients of the first-order quantities:

\begin{eqnarray}
\label{eq:vectorkinD} \mathcal{D}_{(k)} & = & 6 (1+w) H \ell \mathcal{W}_{(k)}, \\
\label{eq:vectorkinZ} \mathcal{Z}_{(k)} & = & - 3 \ell^{-1} \left[ 2 K - (1+w) \mu \ell^2 \right] \mathcal{W}_{(k)}, \\
\label{eq:vectorkinS} \sigma_{(k)} & = & \frac{2 k}{k^2 - 2K} \ell^{-1} \left[ (k^2 - 2K) + 2 (1+w) \mu_0 \ell^2 \right] \mathcal{W}_{(k)}, \\
\label{eq:vectorkinE} E_{(k)} & = & \frac{4 k}{k^2 - 2K} (1 + w) \mu H \ell \mathcal{W}_{(k)}, \\
\label{eq:vectorkinH} H_{(k)} & = & \frac{2 k}{k^2 - 2K} (1 + w) \mu \mathcal{W}_{(k)}.
\end{eqnarray}

At this stage, we have satisfied all of the linearized evolution equations and constraint equations except for (\ref{lineq:sigmaevolution}), (\ref{lineq:weylEevolution}), (\ref{lineq:weylHevolution}) and (\ref{lineq:curlsigmaconstraint}).  We can now verify that $\mathcal{D}_{a}$, $\mathcal{Z}_{a}$, $\sigma_{ab}$, $\omega_{a}$, $E_{ab}$ and $H_{ab}$, as given by (\ref{eq:quantitiesvectorharmonics}) and (\ref{eq:vectorkinD})-(\ref{eq:vectorkinH}) satisfy (\ref{lineq:sigmaevolution}), (\ref{lineq:weylEevolution}), (\ref{lineq:weylHevolution}) and (\ref{lineq:curlsigmaconstraint}) (via Maple).  We have thus established the following result:

\begin{proposition} \label{prop:VectorPerturbations}  For any solution $\mathcal{W}_{(k)}$ of the governing DE (\ref{eq:vectorperturbationmasterequation}), the first order quantities given by (\ref{eq:quantitiesvectorharmonics}) and (\ref{eq:vectorkinD})-(\ref{eq:vectorkinH}) satisfy the complete set of linearized evolution and constraint equations (\ref{lineq:Xevolution})-(\ref{lineq:Hweylconstraint}).
\end{proposition}

\section{Tensor Perturbations} \label{sec:TensorPerturbations}

In this section we focus on \textit{tensor perturbations}, defined as solutions of the linearized equations (\ref{lineq:zeroorder1})-(\ref{lineq:Hweylconstraint}) that can be expanded in terms of the tensor harmonics $Q^{(2)}_{ab}$.  Tensor perturbations are interpreted as gravitational waves, described by a coupling of the electric and magnetic Weyl curvature.

As regards to the choice of basic variable, we shall see that for tensor perturbations the only non-zero first order quantities are $\sigma_{ab}$, $E_{ab}$ and $H_{ab}$.  On the grounds of mathematical simplicity we choose the comoving shear tensor as the basic variable, denoted $\mathcal{X}_{ab}$ and defined by
\begin{equation} \label{eq:dimensionlessshear}
\mathcal{X}_{ab} = \ell \sigma_{ab}.
\end{equation}

\subsection{The Governing DE}

In this section, we derive the governing DE for tensor perturbations.\footnote{A similar derivation was given by Dunsby et al. (1997).}  It is a consequence of applying tensor harmonics that rank two tensors have zero spatial divergence, \textit{i.e.}
\begin{equation} \label{eq:tensorpertdivergencezero}
\hat{\nabla}^{b} \sigma_{ab} = 0, \qquad \hat{\nabla}^{b} E_{ab} = 0, \qquad \hat{\nabla}^{b} H_{ab} = 0.
\end{equation}  It then follows from the constraints (\ref{lineq:sigmaconstraint}), (\ref{lineq:Eweylconstraint}) and (\ref{lineq:Hweylconstraint}) that all vector quantities are zero, \textit{i.e.}
\begin{equation} \label{eq:tensorpertvectorszero}
\mathcal{D}_{a} = 0, \quad \mathcal{Z}_{a} = 0, \quad \mbox{and} \quad \omega_{a} = 0.
\end{equation}

Using (\ref{eq:tensorpertvectorszero}), the shear evolution equation (\ref{lineq:sigmaevolution}) can be rewritten in terms of $\mathcal{X}_{ab}$ as
\begin{equation} \label{eq:tensorperturbationdimsigmaevol}
\dot{\mathcal{X}}_{\langle ab \rangle} = - H \mathcal{X}_{ab} - \ell E_{ab}.
\end{equation}  We now use (\ref{lineq:weylEevolution}) to calculate $(\ell E_{ab})\dot{\ }_{\perp}$, noting that $\dot{\ell} = H \ell$.  We use the constraint (\ref{lineq:curlsigmaconstraint}) to eliminate $H_{ab}$ and then apply (\ref{adiff:curl5}) to the ``$\mathsf{curl}\ \mathsf{curl}\ \mathcal{X}_{ab}$'' term.  The result is
\begin{equation} \label{eq:tensorperturbationdimWeylEevol}
(\ell E_{ab} )\dot{\ }_{\perp} = - 2 H (\ell E_{ab}) + \left( - \tfrac{1}{2} (1 + w) \mu + \frac{3K}{\ell^2} \right) \mathcal{X}_{ab} - \hat{\nabla}^2 \mathcal{X}_{ab}.
\end{equation}  We can now combine these equations to give a second order DE for $\mathcal{X}_{ab}$ (differentiate (\ref{eq:tensorperturbationdimsigmaevol}) using (\ref{lineq:zeroorder1}) and (\ref{eq:tensorperturbationdimWeylEevol}), and then use (\ref{eq:tensorperturbationdimsigmaevol}) to eliminate $E_{ab}$ from the resulting equation).  After expressing the coefficients of $\mathcal{X}_{ab}$ in terms of the density parameters $\Omega$ and $\Omega_\Lambda$ (see \ref{eq:FLdensityparameter}), we obtain\footnote{It should be noted that the evolution equation (22) given in Dunsby et al. (1997, p1219) contains two errors.  The original equation should instead be written as
\begin{displaymath}
\Delta \sigma_{ab} + \tfrac{5}{3} \Theta \dot{\sigma}_{ab} + \left[ \tfrac{1}{9} \Theta^2 - \tfrac{1}{6} \mu (9 \gamma - 10) \right] \sigma_{ab} = 0,
\end{displaymath} where $\Delta \sigma_{ab} = \ddot{\sigma}_{\perp ab} - \hat{\nabla}^2 \sigma_{ab}$, so as to be consistent with the remainder of that paper.  This equation is then equivalent to (\ref{eq:tensorperturbationmasterequation_a}) for $\Lambda = 0$ on recalling that $\mathcal{X}_{ab} = \ell \sigma_{ab}$ and $\Theta = 3 H$.}
\begin{eqnarray} \label{eq:tensorperturbationmasterequation_a}
\ddot{\mathcal{X}}_{\langle ab \rangle} + 3 H \dot{\mathcal{X}}_{\langle ab \rangle} + \left[ - (3 w + 1) \Omega + 2 \Omega_\Lambda + \frac{2 K}{H^2 \ell^2} \right] H^2 \mathcal{X}_{ab} - \hat{\nabla}^2 \mathcal{X}_{ab} = 0.
\end{eqnarray}

We now expand $\mathcal{X}_{ab}$ in terms of tensor harmonics according to
\begin{equation} \label{eq:dimlesssheartensorharmonics}
\mathcal{X}_{ab} = \sum_{k} \mathcal{X}_{(k)} Q^{(2,k)}_{ab}.
\end{equation}  Substituting (\ref{eq:dimlesssheartensorharmonics}) into (\ref{eq:tensorperturbationmasterequation_a}) and using (\ref{ahaeq:tenharmdefinition}) yields
\begin{equation} \label{eq:tensorperturbationmasterequation}
\ddot{\mathcal{X}}_{(k)} + 3 H \dot{\mathcal{X}}_{(k)} + \left[ - (3 w + 1) \Omega + 2 \Omega_\Lambda + \frac{k^2 + 2K}{H^2 \ell^2} \right] H^2 \mathcal{X}_{(k)} = 0.
\end{equation}  This is the governing equation for tensor perturbations, in terms of clock time $t$.

\subsection{Kinematic Quantities and Weyl Curvature} \label{ssec:TensorPertQuantities}

We now write the kinematic quantities and Weyl curvature in terms of the dimensionless shear coefficients $\mathcal{X}_{(k)}$, tensor harmonics $Q^{(2)}_{ab}$ and associated tensor harmonics $P^{(2)}_{ab}$.  Since (\ref{eq:tensorpertdivergencezero}) confirms that $\sigma_{ab}$, $E_{ab}$ and $H_{ab}$ can be expanded in terms of tensor harmonics, we can write
\begin{equation} \label{eq:weyltensorexpansion}
\sigma_{ab} = \sum_{k} \sigma_{(k)} Q^{(2,k)}_{ab}, \quad E_{ab} = \sum_{k} E_{(k)} Q^{(2,k)}_{ab}, \quad H_{ab} = \sum_{k} H_{(k)} P^{(2,k)}_{ab},
\end{equation} where \footnote{See definition in section \ref{acovsec:tensorharmonics}.}
\begin{equation}
P_{ab} = \ell \mathsf{curl}(Q_{ab}).
\end{equation}  Then using (\ref{eq:dimensionlessshear}), (\ref{lineq:sigmaevolution}) and (\ref{lineq:curlsigmaconstraint}) we can write $\sigma_{ab}$, $E_{ab}$ and $H_{ab}$ in terms of $\mathcal{X}_{(k)}$.  We thus obtain the following expressions for the harmonic coefficients of the non-zero first-order quantities:

\begin{eqnarray}
\label{eq:tensorkinS} \sigma_{(k)} & = & \ell^{-1} \mathcal{X}_{(k)}, \\
\label{eq:tensorkinE} E_{(k)} & = & - \ell^{-1} \left( \dot{\mathcal{X}}_{(k)} + H \mathcal{X}_{(k)} \right), \\
\label{eq:tensorkinH} H_{(k)} & = & \ell^{-2} \mathcal{X}_{(k)}.
\end{eqnarray}

At this stage, we have satisfied all of the linearized evolution equations and constraint equations except for (\ref{lineq:weylEevolution}) and (\ref{lineq:weylHevolution}).  We can now verify that $\sigma_{ab}$, $E_{ab}$ and $H_{ab}$, as given by (\ref{eq:weyltensorexpansion}) satisfy (\ref{lineq:weylEevolution}) and (\ref{lineq:weylHevolution}) (via Maple).  We have thus established the following result:

\begin{proposition} \label{prop:TensorPerturbations}  For any solution $\mathcal{X}_{(k)}$ of the governing DE (\ref{eq:tensorperturbationmasterequation}), the first order quantities given by (\ref{eq:weyltensorexpansion}) and (\ref{eq:tensorkinS})-(\ref{eq:tensorkinH}) satisfy the complete set of linearized evolution and constraint equations (\ref{lineq:Xevolution})-(\ref{lineq:Hweylconstraint}).
\end{proposition}

\section{Dynamics of the Linear Perturbations: General Features}

In this section we discuss the dynamics of the linear perturbations.  We first discuss the importance of conformal time in analyzing perturbations of FL and rewrite the governing DEs for each perturbation type in terms of this time variable.  Finally, we examine the dependence of the first-order quantities on the solutions to the governing DEs.

\subsection{Conformal Time and the Particle Horizon} \label{ssec:ConformalTimeParticleHorizon}

It is a consequence of propositions \ref{prop:ScalarPerturbations}, \ref{prop:VectorPerturbations} and \ref{prop:TensorPerturbations} that the spatial gradients, kinematic quantities and Weyl tensor components can be determined directly from the set of solutions of the governing DEs.  As a consequence, we turn our attention to the behaviour of solutions to these equations.

We make the choice of conformal time $\eta$ as the independent quantity, defined in terms of the length scale $\ell$ and clock time $t$ by
\begin{equation} \label{eq:pertconformaltime}
\frac{d\eta}{dt} = \frac{1}{\ell}.
\end{equation}  Under this choice of time variable, the Roberson-Walker line element (\ref{eq:RWmetric}) can be written in the form
\begin{equation}
ds^2 = \ell^2(\eta) (-d\eta^2 + \tilde{g}_{\alpha \beta} dx^{\alpha} dx^{\beta}),
\end{equation} where $\tilde{g}_{\alpha \beta}$ is a 3-metric of constant curvature.

\begin{figure}[ht]
\begin{center}
\includegraphics[height=160pt]{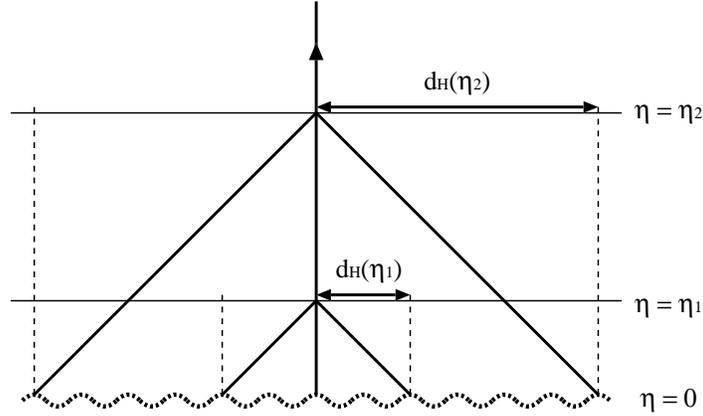}
\end{center}
\caption[Conformal time and the particle horizon.]{The relation between the distance to the particle horizon, denoted $d_H(\eta)$ and the conformal time parameter $\eta$.} \label{fig:particlehorizon}
\end{figure}

Conformal time is directly linked to the notion of the \textit{particle horizon}, defined as the boundary of the part of the universe that is visible to us.  In terms of conformal time, the distance to the particle horizon (see figure \ref{fig:particlehorizon}) is given by
\begin{equation} \label{eq:DistanceToParticleHorizon}
d_H = \ell(\eta) \eta.
\end{equation}  Since the wavelength $\lambda$ for a perturbation of mode $k$ at some fixed time $\eta$ is given by
\begin{equation} \label{eq:PerturbationWavelength}
\lambda(\eta) = \left( \frac{2 \pi}{k} \right) \ell(\eta),
\end{equation} then from (\ref{eq:DistanceToParticleHorizon}) we obtain the key relation
\begin{equation}
\frac{\lambda}{d_H} = \frac{2 \pi}{k \eta}.
\end{equation}  Hence, if $k \eta \ll 1$ the wavelength of the perturbation mode is large compared to the distance to the particle horizon (since $\lambda \gg d_H$).  In this situation it is customary to say that the \textit{perturbation is outside the horizon}.  Conversely, if $k \eta \gg 1$ the wavelength of the perturbation mode is small compared to the distance to the horizon (since $\lambda \ll d_H$).  Similarly, in this situation it is customary to say that the \textit{perturbation is inside the horizon}.

The range of the conformal time variable varies depending on the characteristics of the FL model, as shown in Appendix \ref{app:AlternativeTime}.  A summary of the behaviour of conformal time and clock time for each of the generic classes of FL model with an initial singularity is given below:

\bigskip
\begin{center}
\noindent \begin{tabular}{lcc}
\underline{FL Model} & \underline{Range of $\eta$} & \underline{Range of $t$} \\[0.3ex]
Ever-expanding models, $\Lambda = 0$ & $0 < \eta < +\infty$ & $0 < t < +\infty$ \\[0.3ex]
Ever-expanding models, $\Lambda \neq 0$ & $0 < \eta < \eta_f$ & $0 < t < +\infty$ \\[0.3ex]
Recollapsing models, $\Lambda$ arbitrary & $0 < \eta < \eta_f$ & $0 < t < t_f$
\end{tabular}
\end{center}

\subsection{Governing DEs Using Conformal Time}

We now present the governing DEs for scalar, vector and tensor perturbations in terms of conformal time and specialized to a $\gamma$-law equation of state, \textit{i.e.} $p = (\gamma - 1) \mu$.  It follows from (\ref{eq:soundspeeds}) that
\begin{equation} \label{eq:pertconstanteqstate}
w = c_s^2 = \gamma - 1.
\end{equation}  The three governing DEs, given by (\ref{eq:scalarperturbationmasterequation}), (\ref{eq:vectorperturbationmasterequation}) and (\ref{eq:tensorperturbationmasterequation}) in terms of clock time $t$, assume the following form in terms of conformal time.  The change of variable can be performed simply by using these identities:
\begin{equation} \label{eq:conformaltimechangeofvariable}
\ell \frac{d}{dt} = \frac{d}{d\eta}, \qquad \ell^2 \frac{d^2}{dt^2} = \frac{d^2}{d\eta^2} - H \ell \frac{d}{d\eta},
\end{equation} which follow from (\ref{eq:pertconformaltime}).

\bigskip
\noindent \begin{tabular}{|p{\textwidth}|}
\hline \ \\
\textbf{Governing DEs for Perturbations using Conformal Time} \\ \ \\
\textbf{\ \ \ Scalar}
\begin{equation} \label{eq:scalarperturbationmasterequation_c}
\mathcal{D}_{(k)}^{\prime \prime} + (4 - 3 \gamma) H \ell \mathcal{D}_{(k)}^{\prime} + \left( -\tfrac{3}{2} (2 - \gamma)(3 \gamma - 2) \Omega - 6 (\gamma - 1) \Omega_\Lambda + \frac{c_s^2 k^2}{H^2 \ell^2} \right) H^2 \ell^2 \mathcal{D}_{(k)}  = 0.
\end{equation}
\textbf{\ \ \ Vector}
\begin{equation} \label{eq:vectorperturbationmasterequation_c}
\mathcal{W}_{(k)}^{\prime} - (3 \gamma - 4) H \ell \mathcal{W}_{(k)} = 0.
\end{equation}
\textbf{\ \ \ Tensor}
\begin{equation} \label{eq:tensorperturbationmasterequation_c}
\mathcal{X}_{(k)}^{\prime \prime} + 2 H \ell \mathcal{X}_{(k)}^{\prime} + \left( - (3 \gamma - 2) \Omega + 2 \Omega_\Lambda + \frac{k^2 + 2K}{H^2 \ell^2} \right) H^2 \ell^2 \mathcal{X}_{(k)} = 0.
\end{equation} \\
\hline
\end{tabular}
\bigskip

In the case of vector perturbations, the governing DE (\ref{eq:vectorperturbationmasterequation_c}) can be integrated directly, on applying\footnote{This expression is obtained from (\ref{eq:hubblelengthscale}) on making the change of variable (\ref{eq:conformaltimechangeofvariable}).}
\begin{equation}
H = \frac{\ell^{\prime}}{\ell^2}.
\end{equation}  We then obtain
\begin{equation} \label{eq:vectorperturbationsgeneralsolution}
\mathcal{W}_{(k)} = C (\lambda \ell)^{3 \gamma - 4}.
\end{equation}

In the case of scalar and tensor perturbations, the second-order governing DEs can be transformed into normal form by making a change of dependent variable.  In both cases the resulting DE\footnote{To the best of the author's knowledge, this similarity between tensor and scalar perturbations has not been noted before.} has the following form, in terms of a constant frequency $\omega$ and a potential $U(\eta)$:
\begin{equation}
Y_{(k)}^{\prime \prime} + \left[ \omega^2 - U(\eta) \right] Y_{(k)} = 0.
\end{equation}  The wave number $k$ enters only into the frequency, which also depends on the curvature index $K$, and, in the case of scalar perturbations, on the equation of state.  The potential $U(\eta)$ depends on the background FL model ($H$, $\ell$, $\Omega$ and $\Omega_\Lambda$).  The specific DEs, given in the table below, are obtained by making the indicated change of variable in the governing DEs (\ref{eq:scalarperturbationmasterequation_c}) and (\ref{eq:tensorperturbationmasterequation_c}).  In order to simplify the subsequent analysis we use the notation
\begin{equation} \label{eq:PerturbationBeta}
\beta = \frac{2}{3 \gamma - 2}, \quad \mbox{or, equivalently,} \quad \beta = \frac{2}{3 w + 1},
\end{equation} as in Bardeen (1980).

\bigskip
\noindent \begin{tabular}{|p{\textwidth}|}
\hline \ \\
\textbf{Normal Form of Governing DEs for Scalar and Tensor Perturbations} \\ \ \\
\textbf{\ \ \ Scalar}
\begin{eqnarray}
\label{eq:ScalarMasterDENormalizer} \mbox{Change of variable:} & & \mathcal{D}_{(k)} = (\lambda \ell)^{\frac{1}{\beta} - 1} \hat{\mathcal{D}}_{(k)}, \\
\label{eq:ScalarMasterDENormalForm} \mbox{Governing DE:} & & \hat{\mathcal{D}}_{(k)}^{\prime \prime} + \left[ \omega^2 - U(\eta) \right] \hat{\mathcal{D}}_{(k)} = 0,
\end{eqnarray} \qquad with
\begin{eqnarray}
\label{eq:ScalarMasterDEFrequency} \omega^2 & = & c_s^2 k^2 + \tfrac{1}{\beta^2} \left(\beta - 1\right)^2 K, \\
\label{eq:ScalarMasterDEPotential} U(\eta) & = & \tfrac{1}{\beta^2} \left( 1 + \beta \right) \left( \beta \Omega + \Omega_\Lambda \right) H^2 \ell^2.
\end{eqnarray} \\
\textbf{\ \ \ Tensor}
\begin{eqnarray}
\label{eq:TensorMasterDENormalizer} \mbox{Change of variable:} & & \mathcal{X}_{(k)} = (\lambda \ell)^{-1} \hat{\mathcal{X}}_{(k)}, \\
\label{eq:TensorMasterDENormalForm} \mbox{Governing DE:} & & \hat{\mathcal{X}}_{(k)}^{\prime \prime} + \left[ \omega^2 - U(\eta) \right] \hat{\mathcal{X}}_{(k)} = 0.
\end{eqnarray} \qquad with
\begin{eqnarray}
\label{eq:TensorMasterDEFrequency} \omega^2 & = & k^2 + 3 K, \\
\label{eq:TensorMasterDEPotential} U(\eta) & = & \tfrac{1}{\beta} \left( 1 + \beta \right) \Omega H^2 \ell^2.
\end{eqnarray} \\
\hline
\end{tabular}
\bigskip

\paragraph{Note:}  Here $\lambda$ is the constant conformal parameter introduced in section \ref{ssec:FLIntrinsicParameters}, having dimension $(length)^{-1}$.

\paragraph{Remark:}  Writing the governing DEs in normal form has several advantages.  Firstly, it gives insight into the qualitative behaviour of the solutions.  In particular, if $\omega^2 \gg U(\eta)$ the solutions will be approximated by oscillatory solutions of frequency $\omega$, while if $\omega^2 \ll U(\eta)$, the solutions will be approximated by power law solutions.  Secondly, if the background FL model has $\Lambda = 0$, the general solution can be written in terms of Bessel or Legendre functions, as we now show.

\subsection{General Solutions for Scalar and Tensor Perturbations with $\Lambda = 0$}

In the case $\Lambda = 0$, the potential $U(\eta)$ in the governing DEs is given by
\begin{equation}
U(\eta) = \frac{1}{\beta} ( 1 + \beta ) \Omega H^2 \ell^2
\end{equation} for both scalar and tensor perturbations (see equations (\ref{eq:ScalarMasterDEPotential}) and (\ref{eq:TensorMasterDEPotential})).  The background solution is (\ref{eq:FLsolution-FC-recollapse}):
\begin{equation} \label{eq:PertFCBackgroundSolution}
\ell(\eta) = \lambda^{-1} \left[ S_K(T) \right]^{\beta}, \quad T = \frac{\eta}{\beta},
\end{equation} where
\begin{equation}
S_K(T) = (T, \sinh T, \sin T),
\end{equation} for $K = (0, -1, +1)$.  On noting that $\Omega H^2 \ell^2 = \tfrac{1}{3} \mu \ell^2$, it follows from (\ref{eq:FLsolution-FC-EnergyDensity}) that
\begin{equation} \label{eq:OneFluidBackgroundSolutionsLengthScale}
\Omega H^2 \ell^2 = (\lambda \ell)^{- \frac{2}{\beta}}.
\end{equation}  Thus, using (\ref{eq:PertFCBackgroundSolution}) we obtain
\begin{equation} \label{eq:OneFluidBackgroundSolutions}
\Omega H^2 \ell^2 = \left[ S_K(T) \right]^{-2}.
\end{equation}  Observe now that the governing DEs in normal form (equations (\ref{eq:ScalarMasterDENormalForm}) and (\ref{eq:TensorMasterDENormalForm})) coincide with the Bessel DE or the Legendre DE, as given in Appendix \ref{app:PerturbationSolutions} (see equations (\ref{eqpsol:BesselsDE})-(\ref{eqpsol:ConicalLegendreDESol})), with the parameters determined as follows:

\bigskip

\begin{center}
\begin{tabular}{lclccc}
\underline{Curvature} & \quad & \underline{DE Type} & \quad & \multicolumn{2}{c}{\underline{Parameters}} \\[0.5ex]
$K = 0$ & & Bessel DE & & $\mu = \beta + \tfrac{1}{2}$ & $a = \omega$ \\
$K = -1$ & & Legendre (Toroidal) DE & & $\mu = \beta + \tfrac{1}{2}$ & $\nu = - \tfrac{1}{2} + i \beta \omega$ \\
$K = +1$ & & Legendre (Conical) DE & & $\mu = \beta + \tfrac{1}{2}$ & $\nu = - \tfrac{1}{2} + \beta \omega$ \\
\end{tabular}
\end{center}

\bigskip

\noindent The value of $\omega$, as given by (\ref{eq:ScalarMasterDEFrequency}) and (\ref{eq:TensorMasterDEFrequency}) distinguishes between scalar and tensor perturbations.  The solutions of the three DEs are given by (\ref{eqpsol:BesselsDESol}), (\ref{eqpsol:ToroidalLegendreDESol}) and (\ref{eqpsol:ConicalLegendreDESol}), in terms of Bessel and Legendre functions.  The resulting expressions for $\mathcal{D}_{(k)}$ and $\mathcal{X}_{(k)}$, obtained from (\ref{eq:ScalarMasterDENormalizer}), (\ref{eq:TensorMasterDENormalizer}) and (\ref{eq:PertFCBackgroundSolution}) are listed in the following table

\bigskip
\noindent \begin{tabular}{|p{\textwidth}|}
\hline \ \\
\textbf{General Solution of the Governing DEs for Scalar and Tensor Perturbations When $\Lambda = 0$} \\
\paragraph{\ \ \ Flat FL ($K = 0$)}
\begin{equation} \label{eq:GeneralFlatFLGoverningDESoln}
\left. \begin{array}{ccr}
\mbox{Scalar} & \quad & \mathcal{D}_{(k)}(\eta) = \eta^{\frac{3}{2} - \beta} \\ \mbox{Tensor} & \quad & \mathcal{X}_{(k)}(\eta) = \eta^{\frac{1}{2} - \beta}
\end{array} \right\} \left[ C_{+} J_{\beta + \frac{1}{2}}(a \eta) + C_{-} Y_{\beta + \frac{1}{2}}(a \eta) \right],
\end{equation} \qquad where
\begin{equation}
a = (c_s k, k)
\end{equation} \qquad for scalar and tensor perturbations, respectively.

\paragraph{\ \ \ Open FL ($K = -1$)}
\begin{equation} \label{eq:GeneralOpenFLGoverningDESoln}
\left. \begin{array}{cr}
\mbox{Scalar} & \mathcal{D}_{(k)}(\eta) = \left( \sinh \tfrac{\eta}{\beta} \right)^{\frac{3}{2} - \beta} \\ \mbox{Tensor} & \mathcal{X}_{(k)}(\eta) = \left( \sinh \tfrac{\eta}{\beta} \right)^{\frac{1}{2} - \beta}
\end{array} \right\} \left[ C_{+} Q^\mu_\nu \left( \cosh \tfrac{\eta}{\beta} \right) + C_{-} P^\mu_\nu \left( \cosh \tfrac{\eta}{\beta} \right) \right],
\end{equation} \qquad where
\begin{equation}
\mu = \beta + \tfrac{1}{2}, \qquad \nu + \tfrac{1}{2} = i \beta \omega,
\end{equation} \qquad and the frequency $\omega$ is given by (\ref{eq:ScalarMasterDEFrequency}) and (\ref{eq:TensorMasterDEFrequency}) respectively.

\paragraph{\ \ \ Closed FL ($K = +1$)} \ \\
\qquad Replace hyperbolic functions by trigonometric functions in the solutions (\ref{eq:GeneralOpenFLGoverningDESoln}) for open FL, and use
\begin{equation} \label{eq:GeneralClosedFLGoverningDESoln}
\nu + \tfrac{1}{2} = \beta \omega.
\end{equation} \\
\hline
\end{tabular}
\bigskip

\paragraph{Remark:}  The physically important cases, namely when the matter content is dust (\textit{i.e.} $\gamma = 1$, $\beta = 2$) or radiation (\textit{i.e.} $\gamma = \tfrac{4}{3}$, $\beta = 1$) require comment.  For dust, the scalar solution is not valid since $c_s^2 = \gamma - 1 = 0$.  We will derive the solution in this case in section \ref{sec:DustPerturbations}.  The tensor solution is valid, however, and the relevant Bessel and Legendre functions are in fact elementary and are given in section \ref{ssec:DustTensorPerturbations}.  For radiation, both the scalar and tensor solutions are elementary.  We will give them in section \ref{sec:RadiationPertSolutions}.

One can use asymptotic approximations of the Bessel and Legendre functions (Appendix \ref{app:PerturbationSolutions}) to obtain approximate solutions in restricted regimes.

\subsection{Functional Dependence of First-Order Quantities}

We are interested in describing to what extent a perturbed FL cosmology deviates from an exact FL cosmology.

It is known that cosmological observations place bounds on the first order quantities, specifically on dimensionless quantities formed by dividing the first order quantities by an appropriate power of the Hubble scalar.  The specific quantities are \begin{eqnarray}
\label{eq:HubbleNormalized1} \mbox{Kinematic quantities:} & & \frac{\sigma_{ab}}{H}, \frac{\omega_{a}}{H}, \\
\label{eq:HubbleNormalized2} \mbox{Spatial gradients:} & & \frac{\hat{\nabla}_a \mu}{H \mu}, \frac{\hat{\nabla}_a H}{H^2}, \\
\label{eq:HubbleNormalized3} \mbox{Weyl tensor:} & & \frac{E_{ab}}{H^2}, \frac{H_{ab}}{H^2}.
\end{eqnarray}

The first analysis of this type was due to Sachs and Wolfe (1967) who showed that observations of distant galaxies could, in principle, place bounds on these quantities.  More recently, Maartens et al (1995a,b) have shown that the high isotropy of the CMBR ($\frac{\Delta T}{T} \approx 10^{-5}$) places strong bounds on these quantities.\footnote{The analysis is complicated and subject to assumptions being imposed on the time derivatives of the multipoles of the CMBR.}

It is natural to use these Hubble-normalized quantities as a measure of how close the physical model is to an idealized FL model, and we will thus calculate their time dependence for the three perturbation types.  A growing mode is an indication of an instability in the background FL model.

We now briefly mention the connection between the geometrical approach and Bardeen's gauge-invariant metric approach to cosmological perturbations (Bardeen (1980)).  In Bardeen's approach, the basic variables are gauge-invariant linear combinations of the metric perturbations.  However, it turns out that Bardeen's metric potentials $\Phi_H$, $\Psi$ and $H_T^{(2)}$ have the same time dependence as certain expressions formed from the basic variables in the geometrical approach.  It can be shown that\footnote{This result relies on comparing the governing DEs for $\mathcal{D}_{(k)}$, $\mathcal{W}_{(k)}$ and $\mathcal{X}_{(k)}$ with the governing DEs defined in Bardeen's approach (Bardeen equation (4.9), (4.13) and (4.14)) and noting that the paired quantities ($\mathcal{D}_{(k)}$, $\epsilon_m$), ($\mathcal{W}_{(k)}$, $V_c$) and ($\mathcal{X}_{(k)}$, $\frac{d}{d\eta} H_T^{(2)}$) satisfy the same evolutions equation.  Equations (\ref{eq:BardeenScalarMetricPotential})-(\ref{eq:BardeenVectorMetricPotential}) then follow from Bardeen (4.3) and (4.12).}
\begin{eqnarray}
\label{eq:BardeenScalarMetricPotential} \Phi_H\ \backsim & \Omega H^2 \ell^2 \mathcal{D}_{(k)} & \qquad \mbox{(scalar),} \\
\label{eq:BardeenVectorMetricPotential} \Psi\ \backsim & \Omega H^2 \ell^2 \mathcal{W}_{(k)} & \qquad \mbox{(vector),} \\
\label{eq:BardeenTensorMetricPotential} H_T^{(2)}\ \backsim & 2 H \ell \mathcal{X}_{(k)} + \mathcal{X}_{(k)}^{\prime} & \qquad \mbox{(tensor).}
\end{eqnarray}  We refer the reader to Bruni et al. (1992, p45-9) for an overview of the relation between the geometrical framework and Bardeen's metric approach.

Using the results in sections \ref{ssec:ScalarPertQuantities}, \ref{ssec:VectorPertQuantities} and \ref{ssec:TensorPertQuantities} we can obtain the dependence of the Hubble-normalized quantities (\ref{eq:HubbleNormalized1})-(\ref{eq:HubbleNormalized3}) on the basic variables $\mathcal{D}_{(k)}$, $\mathcal{W}_{(k)}$ and $\mathcal{X}_{(k)}$ in terms of conformal time.  In order to enable comparisons to be made, we also include Bardeen's metric potentials $\Phi_H$ (\ref{eq:BardeenScalarMetricPotential})-(\ref{eq:BardeenTensorMetricPotential}).  These results are provided in table \ref{fig:AsymptoticBehaviourFunctionalDependence}.

\smallskip

\begin{table}[p]
\hspace{-0.5cm} \textbf{Functional dependence of first-order quantities.}\ \\

\hspace{-0.5cm} \begin{tabular}{ccccc}
\underline{Quantity} & & \underline{Scalar Perturbations} & \underline{Vector Perturbations} & \underline{Tensor Perturbations} \\[3.0ex]
$\displaystyle \frac{\sigma_{ab}}{H}$ & $\backsim$ & $\displaystyle -3 (\gamma - 1) \mathcal{D}_{(k)} + \frac{\mathcal{D}_{(k)}^{\prime}}{H \ell}$ & $\displaystyle \left[ \frac{k^2 - 2K}{H^2 \ell^2} + 6 \gamma \Omega \right] H \ell \mathcal{W}_{(k)}$ & $\displaystyle \frac{\mathcal{X}_{(k)}}{H \ell}$ \\[2.0ex]
$\displaystyle \frac{\omega_{a}}{H}$ & $\backsim$ & $0$ & $\displaystyle \frac{\mathcal{W}_{(k)}}{H \ell}$ & $0$ \\[2.0ex]
$\displaystyle \frac{\hat{\nabla}_{a} \mu}{H \mu}$ & $\backsim$ & $\displaystyle \frac{\mathcal{D}_{(k)}}{H \ell}$ & $\displaystyle \mathcal{W}_{(k)}$ & $0$ \\[2.0ex]
$\displaystyle \frac{\hat{\nabla}_{a} H}{H^2}$ & $\backsim$ & $\displaystyle \frac{1}{H \ell} \left[ 3 (\gamma - 1) \mathcal{D}_{(k)} - \frac{\mathcal{D}_{(k)}^{\prime}}{H \ell} \right]$ & $\displaystyle \left[ \frac{2K}{H^2 \ell^2} - 3 \gamma \Omega \right] \mathcal{W}_{(k)}$ & $0$ \\[2.0ex]
$\displaystyle \frac{E_{ab}}{H^2}$ & $\backsim$ & $\displaystyle \Omega \mathcal{D}_{(k)}$ & $\displaystyle \Omega H \ell \mathcal{W}_{(k)}$ & $\displaystyle \frac{\mathcal{X}_{(k)}}{H \ell} + \frac{\mathcal{X}_{(k)}^{\prime}}{H^2 \ell^2}$ \\[2.0ex]
$\displaystyle \frac{H_{ab}}{H^2}$ & $\backsim$ & $0$ & $\displaystyle \Omega \mathcal{W}_{(k)}$ & $\displaystyle \frac{\mathcal{X}_{(k)}}{H^2 \ell^2}$ \\[2.0ex]
\hline \\
$\displaystyle \Phi_H$ & $\backsim$ & $\Omega H^2 \ell^2 \mathcal{D}_{(k)}$ & - & - \\[2.0ex]
$\displaystyle \Psi$ & $\backsim$ & - & $\displaystyle \Omega H^2 \ell^2 \mathcal{W}_{(k)}$ & - \\[2.0ex]
$\displaystyle H_{T}^{(2)}$ & $\backsim$ & - & - & $2 H \ell \mathcal{X}_{(k)} + \mathcal{X}_{(k)}^{\prime}$ \\[2.0ex]
\hline \\
$\displaystyle \frac{\hat{\nabla}_{a}\!\ ^{(3)}\!R}{H^3}$ & $\backsim$ & $\displaystyle \left[ \tfrac{3}{2} \gamma \Omega - 3 (\gamma - 1) \right] \frac{\mathcal{D}_{(k)}}{H \ell} + \frac{\mathcal{D}_{(k)}^{\prime}}{H^2 \ell^2}$ & $\displaystyle \frac{K}{H^2 \ell^2} \mathcal{W}_{(k)}$ & $0$ \\[2.0ex]
$\displaystyle \frac{^{(3)}\!C_{ab}}{H^3}$ & $\backsim$ & $0$ & $0$ & $\displaystyle 2 \frac{\mathcal{X}_{(k)}}{H^2 \ell^2} + \frac{\mathcal{X}_{(k)}^{\prime}}{H^3 \ell^3}$ \\[2.0ex]
\end{tabular}

\hspace{-1.0cm} \caption[Functional dependence of first-order quantities]{Expressions for the first order quantities in terms of the basic variable for each perturbation type.} \label{fig:AsymptoticBehaviourFunctionalDependence}
\end{table}

\clearpage

\section{Perturbations of FL Models with Pressure-Free Matter} \label{sec:DustPerturbations}

For the case where the background FL model contains dust and possibly a cosmological constant, the analysis of scalar perturbations simplifies significantly since one can avoid performing the harmonic decomposition.  The evolution equation (\ref{eq:scalarperturbationmasterequation_a}) for the fractional density gradient $\mathcal{D}_{a}$ simplifies to
\begin{equation} \label{eq:scalarperturbationmasterDEDust_a}
\ddot{\mathcal{D}}_{\langle a \rangle} + 2 H \dot{\mathcal{D}}_{\langle a \rangle} - \tfrac{3}{2} \Omega H^2 \mathcal{D}_{a}  = 0.
\end{equation}  Since (\ref{eq:densitygradientaccelerationrelation}) implies $\dot{u}_a = 0$ for a dust background, a short calculation using (\ref{eq:projectiontensor}) and (\ref{eq:covudecomposition}) reveals $\dot{h}_{ab} = 0$, and hence that
\begin{equation}
\dot{\mathcal{D}}_{\langle a \rangle} = \dot{\mathcal{D}}_{a}, \quad \mbox{and} \quad \ddot{\mathcal{D}}_{\langle a \rangle} = \ddot{\mathcal{D}}_{a}.
\end{equation}  Namely, we use the fact that the fluid 4-velocity is a geodesic in the case of dust.  Thus, we can drop the spatial projections from (\ref{eq:scalarperturbationmasterDEDust_a}) and so obtain
\begin{equation} \label{eq:scalarperturbationmasterDEDust}
\ddot{\mathcal{D}}_{a} + 2 H \dot{\mathcal{D}}_{a} - \tfrac{3}{2} \Omega H^2 \mathcal{D}_{a}  = 0.
\end{equation}  Although it follows that $\mathcal{D}_{a}$ does not depend on the perturbation mode $k$, a harmonic decomposition is required in order to obtain expressions for the shear and electric Weyl tensor (see, for example, (\ref{eq:scalarkinS}) and (\ref{eq:scalarkinE})).

\subsection{The General Solution in Integral Form for Scalar Perturbations}

A second simplification is that the evolution equation (\ref{eq:scalarperturbationmasterDEDust}) has a first integral, which we derive as follows, using the spatial gradients $\mathcal{D}_{a}$, $\mathcal{Z}_{a}$ and $\mathcal{R}_{a}$.

In the case of dust, the evolution equations, given in (\ref{lineq:Devolution}), (\ref{lineq:newZevolution}) and (\ref{lineq:Revolution}) simplify significantly, and can be written as
\begin{eqnarray}
\label{eq:DevolutionDust} \dot{\mathcal{D}}_{a} & = & - \mathcal{Z}_{a}, \\
\label{eq:ZevolutionDust} ( \ell^2 \mathcal{Z}_{a} )\dot{\ } & = & - \tfrac{1}{2} \mu \ell^2 \mathcal{D}_{a}, \\
\label{eq:RevolutionDust} ( \ell^2 \mathcal{R}_{a} )\dot{\ } & = & - 4 K \mathcal{Z}_{a}.
\end{eqnarray}  Note that these DEs are not independent, since the 3 spatial gradients are linearly dependent.  This result can be seen upon taking the spatial gradient of (\ref{a2eq:ricci3scalarB}) and linearizing so as to obtain
\begin{equation} \label{eq:spatialgradientrelation}
\mathcal{R}_a = -4 H \mathcal{Z}_a + 2 \mu \mathcal{D}_a.
\end{equation}

Observe that the quantity $\tilde{C}_{a}$, defined by\footnote{Also see Bruni et al. (1992), eq (38).}
\begin{equation} \label{eq:TildeCDefinition}
\tilde{C}_a = \ell^2 \mathcal{R}_{a} - 4 K \mathcal{D}_{a}.
\end{equation} satisfies
\begin{equation} \label{eq:TildeCZeroDeriv}
\frac{d}{dt} \tilde{C}_{a} = 0.
\end{equation}  We can rewrite (\ref{eq:TildeCDefinition}) in terms of $\mathcal{D}_{a}$ $\dot{\mathcal{D}}_{a}$ using (\ref{eq:DevolutionDust}), (\ref{eq:spatialgradientrelation}), (\ref{eq:FLdensityparameter}) and (\ref{eq:FLdynamicalcurvature}), so as to obtain
\begin{equation} \label{eq:TildeCitoD}
\tilde{C}_{a} = 4 \ell^2 (H \dot{\mathcal{D}}_{a} + ( \tfrac{3}{2} \Omega + \Omega_k ) H^2 \mathcal{D}_{a} ).
\end{equation}  This equation is in fact a first integral of the evolution equation (\ref{eq:scalarperturbationmasterDEDust}), since (\ref{eq:scalarperturbationmasterDEDust}) is a consequence of (\ref{eq:DevolutionDust}) and (\ref{eq:ZevolutionDust}).

On noting that, in the case of dust, (\ref{lineq:zeroorder1}) can be rewritten as
\begin{equation} \label{eq:HevolutionDust}
\dot{H} = - (\tfrac{3}{2} \Omega + \Omega_k) H^2,
\end{equation} we can further simplify (\ref{eq:TildeCitoD}) to read
\begin{equation} \label{eq:spatialgradientfirstintegral}
\frac{d}{dt} \left( \frac{\mathcal{D}_{a}}{H} \right) = \frac{\tilde{C}_{a}}{4 H^2 \ell^2}.
\end{equation}  The solution of (\ref{eq:spatialgradientfirstintegral}) is thus
\begin{equation} \label{eq:GeneralScalarSolutionScalarDust}
\mathcal{D}_{a}(t) = C^{(+)}_{a} H \int^t_0 \frac{d\tilde{t}}{H^2 \ell^2} + C^{(-)}_{a} \frac{H}{\lambda},
\end{equation} for time-independent vectors $C^{(-)}_{a}$ and $C^{(+)}_{a}$, with $C^{(+)}_{a} = \tilde{C}_{a}$.  For simplicity, we define
\begin{equation} \label{eq:GeneralScalarSolutionScalarDustGrowing}
\mathcal{D}^{(+)}(t) = H \int_0^t \frac{d\tilde{t}}{H^2 \ell^2}, \qquad \mathcal{D}^{(-)}(t) = \frac{H}{\lambda},
\end{equation} and so write (\ref{eq:GeneralScalarSolutionScalarDust}) as
\begin{equation}
\mathcal{D}_{a}(t) = C^{(+)}_{a} \mathcal{D}^{(+)}(t) + C^{(-)}_{a} \mathcal{D}^{(-)}(t).
\end{equation}  In terms of conformal time $\eta$, the growing mode $\mathcal{D}^{(+)}(\eta)$, as given by (\ref{eq:GeneralScalarSolutionScalarDustGrowing}), reads
\begin{equation} \label{eq:GeneralScalarSolutionScalarDust_c}
\mathcal{D}^{(+)}(\eta) = H \int_0^\eta \frac{d\tilde{\eta}}{H^2 \ell}.
\end{equation}

Equation (\ref{eq:GeneralScalarSolutionScalarDust_c}) can often be used to gain insight into the asymptotic behaviour of dust perturbations.  If $H \to 0$ or $H \to \infty$ as $\eta \to \eta^{\ast}$ (where $\eta^{\ast} = \pm \infty$ is allowed), we can rewrite $\mathcal{D}^{(+)}$ as
\begin{equation}
\mathcal{D}^{(+)} \sim \frac{\left(\int^{\eta}_{0} \frac{d\tilde{\eta}}{H^2 \ell} \right)}{\left( \frac{1}{H} \right)}.
\end{equation} On applying L'H\^{o}pital's rule, in the limit as $\eta \to \eta^{\ast}$, we obtain
\begin{equation}
\mathcal{D}^{(+)} \sim - \frac{1}{H^{\prime} \ell}.
\end{equation}  Then (\ref{eq:HevolutionDust}) gives an expression for the asymptotic behaviour of the growing mode in terms of zero-order quantities:
\begin{equation} \label{eq:ScalarPerturbationGrowingModeDust}
\mathcal{D}^{(+)}(\eta) \sim \frac{1}{\frac{3}{2} \Omega H^2 \ell^2 - K}, \qquad \mbox{as $\eta \to \eta^{\ast}$}.
\end{equation}

Although the general solution (\ref{eq:GeneralScalarSolutionScalarDustGrowing}) cannot always be obtained in closed form, one can analyze the asymptotic behaviour of these solutions.  We are primarily interested in three asymptotic epochs, namely flat FL ($\eta \approx 0$), Milne ($\eta \gg 1$, $\Omega_k \approx 1$) and de Sitter ($\eta \approx \eta_f$, $\Omega_\Lambda \approx 1$).  The general behaviour of perturbations within these epochs shall be discussed in detail in section \ref{sec:AsymptoticBehaviourMasterDEs}.

\subsection{Scalar Perturbations with $\Lambda = 0$}

We now use the general solution (\ref{eq:GeneralScalarSolutionScalarDust_c}) to obtain the solution for a dust background with $\Lambda = 0$.  The length scale (\ref{eq:PertFCBackgroundSolution}), when specialized to the case of dust, is given by
\begin{equation} \label{eq:GeneralLengthScaleDustPert}
\lambda \ell = \left( S_K(\eta / 2) \right)^{2},
\end{equation} and the Hubble parameter by $H = \ell^{\prime} / \ell^2$ and (\ref{eq:GeneralLengthScaleDustPert}),
\begin{equation} \label{eq:GeneralHubbleParameterDustPert}
H \ell = \frac{C_K(\eta/2)}{S_K(\eta/2)},
\end{equation} where $C_K(T)$ is defined analogous to $S_K(T)$ to be
\begin{equation} \label{eq:FLsolutionC_K}
C_K(T) = (1, \cosh T, \cos T),
\end{equation} for $K = (0, -1, +1)$.

We now consider the case of $K \neq 0$ separately.  In this case, the growing mode can be determined from (\ref{eq:GeneralScalarSolutionScalarDust_c}), (\ref{eq:GeneralLengthScaleDustPert}) and (\ref{eq:GeneralHubbleParameterDustPert}), on integrating
\begin{equation} \label{eq:PertDustZeroLambdaIntegral}
\mathcal{D}^{(+)}(\eta) = \frac{H}{\lambda} \int \frac{S_K(\eta/2)^4}{C_K(\eta/2)^2} d \eta.
\end{equation}  This integral is elementary and can be expressed in terms of $H$ and $\ell$ after a short calculation:
\begin{equation} \label{eq:PertDustUnsimplifiedSolution}
\mathcal{D}^{(+)} = \frac{1}{\lambda \ell} (2 - \tfrac{3}{2} \eta H \ell) + H^2 \ell^2.
\end{equation}

It follows from (\ref{eq:OneFluidBackgroundSolutionsLengthScale}) with $\beta = 2$ that
\begin{equation}
\Omega H^2 \ell^2 = \frac{1}{\lambda \ell},
\end{equation} which allows us to rewrite the Friedmann equation (\ref{eq:FLomegarelation}) as
\begin{equation} \label{eq:PertDustSimplifier}
H^2 \ell^2 = - K + \frac{1}{\lambda \ell}.
\end{equation}  We can now eliminate the $H^2 \ell^2$ term in (\ref{eq:PertDustUnsimplifiedSolution}), obtaining
\begin{equation} \label{eq:PertDustZeroLambdaGrowingMode}
\mathcal{D}^{(+)} = \frac{3}{\lambda \ell} (1 - \tfrac{1}{2} \eta H \ell) - K.
\end{equation}  The solution of (\ref{eq:scalarperturbationmasterDEDust}) for $K \neq 0$ is then
\begin{equation} \label{eq:PertDustZeroLambdaKneq0Soln}
\mathcal{D}_{a} = C^{(+)}_{a} \left[ \frac{3}{\lambda \ell} (1 - \tfrac{1}{2} \eta H \ell) - K \right] + C^{(-)}_{a} \left( \frac{H}{\lambda} \right),
\end{equation} where $H$ and $\ell$ are given by (\ref{eq:GeneralLengthScaleDustPert}) and (\ref{eq:GeneralHubbleParameterDustPert}).

\begin{figure}[htb!]
\begin{center}
\begin{tabular}{ccc}
\includegraphics[width=100pt]{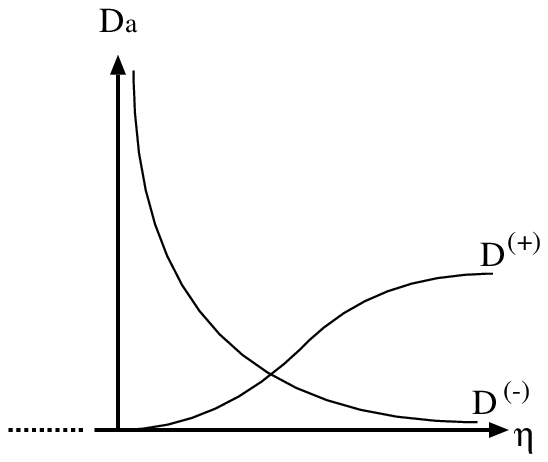} &
\includegraphics[width=100pt]{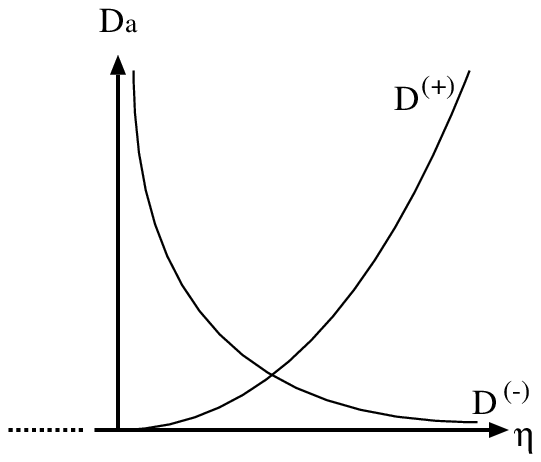} &
\includegraphics[width=100pt]{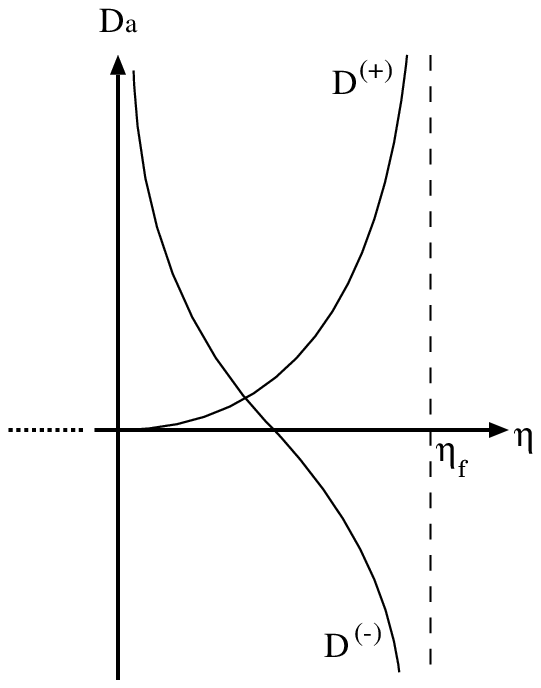}
\end{tabular}
\end{center}
\caption[Scalar perturbation modes for dust and spatial curvature.]{A depiction of the qualitative behaviour of the growing and decaying mode solutions of (\ref{eq:scalarperturbationmasterDEDust}) when $\Lambda = 0$.  From left to right, the plots depict negative ($K = -1$), flat ($K = 0$) and positive ($K = +1$) curvature.} \label{fig:PerturbationScalarModes-DC}
\end{figure}

In the case of $K = 0$, the integral (\ref{eq:PertDustZeroLambdaIntegral}) evaluates to
\begin{equation}
\mathcal{D}^{(+)} = \tfrac{1}{10} \eta^{2},
\end{equation} where the coefficient $\tfrac{1}{10}$ is chosen so as to be asymptotic to the growing mode solution for $K \neq 0$ when $\eta \to 0$.  The solution for a flat FL background is then
\begin{equation} \label{eq:PertDustZeroLambdaKeq0Soln}
\mathcal{D}_{a} = C^{(+)}_{a} \left( \tfrac{1}{10} \eta^{2} \right) + C^{(-)}_{a} (8 \eta^{-3}).
\end{equation}  We depict the growing and decaying mode solutions in figure \ref{fig:PerturbationScalarModes-DC}.

We note that the solution (\ref{eq:PertDustZeroLambdaKneq0Soln}) can also be derived from the governing DE (\ref{eq:scalarperturbationmasterDEDust}), upon applying the general solutions (\ref{eqpsol:BesselsDESol}), (\ref{eqpsol:ResonantLegendre5a}) and (\ref{eqpsol:ResonantLegendre5b}).

\bigskip

\subsection{Scalar Perturbations with $\Lambda > 0$, $K = 0$}

\begin{figure}[htb!]
\begin{center}
\begin{tabular}{ccc}
\includegraphics[height=120pt]{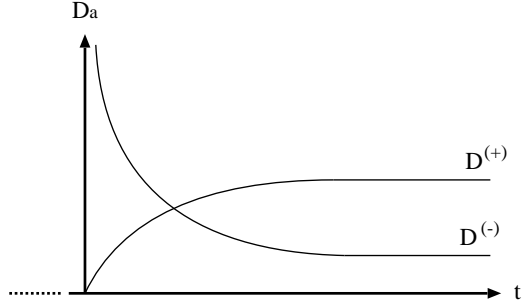}
\end{tabular}
\end{center}
\caption[Scalar perturbation modes for dust and cosmological constant.]{A depiction of the qualitative behaviour of the growing and decaying modes in an FL cosmology for density perturbations with dust and cosmological constant.} \label{fig:PerturbationScalarModes-FL}
\end{figure}

In the case $\Lambda > 0$ with flat spatial geometry ($K = 0$), the general solution (\ref{eq:GeneralScalarSolutionScalarDustGrowing}) can be given in terms of toroidal Legendre functions with clock time as the time variable, as follows.  We transform the governing DE (\ref{eq:scalarperturbationmasterDEDust}) to normal form by means of the change of variable
\begin{equation} \label{eq:FLFlatCosmoCOV}
\mathcal{D}_{a} = (\lambda \ell)^{-1} \hat{\mathcal{D}}_{a}.
\end{equation}  The result is
\begin{equation} \label{eq:DLambdaBackgroundScalarDE}
\hat{\mathcal{D}}_{a}\!\ddot{\ }\ - H^2 \hat{\mathcal{D}}_{a} = 0, \qquad \hat{\mathcal{D}}_{a} = (\lambda \ell)^{-1} \mathcal{D}_{a}.
\end{equation}  The length scale of the background solution is given by (\ref{eq:FLFlatCosmoSoln}).  We calculate $H$ using \mbox{$H = \dot{\ell} / \ell$}.  The results are
\begin{equation} \label{eq:FLFlatCosmoSoln2}
\ell = \lambda^{-1} (\sinh T)^{2/3}, \qquad H = \lambda \coth T,
\end{equation} where
\begin{equation} \label{eq:FLFlatCosmoSoln3}
T = \tfrac{3}{2} \lambda t, \quad \mbox{and} \quad \lambda = \sqrt{\frac{\Lambda}{3}}.
\end{equation}  We substitute (\ref{eq:FLFlatCosmoSoln2}) into (\ref{eq:DLambdaBackgroundScalarDE}) and use $T$ as the independent variable.  After applying a standard identity, we obtain
\begin{equation} \label{eq:DLambdaBackgroundDE}
\frac{d^2}{dT^2} \hat{\mathcal{D}}_{a} - \tfrac{4}{9} \left[ 1 + \frac{1}{\sinh^2 T} \right] \hat{\mathcal{D}}_{a} = 0, \qquad \mathcal{D}_{a} = \left[ \sinh \left( \tfrac{1}{2} \sqrt{3 \Lambda} t \right) \right]^{-2/3} \hat{\mathcal{D}}_{a}.
\end{equation}  This DE is the toroidal Legendre DE (\ref{eqpsol:ToroidalLegendreDE}) with $\mu = \tfrac{5}{6}$ and $\nu = \tfrac{1}{6}$ whose solution is given by (\ref{eqpsol:ToroidalLegendreDESol}).  Thus, we have
\begin{equation} \label{eq:DLambdaBackgroundSolution_a}
\mathcal{D}_{a}(t) = (\sinh T)^{- \frac{1}{6}} \left[ C^{(+)}_a Q^{5/6}_{1/6}(\cosh T) + C^{(-)}_a P^{5/6}_{1/6}(\cosh T) \right].
\end{equation}
The second term of this solution is in fact an elementary function due to the identity\footnote{This result is obtained from Abramowitz and Stegun (1970), equations (8.5.1) and (8.6.16).}
\begin{equation}
(\sinh T)^{-\frac{1}{6}} P^{5/6}_{1/6}(\cosh T) = A \coth(T),
\end{equation} for $A$ constant.  It follows from (\ref{eq:FLFlatCosmoSoln2}) that (\ref{eq:DLambdaBackgroundSolution_a}) can be rewritten as
\begin{equation} \label{eq:DLambdaBackgroundSolution}
\mathcal{D}_{a}(t) = C^{(+)}_a (\sinh T)^{- \frac{1}{6}} Q^{5/6}_{1/6}(\cosh T) + C^{(-)}_a \frac{H}{\lambda},
\end{equation} where $T$ is given by (\ref{eq:FLFlatCosmoSoln3}).  Observe that the second term corresponds to the second term in the general solution (\ref{eq:GeneralScalarSolutionScalarDust}).  However, the growing mode in (\ref{eq:DLambdaBackgroundSolution}) does not correspond to the growing mode given in the general solution (\ref{eq:GeneralScalarSolutionScalarDust}), since it diverges as $t \to 0^{+}$ (see (\ref{eqpsol:AsymptAssocQT1})).  Thus the growing mode $\mathcal{D}^{(+)}(t)$ in (\ref{eq:GeneralScalarSolutionScalarDust}) must be a linear combination of the two solutions in (\ref{eq:DLambdaBackgroundSolution}):
\begin{equation} \label{eq:DLambdaBackgroundAlternateGrowingMode}
\mathcal{D}^{(+)} = H \int_0^t \frac{dt}{H^2 \ell^2} = a_1 (\sinh T)^{- \frac{1}{6}} Q^{5/6}_{1/6}(\cosh T) + a_2 \frac{H}{\lambda}.
\end{equation}  The growing mode defined in this way will satisfy
\begin{equation}
\lim_{t \to 0} \mathcal{D}^{(+)} = 0.
\end{equation}  We depict the growing and decaying mode solutions in figure \ref{fig:PerturbationScalarModes-FL}.

\subsection{Tensor Perturbations for $\Lambda = 0$} \label{ssec:DustTensorPerturbations}

The solution for $\mathcal{X}_{(k)}(\eta)$ is obtained in terms of Bessel and Legendre functions by choosing $\beta = 2$ in the general solution (\ref{eq:GeneralFlatFLGoverningDESoln}), (\ref{eq:GeneralOpenFLGoverningDESoln}) and (\ref{eq:GeneralClosedFLGoverningDESoln}).  Since the index has the value $\mu = \tfrac{5}{2}$, these solutions are elementary functions, given by (\ref{eqpsol:SpecialSolution5a})-(\ref{eqpsol:SpecialSolution5f}).  Upon comparing the functional dependence in each case, we obtain the following unified form of the solution:
\begin{equation} \label{eq:DustTensorPertrubationsSolution}
\mathcal{X}_{(k)}(\eta) = \frac{1}{\lambda \ell} \big[c_{+} \mathcal{X}_{(k)}^{+}(\eta) + c_{-} \mathcal{X}_{(k)}^{-}(\eta)\big],
\end{equation} where
\begin{eqnarray}
\mathcal{X}_{(k)}^{+}(\eta) & = & (4 n^2 - K - 3 H^2 \ell^2) \sin(n \eta) + 6 n H \ell \cos(n \eta), \\
\mathcal{X}_{(k)}^{-}(\eta) & = & (4 n^2 - K - 3 H^2 \ell^2) \cos(n \eta) - 6 n H \ell \sin(n \eta),
\end{eqnarray} and $n^2 = k^2 + 3 K$.  Here, the coefficients $c_{+}$ and $c_{-}$ are constants that depend on $k$.

Tensor perturbations describe the propagation of gravitational waves, which are represented by the electric and magnetic Weyl tensor, $E_{ab}$ and $H_{ab}$.  We specialize the solution (\ref{eq:DustTensorPertrubationsSolution}) to $K = -1$ and use (\ref{eq:FLsolution-FC-recollapse}) and $H = \ell^{\prime} / \ell^2$ to determine $\ell$ and $H$.  On applying (\ref{eq:tensorkinE}) and (\ref{eq:tensorkinH}), we can determine the asymptotic behaviour of $E_{ab}$ and $H_{ab}$ in three asymptotic regimes:  The first regime ($\eta \ll 1$ and $k \eta \ll 1$) describes long wavelength perturbations that occur outside the particle horizon close to the singularity; the second regime ($\eta \ll 1$ and $k \eta \gg 1$) describes perturbations within the particle horizon close to the singularity; lastly, the long term behaviour of the perturbation, when spatial curvature is significant, is described in the third regime ($\eta \gg 1$).

\bigskip

\begin{center}
\begin{tabular}{c|c|c|c}
& $\eta \ll 1$ and $k \eta \ll 1$ & $\eta \ll 1$ and $k \eta \gg 1$ & $\eta \gg 1$ \\[0.3ex]
\hline $\displaystyle E_{ab} / H^2$ & $c_{+} \eta^2 + c_{-} \eta^{-3}$ & $A \cos (n \eta + \phi)$ & $A e^{- \eta} \cos(n \eta + \phi)$ \\[1.6ex]
$\displaystyle H_{ab} / H^2$ & $c_{+} \eta^3 + c_{-} \eta^{-2}$ & $A \sin (n \eta + \phi)$ & $A e^{- \eta} \sin(n \eta + \phi)$ \\[1.6ex]
\hline
\end{tabular} \\
\end{center}

\bigskip

\begin{figure}[p]
\begin{center}
\includegraphics[width=\textwidth]{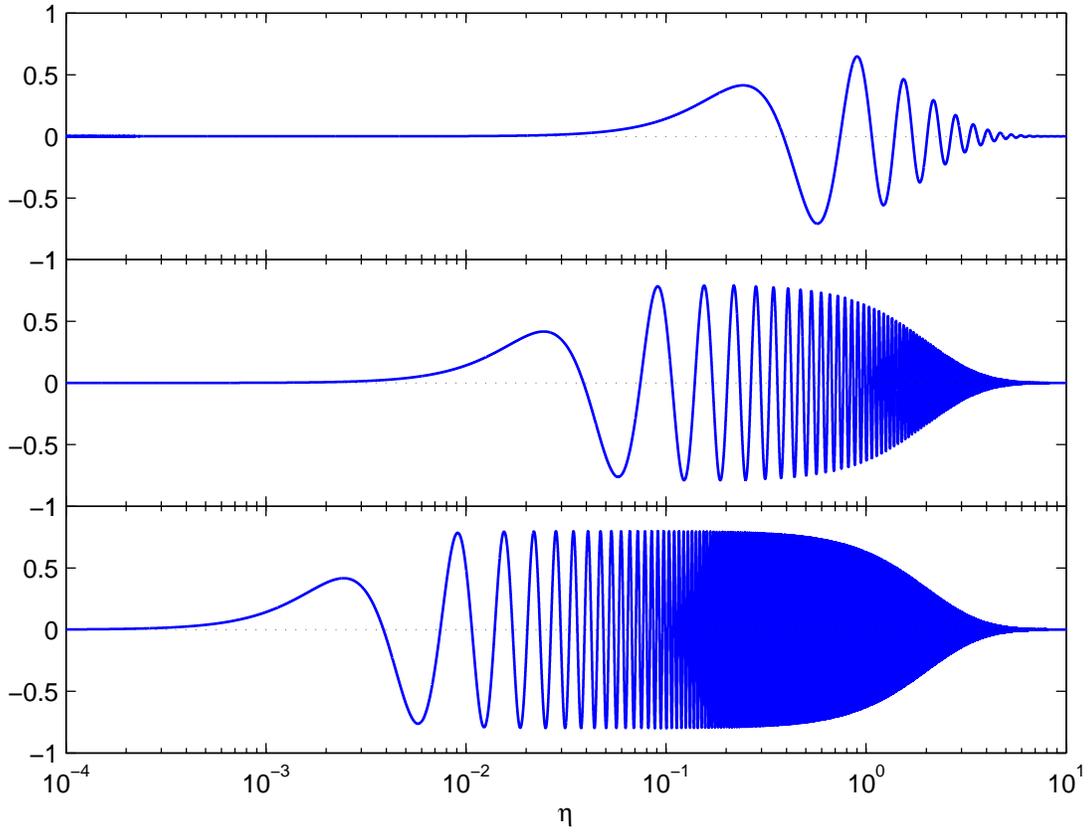}
\end{center}
\caption[Tensor modes of the Hubble-normalized electric Weyl tensor in open FL with dust]{Tensor modes of the Hubble-normalized electric Weyl tensor $E_{(k)} / H^2$ for an open FL background ($K = -1$) with dust.  From top to bottom, the eigenmodes in each plot are $n=10$, $n=100$ and $n=1000$.} \label{fig:DustTensorEplot}
\end{figure}

Here $c_{\pm}$, $A$ and $\phi$ are arbitrary constants that depend on $k$.  In the first regime we can see that the tensor perturbations do not represent gravitational waves since they have a power law dependence on $\eta$.  We depict the behaviour of $E_{(k)} / H^2$ in figure \ref{fig:DustTensorEplot}, for three given choices of $n$.

\section{Perturbations of Radiation-Filled FL Models} \label{sec:RadiationPertSolutions}

We now derive the general solution of the governing DEs for a radiation fluid with zero cosmological constant ($\Lambda = 0$) in terms of elementary functions.

The solution for $\mathcal{D}_{(k)}(\eta)$ is obtained by choosing $\beta = 1$ in the general solution (\ref{eq:GeneralFlatFLGoverningDESoln}), (\ref{eq:GeneralOpenFLGoverningDESoln}) and (\ref{eq:GeneralClosedFLGoverningDESoln}).  Since the index has the value $\mu = \tfrac{3}{2}$, these solutions are elementary functions, given by (\ref{eqpsol:SpecialSolution3a})-(\ref{eqpsol:SpecialSolution3f}).  Upon comparing the functional dependence in each case, we obtain the following unified form of the solution:
\begin{equation} \label{eq:RadiationTensorPertrubationsSolution}
\mathcal{D}_{(k)}(\eta) = c_{+} \mathcal{D}^{(+)}(\eta) + c_{-} \mathcal{D}^{(-)}(\eta),
\end{equation} where
\begin{eqnarray}
\mathcal{D}^{(+)}(\eta) & = & H \ell \sin(c_s k \eta) - c_s k \cos(c_s k \eta), \\
\mathcal{D}^{(-)}(\eta) & = & H \ell \cos(c_s k \eta) + c_s k \sin(c_s k \eta),
\end{eqnarray} and
\begin{equation}
H \ell = \left\{ \begin{array}{ll} \coth \eta & \quad \mbox{Open FL ($K = -1$),} \\[0.5ex] \eta^{-1} & \quad \mbox{Flat FL ($K = 0$),} \\[0.5ex] \cot \eta & \quad \mbox{Closed FL ($K = +1$).} \end{array} \right.
\end{equation}  The coefficients $c_{+}$ and $c_{-}$ are constants that depend on $k$.

The solution for $\mathcal{X}_{(k)}(\eta)$ is obtained by choosing $\beta = 1$ in the general solution (\ref{eq:GeneralFlatFLGoverningDESoln}), (\ref{eq:GeneralOpenFLGoverningDESoln}) and (\ref{eq:GeneralClosedFLGoverningDESoln}).  Since the index has the value $\mu = \tfrac{3}{2}$, these solutions are elementary functions, given by (\ref{eqpsol:SpecialSolution3a})-(\ref{eqpsol:SpecialSolution3f}).  Upon comparing the functional dependence in each case, we obtain the following unified form of the solution:
\begin{equation}
\mathcal{X}_{(k)}(\eta) = \frac{1}{\lambda \ell} \big[ c_{+} \mathcal{X}_{(k)}^{+}(\eta) + c_{-} \mathcal{X}_{(k)}^{-}(\eta) \big],
\end{equation} where
\begin{eqnarray}
\mathcal{X}_{(k)}^{+}(\eta) & = & H \ell \sin(n \eta) - n \cos(n \eta), \\
\mathcal{X}_{(k)}^{-}(\eta) & = & H \ell \cos(n \eta) + n \sin(n \eta), \\
\end{eqnarray} and $n^2 = k^2 + 3 K$.  The coefficients $c_{+}$ and $c_{-}$ are constants that depend on $k$.

As with perturbations of a dust background, we specialize the solution (\ref{eq:RadiationTensorPertrubationsSolution}) to $K = -1$ and use (\ref{eq:FLsolution-FC-recollapse}) and $H = \ell^{\prime} / \ell^2$ to determine $\ell$ and $H$.  On applying (\ref{eq:tensorkinE}) and (\ref{eq:tensorkinE}), we can determine the asymptotic behaviour of $E_{ab}$ and $H_{ab}$ in three asymptotic regimes.

\bigskip

\begin{center}
\begin{tabular}{c|c|c|c}
& $\eta \ll 1$ and $k \eta \ll 1$ & $\eta \ll 1$ and $k \eta \gg 1$ & $\eta \gg 1$ \\[0.3ex]
\hline $\displaystyle E_{ab} / H^2$ & $c_{+} \eta^2 + c_{-} \eta^{-1}$ & $A \eta \cos (n \eta + \phi)$ & $A e^{- \eta} \cos(n \eta + \phi)$ \\[1.6ex]
$\displaystyle H_{ab} / H^2$ & $c_{+} \eta^3 + c_{-}$ & $A \eta \sin (n \eta + \phi)$ & $A e^{- \eta} \sin(n \eta + \phi)$ \\[1.6ex]
\hline
\end{tabular} \\
\end{center}

\bigskip

\begin{figure}[p]
\begin{center}
\includegraphics[width=\textwidth]{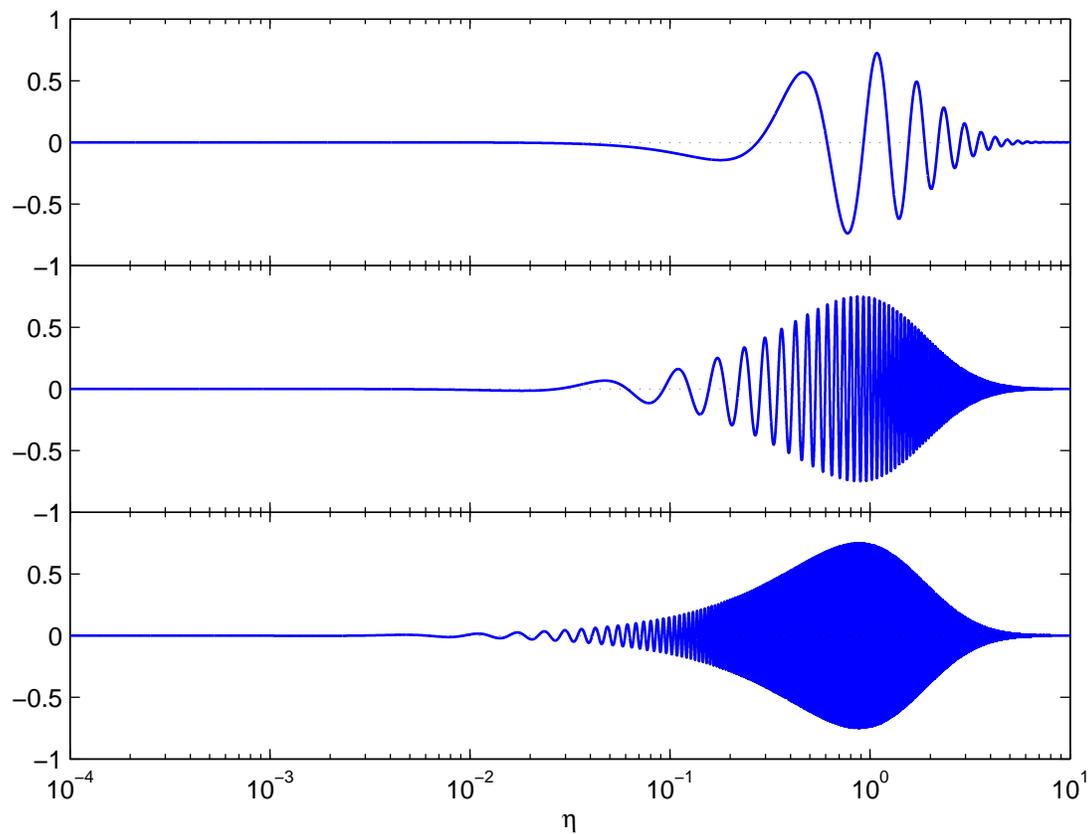}
\end{center}
\caption[Tensor modes of the Hubble-normalized electric Weyl tensor in open FL with radiation]{Tensor modes of the Hubble-normalized electric Weyl tensor $E_{(k)} / H^2$ for an open FL background ($K = -1$) with radiation.  From top to bottom, the eigenmodes in each plot are $n=10$, $n=100$ and $n=1000$.} \label{fig:RadiationTensorEplot}
\end{figure}

Here $c_{\pm}$, $A$ and $\phi$ are arbitrary constants that depend on $k$.  We depict the behaviour of $E_{(k)} / H^2$ in figure \ref{fig:RadiationTensorEplot}, for three given choices of $n$.

\clearpage

\section{Perturbations in Asymptotic Epochs of FL Models} \label{sec:AsymptoticBehaviourMasterDEs}

In this section we give approximate solutions of the governing ODEs for scalar and tensor perturbations in three asymptotic epochs for FL cosmologies.  We also give the behaviour of the spatial gradients, kinematic quantities and Weyl tensor components for these approximate solutions.  Further, to facilitate comparisons, we give the Bardeen (1980) gauge-invariant metric potentials for the approximate solutions.

The three asymptotic epochs are as follows:

\paragraph{Flat FL.}  The flat FL background is described by
\begin{equation} \label{eq:AsymptoticFlatFLBackground}
\Omega = 1, \quad \Omega_\Lambda = 0, \quad \Omega_k = 0, \quad \lambda \ell = \left( \frac{\eta}{\beta} \right)^{\beta}, \quad H \ell = \frac{\beta}{\eta}.
\end{equation}  This solution approximates the dynamics of any FL cosmology with an initial singularity with cosmological constant and arbitrary curvature at early times, since these models satisfy
\begin{equation} \label{eq:AsymptoticFlatVariables}
\lim_{\eta \to 0} (\Omega, \Omega_\Lambda, \Omega_k) = (1, 0, 0).
\end{equation} (see table \ref{table:FLHubbleAsymptoticBehaviour}).

\paragraph{Milne.}  The Milne background is described by
\begin{equation} \label{eq:AsymptoticMilneBackground}
\Omega = 0, \quad \Omega_\Lambda = 0, \quad \Omega_k = 1, \quad \lambda \ell = e^{\eta}, \quad H \ell = 1.
\end{equation}  This solution approximates the dynamics of any open FL cosmology with zero cosmological constant at late times, since these models satisfy
\begin{equation}
\lim_{\eta \to \infty} (\Omega, \Omega_\Lambda, \Omega_k) = (0, 0, 1).
\end{equation} (see table \ref{table:FLHubbleAsymptoticBehaviour}).

\paragraph{de Sitter.}  The de Sitter background is described by
\begin{equation} \label{eq:AsymptoticDeSitterBackground}
\Omega = 0, \quad \Omega_\Lambda = 1, \quad \Omega_k = 0, \quad \lambda \ell = \frac{1}{\tilde{\eta}}, \quad H \ell = \frac{1}{\tilde{\eta}},
\end{equation} where $\tilde{\eta}$ is defined by (\ref{eq:FLdeSitterTildeEta}):
\begin{equation}
\tilde{\eta} = \eta - \eta_f.
\end{equation}  This solution approximates the dynamics of any ever-expanding FL cosmology with cosmological constant and arbitrary curvature at late times, since these models satisfy
\begin{equation}
\lim_{\eta \to \eta_f} (\Omega, \Omega_\Lambda, \Omega_k) = (0, 1, 0).
\end{equation} (see table \ref{table:FLHubbleAsymptoticBehaviour}).

\paragraph{Remark:}  The approximate solution for the case of flat FL with $k \eta \ll 1$ (Table \ref{fig:AsymptoticBehaviourFlatFLOutside}) have been given by Goode (1989).  To the best of the author's knowledge, the other results are new.

\clearpage

\subsection{The Flat FL Asymptotic Epoch}

The solutions of the governing equations in the flat FL background are obtained by substituting (\ref{eq:AsymptoticFlatFLBackground}) into (\ref{eq:GeneralFlatFLGoverningDESoln}) and (\ref{eq:vectorperturbationsgeneralsolution}) and are given by
\begin{eqnarray}
\label{eq:AsymptoticFlatFLScalarSoln} \mathcal{D}_{(k)}(\eta) & = & \eta^{\frac{3}{2} - \beta} \left[ c_{+} J_{\beta + \frac{1}{2}}(c_s k \eta) + c_{-} Y_{\beta + \frac{1}{2}} (c_s k \eta) \right], \\
\label{eq:AsymptoticFlatFLVectorSoln} \mathcal{W}_{(k)}(\eta) & = & C \eta^{2(1 - \beta)}, \\
\label{eq:AsymptoticFlatFLTensorSoln} \mathcal{X}_{(k)}(\eta) & = & \eta^{\frac{1}{2} - \beta} \left[ c_{+} J_{\beta + \frac{1}{2}}(k \eta) + c_{-} Y_{\beta + \frac{1}{2}}(k \eta) \right].
\end{eqnarray}

In the expressions for $\mathcal{D}_{(k)}(\eta)$ and $\mathcal{X}_{(k)}(\eta)$ we can identify two asymptotic regimes, namely $k \eta \ll 1$ and $k \eta \gg 1$.  The restriction $k \eta \ll 1$ means that the perturbation mode corresponding to $k$ is outside the particle horizon, while $k \eta \gg 1$ means the perturbation mode is inside the particle horizon (see section \ref{ssec:ConformalTimeParticleHorizon}).  The approximate solutions subject to these restrictions can be obtained from the asymptotic expressions for the Bessel functions in Appendix \ref{asec:AsymptoticBehaviourBesselLegendre}.

These approximate solutions will also be valid for an open or closed FL background, for the following reason.  If $K = \pm 1$ and $k \eta \ll 1$, it follows that $\eta$ must satisfy $\eta \ll 1$ since $k > 1$ -- see (\ref{eq:scalareigenvalues}), (\ref{eq:vectoreigenvalues}) and (\ref{eq:tensoreigenvalues}).  It thus follows from (\ref{eq:AsymptoticFlatVariables}) that
\begin{equation} \label{eq:FlatFLAsymptoticCurvatureCondition}
\frac{\Omega_k}{\Omega} \ll 1,
\end{equation} \textit{i.e.} the spatial curvature is dynamically negligible in the regime $k \eta \ll 1$.

The approximate solutions and the resulting time dependence of the first order quantities are given in table \ref{fig:AsymptoticBehaviourFlatFLOutside} ($k \eta \ll 1$) and table \ref{fig:AsymptoticBehaviourFlatFLInside} ($k \eta \gg 1$).

\begin{table}[tbp]
\textbf{Perturbations of FL models subject to $k \eta \ll 1$ and $\Omega_k / \Omega_m \ll 1$.}\ \\

\hspace{0.5cm} \begin{tabular}{ccccc}
\underline{Quantity} & & \underline{Scalar Perturbations} & \underline{Vector Perturbations} & \underline{Tensor Perturbations} \\[1.5ex]
$\displaystyle \mathcal{D}_{(k)}$ & $\backsim$ & $\displaystyle \{\eta^2, \eta^{1 - 2 \beta}\}$ & - & - \\[1.5ex]
$\displaystyle \mathcal{W}_{(k)}$ & $\backsim$ & - & $\displaystyle \{\eta^{2 - 2 \beta}\}$ & - \\[1.5ex]
$\displaystyle \mathcal{X}_{(k)}$ & $\backsim$ & - & - & $\displaystyle \{\eta, \eta^{- 2 \beta}\}$ \\[1.5ex]
\hline \\
$\displaystyle \frac{\sigma_{ab}}{H}$ & $\backsim$ & $\displaystyle \{\eta^2, \eta^{1 - 2 \beta}\}$ & $\displaystyle \{\eta^{1 - 2 \beta}\}$ & $\displaystyle \{\eta^2, \eta^{1 - 2 \beta}\}$ \\[1.5ex]
$\displaystyle \frac{\omega_{a}}{H}$ & $\backsim$ & $0$ & $\displaystyle \{\eta^{3 - 2 \beta}\}$ & $0$ \\[1.5ex]
$\displaystyle \frac{\hat{\nabla}_{a} \mu}{H \mu} $ & $\backsim$ & $\displaystyle \{\eta^3, \eta^{2 - 2 \beta}\}$ & $\displaystyle \{\eta^{2 - 2 \beta}\}$ & $0$ \\[1.5ex]
$\displaystyle \frac{\hat{\nabla}_{a} H}{H^2}$ & $\backsim$ & $\displaystyle \{\eta^3, \eta^{2 - 2 \beta}\}$ & $\displaystyle \{\eta^{2 - 2 \beta}\}$ & $0$ \\[1.5ex]
$\displaystyle \frac{\hat{\nabla}_{a}\!\ ^{(3)}\!R}{H^3}$ & $\backsim$ & $\displaystyle \{\eta^3, \eta^{4 - 2\beta}\}^{\dagger}$ & $0$ & $0$ \\[1.5ex]
$\displaystyle \frac{E_{ab}}{H^2}$ & $\backsim$ & $\displaystyle \{\eta^2, \eta^{1 - 2 \beta}\}$ & $\displaystyle \{\eta^{1 - 2 \beta}\}$ & $\displaystyle \{\eta^2, \eta^{1 - 2 \beta}\}$ \\[1.5ex]
$\displaystyle \frac{H_{ab}}{H^2}$ & $\backsim$ & $0$ & $\displaystyle \{\eta^{2 - 2 \beta}\}$ & $\displaystyle \{\eta^{3}, \eta^{2 - 2 \beta}\}$ \\[1.5ex]
\hline \\
$\displaystyle \Phi_H$ & $\backsim$ & $\displaystyle \{1, \eta^{-1 - 2 \beta}\}$ & - & - \\[1.5ex]
$\displaystyle \Psi$ & $\backsim$ & - & $\displaystyle \{\eta^{-2 \beta}\}$ & - \\[1.5ex]
$\displaystyle H_{T}^{(2)}$ & $\backsim$ & - & - & $\displaystyle \{1, \eta^{1 - 2 \beta}\}^{\ddagger}$
\end{tabular}

\bigskip
\caption[Perturbations of flat FL with $k \eta \ll 1$]{This table describes perturbations of any FL model subject to \mbox{$k \eta \ll 1$} and \mbox{$\Omega_k / \Omega_m \ll 1$}.  For open and closed FL models the second restriction requires that \mbox{$\eta \ll 1$}, which is necessarily satisfied on account of the first restriction, since $k > 1$.  Recall that $\beta = 2/(3\gamma-2)$, where $\gamma$ is the equation of state parameter (see (\ref{eq:PerturbationBeta})). \vspace{0.1in} \newline \footnotesize{$\ ^\dagger$\ \ The second mode in this case is obtained directly from the exact solution and is zero in the case of dust. \newline $\ ^\ddagger$\ \ The second mode in this case cannot be obtained from the asymptotic behaviour of $\mathcal{X}_{(k)}$.}} \label{fig:AsymptoticBehaviourFlatFLOutside}
\end{table}

\begin{table}[tbp]
\textbf{Perturbations of FL models subject to $k \eta \gg 1$ and $\Omega_k / \Omega_m \ll 1$.}\ \\

\hspace{0.5cm} \begin{tabular}{ccccc}
\underline{Quantity} & & \underline{Scalar Perturbations $(\gamma \neq 1)^{\dagger}$} & \underline{Vector Perturbations} & \underline{Tensor Perturbations} \\[1.5ex]
$\displaystyle \mathcal{D}_{(k)}$ & $\backsim$ & $\displaystyle \eta^{1 - \beta} \sin(c_s k \eta + \phi)$ & - & - \\[1.5ex]
$\displaystyle \mathcal{W}_{(k)}$ & $\backsim$ & - & $\displaystyle \eta^{2 - 2 \beta}$ & - \\[1.5ex]
$\displaystyle \mathcal{X}_{(k)}$ & $\backsim$ & - & - & $\displaystyle \eta^{-\beta} \sin(k \eta + \phi)$ \\[1.5ex]
\hline \\
$\displaystyle \frac{\sigma_{ab}}{H}$ & $\backsim$ & $\displaystyle \eta^{2 - \beta} \cos(c_s k \eta + \phi)$ & $\displaystyle \eta^{3 - 2 \beta}$ & $\displaystyle \eta^{1 - \beta} \sin(k \eta + \phi)$ \\[1.5ex]
$\displaystyle \frac{\omega_{a}}{H}$ & $\backsim$ & $0$ & $\displaystyle \eta^{3 - 2 \beta}$ & $0$ \\[1.5ex]
$\displaystyle \frac{\hat{\nabla}_{a} \mu}{H \mu}$ & $\backsim$ & $\displaystyle \eta^{2 - \beta} \sin(c_s k \eta + \phi)$ & $\displaystyle \eta^{2 - 2 \beta}$ & $0$ \\[1.5ex]
$\displaystyle \frac{\hat{\nabla}_{a} H}{H^2}$ & $\backsim$ & $\displaystyle \eta^{3 - \beta} \cos(c_s k \eta + \phi)$ & $\displaystyle \eta^{2 - 2 \beta}$ & $0$ \\[1.5ex]
$\displaystyle \frac{\hat{\nabla}_{a}\!\ ^{(3)}\!R}{H^3}$ & $\backsim$ & $\displaystyle \eta^{3 - \beta} \sin(c_s k \eta + \tilde{\phi})$ & $0$ & $0$ \\[1.5ex]
$\displaystyle \frac{E_{ab}}{H^2}$ & $\backsim$ & $\displaystyle \eta^{1 - \beta} \sin(c_s k \eta + \phi)$ & $\displaystyle \eta^{1 - 2 \beta}$ & $\displaystyle \eta^{2 - \beta} \cos(k \eta + \phi)$ \\[1.5ex]
$\displaystyle \frac{H_{ab}}{H^2}$ & $\backsim$ & $0$ & $\displaystyle \eta^{2 - 2 \beta}$ & $\displaystyle \eta^{2 - \beta} \sin(k \eta + \phi)$ \\[1.5ex]
\hline \\
$\displaystyle \Phi_H$ & $\backsim$ & $\displaystyle \eta^{- 1 - \beta} \sin(c_s k \eta + \phi)$ & - & - \\[1.5ex]
$\displaystyle \Psi$ & $\backsim$ & - & $\displaystyle \eta^{-2 \beta}$ & - \\[1.5ex]
$\displaystyle H_{T}^{(2)}$ & $\backsim$ & - & - & $\eta^{-\beta} \cos(k \eta + \phi)$
\end{tabular}

\bigskip
\caption[Perturbations of flat FL, with $k \eta \gg 1$]{This table describes perturbations of flat FL with $k \eta \gg 1$, and open and closed FL with $k \eta \gg 1$ and $\eta \ll 1$.  For open and closed FL models the second restriction requires that $\frac{\Omega_k}{\Omega} \ll 1$.  The first restriction then requires that $k$ is sufficiently large.  Recall that $\beta = 2/(3\gamma-2)$, where $\gamma$ is the equation of state parameter (see (\ref{eq:PerturbationBeta})).  \vspace{0.1in} \newline \footnotesize{$\ ^\dagger$\ \ For the case of $\gamma = 1$ see table \ref{fig:AsymptoticBehaviourFlatFLOutside}.}} \label{fig:AsymptoticBehaviourFlatFLInside}
\end{table}

\clearpage

\subsection{The Milne Asymptotic Epoch}

The governing DEs in the Milne background are obtained by substituting (\ref{eq:AsymptoticMilneBackground}) into (\ref{eq:ScalarMasterDENormalForm}), (\ref{eq:vectorperturbationmasterequation_c}) and (\ref{eq:TensorMasterDENormalForm}) and are given by
\begin{align}
\hat{\mathcal{D}}_{(k)}^{\prime \prime} + \omega^2 \hat{\mathcal{D}}_{(k)} = 0, & & \mathcal{D}_{(k)} &= e^{- (1 - \frac{1}{\beta}) \eta} \hat{\mathcal{D}}_{(k)}, \\
\mathcal{W}_{(k)}^{\prime} + \tfrac{2}{\beta} (\beta - 1) \mathcal{W}_{(k)} = 0, & \\
\hat{\mathcal{X}}_{(k)}^{\prime \prime} + n^2 \hat{\mathcal{X}}_{(k)} = 0, & & \mathcal{X}_{(k)} &= e^{- \eta} \hat{\mathcal{X}}_{(k)}.
\end{align} where
\begin{equation}
\omega^2 = c_s^2 k^2 - (1 - \tfrac{1}{\beta})^2
\end{equation} holds for scalar perturbations and
\begin{equation}
n^2 = k^2 - 3
\end{equation} holds for tensor perturbations.  For scalar perturbations there are two cases, namely $\omega^2 > 0$ and $\omega^2 \leq 0$.  The second case includes dust (\textit{i.e.} \mbox{$c_s^2 = \gamma - 1 = 0$}, \mbox{$\beta = 2$}), while the first case includes radiation (\textit{i.e.} \mbox{$\gamma = \tfrac{4}{3}$}, \mbox{$\beta = 1$}).  In addition, the condition $\omega^2 > 0$ is satisfied for all perfect fluids other than dust (\textit{i.e.} \mbox{$1 < \gamma < 2$}) provided that the wave number $k$ is sufficiently large, \textit{i.e.} $k > k_{crit}$.  We will restrict our considerations to the cases $\omega^2 > 0$ and dust.

The solutions are
\begin{eqnarray}
\label{eq:AsymptoticMilneScalarSoln} \mathcal{D}_{(k)}(\eta) & = & \left\{ \begin{array}{ll} A e^{- (1 - \frac{1}{\beta}) \eta} \sin (\omega \eta + \phi), & \mbox{if\ $\gamma > 1, k > k_{crit}$,} \\ c_{+} + c_{-} e^{- \eta}, & \mbox{if\ $\gamma = 1$,} \end{array} \right.  \\
\label{eq:AsymptoticMilneVectorSoln} \mathcal{W}_{(k)}(\eta) & = & C\ e^{-2 (1 - \frac{1}{\beta}) \eta}, \\
\label{eq:AsymptoticMilneTensorSoln} \mathcal{X}_{(k)}(\eta) & = & A e^{-\eta} \sin(n \eta + \phi),
\end{eqnarray} where $A$ is an arbitrary constant amplitude.

In order to calculate the approximate expression for $E_{ab}$ we need an approximate expression for $\Omega$ subject to $\eta \gg 1$.  A simple calculation using (\ref{eq:FLsolution-FC-recollapse}) and (\ref{eq:FLsolution-FC-EnergyDensity}) yields the approximate asymptotic behaviour
\begin{equation}
\Omega = \exp \left( - \tfrac{2}{\beta} \eta \right) \ll 1.
\end{equation}

The approximate solutions and the resulting time dependence of the first order quantities are given in table \ref{fig:AsymptoticBehaviourMilne}.

\begin{table}[tbp]
\textbf{Perturbations of Milne.}\ \\

\hspace{-0.5cm} \begin{tabular}{cccccc}
\underline{Quantity} & & \multicolumn{2}{c}{\underline{Scalar Perturbations}} & \underline{Vector Perturbations} & \underline{Tensor Perturbations} \\[1.5ex]
& & \underline{$\gamma = 1$} & \underline{$\gamma > 1, k^2 > k^2_{crit}$} & & \\[1.5ex]
$\displaystyle \mathcal{D}_{(k)}$ & $\backsim$ & $\displaystyle \{1, e^{-\eta}\}$ & $\displaystyle e^{- (1 - \frac{1}{\beta}) \eta} \sin(\omega \eta + \phi)$ & - & - \\[1.5ex]
$\displaystyle \mathcal{W}_{(k)}$ & $\backsim$ & - & - & $\displaystyle e^{-2 (1 - \frac{1}{\beta}) \eta}$ & - \\[1.5ex]
$\displaystyle \mathcal{X}_{(k)}$ & $\backsim$ & - & - & - & $\displaystyle e^{-\eta} \sin(n \eta + \phi)$ \\[1.5ex]
\hline \\
$\displaystyle \frac{\sigma_{ab}}{H}$ & $\backsim$ & $\displaystyle \{\eta e^{-\eta}, e^{-\eta}\}^\dagger$ & $\displaystyle e^{- (1 - \frac{1}{\beta}) \eta} \sin(\omega \eta + \tilde{\phi})$ & $\displaystyle e^{-2 (1 - \frac{1}{\beta}) \eta}$ & $\displaystyle e^{-\eta} \sin(n \eta + \phi)$ \\[1.5ex]
$\displaystyle \frac{\omega_{a}}{H}$ & $\backsim$ & $0$ & $0$ & $\displaystyle e^{- 2 (1 - \frac{1}{\beta}) \eta}$ & $0$ \\[1.5ex]
$\displaystyle \frac{\hat{\nabla}_{a} \mu}{H \mu}$ & $\backsim$ & $\displaystyle \{1, e^{-\eta}\}$ & $\displaystyle e^{- (1 - \frac{1}{\beta}) \eta} \sin(\omega \eta + \phi)$ & $\displaystyle e^{-2 (1 - \frac{1}{\beta}) \eta}$ & $0$ \\[1.5ex]
$\displaystyle \frac{\hat{\nabla}_{a} H}{H^2}$ & $\backsim$ & $\displaystyle \{\eta e^{-\eta}, e^{-\eta}\}^\dagger$ & $\displaystyle e^{- (1 - \frac{1}{\beta}) \eta} \sin(\omega \eta + \tilde{\phi})$ & $\displaystyle e^{-2 (1 - \frac{1}{\beta}) \eta}$ & $0$ \\[1.5ex]
$\displaystyle \frac{\hat{\nabla}_{a}\!\ ^{(3)}\!R}{H^3}$ & $\backsim$ & $\displaystyle \{e^{-\eta},e^{-\eta}\}$ & $\displaystyle e^{- (1 - \frac{1}{\beta}) \eta} \sin(\omega \eta + \tilde{\phi})$ & $\displaystyle e^{-2 (1 - \frac{1}{\beta}) \eta}$ & $0$ \\[1.5ex]
$\displaystyle \frac{E_{ab}}{H^2}$ & $\backsim$ & $\displaystyle \{e^{-\eta}, e^{-2 \eta}\}$ & $\displaystyle e^{- (1 + \frac{1}{\beta}) \eta} \sin(\omega \eta + \phi)$ & $\displaystyle e^{-2 \eta}$ & $\displaystyle e^{-\eta} \cos(n \eta + \phi)$ \\[1.5ex]
$\displaystyle \frac{H_{ab}}{H^2}$ & $\backsim$ & $0$ & $0$ & $\displaystyle e^{-2 \eta}$ & $\displaystyle e^{-\eta} \sin(n \eta + \phi)$ \\[1.5ex]
\hline \\
$\displaystyle \Phi_H$ & $\backsim$ & $\displaystyle \{e^{-\eta}, e^{-2 \eta}\}$ & $\displaystyle e^{- (1 + \frac{1}{\beta}) \eta} \sin(\omega \eta + \phi)$ & - & - \\[1.5ex]
$\displaystyle \Psi$ & $\backsim$ & - & - & $\displaystyle e^{-2 \eta}$ & - \\[1.5ex]
$\displaystyle H_{T}^{(2)}$ & $\backsim$ & - & - & - & $e^{-\eta} \sin(n \eta + \phi)$
\end{tabular}

\bigskip
\caption[Perturbations of Milne]{This table describes perturbations of open FL models with $\eta \gg 1$ and $\Omega_k/\Omega_m \gg 1$ and hence $k \eta \gg 1$.  Recall that $\beta = 2/(3\gamma-2)$, where $\gamma$ is the equation of state parameter (see (\ref{eq:PerturbationBeta})).  \vspace{0.1in} \newline \footnotesize{$\ ^\dagger$\ \ In the case of $\gamma = 1$ the growing mode is obtained from the growing mode for dust in an open FL background, given by (\ref{eq:PertDustZeroLambdaGrowingMode}).}} \label{fig:AsymptoticBehaviourMilne}
\end{table}

\clearpage

\subsection{The de Sitter Asymptotic Epoch}

The governing DEs in the flat de Sitter background are obtained by substituting (\ref{eq:AsymptoticDeSitterBackground}) into (\ref{eq:ScalarMasterDENormalForm}), (\ref{eq:vectorperturbationmasterequation_c}) and (\ref{eq:TensorMasterDENormalForm}) and are given by
\begin{align}
\label{eq:AsymptoticDeSitterScalarEq} \hat{\mathcal{D}}_{(k)}^{\prime \prime} + \left[ c_s^2 k^2 - \tfrac{1}{\beta} \left( 1 + \tfrac{1}{\beta} \right) \frac{1}{\tilde{\eta}^2} \right] \hat{\mathcal{D}}_{(k)} &= 0, & \mathcal{D}_{(k)} &= \tilde{\eta}^{1 - \frac{1}{\beta}} \hat{\mathcal{D}}_{(k)}, \\
\mathcal{W}_{(k)}^{\prime} + 2 \left( \tfrac{1}{\beta} - 1 \right) \frac{1}{\tilde{\eta}} \mathcal{W}_{(k)} &= 0, \\
\hat{\mathcal{X}}_{(k)}^{\prime \prime} + n^2 \hat{\mathcal{X}}_{(k)} &= 0, & \mathcal{X}_{(k)} &= \tilde{\eta} \hat{\mathcal{X}}_{(k)},
\end{align} where the prime $(^\prime)$ denotes differentiation with respect to $\tilde{\eta}$ and \mbox{$n^2 = k^2 + 3$} -- see (\ref{eq:tensoreigenvalues}).  The solutions to these equations are given by\footnote{Observe that (\ref{eq:AsymptoticDeSitterScalarEq}) is Bessel's DE (\ref{eqpsol:BesselsDE}) in normal form with $a = c_s k$ and $\mu^2 - \tfrac{1}{4} = \tfrac{1}{\beta}(1 + \tfrac{1}{\beta})$, \textit{i.e.} $\mu = \tfrac{1}{2} + \tfrac{1}{\beta}$.}
\begin{eqnarray}
\label{eq:AsymptoticDeSitterScalarSoln} \mathcal{D}_{(k)}(\eta) & = & \left\{ \begin{array}{ll} \tilde{\eta}^{\frac{3}{2} - \frac{1}{\beta}} \left[ c_{-} J_{\frac{1}{2} + \frac{1}{\beta}}(c_s k \tilde{\eta}) + c_{+} Y_{\frac{1}{2} + \frac{1}{\beta}} (c_s k \tilde{\eta}) \right], & \mbox{if\ $\gamma \neq 1$} \\[1.5ex] c_{-} \tilde{\eta}^2 + c_{+}, & \mbox{if\ $\gamma = 1$} \end{array} \right.  \\
\label{eq:AsymptoticDeSitterVectorSoln} \mathcal{W}_{(k)}(\eta) & = & c\ \tilde{\eta}^{2 - \frac{2}{\beta}}, \\
\label{eq:AsymptoticDeSitterTensorSoln} \mathcal{X}_{(k)}(\eta) & = & c_{-} \tilde{\eta} \sin(n \tilde{\eta}) + c_{+} \tilde{\eta} \cos(n \tilde{\eta}).
\end{eqnarray}

As in the case of the flat FL background, we can distinguish two regimes, depending on the wave number, \textit{i.e.} $k \tilde{\eta} \ll 1$ and $k \tilde{\eta} \gg 1$.  These results are valid also for $K = \pm 1$, but require that $\Omega_k / \Omega_\Lambda \ll 1$, which holds for $\tilde{\eta} \ll 1$ on account of (\ref{eq:AsymptoticDeSitterBackground}).

In order to calculate the approximate expression for $E_{ab}$ we need an approximate expression for $\Omega$ subject to $\tilde{\eta} \ll 1$.  A simple calculation using (\ref{eq:FLFlatCosmoSoln}) and (\ref{eq:FLFlatCosmoEnergyDensity}) yields the approximate asymptotic behaviour
\begin{equation}
\Omega \sim \tilde{\eta}^{3 \gamma}.
\end{equation}

The approximate solutions and the resulting time dependence of the first order quantities are given in table \ref{fig:AsymptoticBehaviourDeSitter2} ($k \tilde{\eta} \ll 1$) and table \ref{fig:AsymptoticBehaviourDeSitter} ($k \tilde{\eta} \gg 1$).

\begin{table}[tbp]
\textbf{Perturbations of de Sitter subject to $k \tilde{\eta} \ll 1$ and $\Omega_k / \Omega_\Lambda \ll 1$.}\ \\

\hspace{0.5cm} \begin{tabular}{ccccc}
\underline{Quantity} & & \underline{Scalar Perturbations} & \underline{Vector Perturbations} & \underline{Tensor Perturbations} \\[1.5ex]
$\displaystyle \mathcal{D}_{(k)}$ & $\backsim$ & $\displaystyle \{\tilde{\eta}^{1 - \frac{2}{\beta}}, \tilde{\eta}^2\}$ & - & - \\[1.5ex]
$\displaystyle \mathcal{W}_{(k)}$ & $\backsim$ & - & $\displaystyle \tilde{\eta}^{2 - \frac{2}{\beta}}$ & - \\[1.5ex]
$\displaystyle \mathcal{X}_{(k)}$ & $\backsim$ & - & - & $\displaystyle \{\tilde{\eta}, \tilde{\eta}^2\}$ \\[1.5ex]
\hline \\
$\displaystyle \frac{\sigma_{ab}}{H}$ & $\backsim$ & $\displaystyle \{\tilde{\eta}^{3 - \frac{2}{\beta}},\tilde{\eta}^2\}^{\dagger}$ & $\displaystyle \tilde{\eta}^{3 - \frac{2}{\beta}}$ & $\displaystyle \{\tilde{\eta}^2, \tilde{\eta}^3\}$ \\[1.5ex]
$\displaystyle \frac{\omega_{a}}{H}$ & $\backsim$ & $0$ & $\displaystyle \tilde{\eta}^{3 - \frac{2}{\beta}}$ & $0$ \\[1.5ex]
$\displaystyle \frac{\hat{\nabla}_{a} \mu}{H \mu}$ & $\backsim$ & $\displaystyle \{\tilde{\eta}^{2 - \frac{2}{\beta}}, \tilde{\eta}^3\}$ & $\displaystyle \tilde{\eta}^{2 - \frac{2}{\beta}}$ & $0$ \\[1.5ex]
$\displaystyle \frac{\hat{\nabla}_{a} H}{H^2}$ & $\backsim$ & $\displaystyle \{\tilde{\eta}^{4 - \frac{2}{\beta}}, \tilde{\eta}^3\}^{\dagger}$ & $\displaystyle \tilde{\eta}^{4}$ & $0$ \\[1.5ex]
$\displaystyle \frac{\hat{\nabla}_{a}\!\ ^{(3)}\!R}{H^3}$ & $\backsim$ & $\displaystyle \{\tilde{\eta}^{4 - \frac{2}{\beta}}, \tilde{\eta}^3\}$ & $0$ & $0$ \\[1.5ex]
$\displaystyle \frac{E_{ab}}{H^2}$ & $\backsim$ & $\displaystyle \{\tilde{\eta}^{3}, \tilde{\eta}^{2 (2 + \frac{1}{\beta})}\}$ & $\displaystyle \tilde{\eta}^{3}$ & $\displaystyle \{\tilde{\eta}^4, \tilde{\eta}^3\}$ \\[1.5ex]
$\displaystyle \frac{H_{ab}}{H^2}$ & $\backsim$ & $0$ & $\displaystyle \tilde{\eta}^{4}$ & $\displaystyle \{\tilde{\eta}^3, \tilde{\eta}^4\}$ \\[1.5ex]
\hline \\
$\displaystyle \Phi_H$ & $\backsim$ & $\displaystyle \{\tilde{\eta}, \tilde{\eta}^{2 + \frac{2}{\beta}}\}$ & - & - \\[1.5ex]
$\displaystyle \Psi$ & $\backsim$ & - & $\displaystyle \tilde{\eta}^{2}$ & - \\[1.5ex]
$\displaystyle H_{T}^{(2)}$ & $\backsim$ & - & - & $\{1, \tilde{\eta}^3\}$
\end{tabular}

\bigskip
\caption[Perturbations of de Sitter ($k \tilde{\eta} \ll 1$)]{This table describes perturbations of any FL model that is future asymptotic to de Sitter, subject to $k \tilde{\eta} \ll 1$.  Recall that $\beta = 2/(3\gamma-2)$, where $\gamma$ is the equation of state parameter (see (\ref{eq:PerturbationBeta})).  \vspace{0.1in} \newline \footnotesize{$\ ^\dagger$\ \ Due to cancellation of higher order terms, the growing mode must be obtained from the exact solution (\ref{eq:AsymptoticDeSitterScalarSoln}).}} \label{fig:AsymptoticBehaviourDeSitter2}
\end{table}

\begin{table}[tbp]
\textbf{Perturbations of de Sitter subject to $k \tilde{\eta} \gg 1$ and $\Omega_k / \Omega_\Lambda \ll 1$.}\ \\

\hspace{0.5cm} \begin{tabular}{ccccc}
\underline{Quantity} & & \underline{Scalar Perturbations ($\gamma \neq 1$)} & \underline{Tensor Perturbations} \\[1.5ex]
$\displaystyle \mathcal{D}_{(k)}$ & $\backsim$ & $\displaystyle \tilde{\eta}^{1 - \frac{1}{\beta}} \sin(c_s k \tilde{\eta} + \phi)$ & - \\[1.5ex]
$\displaystyle \mathcal{X}_{(k)}$ & $\backsim$ & - & $\displaystyle \tilde{\eta} \sin(n \tilde{\eta} + \phi)$ \\[1.5ex]
\hline \\
$\displaystyle \frac{\sigma_{ab}}{H}$ & $\backsim$ & $\displaystyle \tilde{\eta}^{2 - \frac{1}{\beta}} \cos(c_s k \tilde{\eta} + \phi)$ & $\displaystyle \tilde{\eta}^2 \sin(n \tilde{\eta} + \phi)$ \\[1.5ex]
$\displaystyle \frac{\omega_{a}}{H}$ & $\backsim$ & $0$ & $0$ \\[1.5ex]
$\displaystyle \frac{\hat{\nabla}_{a} \mu}{H \mu}$ & $\backsim$ & $\displaystyle \tilde{\eta}^{2 - \frac{1}{\beta}} \sin(c_s k \tilde{\eta} + \phi)$ & $0$ \\[1.5ex]
$\displaystyle \frac{\hat{\nabla}_{a} H}{H^2}$ & $\backsim$ & $\displaystyle \tilde{\eta}^{3 - \frac{1}{\beta}} \cos(c_s k \tilde{\eta} + \phi)$ & $0$ \\[1.5ex]
$\displaystyle \frac{\hat{\nabla}_{a}\!\ ^{(3)}\!R}{H^3}$ & $\backsim$ & $\displaystyle \tilde{\eta}^{3 - \frac{1}{\beta}} \cos(c_s k \tilde{\eta} + \phi)$ & $0$ \\[1.5ex]
$\displaystyle \frac{E_{ab}}{H^2}$ & $\backsim$ & $\displaystyle \tilde{\eta}^{3 + \frac{1}{\beta}} \sin(c_s k \tilde{\eta} + \phi)$ & $\displaystyle \tilde{\eta}^3 \cos(n \tilde{\eta} + \phi)$ \\[1.5ex]
$\displaystyle \frac{H_{ab}}{H^2}$ & $\backsim$ & $0$ & $\displaystyle \tilde{\eta}^3 \sin(n \tilde{\eta} + \phi)$ \\[1.5ex]
\hline \\
$\displaystyle \Phi_H$ & $\backsim$ & $\displaystyle \tilde{\eta}^{1 + \frac{1}{\beta}} \sin(c_s k \tilde{\eta} + \phi)$ & - \\[1.5ex]
$\displaystyle H_{T}^{(2)}$ & $\backsim$ & - & $\displaystyle \tilde{\eta} \cos(n \tilde{\eta} + \phi)$
\end{tabular}

\bigskip
\caption[Perturbations of de Sitter ($k \tilde{\eta} \gg 1$)]{This table describes perturbations of any FL model that is future asymptotic to de Sitter, with $k \tilde{\eta} \gg 1$ and $\tilde{\eta} \ll 1$.  In the case of scalar perturbations with $\gamma = 1$ or vector perturbations, the behaviour is identical to that given in table \ref{fig:AsymptoticBehaviourDeSitter2}.  Recall that $\beta = 2/(3\gamma-2)$, where $\gamma$ is the equation of state parameter (see (\ref{eq:PerturbationBeta})).} \label{fig:AsymptoticBehaviourDeSitter}
\end{table}

\clearpage

\section{Discussion}

We now summarize the principal features of linear perturbations of FL cosmologies.

\begin{enumerate}
\item[i)] For each perturbation type the complete system of linearized evolution and constraint equations is reduced to a single ODE for the harmonic modes, together with expressions for the Hubble-normalized quantities (\ref{eq:HubbleNormalized1})-(\ref{eq:HubbleNormalized3}).  Solutions of the governing DE for vector perturbations can always be determined explicitly in terms of $\ell$ according to (\ref{eq:vectorperturbationsgeneralsolution}).

\item[ii)] When written in normal form using conformal time, the governing DEs for scalar and tensor perturbations are essentially the same if $\Lambda = 0$, \textit{i.e.} they are governed by the same potential $U$ with a different frequency $\omega$, but differ significantly if $\Lambda > 0$.

\item[iii)] We have given a unified derivation of the known explicit solutions of the governing DEs for scalar and tensor perturbations in (\ref{eq:GeneralFlatFLGoverningDESoln})-(\ref{eq:GeneralClosedFLGoverningDESoln}).  Solutions are given in terms of Bessel functions if $\Lambda = 0$ and $K = 0$ with $\gamma$ arbitrary.  Solutions are given in terms of Legendre functions if $\Lambda = 0$ and $K = \pm 1$ with $\gamma$ arbitrary.  In the case of dust ($\gamma = 1$) and radiation ($\gamma = \tfrac{4}{3}$), solutions for all perturbation modes can be written in terms of elementary functions.

\end{enumerate}

We now give a brief overview of the behaviour of scalar and tensor perturbations when the matter content is dust or radiation and possibly with a cosmological constant.

\textit{Scalar perturbations}, as described by the spatial density gradient $\mathcal{D}_{a}$, have a growing mode and a decaying mode in the regime $k \eta \ll 1$.  For \textit{pressure-free matter} these two modes persist for all $\eta$.  The growing mode is bounded in the Milne regime, \textit{i.e.} if $\Lambda = 0$ and $K = -1$, and in the de Sitter regime, \textit{i.e.} if $\Lambda > 0$.

For \textit{radiation} (or more generally if the barotropic index satisfies $w > 1$), scalar perturbations are wavelike in the regime $c_s k \eta \gg 1$ (interpreted as sound waves with speed of propagation $c_s$).  The spatial density gradient is wavelike with constant amplitude in the Milne regime (unbounded if $w > \tfrac{1}{3}$) and in the de Sitter regime (with $k \tilde{\eta} \ll 1$) it has a growing and unbounded mode and a decaying mode (more generally this occurs if $w > 0$).

\textit{Tensor perturbations}, as described by the electric and magnetic Weyl curvature, have a growing mode and a decaying mode in the regime $k \eta \ll 1$.  In the regime $k \eta \gg 1$ where $\Omega_k / \Omega_m \ll 1$ the behaviour is sinusoidal, with power-law amplitude.

In tables \ref{fig:AsymptoticBehaviourFlatFLOutside}-\ref{fig:AsymptoticBehaviourDeSitter} we summarize the time-dependence of the Hubble-normalized kinematic quantities, Hubble-normalized Weyl tensor components and metric potentials in three asymptotic epochs for FL cosmologies.  Tables \ref{fig:AsymptoticBehaviourFlatFLOutside} and \ref{fig:AsymptoticBehaviourFlatFLInside} show clearly that the flat FL model is unstable, both with respect to long wavelength and short wavelength perturbations.  This instability is not revealed by the gauge invariant metric potentials.  Tables \ref{fig:AsymptoticBehaviourMilne} - \ref{fig:AsymptoticBehaviourDeSitter} show that spatial curvature has a stabilizing effect, as does the presence of a cosmological constant.

\appendix

\chapter{Evolution and Constraint Equations} \label{app:evolutionandconstrainteqs}

As discussed in section \ref{sec:evolutionandconstraintequations}, when the Ricci identities (\ref{eq:ricciidentities}) and Bianchi identities (\ref{eq:bianchiidentities}) are applied to the Einstein field equations (\ref{eq:einsteinfieldequations}), we obtain a set of evolution and constraint equations for the kinematic quantities and Weyl tensor components.  For an outline of the derivation of these equations see Ellis (1973, p24-30) and Ellis and van Elst (1998, p10-12).

\paragraph{Projected Field Equations.}  The Einstein field equations (\ref{eq:einsteinfieldequations}) can be rewritten as
\begin{equation} \label{a1eq:EinsteinFieldEquations2}
R_{ab} = T_{ab} - \tfrac{1}{2} T g_{ab}.
\end{equation}  We contract (\ref{a1eq:EinsteinFieldEquations2}) with $h_{ab}$ and $u_a$ and use (\ref{eq:energymomentumtensor}) in order to obtain the following relations:
\begin{eqnarray}
\label{a1eq:contractedefes1} R & = & \mu - 3 p, \\
\label{a1eq:contractedefes2} R_{ab} u^a u^b & = & \tfrac{1}{2} (\mu + 3 p), \\
\label{a1eq:contractedefes3} R_{ab} u^a h^b_{\ c} & = & - q_c, \\
\label{a1eq:contractedefes4} R_{\langle ab \rangle} & = & \pi_{ab},
\end{eqnarray} (see Ellis (1973), p24).  Using (\ref{eq:weyltensor}) we also obtain two important relations relating the Riemann curvature tensor and Weyl tensor components:
\begin{eqnarray}
R_{acbd} u^c u^d & = & E_{ab} - \tfrac{1}{2} R_{\langle ab \rangle} + \tfrac{1}{3} h_{ab} R_{cd} u^{c} u^{d}, \\
\tfrac{1}{2} \eta_{ac}^{\phantom{ac}ef} R_{efbd} u^{c} u^{d} & = & H_{ab} + \tfrac{1}{2} \eta_{abcd} R^{c}_{\ e} u^{d} u^{e}.
\end{eqnarray}

\bigskip
\noindent \begin{tabular}{|p{\textwidth}|}
\hline \ \\
\textbf{Ricci Identities}
\begin{eqnarray}
\label{a1eq:ricciidentity1} \dot{H} & = & -H^2 + \tfrac{1}{3}( \hat{\nabla}_a \dot{u}^{a} + \dot{u}_a \dot{u}^a - 2 \sigma^2 + 2 \omega^2) - \tfrac{1}{6} (\mu + 3p), \\
\label{a1eq:ricciidentity2} \dot{\sigma}_{\langle ab \rangle} & = & -2 H \sigma_{ab} + \hat{\nabla}_{\langle a} \dot{u}_{b \rangle} + \dot{u}_{\langle a} \dot{u}_{b \rangle} - \sigma_{\langle a}^{\ \ c} \sigma_{b \rangle c} - \omega_{\langle a} \omega_{b \rangle} - (E_{ab} - \tfrac{1}{2} \pi_{ab}), \\
\label{a1eq:ricciidentity3} \dot{\omega}_{\langle a \rangle} & = & - 2 H \omega_a + \sigma_{ab} \omega^b - \tfrac{1}{2} \mathsf{curl}(\dot{u}_a), \\
\label{a1eq:ricciidentity4} 0 & = & \hat{\nabla}^b \sigma_{ab} - 2 \hat{\nabla}_a H - \mathsf{curl}(\omega_a) - 2 \epsilon_{abc} \dot{u}^b \omega^c + q_a, \\
\label{a1eq:ricciidentity5} 0 & = & \hat{\nabla}^a \omega_a - \dot{u}^a \omega_a, \\
\label{a1eq:ricciidentity6} 0 & = & H_{ab} - 2 \dot{u}_{\langle a} \omega_{b \rangle} - \hat{\nabla}_{\langle a} \omega_{b \rangle} - \mathsf{curl}(\sigma_{ab}).
\end{eqnarray} \\
\hline
\end{tabular}
\bigskip

The twice-contracted Bianchi identities (\ref{eq:bianchiidentities_b}) lead to the following evolution equations:
\begin{eqnarray}
\label{a1eq:contractedbianchi1} \dot{\mu} & = & -3 H (\mu + p) - \hat{\nabla}_{a} q^a - 2 \dot{u}_a q^a - \sigma^a_{\ b} \pi^{b}_{\ a}, \\
\label{a1eq:contractedbianchi2} \dot{q}_{\langle a \rangle} & = & -4 H q_a - \hat{\nabla}_a p - (\mu + p) \dot{u}_a - \hat{\nabla}^b \pi_{ab} \nonumber \\ & & - \dot{u}^b \pi_{ab} - \sigma^{ab} q_{b} + \epsilon_{abc} \omega^b q^c.
\end{eqnarray}

We are primarily interested in the case of a perfect fluid, where these equations reduce to

\bigskip
\noindent \begin{tabular}{|p{\textwidth}|}
\hline \ \\
\textbf{Twice-Contracted Bianchi Identities (Perfect Fluid Source)}
\begin{eqnarray}
\label{a1eq:contractedbianchi3} \dot{\mu} & = & - 3 H (\mu + p), \\
\label{a1eq:contractedbianchi4} 0 & = & (\mu + p) \dot{u}_a + \hat{\nabla}_{a} p.
\end{eqnarray} \\
\hline
\end{tabular}
\bigskip

For simplicity, the equations obtained from the full Bianchi identities (\ref{eq:bianchiidentities}) are presented here in the case of a perfect fluid source.

\bigskip
\noindent \begin{tabular}{|p{\textwidth}|}
\hline \ \\
\textbf{Bianchi Identities (Perfect Fluid Source)}
\begin{eqnarray}
\label{a1eq:fullbianchi1} \dot{E}_{\langle ab \rangle} & = & - 3 H E_{ab} + 3 \sigma_{\langle a}^{\ \ c} E_{b \rangle c}^{\phantom{c}} - \tfrac{1}{2} (\mu + p) \sigma_{ab} + \mathsf{curl}(H_{ab}) \nonumber \\ & & + \epsilon_{ef(a}^{\phantom{f}} (2 \dot{u}^e H_{b)}^{\ f} - \omega^e E_{b)}^{\ f}), \\
\label{a1eq:fullbianchi2} \dot{H}_{\langle ab \rangle} & = & -3 H H_{ab} + 3 \sigma_{\langle a}^{\ \ c} H_{b \rangle c}^{\phantom{c}} - \mathsf{curl}(E_{ab})\nonumber \\ & & - \epsilon_{ef(a}^{\phantom{f}} (2 \dot{u}^e E_{b)}^{\ f} + \omega^e H_{b)}^{\ f}), \\
\label{a1eq:fullbianchi3} 0 & = & \hat{\nabla}^b E_{ab} + 3 \omega^b H_{ab} + \epsilon_{abc} \sigma^{be} H_{e}^{\ c} - \tfrac{1}{3} \hat{\nabla}_a \mu, \\
\label{a1eq:fullbianchi4} 0 & = & \hat{\nabla}^b H_{ab} - 3 \omega^b E_{ab} - \epsilon_{abc} \sigma^{be} E_{e}^{\ c} - (\mu + p) \omega_a.
\end{eqnarray} \\
\hline
\end{tabular}
\bigskip

\paragraph{Spatial Gradients of Zero-Order Quantities.}  In order to derive the governing equations for linear perturbations of FL, it is necessary to derive evolution equations for the spatial gradients of the matter density $\mu$ and the Hubble parameter $H$, defined by (\ref{eq:fractionaldensitygradient}) and (\ref{eq:spatialgradienthubblescalar}).  By propositions \ref{prop:energymomentumisotropy} and \ref{prop:kinematiccharacterization1}, these quantities are zero in the background and hence gauge-invariant by the Stewart-Walker lemma.

We make use of the fact that the Raychaudhuri equation (\ref{a1eq:ricciidentity1}) can be rewritten as
\begin{equation} \label{eq:raychaudhuriwithR}
\dot{H} = \tfrac{1}{3} \mathcal{T} - \tfrac{1}{2} (\mu + p),
\end{equation} where
\begin{equation}
\mathcal{T} = -3 H^2 + \mu + \Lambda + \hat{\nabla}_a \dot{u}^a + \dot{u}^{a} \dot{u}_{a} - 2 \sigma^2 + 2 \omega^2.
\end{equation}

In order to proceed, we require an identity which permits the interchange of the $\dot{\ }$ derivative and the spatial derivative $\hat{\nabla}_{a}$ when applied to a scalar field $f$.  A straightforward calculation using (\ref{eq:covudecomposition}) yields
\begin{equation} \label{eq:timespacederivativeinterchange1}
(\hat{\nabla}_a f)\dot{\ }_{\perp} = - H \hat{\nabla}_a f + ( \hat{\nabla}_a + \dot{u}_a ) \dot{f} - (\sigma_{a}^{\ b} + \omega_{a}^{\ b}) \hat{\nabla}_b f,
\end{equation} for any scalar $f$ (also see equation (92) in Ellis et al. (1990)).

We now derive evolution equations for $X_a$ and $Z_a$, following Ellis and van Elst (1998, p56).  Substituting $f = \mu$ into (\ref{eq:timespacederivativeinterchange1}) and using (\ref{eq:fractionaldensitygradient}), (\ref{eq:spatialgradienthubblescalar}), (\ref{a1eq:contractedbianchi3}) and (\ref{a1eq:contractedbianchi4}) yields the evolution equation for $X_{a}$.  If we instead choose $f = H$ in (\ref{eq:timespacederivativeinterchange1}) and use (\ref{eq:fractionaldensitygradient}), (\ref{eq:spatialgradienthubblescalar}), (\ref{eq:raychaudhuriwithR}), (\ref{a1eq:contractedbianchi3}) and (\ref{a1eq:contractedbianchi4}) we obtain the evolution equation for $Z_{a}$.  The results of these derivations are given as follows:

\bigskip
\noindent \begin{tabular}{|p{\textwidth}|}
\hline \ \\
\textbf{Spatial Gradients of Zero-Order Quantities}
\begin{eqnarray}
\label{eq:fractionaldensitygradient_evol} \dot{X}_{\langle a \rangle} & = & -4 H X_{a} - (\sigma_{a}^{\ b} + \omega_{a}^{\ b}) X_{b} - (1 + w) \mu Z_{a}, \\
\label{eq:spatialgradienthubblescalar_evol} \dot{Z}_{\langle a \rangle} & = & - 3 H \mathcal{Z}_a - (\sigma_a^{\ b} + \omega_a^{\ b}) Z_b - \tfrac{1}{2} X_a \nonumber \\ & & + \mathcal{T} \dot{u}_a + \hat{\nabla}_a (\hat{\nabla}^b \dot{u}_{b} + \dot{u}^{a} \dot{u}_{a} - 2 \sigma^2 + 2 \omega^2).
\end{eqnarray} \\
\hline
\end{tabular}
\bigskip

\chapter{Spatial Curvature}

In this appendix we introduce the 3-curvature when the fundamental congruence has non-zero vorticity.  This analysis leads to a generalization of the Gauss equation (\ref{eq:gaussequation}), the 3-Ricci tensor (\ref{eq:spatialricci3tensor}) and 3-Ricci scalar.

\section{The Generalized Gauss Equation}

We note that the projected covariant derivative $\hat{\nabla}_{a}$, defined in (\ref{eq:orthogonalcovariantderiv}), is a non-commutative operator on the space-time manifold when vorticity is present, \textit{i.e.} $\omega_{a} \neq 0$.  After a short calculation using (\ref{eq:projectiontensor}), (\ref{eq:orthogonalcovariantderiv}) and (\ref{eq:covudecomposition}), we obtain
\begin{equation} \label{eq:covariantderivativecommutator1}
\hat{\nabla}_{a} \hat{\nabla}_{b} f - \hat{\nabla}_{b} \hat{\nabla}_{a} f = - 2 \omega_{ab} \dot{f}.
\end{equation}  We now derive a similar identity for tensors with one or more index.

On projecting (\ref{eq:covudecomposition}) orthogonal to $u^{a}$ and applying (\ref{eq:orthogonalcovariantderiv}), we obtain
\begin{equation}
\hat{\nabla}_{b} u_{a} = k_{ab},
\end{equation} where $k_{ab}$ is defined by
\begin{equation}
k_{ab} = \omega_{ab} + \sigma_{ab} + H h_{ab},
\end{equation} (see Ellis et al. (1990), eq. (76)).

Upon performing a similar computation as in (\ref{eq:covariantderivativecommutator1}) and using (\ref{eq:ricciidentities}), we obtain the expression
\begin{equation} \label{eq:Generalized3RicciIdentity}
\hat{\nabla}_{a} \hat{\nabla}_{b} X_c - \hat{\nabla}_{b} \hat{\nabla}_{a} X_c = - 2 \omega_{ab} \dot{X}_{\langle c \rangle} + ^{(3)}\!R_{abcd} X^d,
\end{equation} where $^{(3)}\!R_{abcd}$ is defined by
\begin{equation} \label{eq:GeneralizedGaussEquation}
^{(3)}\!R_{abcd} = (R_{abcd})_{\perp} - k_{ac} k_{bd} + k_{ad} k_{bc},
\end{equation} (see Ellis et al. (1990), eq. (79)).  We note that this definition reduces to the 3-Riemann tensor given in (\ref{eq:gaussequation}) when $u^a$ is irrotational.  A similar identity to (\ref{eq:Generalized3RicciIdentity}) can be obtained for tensors with two or more indices, generalized according to
\begin{equation} \label{adiff:comm4}
\hat{\nabla}_{[a} \hat{\nabla}_{b]} S_{cd \cdots ef} = - \omega_{ab} \dot{S}_{\perp cd \cdots ef} + \tfrac{1}{2} (\ ^{(3)}\!R_{scba} T^{s}_{\ d \cdots ef} + \cdots +\ ^{(3)}\!R_{sfba} T^{\ \ \ \ \ \ s}_{cd \cdots e} ).
\end{equation}

On contracting (\ref{eq:GeneralizedGaussEquation}), we can obtain an expression for the spatial Ricci scalar and trace-free spatial Ricci tensor in terms of the spacetime quantities from (\ref{eq:GeneralizedGaussEquation}), incorporating the Weyl tensor through (\ref{eq:weyltensor}) and (\ref{eq:weyldecomposition}):
\begin{eqnarray}
\label{asceq:Generalricci3scalar} ^{(3)}\!R & = & R + 2 R_{bd} u^b u^d - 6 H^2 + 2 \sigma^2 - 2 \omega^2, \\
\label{asceq:Generalspatialricci3tensorfull} ^{(3)}\!S_{ab} & = & E_{ab} + \tfrac{1}{2} (h_{a}^{\ c} h_{b}^{\ d} - \tfrac{1}{3} h^{cd} h_{ab}) R_{cd} - H \sigma_{ab} - H \omega_{ab} \nonumber \\ & & - \tfrac{2}{3} (\sigma^2 - \omega^2) h_{ab} + \sigma^{\ c}_{a} \sigma_{cb} + \omega^{\ c}_{a} \omega_{cb} + \omega^{\ c}_{a} \sigma_{cb} + \sigma^{\ c}_{a} \omega_{cb}.
\end{eqnarray}  Using the field equations (\ref{eq:einsteinfieldequations}) in the form (\ref{a1eq:contractedefes1})-(\ref{a1eq:contractedefes4}) to eliminate $R$ and $R_{ab}$, we obtain
\begin{eqnarray}
\label{a2eq:ricci3scalarW} ^{(3)}\!R & = & -6 H^2 + 2 \mu + 2 \sigma^2 - 2 \omega^2, \\
\label{a2eq:spatialricci3tensorW} ^{(3)}\!S_{ab} & = & E_{ab} + \tfrac{1}{2} \pi_{ab} - H \sigma_{ab} - H \omega_{ab} - \tfrac{2}{3} (\sigma^2 - \omega^2) h_{ab} \nonumber \\ & & + \sigma^{\ c}_{a} \sigma_{cb} + \omega^{\ c}_{a} \omega_{cb} + \omega^{\ c}_{a} \sigma_{cb} + \sigma^{\ c}_{a} \omega_{cb}.
\end{eqnarray}

\section{The Spatial Gradient of the 3-Ricci Scalar}

The \textit{spatial gradient of the 3-Ricci scalar} is defined by
\begin{equation}
\mathcal{R}_a = \ell\ \hat{\nabla}_a\ ^{(3)}\!R.
\end{equation}  An evolution equation for $\mathcal{R}_{a}$ is obtained in the same manner as for $\mathcal{D}_{a}$ and $\mathcal{Z}_{a}$, defined in (\ref{eq:fractionaldensitygradientDC}) and (\ref{eq:fractionaldensitygradientZC}), respectively.  If we choose $f =\ ^{(3)}\!R$ in (\ref{eq:timespacederivativeinterchange1}) and use (\ref{eq:spatialgradienthubblescalar}), (\ref{eq:raychaudhuriwithR}), (\ref{a1eq:contractedbianchi3}) and (\ref{a1eq:contractedbianchi4}) we obtain
\begin{eqnarray} \label{eq:spatialgradientcurvature_evol}
\dot{\mathcal{R}}_{\langle a \rangle} & = & - 2 H \mathcal{R}_a - (\sigma_a^{\ b} + \omega_a^{\ b}) \mathcal{R}_b - \tfrac{4}{3} \mathcal{T} \mathcal{Z}_{a} \nonumber \\ & & - 4 \ell \mathcal{T} \dot{u}_a + \ell \hat{\nabla}_a (\hat{\nabla}^b \dot{u}_{b} - 2 \sigma^2 + 2 \omega^2).
\end{eqnarray}

Further, we can obtain an auxiliary equation describing the evolution of $\mathcal{R}_a$ from (\ref{eq:spatialgradientcurvature_evol}) and (\ref{eq:accelerationdoubledivergenceterm}), presented here for completeness:
\begin{equation} \label{lineq:Revolution}
\dot{\mathcal{R}}_{\langle a \rangle} = -2 H \mathcal{R}_a - \frac{4 K}{\ell^2} \mathcal{Z}_a + 4 H \frac{c_s^2}{(1+w)} \left( \hat{\nabla}^2 + \frac{K}{\ell^2} \right) \mathcal{D}_{a} + 24 c_s^2 H^2 \ell\ \mathsf{curl}(\omega_a).
\end{equation}

\chapter{Solutions of the Friedmann Equation} \label{app:FriedmannSolutions}

In this appendix we list the solutions of the three first order quadratic DEs that arise by transforming the Friedmann DE.  In the first two cases, the solution is uniquely determined by requiring $\frac{dL}{dT} > 0$ and the initial condition
\begin{equation}
L(0) = 0.
\end{equation}  In the third case we require that
\begin{equation}
L(0) = 1, \quad \mbox{and} \quad L^{\prime}(T) > 0\ \mbox{for $T > 0$.}
\end{equation}

\begin{align}
& \mbox{\underline{Differential Equation}} & & \mbox{\underline{Solution}} \nonumber \\
\label{aFLsol:case1} \mbox{\textbf{Case 1}} \quad & \left( \frac{dL}{dT} \right)^2 = 1 + 2 \alpha L + L^2, & & L = \sinh T + 2 \alpha \sinh^2( \tfrac{1}{2} T ). \\
\label{aFLsol:case2} \mbox{\textbf{Case 2}} \quad & \left( \frac{dL}{dT} \right)^2 = 1 + 2 \alpha L - L^2, & & L = \sin T + 2 \alpha \sin^2( \tfrac{1}{2} T ). \\
\label{aFLsol:case3} \mbox{\textbf{Case 3}} \quad & \left( \frac{dL}{dT} \right)^2 = L^2 - 1, & & L = \cosh T.
\end{align}

\pagebreak

There are four known cases where the Friedmann equation can be transformed into the quadratic form (\ref{eq:FLquadraticform}) via a change of variable:

\begin{enumerate}
\item[i)] 1-fluid plus spatial curvature.
\item[ii)] 2-fluid flat FL.
\item[iii)] 2-fluid FL plus arbitrary curvature.  The equation of state parameters must satisfy
\begin{equation}
\gamma_1 = \tfrac{2}{3} + 2 \Delta \gamma, \qquad \gamma_2 = \tfrac{2}{3} + \Delta \gamma,
\end{equation} where $\Delta \gamma$ is constant.
\item[iv)] 3-fluid FL.  The equation of state parameters must satisfy
\begin{equation}
\gamma_1 > \gamma_2 > \gamma_3, \qquad \gamma_1 - \gamma_2 = \gamma_2 - \gamma_3 = \Delta \gamma,
\end{equation} where $\Delta \gamma$ is constant.
\end{enumerate}

\section{Solutions with a Stiff Fluid}

There are three known cases where we can obtain a solution to the Friedmann equation where one fluid is a so-called stiff fluid, \textit{i.e.} $\gamma_s = 2$.  In this section we briefly present each of these solutions.

\paragraph{$SD$-Universes.}  We choose the conformal parameter to be $\lambda = \lambda_d$, which, on using (\ref{eq:FLM}), leads to a single mass parameter defined by
\begin{equation}
m_s = \left( \frac{\lambda_s}{\lambda} \right)^{-4}.
\end{equation}  In this case the Friedmann equation reads
\begin{equation}
\left( \frac{d\ell}{dt} \right)^2 = (\lambda_s \ell)^{-4} + (\lambda_d \ell)^{-1}.
\end{equation}  We apply the change of variable
\begin{equation}
\mathcal{L} = \frac{\lambda_s^4}{\lambda} \ell^3, \qquad \mathcal{T} = 3 \frac{\lambda_s^2}{\lambda} t,
\end{equation} obtaining
\begin{equation}
\left( \frac{d\mathcal{L}}{d\mathcal{T}} \right)^2 = 1 + \mathcal{L}.
\end{equation}  Subject to the IC $\mathcal{L}(0) = 0$, the solution is
\begin{equation}
\mathcal{L}(\mathcal{T}) = \tfrac{1}{4} \mathcal{T}^2 + \mathcal{T},
\end{equation} or equivalently
\begin{equation}
\ell(t) = \lambda^{-1} \left[ \tfrac{9}{4} (\lambda t)^2 + 3 \sqrt{m_s} (\lambda t) \right]^{1/3}.
\end{equation}

\paragraph{$SRC$-Universes.}  We choose the conformal parameter to be $\lambda = \lambda_r$, which, on using (\ref{eq:FLM}), leads to a single mass parameter defined by
\begin{equation}
m_s = \left( \frac{\lambda_s}{\lambda} \right)^{-4}.
\end{equation}  In this case the Friedmann equation reads
\begin{equation}
\left( \frac{d\ell}{d\eta} \right)^2 = (\lambda_s \ell)^{-4} + (\lambda_r \ell)^{-2} - K.
\end{equation}  We apply the change of variable
\begin{equation}
\mathcal{L} = \lambda_s^2 \ell^2, \qquad \mathcal{T} = 2 \eta,
\end{equation} obtaining
\begin{equation}
\left( \frac{d\mathcal{L}}{d\mathcal{T}} \right)^2 = 1 + \frac{1}{\sqrt{m_s}} \mathcal{L} - K \mathcal{L}^2.
\end{equation}  Subject to the IC $\mathcal{L}(0) = 0$, the solution is
\begin{equation}
\ell(t) = \lambda^{-1} \left[ \sqrt{m_s} S_K(2 \eta) + S_K(\eta)^2 \right]^{1/2}.
\end{equation}

\paragraph{$SD\Lambda$-Universes.}  We choose the conformal parameter to be $\lambda = \lambda_\Lambda$, which, on using (\ref{eq:FLM}), leads to two mass parameters defined by
\begin{equation}
m_s = \left( \frac{\lambda_s}{\lambda} \right)^{-4}, \qquad m_d = \left( \frac{\lambda_d}{\lambda} \right)^{-1}.
\end{equation}  In this case the Friedmann equation reads
\begin{equation}
\left( \frac{d\ell}{d\eta} \right)^2 = (\lambda_s \ell)^{-4} + (\lambda_d \ell)^{-1} + (\lambda \ell)^2.
\end{equation}  We apply the change of variable
\begin{equation}
\mathcal{L} = \lambda_s^2 \lambda \ell^3, \qquad \mathcal{T} = 3 \lambda t,
\end{equation} obtaining
\begin{equation}
\left( \frac{d\mathcal{L}}{d\mathcal{T}} \right)^2 = 1 + \frac{m_d}{\sqrt{m_s}} \mathcal{L} + \mathcal{L}^2.
\end{equation}  Subject to the IC $\mathcal{L}(0) = 0$, the solution is
\begin{equation}
\ell(t) = \lambda^{-1} \left[ \sqrt{m_s} \sinh(3 \lambda t) + m_d \sinh^2(\tfrac{3}{2} \lambda t) \right]^{1/3}.
\end{equation}

\section{Existence-Uniqueness for the Friedmann Equation} \label{sec:FLexistenceuniqueness}

In this section we prove existence-uniqueness of solutions to the Friedmann DE (\ref{eq:FLDimlessLDE}), given by
\begin{equation} \label{FLEUeq:FLDimlessLDE}
\left( \frac{dL}{dT} \right)^2 = \sum_{i=1}^{n} m_i L^{-3 \gamma_i + 2} - K,
\end{equation} for any choice of parameters $m_i > 0$, $i = 1, \ldots, n-1$ and subject to the conditions
\begin{equation}
L(0) = 0, \quad \mbox{and} \quad \frac{dL}{dT} > 0,
\end{equation}

We can rewrite (\ref{FLEUeq:FLDimlessLDE}) as
\begin{equation}
\left( \frac{dL}{dT} \right)^2 = L^{-(3 \gamma_1 - 2)} \left[ \sum_{i = 1}^{n} m_i L^{3 (\gamma_1 - \gamma_i)} - K L^{3 \gamma_1 - 2} \right].
\end{equation}  Since $\gamma_1 > \gamma_2 > \cdots > \gamma_n$ and $\gamma_1 > \tfrac{2}{3}$, this DE is of the form
\begin{equation} \label{FLEUeq:FLDimlessLDE2}
\left( \frac{dL}{dT} \right)^2 = L^{- (3 \gamma_1 - 2)} F(L),
\end{equation} with
\begin{equation}
F(0) = m_1 > 0.
\end{equation}  Since $\frac{dL}{dT} > 0$, we can rewrite (\ref{FLEUeq:FLDimlessLDE2}) as
\begin{equation} \label{FLEUeq:FLDimlessLDE3}
\frac{dT}{dL} = \frac{L^{\frac{1}{2} (3 \gamma_1 - 2)}}{F(L)^{1/2}}, \qquad L \geq 0.
\end{equation}  The general solution to (\ref{FLEUeq:FLDimlessLDE3}) is then, for all $m_i > 0$,
\begin{equation}
T = G(L) + C,
\end{equation} where $G(L)$ is the antiderivative of the RHS of (\ref{FLEUeq:FLDimlessLDE3}).  The requirement that $T = 0$ when $L = 0$ fixes $C$, and hence gives a unique solution.

\chapter{Conformal Time} \label{app:AlternativeTime}

The conformal time variable $\eta$ is defined in terms of clock time $t$ and the length scale $\ell$ via
\begin{equation} \label{eq:conformaltime}
\frac{d\eta}{dt} = \frac{1}{\ell}.
\end{equation}

Unlike clock time, conformal time is not necessarily unbounded into the future for ever-expanding models.  In fact, the presence or absence of certain types of fluids determines the behaviour of the conformal time variable.  We present the following theorem:

\renewcommand{\thetheorem}{\thechapter.1}

\begin{theorem} Let ($\mathcal{M}$, $\mathbf{g}$, $\mathbf{u}$) be an ever-expanding $n$-fluid FL cosmology whose length scale satisfies $\ell(0) = 0$ and where each fluid satisfies a linear equation of state (\ref{eq:FLequationofstateNFluid}) with equation of state parameters $\gamma_i$ ordered according to \mbox{$\gamma_1 > \gamma_2 > \cdots > \gamma_n$}.  It follows that $\eta$ satisfies the following constraints:
\begin{enumerate}
\item[(a)] $\eta$ is bounded below iff $\gamma_1 > \tfrac{2}{3}$.
\item[(b)] $\eta$ is bounded above iff $\gamma_2 < \tfrac{2}{3}$.
\end{enumerate}
\end{theorem}

\paragraph{Proof.}  It follows from the definition of an ever-expanding model (see section \ref{ssec:FLClassificationTheorem}) that
\begin{equation} \label{eqconf:EllEvolution}
\frac{d\ell}{dt} > 0\ \forall\ t > 0.
\end{equation}  Using the chain rule we expand (\ref{eq:conformaltime}) so as to obtain
\begin{equation} \label{eqconf:ConformalTimeLS}
\frac{d\eta}{d\ell} = \left[ \ell \frac{d\ell}{dt} \right]^{-1}.
\end{equation}  It then follows from (\ref{eqconf:EllEvolution}) and (\ref{eqconf:ConformalTimeLS}) that $\eta$ is a strictly increasing function of $\ell$ for $t > 0$.

Let $\eta_0$ and $\eta_{\ast}$ be the value of conformal time at $\ell_0$ and $\ell_{\ast}$, respectively.  We now integrate (\ref{eqconf:ConformalTimeLS}) over $(\eta_0, \eta_{\ast})$ so as to obtain
\begin{equation} \label{eqconf:ConformalTime}
\eta_{\ast} - \eta_0 = F(\ell_{\ast}) = \int_{\ell_0}^{\ell_{\ast}} \left[ \ell \left(\frac{d\ell}{dt}\right) \right]^{-1} d\ell.
\end{equation}
Since $\eta$ is strictly increasing, we note that $\eta_{\ast}$ is bounded above if and only if $F(\ell_{\ast})$ is bounded as $\ell_{\ast} \rightarrow +\infty$.  Similarly, we note that $\eta_{\ast}$ is bounded below if and only if $F(\ell_{\ast})$ is bounded as $\ell_{\ast} \rightarrow 0^{+}$.  We now prove the two results separately:

\paragraph{(a)} Assume that condition (a) is not met.  It then follows from (\ref{eq:FLfriedmannDE}) that $\exists$ a constant $C > 0$ such that $\frac{d\ell}{dt} < C\ \forall\ t < t_0$.  The integral (\ref{eqconf:ConformalTime}) then takes the form
\begin{equation}
\eta_{\ast} = \eta_0 - \int_{0}^{\ell_0} \left[ \ell \left(\frac{d\ell}{dt}\right) \right]^{-1} d\ell < \eta_0 - \frac{1}{C} \int_{0}^{\ell_0} \frac{d\ell}{\ell},
\end{equation} which is unbounded.

If condition (a) is met it then follows from (\ref{eq:FLfriedmannDE}) that $\exists$ constants $C > 0 $ and $\epsilon > 0$ such that $\frac{d\ell}{dt} > C \ell^{-\epsilon} \ \forall\ t < t_0$.  The integral (\ref{eqconf:ConformalTime}) then takes the form
\begin{equation}
\eta_{\ast} = \eta_0 - \int_{0}^{\ell_0} \left[ \ell \left(\frac{d\ell}{dt}\right) \right]^{-1} d\ell > \eta_0 - \frac{1}{C} \int_{0}^{\ell_0} \frac{d\ell}{\ell^{1-\epsilon}},
\end{equation} which is bounded.

\paragraph{(b)} Assume that condition (b) is not met.  It then follows from (\ref{eq:FLfriedmannDE}) that $\exists$ a constant $C > 0$ such that $\frac{d\ell}{dt} < C\ \forall\ t > t_0$.  The integral (\ref{eqconf:ConformalTime}) then takes the form
\begin{equation}
\eta_{\ast} = \eta_0 + \lim_{\ell_{\ast} \rightarrow +\infty} \int_{\ell_0}^{\ell_{\ast}} \left[ \ell \left(\frac{d\ell}{dt}\right) \right]^{-1} d\ell > \eta_0 + \lim_{\ell_{\ast} \rightarrow +\infty} \frac{1}{C} \int_{\ell_0}^{\ell_{\ast}} \frac{d\ell}{\ell},
\end{equation} which is unbounded.

If condition (b) is met it then follows from (\ref{eq:FLfriedmannDE}) that $\exists$ constants $C > 0 $ and $\epsilon > 0$ such that $\frac{d\ell}{dt} > C \ell^{\epsilon} \ \forall\ t < t_0$.  The integral (\ref{eqconf:ConformalTime}) then takes the form
\begin{equation}
\eta_{\ast} = \eta_0 + \lim_{\ell_{\ast} \rightarrow +\infty} \int_{\ell_0}^{\ell_{\ast}} \left[ \ell \left(\frac{d\ell}{dt}\right) \right]^{-1} d\ell > \eta_0 + \lim_{\ell_{\ast} \rightarrow +\infty} \frac{1}{C} \int_{\ell_0}^{\ell_{\ast}} \frac{d\ell}{\ell^{1+\epsilon}},
\end{equation} which is bounded.

Thus, we conclude that the constraints hold, completing the proof. $\square$

\paragraph{} We note that this theorem also holds in the case of closed FL as long as $\frac{d\ell}{dt} > C$, where $C$ is a constant satisfying $C > 0$.  This constraint prevents models which recollapse or whose length scale tend to a constant as $t \rightarrow +\infty$.

\chapter{Covariant Differential Identities} \label{app:covdiffident}

In this appendix, we present a list of useful differential identities involving the derivative operators used in this thesis.  This list represents an extension of the list given in Ellis et al. (1990) Appendix A by incorporating identities involving the Laplacian and curl operators.

\paragraph{Zero-order 3-curvature quantities for FL models:}

Before proceeding, we must first obtain zero order expressions for the background curvature.  It quickly follows from (\ref{eq:curvatureparameter}), (\ref{a2eq:spatialricci3tensor}) and (\ref{a2eq:riemann3tensorbyricci3tensor}) that, to zero order, the background 3-curvature quantities satisfy the following relationships:

\begin{eqnarray}
\label{adiff:curv1} \ell^2\ ^{(3)}\!R_{abcd} & = & K\ (h_{ac} h_{bd} - h_{ad} h_{bc}), \\
\label{adiff:curv2} \ell^2\ ^{(3)}\!R_{ac} & = & 2 K\ h_{ac}, \\
\label{adiff:curv3} \ell^2\ ^{(3)}\!R & = & 6 K.
\end{eqnarray}

We now present a list of linearized identities, derived using (\ref{eq:fundamentalcongderiv}), (\ref{eq:orthogonalcovariantderiv}), (\ref{eq:covariantvectorcurl})-(\ref{eq:covarianttensorcurl}), (\ref{eq:covariantderivativecommutator1}) and (\ref{adiff:comm4}).  In all of the following relations, we shall assume that $\hat{\nabla}_{a} f$, $X_{a}$ and $T_{ab}$ and all their derivatives are first order quantities.  Further, we assume that $T_{ab}$ is symmetric and trace-free.

\paragraph{First-order time derivatives:}

The following set of linearized identities are used when interchanging the derivative along the fundamental congruence ($\dot{\ }$) with $\hat{\nabla}$, $\mathsf{curl}(\cdot)$ or $\hat{\nabla}^2$:

\begin{eqnarray}
\label{adiff:time1a} \hat{\nabla}_{a}(\dot{f}) - (\hat{\nabla}_{a} f)\dot{\ }_{\!\perp} & = & H \hat{\nabla}_{a} f - \dot{f} \dot{u}_{a}, \\
\label{adiff:time2a} \hat{\nabla}_{a}(\dot{X}_{b}) - (\hat{\nabla}_{a} X_{b})\dot{\ }_{\!\perp} & = & H \hat{\nabla}_{a} X_{b}, \\
\label{adiff:time3a} \hat{\nabla}_{a}(\dot{T}_{bc}) - (\hat{\nabla}_{a} T_{bc})\dot{\ }_{\!\perp} & = & H \hat{\nabla}_{a} T_{bc}.
\end{eqnarray}
\begin{eqnarray}
\label{adiff:time1b} \ell \hat{\nabla}_{a}(\dot{f}) & = & (\ell \hat{\nabla}_{a} f)\dot{\ }_{\!\perp} - \ell \dot{f} \dot{u}_{a}, \\
\label{adiff:time2b} \ell \hat{\nabla}_{a}(\dot{X}_{b}) & = & (\ell \hat{\nabla}_{a} X_{b})\dot{\ }_{\!\perp}, \\
\label{adiff:time3b} \ell \hat{\nabla}_{a}(\dot{T}_{bc}) & = & (\ell \hat{\nabla}_{a} T_{bc})\dot{\ }_{\!\perp}.
\end{eqnarray}
\begin{eqnarray}
\label{adiff:time4a} \mathsf{curl}(\dot{X}_{a}) - (\mathsf{curl}(X_{a}))\dot{\ } & = & H \mathsf{curl}(X_{a}), \\
\label{adiff:time5a} \mathsf{curl}(\dot{T}_{ab}) - (\mathsf{curl}(T_{ab}))\dot{\ } & = & H \mathsf{curl}(T_{ab}).
\end{eqnarray}
\begin{eqnarray}
\label{adiff:time4b} \ell \mathsf{curl}(\dot{X}_{a}) & = & (\ell \mathsf{curl}(X_{a}))\dot{\ }_{\!\perp}, \\
\label{adiff:time5b} \ell \mathsf{curl}(\dot{T}_{ab}) & = & (\ell \mathsf{curl}(T_{ab}))\dot{\ }_{\!\perp}.
\end{eqnarray}
\begin{eqnarray}
\label{adiff:time6a} \hat{\nabla}^2(\dot{X}_{a}) = (\hat{\nabla}^2 X_{a})\dot{\ }_{\!\perp} + 2 H \hat{\nabla}^2 X_{a}, \\
\label{adiff:time7a} \hat{\nabla}^2(\dot{T}_{ab}) = (\hat{\nabla}^2 T_{ab})\dot{\ }_{\!\perp} + 2 H \hat{\nabla}^2 T_{ab}.
\end{eqnarray}
\begin{eqnarray}
\label{adiff:time6b} \ell^2 \hat{\nabla}^2(\dot{X}_{a}) = (\ell^2 \hat{\nabla}^2 X_{a})\dot{\ }_{\!\perp}, \\
\label{adiff:time7b} \ell^2 \hat{\nabla}^2(\dot{T}_{ab}) = (\ell^2 \hat{\nabla}^2 T_{ab})\dot{\ }_{\!\perp}.
\end{eqnarray}

\paragraph{Laplacian identities:}

The following linearized identities are used when interchanging the Laplacian operator $\hat{\nabla}^2$ with $\hat{\nabla}$ or $\mathsf{curl}(\cdot)$:

\begin{eqnarray}
\label{adiff:laplace1} \ell^2 \hat{\nabla}^{a} \hat{\nabla}^2 f & = & \ell^2 \hat{\nabla}^2 \hat{\nabla}^{a} f - 2 K \hat{\nabla}^{a} f - 2 \dot{f} \mathsf{curl}(\omega_{a}), \\
\label{adiff:laplace2} \ell^2 \hat{\nabla}^{a} \hat{\nabla}^2 X_{b} & = & \ell^2 \hat{\nabla}^2 \hat{\nabla}^{a} X_{b} - 2 K \big[\hat{\nabla}^{a} X_{b} + \hat{\nabla}^{b} X_{a} - h_{\ b}^{a} \hat{\nabla}^{c} X_{c} \big], \\
\label{adiff:laplace2b} \ell^2 \hat{\nabla}^{a} \hat{\nabla}^2 T_{bc} & = & \ell^2 \hat{\nabla}^2 \hat{\nabla}^{a} T_{bc} - 2 K \left[\hat{\nabla}^{a} T_{bc} + \hat{\nabla}_b T^{a}_{\ c} + \hat{\nabla}_{c} T^{\ a}_{b} - h^{a}_{\ b} \hat{\nabla}^s T_{sc} - h^{a}_{\ c} \hat{\nabla}^s T_{bs} \right], \nonumber \\ & &
\end{eqnarray}
\begin{eqnarray}
\label{adiff:laplace3} \ell^2 \hat{\nabla}^{a} \hat{\nabla}^2 X_{a} & = & \ell^2 \hat{\nabla}^2 \hat{\nabla}^{a} X_{a} + 2 K \hat{\nabla}^{a} X_{a}, \\
\label{adiff:laplace4} \ell^2 \hat{\nabla}^{b} \hat{\nabla}^2 T_{ab} & = & \ell^2 \hat{\nabla}^2 \hat{\nabla}^{b} T_{ab} + 4 K \hat{\nabla}^{b} T_{ab},
\end{eqnarray}
\begin{eqnarray}
\label{adiff:curl6} \hat{\nabla}^2 \mathsf{curl}\ X_a & = & \mathsf{curl} ( \hat{\nabla}^2 X_a ), \\
\label{adiff:curl7} \hat{\nabla}^2 \mathsf{curl}\ T_{ab} & = & \mathsf{curl} ( \hat{\nabla}^2 T_{ab} ).
\end{eqnarray}

\paragraph{Curl identities:}

Finally, the following linearized identities are used when interchanging the curl operator $\mathsf{curl}(\cdot)$ with $\hat{\nabla}$:

\begin{eqnarray}
\label{adiff:curl1} \mathsf{curl}(\hat{\nabla}_{a} f) & = & - 2 \omega_a \dot{f}, \\
\label{adiff:curl2} \mathsf{curl}(\hat{\nabla}_{a} X_{b}) & = & \hat{\nabla}_{(a} \mathsf{curl}(X_{b)}), \\
\label{adiff:curl2a} \mathsf{curl}(\hat{\nabla}_{(a} X_{b)}) & = & \tfrac{1}{2} \hat{\nabla}_{(a} \mathsf{curl}(X_{b)}), \\
\label{adiff:curl2b} \hat{\nabla}^{a} \mathsf{curl}(X_{a}) & = & 0, \\
\label{adiff:curl3} \mathsf{curl}(\hat{\nabla}^{b} T_{ab}) & = & 2 \hat{\nabla}^{b} \mathsf{curl}(T_{ab}),
\end{eqnarray}

Further, we obtain the following linearized identities for the curl of a curl, in the linearized context:

\begin{eqnarray}
\label{adiff:curl4} \ell^2 \mathsf{curl}\ \mathsf{curl}\ X_{a} & = & - \ell^2 \hat{\nabla}^2 X_{a} + \ell^2 \hat{\nabla}_{a} \hat{\nabla}^{b} X_{b} + 2 K X_{a}, \\
\label{adiff:curl5} \ell^2 \mathsf{curl}\ \mathsf{curl}\ T_{ab} & = & - \ell^2 \hat{\nabla}^2 T_{ab} + \tfrac{3}{2} \ell^2 \hat{\nabla}_{\langle a} \hat{\nabla}^{c} T_{b \rangle c} + 3 K T_{ab},
\end{eqnarray}

\chapter{Covariant Harmonics} \label{app:covariantharmonics}

In this appendix, we present an overview of general identities and results related to the covariant harmonics.  We extend the results of Bruni et al. (1992) Appendix B by introducing the notion of associated vector and tensor harmonics.

As in Hawking (1966), we define the \textit{spatial harmonics} $Q^{(n)}_{ab \cdots c}$ as eigenfunctions of the spatial Laplace-Beltrami operator:
\begin{equation} \label{eq:spatialharmonics}
\hat{\nabla}^2 Q^{(k)}_{a b \cdots c} + \frac{k^2}{\ell^2} Q^{(k)}_{a b \cdots c} = 0,
\end{equation} where $k$ is known as the \textit{wavenumber} of the harmonic.  Further, using (\ref{eq:spatialharmonics}) we impose that the $Q^{(k)}$ quantities are constant in time:
\begin{equation} \label{eq:spatialharmonicscomoving}
\dot{Q}^{(k)}_{a b \cdots c} = 0.
\end{equation}  See Bruni et al. (1992, Appendix B) for the relation between the metric harmonics $Y$ and the covariant harmonics $Q$.  Hereafter, as we are working in the geometrical formalism, the use of the term \textit{harmonics} will refer to the covariant harmonics defined by (\ref{eq:spatialharmonics}) and (\ref{eq:spatialharmonicscomoving}).  For the problem at hand, we only require scalar harmonics, vector harmonics and two-index tensor harmonics (which, for purposes of brevity will be referred to as tensor harmonics).

We briefly discuss notation for harmonics and quantities derived from harmonics.  The $0$ superscript of the harmonic quantity $Q^{(0,k)}$ denotes that it is a scalar harmonic.  Similarly, we use $1$ and $2$ to denote vector and tensor harmonics, respectively.  It is common in the literature to drop the wavenumber $k$ when writing a harmonic quantity, so writing $Q^{(0,k)}$ as $Q^{(0)}$ instead.  The wavenumber of the harmonic being used is instead implicitly defined by the equation where the harmonic appears.  Further, here and elsewhere, we shall use $Q$ to denote the whole set of harmonics and the standard quantities constructed from $Q$: we have scalar harmonics $Q^{(0)}$, $Q^{(0)}_{a}$, $Q^{(0)}_{ab}$, vector harmonics $Q^{(1)}_{a}$, $Q^{(1)}_{ab}$ and tensor harmonics $Q^{(2)}_{ab}$.

We now give the definitions and properties of each of these three classes of harmonic types.

\section{Scalar Harmonics} \label{acovsec:scalarharmonics}

Scalar harmonics are defined as solutions of
\begin{equation} \label{ahaeq:scaharmdefinition}
\hat{\nabla}^2 Q = - \frac{k^2}{\ell^2} Q,
\end{equation} that satisfy the time-independence constraint
\begin{equation} \label{ahaeq:scatimeindependence}
\dot{Q} = 0.
\end{equation}  Each scalar harmonics $Q$ defines a vector and a trace-free symmetric tensor via
\begin{equation} \label{ahaeq:scavecdef}
Q_{a} = - \frac{\ell}{k} \hat{\nabla}_{a} Q,
\end{equation} and
\begin{equation} \label{ahaeq:scatensdef}
Q_{ab} = \frac{\ell^2}{k^2} \hat{\nabla}_{b} \hat{\nabla}_{a} Q + \tfrac{1}{3} h_{ab} Q.
\end{equation}  Since it follows from (\ref{ahaeq:scatimeindependence}) and (\ref{eq:covariantderivativecommutator1}) that $Q$ satisfies
\begin{equation}
\hat{\nabla}_{[a} \hat{\nabla}_{b]} Q = 0,
\end{equation} it can be quickly shown that $Q_{a}$ and $Q_{ab}$ satisfy the curl-free constraint,
\begin{equation} \label{ahaeq:scaharmzerocurl}
\mathsf{curl}(Q_{a}) = 0, \qquad \mathsf{curl}(Q_{ab}) = 0,
\end{equation} congruence-orthogonality,
\begin{equation}
u^a Q_{a} = 0, \qquad u^a Q_{ab} = 0, \qquad u^b Q_{ab} = 0,
\end{equation} and time invariance,
\begin{equation} \label{ahaeq:scaharmtimeinvariance}
\dot{Q}_{a} = 0, \qquad \dot{Q}_{ab} = 0.
\end{equation}

We are restricted in our choice of $k^2$ to the eigenvalues determined by Lifshitz (1946), which in turn are determined by the spatial curvature according to
\begin{equation} \label{eq:scalareigenvalues}
k^2 = n^2 - K \quad \left\{\begin{array}{ll}n > 0 & \mbox{if $K = -1,0$,} \\ n = 3, 4, \ldots & \mbox{if $K = +1$.} \end{array} \right.
\end{equation}  We note that it is common to include $n = 2$ as a possibly harmonic mode in the case of $K = +1$, but it can be shown via the constraint equation (\ref{lineq:Eweylconstraint}) and identity (\ref{ahaeq:scatenprop1}) that this mode is unphysical.  In fact, this mode is gauge-dependent and associated with the freedom of choice of the velocity field $u^a$.

Some important identities involving the vector quantity $Q_a$ can be derived using the differential identities in Appendix \ref{app:covdiffident} and (\ref{ahaeq:scaharmdefinition})-(\ref{ahaeq:scavecdef}):
\begin{eqnarray}
\label{ahaeq:scavecprop1} \ell \hat{\nabla}_{a} Q^a & = & k Q, \\
\label{ahaeq:scavecprop2} \ell^2 \hat{\nabla}^2 Q_a & = & - (k^2 - 2K) Q_{a}, \\
\label{ahaeq:scavecprop3} \ell \hat{\nabla}_{b} Q_a & = & - k (Q_{ab} - \tfrac{1}{3} h_{ab} Q), \\
\label{ahaeq:scavecprop4} \ell^2 \hat{\nabla}_{[a} \hat{\nabla}_{b]} Q_c & = & \tfrac{1}{2} K (h_{ac} Q_{b} - h_{bc} Q_{a}).
\end{eqnarray}

Similarly, some important identities involving the tensor quantity $Q_{ab}$ can be derived using the differential identities in appendix \ref{app:covdiffident} and (\ref{ahaeq:scaharmdefinition})-(\ref{ahaeq:scatensdef}):
\begin{eqnarray}
\label{ahaeq:scatenprop1} \ell \hat{\nabla}^{b} Q_{ab} & = & \tfrac{2}{3} k^{-1} (k^2 - 3 K) Q_{a}, \\
\label{ahaeq:scatenprop2} \ell^2 \hat{\nabla}^{a} \hat{\nabla}^{b} Q_{ab} & = & \tfrac{2}{3} (k^2 - 3 K) Q, \\
\label{ahaeq:scatenprop3} \ell^2 \hat{\nabla}_{b} \hat{\nabla}^{c} Q_{ac} & = & - \tfrac{2}{3} (k^2 - 3 K)(Q_{ab} - \tfrac{1}{3} h_{ab} Q), \\
\label{ahaeq:scatenprop4} \ell^2 \hat{\nabla}^2 Q_{ab} & = & - (k^2 - 6 K) Q_{ab}, \\
\label{ahaeq:scatenprop5} \ell^2 \hat{\nabla}_{[a} \hat{\nabla}_{b]} Q_{cd} & = & \tfrac{1}{2} K \big[(h_{ac} Q_{bd} - h_{bc} Q_{ad}) + (h_{ad} Q_{bc} - h_{bd} Q_{ac}) \big].
\end{eqnarray}

\section{Vector Harmonics} \label{acovsec:vectorharmonics}

Vector harmonics are defined as solutions of
\begin{equation} \label{ahaeq:vecharmdefinition}
\hat{\nabla}^2 Q_{a} = - \frac{k^2}{\ell^2} Q_{a},
\end{equation} where $Q_{a}$ is divergence-free and time-independent,
\begin{equation} \label{ahaeq:vecharmtimeindependence}
\hat{\nabla}^a Q_{a} = 0, \qquad \dot{Q}_a = 0.
\end{equation}  Every vector harmonic $Q_{a}$ defines an \textit{associated vector harmonic} $P_{a}$ via
\begin{equation} \label{ahaeq:associatedvecharm}
P_a = \ell\ \mathsf{curl}(Q_a),
\end{equation} which can be shown to satisfy (\ref{ahaeq:vecharmdefinition}) and (\ref{ahaeq:vecharmtimeindependence}) using (\ref{adiff:curl6}), (\ref{adiff:curl2b}) and (\ref{adiff:time4b}).  On taking the curl of (\ref{ahaeq:associatedvecharm}) and applying (\ref{ahaeq:vecharmdefinition}) and (\ref{adiff:curl4}), we obtain $Q_{a}$ in terms of $P_{a}$,
\begin{equation}
Q_a = (k^2 + 2 K)^{-1}\ \ell\ \mathsf{curl}(P_a).
\end{equation}  The vector harmonics $Q$ and associated vector harmonics $P$ then define trace-free symmetric tensor quantities via
\begin{equation} \label{ahaeq:vecharmtensor}
Q_{ab} = - \frac{\ell}{k} \hat{\nabla}_{(a} Q_{b)},
\end{equation} and
\begin{equation} \label{ahaeq:vecharmtensorassoc}
P_{ab} = - \frac{\ell}{k} \hat{\nabla}_{(a} P_{b)}.
\end{equation}  These tensor-valued vector harmonics are defined in order to satisfy zero second divergence,
\begin{equation}
\hat{\nabla}^{a} \hat{\nabla}^{b} Q_{ab} = 0, \qquad \hat{\nabla}^{a} \hat{\nabla}^{b} P_{ab} = 0,
\end{equation} congruence orthogonality,
\begin{equation}
\qquad u^a Q_{ab} = 0, \qquad u^b Q_{ab} = 0, \qquad u^a P_{ab} = 0, \qquad u^b P_{ab} = 0,
\end{equation} and time invariance,
\begin{equation}
\dot{Q}_{ab} = 0, \qquad \dot{P}_{ab} = 0.
\end{equation}  The tensor-valued vector harmonics are related using (\ref{ahaeq:vecharmtensor}), (\ref{ahaeq:vecharmtensorassoc}) and (\ref{adiff:curl7}), as follows:
\begin{eqnarray}
\ell\ \mathsf{curl}(P_{ab}) & = & \tfrac{1}{2} (k^2 + 2 K) Q_{ab}, \\
\ell\ \mathsf{curl}(Q_{ab}) & = & \tfrac{1}{2} P_{ab}.
\end{eqnarray}

As with scalar harmonics, we are restricted in our choice of $k^2$ to the eigenvalues determined by Lifshitz (1946), according to
\begin{equation} \label{eq:vectoreigenvalues}
k^2 = n^2 - 2 K \quad \left\{\begin{array}{ll}n > 0 & \mbox{if $K = -1,0$} \\ n = 3, 4, \ldots & \mbox{if $K = +1$} \end{array} \right. .
\end{equation}  Again, it is common to include $n = 2$ as a possibly harmonic mode in the case of $K = +1$, but it can be shown via the constraint equation (\ref{a1eq:fullbianchi4}) that this mode is unphysical.

Using the differential identities in Appendix \ref{app:covdiffident} and (\ref{ahaeq:vecharmdefinition})-(\ref{ahaeq:associatedvecharm}) we can derive the following identity for the vector harmonics $Q_a$ and associated vector harmonics $P_a$ (in this case, we may replace $Q_a$ with $P_a$ to obtain a second identity):
\begin{eqnarray}
\label{ahaeq:vecvecprop1} \ell^2 \hat{\nabla}_{[a} \hat{\nabla}_{b]} Q_c & = & \tfrac{1}{2} K (h_{ac} Q_{b} - h_{bc} Q_{a}).
\end{eqnarray}

Some identities involving the tensor-valued vector harmonics $Q_{ab}$ and associated tensor-valued vector harmonics $P_{ab}$ can be derived using the differential identities in appendix \ref{app:covdiffident} and (\ref{ahaeq:vecharmdefinition})-(\ref{ahaeq:vecharmtensorassoc}).  In each of the following identities, the pair ($Q_{ab}$, $Q_{a}$) may be replaced with ($P_{ab}$, $P_{a}$):
\begin{eqnarray}
\label{ahaeq:tenvecprop1} \ell \hat{\nabla}^{b} Q_{ab} & = & \tfrac{1}{2} k^{-1} (k^2 - 2 K) Q_{a}, \\
\label{ahaeq:tenvecprop2} \ell^2 \hat{\nabla}_{(a} \hat{\nabla}^{c} Q_{a)c} & = & - \tfrac{1}{2} (k^2 - 2 K) Q_{ab}, \\
\label{ahaeq:tenvecprop3} \ell^2 \hat{\nabla}^2 Q_{ab} & = & - (k^2 - 4 K) Q_{ab}, \\
\label{ahaeq:tenvecprop4} \ell^2 \hat{\nabla}_{[a} \hat{\nabla}_{b]} Q_{cd} & = & \tfrac{1}{2} K \big[(h_{ac} Q_{bd} - h_{bc} Q_{ad}) + (h_{ad} Q_{bc} - h_{bd} Q_{ac}) \big].
\end{eqnarray}

\section{Tensor Harmonics} \label{acovsec:tensorharmonics}

Tensor harmonics are defined as solutions of
\begin{equation} \label{ahaeq:tenharmdefinition}
\hat{\nabla}^2 Q_{ab} = - \frac{k^2}{\ell^2} Q_{ab},
\end{equation} where $Q_{ab}$ is divergence-free, trace-free and time independent:
\begin{equation} \label{ahaeq:tenharmtimeindependence}
\hat{\nabla}^b Q_{ab} = 0, \qquad Q^{a}_{\ a} = 0, \qquad \dot{Q}_{ab}  \\= 0.
\end{equation}  As with vector harmonics, every tensor harmonic $Q_{ab}$ defines an \textit{associated tensor harmonic} $P_{ab}$ via
\begin{equation} \label{ahaeq:associatedtenharm}
P_{ab} = \ell \mathsf{curl}(Q_{ab}).
\end{equation}  It can be shown that $P_{ab}$ satisfies the definition of a tensor harmonic (\ref{ahaeq:tenharmdefinition})-(\ref{ahaeq:tenharmtimeindependence}) using (\ref{adiff:curl7}), (\ref{adiff:curl3}) and (\ref{adiff:time5b}).  It then follows immediately from (\ref{ahaeq:tenharmdefinition}) that the quantities $Q_{ab}$ and $P_{ab}$ satisfy congruence orthogonality:
\begin{equation}
u^a Q_{ab} = 0, \qquad u^b Q_{ab} = 0, \qquad u^a P_{ab} = 0, \qquad u^b P_{ab} = 0.
\end{equation}  Upon taking the curl of (\ref{ahaeq:associatedtenharm}) and applying (\ref{ahaeq:tenharmdefinition}) and (\ref{adiff:curl5}), we obtain $Q_{ab}$ in terms of $P_{ab}$:
\begin{equation}
Q_{ab} = (k^2 + 3K)^{-1} \ell \mathsf{curl}(P_{ab}).
\end{equation}

As with scalar and vector harmonics, we are restricted in our choice of $k^2$ to the eigenvalues determined by Lifshitz (1946):
\begin{equation} \label{eq:tensoreigenvalues}
k^2 = n^2 - 3 K \quad \left\{\begin{array}{ll}n > 0 & \mbox{if $K = -1,0$} \\ n = 3, 4, \ldots & \mbox{if $K = +1$} \end{array} \right. .
\end{equation}

We can obtain an important identity for $Q_{ab}$ and $P_{ab}$ using (\ref{adiff:comm4}) and (\ref{adiff:curv1}), as follows (the identity for $P_{ab}$ is identical, except with $Q_{ab}$ replaced with $P_{ab}$):
\begin{equation}
\ell^2 \hat{\nabla}_{[a} \hat{\nabla}_{b]} Q_{cd} = \tfrac{1}{2} K \big[(h_{ac} Q_{bd} - h_{bc} Q_{ad}) + (h_{ad} Q_{bc} - h_{bd} Q_{ac}) \big].
\end{equation}
\chapter{The 3-Cotton-York Tensor}

In this appendix we introduce the 3-Cotton-York tensor, which is useful in the analysis of tensor perturbations.  Specifically, we are interested in using the 3-Cotton-York tensor in place of the co-moving shear $\mathcal{X}_{ab}$ in developing a governing DE for tensor perturbations.  It is of interest in this regard due to its relation with Bardeen's metric potential $H_T^{(2)}$.  We will further show that the coefficients of the shear and Weyl tensor components can be rewritten in terms of $\mathcal{C}_{(k)}$.  To the best of the author's knowledge, this analysis of tensor perturbations in terms of the 3-Cotton-York tensor is new.

In terms of the trace-free spatial Ricci tensor (\ref{a2eq:spatialricci3tensor}), the \textit{3-Cotton-York tensor} is defined as (see, for example, van Elst and Uggla (1997, p2680-1))
\begin{equation} \label{eq:cottonyorktensorA}
^{(3)}\!C_{ab} = \mathsf{curl}(^{(3)}\!S_{ab}),
\end{equation} and hence is symmetric and tracefree.  In the linearized context, we can use (\ref{a2eq:spatialricci3tensorW}) and (\ref{lineq:curlsigmaconstraint}) to obtain an alternative first-order expression for $^{(3)}\!C_{ab}$ in terms of the Weyl tensor components, namely
\begin{equation} \label{lineq:cottonyorktensor}
^{(3)}\!C_{ab} = - H H_{ab} + \mathsf{curl}(E_{ab}).
\end{equation}  We note that an important property of this quantity is that it has zero spatial divergence in the linear theory.  Indeed, upon taking the divergence of (\ref{lineq:cottonyorktensor}) and applying (\ref{eq:densitygradientvorticityrelation}), (\ref{lineq:Eweylconstraint}), (\ref{lineq:Hweylconstraint}) and (\ref{adiff:curl3}), we obtain, to first order,
\begin{equation} \label{eq:cottonyorktensordivzero}
\hat{\nabla}^{b}\ ^{(3)}\!C_{ab} = 0.
\end{equation}  The evolution equation for $^{(3)}\!C_{ab}$ then follows upon taking the time derivative of (\ref{lineq:cottonyorktensor}) and applying (\ref{lineq:zeroorder1}), (\ref{lineq:weylEevolution}), (\ref{lineq:weylHevolution}) and (\ref{adiff:time5a}), giving
\begin{equation} \label{eq:cottonyorktensorevol}
^{(3)}\!\dot{C}_{ab} = -3 H\ ^{(3)}\!C_{ab} + \frac{2K}{\ell^2} H_{ab} - \hat{\nabla}^2 H_{ab} + 2 \hat{\nabla}_{\langle a} \hat{\nabla}^{c} H_{b \rangle c}.
\end{equation}

\section{Evolution Equation for the 3-Cotton-York Tensor}

Instead of choosing the co-moving shear to be the basic quantity for tensor perturbations (as in section \ref{sec:TensorPerturbations}), we can instead choose the basic quantity to be the \textit{dimensionless 3-Cotton-York tensor}, defined by
\begin{equation} \label{eq:dimensionlesscottonyork}
\mathcal{C}_{ab} = \ell^3\ ^{(3)}\!C_{ab}.
\end{equation}

Upon recalling from (\ref{eq:cottonyorktensordivzero}) that the 3-Cotton-York tensor has zero divergence,  it quickly follows that $\mathcal{C}_{ab}$ also has zero divergence, \textit{i.e.}
\begin{equation}
\hat{\nabla}^{b} \mathcal{C}_{ab} = 0.
\end{equation}  Hence, $\mathcal{C}_{ab}$ can be expanded purely in terms of associated tensor harmonics, according to
\begin{equation} \label{eq:cottonyorktensorharmonics}
\mathcal{C}_{ab} = \sum_k \mathcal{C}_{(k)} P^{(2)}_{ab},
\end{equation}  The derivatives of $\mathcal{C}_{ab}$ along the fundamental congruence are obtained by differentiating (\ref{eq:cottonyorktensorharmonics}), giving
\begin{equation} \label{eq:cottonyorkderivs}
\dot{\mathcal{C}}_{\langle ab \rangle} = \sum_k \dot{\mathcal{C}}_{(k)} P^{(2,k)}_{ab}, \qquad \ddot{\mathcal{C}}_{\langle ab \rangle} = \sum_k \ddot{\mathcal{C}}_{(k)} P^{(2,k)}_{ab}.
\end{equation}

Upon differentiating (\ref{eq:dimensionlesscottonyork}) and using (\ref{eq:hubblelengthscale}) and (\ref{eq:cottonyorktensorevol}) we obtain
\begin{equation} \label{eq:dimensionlesscottonyorktensorevol}
\dot{\mathcal{C}}_{ab} = \ell^3 \left(\frac{2 K}{\ell^2} - \hat{\nabla}^2 \right) H_{ab}.
\end{equation}  Differentiating again and applying (\ref{eq:dimensionlesscottonyork}) and (\ref{adiff:time7b}) yields
\begin{equation} \label{eq:dimensionlesscottonyorktensorevol_a}
\ddot{\mathcal{C}}_{ab} + 3 H \dot{\mathcal{C}}_{ab} + \left(\frac{2 K}{\ell^2} - \hat{\nabla}^2 \right) \mathcal{C}_{ab} = 0.
\end{equation}  We now expand each term of this equation in terms of tensor harmonics using (\ref{eq:cottonyorktensorharmonics}) and (\ref{eq:cottonyorkderivs}), applying (\ref{ahaeq:tenharmdefinition}) to remove the Laplacian term, obtaining
\begin{equation} \label{eq:dimensionlesscottonyorktensorevol_b}
\ddot{\mathcal{C}}_{(k)} + 3 H \dot{\mathcal{C}}_{(k)} + \frac{1}{\ell^2} \left(k^2 + 2 K \right) \mathcal{C}_{(k)} = 0.
\end{equation}  This is then the second-order DE governing the evolution of the 3-Cotton-York tensor.  In terms of conformal time, this DE reads
\begin{equation}
\mathcal{C}_{(k)}^{\prime \prime} + 2 H \ell \mathcal{C}_{(k)}^{\prime} + (k^2 + 2 K) \mathcal{C}_{(k)} = 0,
\end{equation} which is identical to Bardeen's DE for the metric potential $H_T^{(2)}$, given in Bardeen (1980) equation (4.14).

\section{Tensor Perturbations}

We now show the relation between (\ref{eq:dimensionlesscottonyorktensorevol_b}) and the governing DE for tensor perturbations (\ref{eq:tensorperturbationmasterequation}).  We note that it follows by (\ref{eq:cottonyorktensordivzero}) that the 3-Cotton-York tensor is purely tensorial, \textit{i.e.} the scalar and vector contributions to $^{(3)}\!C_{(k)}$ are exactly zero.  Expanding $^{(3)}\!C_{(k)}$ in terms of associated tensor harmonics yields
\begin{equation}
^{(3)}\!C_{ab} = \sum_k\ ^{(3)}\!C_{(k)} P^{(2)}_{ab}.
\end{equation}  The coefficients $^{(3)}\!C_{(k)}$ can then be determined in terms of $\mathcal{X}_{(k)}$ and the tensor harmonics via (\ref{lineq:cottonyorktensor}) upon applying (\ref{eq:weyltensorexpansion}), (\ref{eq:tensorkinE}), (\ref{eq:tensorkinH}) and (\ref{ahaeq:associatedtenharm}).  We obtain
\begin{equation} \label{eq:tensorkinC}
^{(3)}C_{(k)} \backsim - \frac{1}{H^2 \ell^2} \left( \dot{\mathcal{X}}_{(k)} + 2 H \mathcal{X}_{(k)} \right).
\end{equation}

If we expand (\ref{eq:dimensionlesscottonyorktensorevol}) in terms of tensor harmonics using (\ref{eq:weyltensorexpansion}), (\ref{ahaeq:tenharmdefinition}) and (\ref{eq:cottonyorkderivs}), we obtain
\begin{equation}
\dot{\mathcal{C}}_{(k)} = \ell \left(k^2 + 2 K\right) H_{(k)},
\end{equation} and hence using (\ref{eq:tensorkinH}) that
\begin{equation} \label{eq:BasicQuantityXCRelation}
\mathcal{X}_{(k)} = \frac{1}{k^2 + 2K} \ell \dot{\mathcal{C}}_{(k)}.
\end{equation}  Now using (\ref{eq:BasicQuantityXCRelation}), we can rewrite (\ref{eq:tensorkinS})-(\ref{eq:tensorkinH}) as
\begin{eqnarray}
\label{SCeq:SCtensorkinS} \sigma_{(k)} & = & \frac{1}{k^2 + 2K} \dot{\mathcal{C}}_{(k)}, \\
\label{SCeq:SCtensorkinE} E_{(k)} & = & - \frac{1}{k^2 + 2K} \left[ \ddot{\mathcal{C}}_{(k)} + 2 H \dot{\mathcal{C}}_{(k)} \right], \\
\label{SCeq:SCtensorkinH} H_{(k)} & = & \frac{1}{k^2 + 2K} \ell^{-1} \dot{\mathcal{C}}_{(k)}.
\end{eqnarray}  Thus, we obtain the coefficients of the shear and Weyl tensor components in terms of the coefficients of the 3-Cotton-York tensor $\mathcal{C}_{(k)}$.
\chapter{Bessel and Associated Legendre Functions} \label{app:PerturbationSolutions}

In this appendix we present solutions of the Bessel DE, toroidal Legendre DE and conical Legendre DE, which are used in solving the governing DEs for scalar and tensor perturbations.  In section \ref{asec:NormalFormsBesselLegendreDEs} we present the normal forms of the Bessel and associated Legendre DEs and their general solutions in terms of Bessel and associated Legendre functions.  As discussed in section \ref{asec:SpecialGoverningDESolutions}, these solutions can be written in terms of elementary functions in certain special cases.  Further, resonant solutions of the Legendre DE appear in certain cases, as discussed in section \ref{asec:ResonantSolutionsGoverningDE}.  Finally, the asymptotic behaviour of the general solutions is given in section \ref{asec:AsymptoticBehaviourBesselLegendre}.

\section{Normal Forms of Bessel and Associated Legendre DEs} \label{asec:NormalFormsBesselLegendreDEs}

In this section we present the normal forms of Bessel's DE, the toroidal Legendre DE and the conical Legendre DE along with their general solution in terms of Bessel functions and associated Legendre functions.  Formally, the toroidal and conical Legendre functions are defined as the composition of the associated Legendre functions with $\cosh$ and $\cos$, respectively.

\paragraph{Bessel's DE}
Normal Form:
\begin{equation} \label{eqpsol:BesselsDE}
\frac{d^2 X}{d T^2} + \left[ a^2 - \frac{(\mu^2 - \tfrac{1}{4})}{T^2} \right] X = 0.
\end{equation}
Solution:
\begin{equation} \label{eqpsol:BesselsDESol}
X(T) = T^{\frac{1}{2}} \left[ c_1 J_{\mu} (a T) + c_2 Y_{\mu} (a T) \right].
\end{equation}

\paragraph{Legendre DE (Toroidal)}
Normal Form:
\begin{equation} \label{eqpsol:ToroidalLegendreDE}
\frac{d^2 X}{d T^2} + \left[ - (\nu + \tfrac{1}{2})^2 - \frac{(\mu^2 - \tfrac{1}{4})}{\sinh^2 T} \right] X = 0.
\end{equation}
Solution:
\begin{equation} \label{eqpsol:ToroidalLegendreDESol}
X(T) = (\sinh T)^{\frac{1}{2}} \left[ c_1 P_{\nu}^{\mu} (\cosh T) + c_2 Q_{\nu}^{\mu} (\cosh T) \right].
\end{equation}

\paragraph{Legendre DE (Conical)}
Normal Form:
\begin{equation} \label{eqpsol:ConicalLegendreDE}
\frac{d^2 X}{d T^2} + \left[ (\nu + \tfrac{1}{2})^2 - \frac{(\mu^2 - \tfrac{1}{4})}{\sin^2 T} \right] X = 0.
\end{equation}
Solution:
\begin{equation} \label{eqpsol:ConicalLegendreDESol}
X(T) = (\sin T)^{\frac{1}{2}} \left[ c_1 P_{\nu}^{\mu} (\cos T) + c_2 Q_{\nu}^{\mu} (\cos T) \right].
\end{equation}

\section{Special Solutions of the Governing DEs} \label{asec:SpecialGoverningDESolutions}

For certain values of $\gamma$, $\mu$ and $\nu$, the Bessel functions and associated Legendre $P$ functions reduce to closed-form trigonmetric expressions.  Further, in each of these cases we can derive a second linearly independent solution $\tilde{Q}$ of the associated Legendre DEs (\ref{eqpsol:ToroidalLegendreDE}) and (\ref{eqpsol:ConicalLegendreDE}) that is defined so as to reduce to $J_{\mu}(bz)$ in the limit $z \to 0$.  This solution is a preferred choice, since it emphasizes the relation between solutions for $K = \pm 1$ and solutions with $K = 0$ in the early time limit.  In this section, we present the first few instances of this behaviour and its importance in modelling perturbations with a given fluid background.

For brevity, we shall make use of the notation
\begin{equation}
C(z) = \cos(z), \qquad S(z) = \sin(z),
\end{equation} and use the tilde $\backsim$ to denote dependence up to a multiplicative constant.

\paragraph{$\mu = \tfrac{1}{2}$, $\nu = - \tfrac{1}{2} + \sqrt{K} b$}

Describes evolution of ($\mathcal{C}$, $\gamma = \tfrac{4}{3}$).
\begin{eqnarray}
z^{1/2} \cdot Y_{\mu}(bz) & \backsim & C(bz), \\
z^{1/2} \cdot J_{\mu}(bz) & \backsim & S(bz), \\
(\sinh z)^{1/2} \cdot P^{\mu}_{\nu}(\cosh z) & \backsim & C(bz), \\
(\sinh z)^{1/2} \cdot \tilde{Q}^{\mu}_{\nu}(\cosh z) & \backsim &  S(bz), \\
(\sin z)^{1/2} \cdot P^{\mu}_{\nu}(\cos z) & \backsim & C(bz), \\
(\sin z)^{1/2} \cdot \tilde{Q}^{\mu}_{\nu}(\cos z) & \backsim & S(bz).
\end{eqnarray}

\paragraph{$\mu = \tfrac{3}{2}$, $\nu = - \tfrac{1}{2} + \sqrt{K} b$}

Describes evolution of ($\mathcal{C}$, $\gamma = 1$), ($\mathcal{X}$, $\gamma = \tfrac{4}{3}$), ($\mathcal{D}$, $\gamma = \tfrac{4}{3}$).
\begin{eqnarray}
\label{eqpsol:SpecialSolution3a} z^{1/2} \cdot Y_{\mu}(bz) & \backsim & z^{-1} C(bz) + b S(bz), \\
\label{eqpsol:SpecialSolution3b} z^{1/2} \cdot J_{\mu}(bz) & \backsim & - z^{-1} S(bz) + b C(bz), \\
\label{eqpsol:SpecialSolution3c} (\sinh z)^{1/2} \cdot P^{\mu}_{\nu}(\cosh z) & \backsim & (\coth z) C(bz) + b S(bz), \\
\label{eqpsol:SpecialSolution3d} (\sinh z)^{1/2} \cdot \tilde{Q}^{\mu}_{\nu}(\cosh z) & \backsim & - (\coth z) S(bz) + b C(bz), \\
\label{eqpsol:SpecialSolution3e} (\sin z)^{1/2} \cdot P^{\mu}_{\nu}(\cos z) & \backsim & (\cot z) C(bz) + b S(bz), \\
\label{eqpsol:SpecialSolution3f} (\sin z)^{1/2} \cdot \tilde{Q}^{\mu}_{\nu}(\cos z) & \backsim & - (\cot z) S(bz) + b C(bz).
\end{eqnarray}

\paragraph{$\mu = \tfrac{5}{2}$, $\nu = - \tfrac{1}{2} + \sqrt{K} b$}

Describes evolution of ($\mathcal{C}$, $\gamma = \tfrac{1}{3}$), ($\mathcal{X}$, $\gamma = 1$).

\begin{eqnarray}
\label{eqpsol:SpecialSolution5a} z^{1/2} \cdot Y_{\mu}(bz) & \backsim & (b^2 - 3 z^{-2}) C(bz) - 3 b z^{-1} S(bz), \\
\label{eqpsol:SpecialSolution5b} z^{1/2} \cdot J_{\mu}(bz) & \backsim & (b^2 - 3 z^{-2}) S(bz) + 3 b z^{-1} C(bz),\\
\label{eqpsol:SpecialSolution5c} (\sinh z)^{1/2} \cdot P^{\mu}_{\nu}(\cosh z) & \backsim & (b^2 - 3 \coth^2 z + 1) C(bz) - 3 b (\coth z) S(bz), \\
\label{eqpsol:SpecialSolution5d} (\sinh z)^{1/2} \cdot \tilde{Q}^{\mu}_{\nu}(\cosh z) & \backsim & (b^2 - 3 \coth^2 z + 1) S(bz) + 3 b (\coth z) C(bz),\\
\label{eqpsol:SpecialSolution5e} (\sin z)^{1/2} \cdot P^{\mu}_{\nu}(\cos z) & \backsim & (b^2 - 3 \cot^2 z - 1) C(bz) - 3 b (\cot z) S(bz), \\
\label{eqpsol:SpecialSolution5f} (\sin z)^{1/2} \cdot \tilde{Q}^{\mu}_{\nu}(\cos z) & \backsim & (b^2 - 3 \cot^2 z - 1) S(bz) + 3 b (\cot z) C(bz).
\end{eqnarray}

\section{Resonant Solutions of the Legendre DE} \label{asec:ResonantSolutionsGoverningDE}

For certain special cases, the associated Legendre functions $P$ and $Q$ are no longer linearly independent (see Abramowitz and Stegun p333, eq. (8.1.8)).  One such example is the case of scalar perturbations with a dust background (\mbox{$\gamma = 1$}), where a second solution must be determined through alternative means.  We now give the first five examples of resonance in the Legendre DE and give the second resonant solution of the Legendre DE for each case, denoted by $R^{\mu}_{\nu}(z)$.  The toroidal solutions are given below.  Conical solutions are obtained by replacing hyperbolic functions with trigonometric functions.

\paragraph{$\mu = \tfrac{1}{2}$, $\nu = - \tfrac{1}{2}$}
\begin{eqnarray}
(\sinh z)^{1/2} \cdot P^{1/2}_{-1/2}(\cosh z) & \backsim & 1, \\
(\sinh z)^{1/2} \cdot R^{1/2}_{-1/2}(\cosh z) & \backsim & z.
\end{eqnarray}

\paragraph{$\mu = \tfrac{3}{2}$, $\nu = -\tfrac{1}{2}$}
\begin{eqnarray}
(\sinh z)^{1/2} \cdot P^{3/2}_{-1/2}(\cosh z) & \backsim & \coth z, \\
(\sinh z)^{1/2} \cdot R^{3/2}_{-1/2}(\cosh z) & \backsim & z \coth z - 1.
\end{eqnarray}

\paragraph{$\mu = \tfrac{5}{2}$, $\nu = -\tfrac{1}{2}$}
\begin{eqnarray}
(\sinh z)^{1/2} \cdot P^{5/2}_{-1/2}(\cosh z) & \backsim & 2 (\coth z)^2 + (\sinh z)^{-2}, \\
(\sinh z)^{1/2} \cdot R^{5/2}_{-1/2}(\cosh z) & \backsim & z \left( 2 (\coth z)^2 + (\sinh z)^{-2} \right) - 3 (\coth z).
\end{eqnarray}

\paragraph{$\mu = \tfrac{3}{2}$, $\nu = \tfrac{1}{2}, - \tfrac{3}{2}$}
\begin{eqnarray}
(\sinh z)^{1/2} \cdot P^{3/2}_{-3/2}(\cosh z) & \backsim & (\sinh z)^{-1}, \\
(\sinh z)^{1/2} \cdot R^{3/2}_{-3/2}(\cosh z) & \backsim & z (\sinh z)^{-1} - \cosh z.
\end{eqnarray}

\paragraph{$\mu = \tfrac{5}{2}$, $\nu = \tfrac{1}{2}, - \tfrac{3}{2}$}
\begin{eqnarray}
\label{eqpsol:ResonantLegendre5a} (\sinh z)^{1/2} \cdot P^{5/2}_{-3/2}(\cosh z) & \backsim & (\coth z) (\sinh z)^{-1}, \\
\label{eqpsol:ResonantLegendre5b} (\sinh z)^{1/2} \cdot R^{5/2}_{-3/2}(\cosh z) & \backsim & 3 z (\coth z) (\sinh z)^{-1} - ((\cosh z)^2 + 2) (\sinh z)^{-1}.
\end{eqnarray}

\section{Asymptotic Behaviour} \label{asec:AsymptoticBehaviourBesselLegendre}

We now present the asymptotic behaviour of the Bessel functions and associated Legendre functions, both for early times and late times.

\paragraph{Bessel Functions}
\begin{align}
J_u(t \rightarrow 0) &\sim \left\{ \begin{array}{ll} t^{|u|} & \mbox{$u$ integer} \\ t^{u} & \mbox{$u$ noninteger} \end{array} \right. &
Y_u(t \rightarrow 0) &\sim \left\{ \begin{array}{ll} t^{-|u|} & \mbox{$u \neq 0$} \\ \ln(t) & \mbox{$u = 0$} \end{array} \right.
\end{align}
\begin{align}
J_u(t \rightarrow +\infty) &\sim t^{-1/2} \sin(t + \phi) & Y_u(t \rightarrow +\infty) &\sim t^{-1/2} \cos(t + \phi)
\end{align}

\paragraph{Associated Legendre Functions ($t \to 1$)}
\begin{eqnarray}
\label{eqpsol:AsymptAssocPT1} P_{\nu}^{\mu} (t \rightarrow 1) & \sim & \left\{ \begin{array}{ll} 1 & \mbox{$\mu = 0$} \\ (t - 1)^{|\mu|/2} & \mbox{$\mu$ integer} \\ (t - 1)^{-|\mu|/2} & \mbox{$\mu$ noninteger} \end{array} \right. \\
\nonumber \vspace{0.5cm} \\
\label{eqpsol:AsymptAssocQT1} Q_{\nu}^{\mu} (t \rightarrow 1) & \sim & \left\{ \begin{array}{ll} \ln(t - 1) & \mbox{$\mu = 0$} \\ (t - 1)^{-|\mu|/2} & \mbox{$\mu$ integer} \\ (t - 1)^{-|\mu|/2} & \mbox{$\mu$ noninteger} \end{array} \right.
\end{eqnarray}

\paragraph{Toroidal Legendre Functions ($t \to \infty$)}\ \\

For the case
\begin{equation}
\nu = - \tfrac{1}{2} + i b, \qquad b > 0,
\end{equation} the asymptotic behaviour of the toroidal Legendre functions is given by
\begin{eqnarray}
P_{\nu}^{\mu} (\mathrm{cosh}(t \rightarrow \infty)) & \sim & \exp(-t/2) \cos(bt + \phi), \\
Q_{\nu}^{\mu} (\mathrm{cosh}(t \rightarrow \infty)) & \sim & \exp(-t/2) \sin(bt + \phi),
\end{eqnarray} as $t \to \infty$.


\begin{thebibliography}{99}

\bibitem{handbookfns} Abramowitz, M. and Stegun, I. (1970).  \textit{Handbook of Mathematical Functions with Formulas, Graphs, and Mathematical Tables, 9th ed.}  New York:  Dover Publications, Inc.

\bibitem{gaugeinvpert} Bardeen, J.M. (1980).  Gauge Invariant Cosmological Perturbations, \textit{Phys. Rev. D} \textbf{22}, p1882-905.

\bibitem{cosmopertub} Bruni, M., Dunsby, P.K.S. and Ellis, G.F.R. (1992).  Cosmological perturbations and the physical meaning of gauge-invariant variables, \textit{Astrophys. J.} \textbf{395}, 34-53.

\bibitem{interactingfluids} Clifton, T. and Barrow, J.D. (2007).  The Ups and Downs of Cyclic Universes, \textit{Phys. Rev. D} \textbf{75}.

\bibitem{coleslucchin} Coles, P. and Lucchin, F. (1995).  \textit{Cosmology: the Origin and Evolution of Cosmic Structure}.  John Wiley and Sons.

\bibitem{roleofshear} Collins, C.B. and Wainwright, J. (1983).  The role of shear in general-relativistic cosmological and stellar models, \textit{Phys. Rev. D} \textbf{27}, 1209-18.

\bibitem{analyticdisFri} Coquereaux, R. and Grossmann, A. (1982).  Analytic Discussion of Spatially Closed Friedman Universes with Cosmological Constant and Radiation Pressure, \textit{Ann. of Phys.} \textbf{143}, 296-356.

\bibitem{analyticsolFri} Dabrowski, M. and Stelmach, J. (1986).  Analytic Solutions of Friedman Equation for Spatially Opened Universes with Cosmological Constant and Radiation Pressure, \textit{Ann. of Phys.} \textbf{166}, 422-42.

\bibitem{evoldensitypert} Demia\'{n}ski, M., Zdzislaw, A.G. and Woszczyna, A. (2005).  Evolution of density perturbations in a realistic universe, \textit{Gen. Relativ. Gravit.} \textbf{37}, 2063-2082.

\bibitem{covgravwaves} Dunsby, P.K.S., Bassett B.A. and Ellis, G.F.R. (1997).  Covariant analysis of gravitational waves in a cosmological context, \textit{Class. Quan. Grav.} \textbf{14}, 1215-22.

\bibitem{phasespaceFL} Ehlers, J. and Rindler, W. (1989).  A phase-space representation of Friedmann-Lema\^{i}tre universes containing both dust and radiation and the inevitability of a big bang, \textit{Mon. Not. R. Astr. Soc.} \textbf{238}, 503-21.

\bibitem{einsteindesit} Einstein, A. and de Sitter, W. (1932).  On the relation between the expansion and the mean density of the universe, \textit{Proc. Natl. Acad. Sci. USA} \textbf{18}, 213-14.

\bibitem{relcosmo} Ellis, G.F.R. (1973).  Relativistic Cosmology.  In \textit{Carg\`ese Lectures in Physics Physics, Volume 6}, ed. E. Schatzmann.  Gordon and Breach.

\bibitem{relhist} Ellis, G.F.R. (1989).  \textit{The Expanding Universe:  A History of Cosmology from 1917 to 1960}, in D. Howard and J. Stachel, ed., \textit{Einstein and the History of General Relativity}, p367-431.  Boston:  The Center for Einstein Studies.

\bibitem{covgauge} Ellis, G.F.R. and Bruni, M. (1989).  Covariant and gauge-invariant approach to cosmological density fluctuations, \textit{Phys. Rev. D} \textbf{40}, 1804-18.

\bibitem{covgaugeperf} Ellis, G.F.R., Hwang, J. and Bruni, M. (1989).  Covariant and gauge-independent perfect-fluid Robertson-Walker perturbations, \textit{Phys. Rev. D} \textbf{40}, 1819-26.

\bibitem{dengradvort} Ellis, G.F.R., Bruni, M. and Hwang, J. (1990).  Density-gradient-vorticity relation in perfect-fluid Robertson-Walker perturbations, \textit{Phys. Rev. D} \textbf{42}, 1035-46.

\bibitem{cosmomodels} Ellis, G.F.R. and van Elst, H. (1998).  Cosmological Models.  In \textit{Carg\`ese Lectures 1998.}

\bibitem{diffforms} Flanders, H. (1989).  \textit{Differential Forms with Applications to the Physical Sciences.}  New York:  Dover Publications, Inc.

\bibitem{spatinhomo} Goode, S.W. (1983).  Spatially inhomogeneous cosmologies and their relation with the Friedmann-Robertson-Walker cosmologies, \textit{PhD Thesis} University of Waterloo.

\bibitem{spathomoper} Goode, S.W. (1989).  Analysis of spatially inhomogeneous perturbations of the FRW cosmologies, \textit{Phys. Rev. D} \textbf{39}, 2882-92.

\bibitem{genrellec} Hall, G.S. and Pulham, J.R. (1996). \textit{General Relativity.}  Bristol: Scottish Universities Summer School in Physics.

\bibitem{classuniform} Harrison, E.R. (1967).  Classification of Uniform Cosmological Models, \textit{Mon. Not. R. Astron. Soc.} \textbf{137}, 69.

\bibitem{perturbexpand} Hawking, S.W. (1966).  Perturbations of an expanding universe, \textit{Astrophys. J.} \textbf{145}, 544-54.

\bibitem{spacetime} Hawking, S.W. and Ellis, G.F.R. (1973).  \textit{The Large Scale Structure of Space-Time.}  Cambridge:  Cambridge University Press.

\bibitem{friedstruct} Jantzen, R.T. and Uggla, C. (1992).  Structure of the generalized Friedmann problem, \textit{Gen. Rel. Grav.} \textbf{24}, 59-85.

\bibitem{flatnessproblem} Lake, K. (2005).  \textit{The Flatness Problem and $\Lambda$.}  Phys. Rev. Lett. \textbf{94}, 201102.

\bibitem{integfriedmann} Lake, K. (2006).  \textit{Integration of the Friedmann equation for universes of arbitrary complexity.}  Preprint, gr-qc/0603028.

\bibitem{lifshitz1} Lifshitz, M. (1946).  On the Gravitational Stability of the Expanding Universe, \textit{J. Phys (Moscow)} \textbf{10}, 116.

\bibitem{lifshitz2} Lifshitz, M. and Khalatnikov, I. (1963).  Investigations in Relativistic Cosmology, \textit{Adv. Phys.} \textbf{12}, 185-249.

\bibitem{lythmukherjee} Lyth, D.H. and Mukherjee, M. (1988).  Fluid flow description of density irregularities in the universe, \textit{Phys. Rev. D} \textbf{38}, 485-9.

\bibitem{krasinski} Krasinski, A. (1997).  \textit{Inhomogeneous Cosmological Models.}  Cambridge:  Cambridge University Press.

\bibitem{limitsanisotropy} Maartens, R., Ellis, G.F.R. and Stoeger, W.R. (1995a).  Limits on anisotropy and inhomogeneity from the cosmic background radiation, \textit{Phys. Rev. D} \textbf{51}, 1525-35.

\bibitem{improvedlimitsisotropy} Maartens, R., Ellis, G.F.R. and Stoeger, W.R. (1995b).  Improved limits on anisotropy and inhomogeneity from the cosmic background radiation, \textit{Phys. Rev. D} \textbf{51}, 5942-45.

\bibitem{lininstab} Maartens, R. (1997).  Linearization instability of gravity waves? \textit{Phys. Rev. D} \textbf{55}, 463-7.

\bibitem{physicalfound} Mukhanov, V. (2005).  \textit{Physical Foundations of Cosmology.}  Cambridge:  Cambridge University Press.

\bibitem{cosmophysics} Peacock, J.A. (1999).  \textit{Cosmological Physics.}  Cambridge:  Cambridge University Press.

\bibitem{diffeqdyn} Perko, L. (1996).  \textit{Differential Equations and Dynamical Systems, 2nd ed.}  New York:  Springer-Verlag.

\bibitem{infopenuni} Ratra, B. and Peebles P.J.E. (1995).  Inflation in an open universe, \textit{Phys. Rev. D} \textbf{52}, 1837-94.

\bibitem{essrel} Rindler, W. (1977).  \textit{Essential Relativity:  Special, General, and Cosmological, revised 2nd ed.}  New York:  Springer-Verlag.

\bibitem{sachswolfe} Sachs, R.K. and Wolfe, A.M. (1967).  Perturbations of a cosmological model and angular variations of the microwave background, \textit{Astrophys. J.} \textbf{147}, 73-90.

\bibitem{WMAP3year} Spergel, D.N. et al. (2006).  Wilkinson Microwave Anisotropy Probe (WMAP) Three Year Results: Implications for Cosmology.  Preprint, astro-ph/0603449.

\bibitem{exactsols} Stephani, H. et al. (2003).  \textit{Exact Solutions to Einstein's Field Equations 2nd ed.}  Cambridge: Cambridge University Press.

\bibitem{stewpert} Stewart, J.M.  Perturbations of Friedmann-Robertson-Walker cosmological models, \textit{Class. Quan. Grav} \textbf{7}, 1169-80.

\bibitem{stewwalker} Stewart, J.M. and Walker, M. (1974).  Perturbations of spacetimes in general relativity, \textit{Proc. Roy. Soc. London A} \textbf{341}, 49-74. [Chapter 14]

\bibitem{cosmoparamSDSSWMAP} Tegmark, M. et al. (2004).  Cosmological parameters from SDSS and WMAP, \textit{Phys. Rev. D} \textbf{69}, 103501.

\bibitem{tolman} Tolman, R.C. (1934).  \textit{Relativity Thermodynamics and Cosmology.}  Oxford:  Clarendon Press.

\bibitem{orthoframe} van Elst, H. and Uggla, C. (1997).  General Relativistic 1 + 3 Orthonormal Frame Approach, \textit{Class. Quan. Grav} \textbf{14}, 2673–95.

\bibitem{compactden} Wainwright, J. (1996) \textit{Relativistic Cosmology}, in Hall, G.S. and Pulham, J.R. General Relativity.  Great Britain: Scotish Universities Summer School in Physics, p107-41.

\bibitem{dynsys} Wainwright, J. and Ellis, G.F.R. (1997).  \textit{Dynamical Systems in Cosmology.}  Cambridge: Cambridge University Press.

\bibitem{genrel} Wald, R. (1984).  \textit{General Relativity.}  Chicago and London: The University of Chicago Press.

\bibitem{gravcosmo} Weinberg, S. (1972).  \textit{Gravitation and Cosmology: Principles and Applications of the General Theory of Relativity}, New York: Wiley.

\end{thebibliography}
\end{document}